\documentclass[final,3p,times]{elsarticle}

\usepackage{amssymb}
\begin{document}

\begin{frontmatter}



\title{Modeling laser wakefield accelerators in a Lorentz boosted frame.}
%\thanks{Work supported by US-DOE Contracts DE-AC02-05CH11231 and DE-AC52-07NA27344, and US-DOE SciDAC program ComPASS. Used resources of NERSC, supported by US-DOE Contract DE-AC02-05CH11231.}}


\author[LBNL]{J.-L. Vay}
\author[LBNL]{C. G. R. Geddes}
\author[TECHX]{E. Cormier-Michel}
\author[LLNL]{D. P. Grote}
\address[LBNL]{LBNL, Berkeley, CA, USA}
\address[TECHX]{Tech-X Corp., Boulder, CO, USA}
\address[LLNL]{LLNL, Livermore, CA, USA\\[-5mm]}

\begin{abstract}
%Paradoxically, the design of plasma wakefield accelerators, which promise to cut the length 
%of accelerators of particles by orders of magnitude for some applications, is in some respect more 
%computationally demanding than its counterpart for conventional accelerators. 
%The scaling of the speedup with regard to the velocity 
%of the frame of reference relative to the laboratory is derived, and applied to a range of acceleration 
%energies between 100 MeV to 100 GeV for a stage design. 
%A speedup reaching as high 
%as 30,000 is predicted for a 1 GeV stage. 
Modeling of laser-plasma wakefield accelerators in an optimal frame of reference \cite{VayPRL07} is shown to produce orders of magnitude speed-up of calculations from first principles.  Obtaining these speedups requires mitigation of a high-frequency instability that otherwise limits effectiveness in addition to solutions for handling data input and output in a relativistically boosted frame of reference. The observed high-frequency instability is mitigated using methods including an electromagnetic solver with tunable coefficients, its extension to accomodate Perfectly Matched Layers and Friedman's damping algorithms, as well as an efficient large bandwidth digital filter. It is shown that choosing the frame of the wake as the frame of reference allows for higher levels of filtering and damping than is possible in other frames for the same accuracy. Detailed testing also revealed serendipitously the existence of a singular time step at which the instability level is minimized, independently of numerical dispersion, thus indicating that the observed instability may not be due primarily to Numerical Cerenkov as has been conjectured. The techniques developed for Cerenkov mitigation prove nonetheless to be very efficient at controlling the instability. Using these techniques, agreement at the percentage level is demonstrated between simulations using different frames of reference, with speedups reaching two orders of magnitude for a 0.1 GeV class stages. The method then allows direct and efficient full-scale modeling of deeply depleted laser-plasma stages of 10 GeV-1 TeV for the first time, verifying the scaling of plasma accelerators to very high energies. Over 4, 5 and 6 orders of magnitude speedup is achieved for the modeling of 10 GeV, 100 GeV and 1 TeV class stages, respectively.
\end{abstract}


\end{frontmatter}

\clearpage
\tableofcontents{}
% \linenumbers

%\section{}
%\label{}



%% References


\section{Introduction}
%Introduce collider design, need for staging, optimal stage length.  Reference Carl's AAC collider paper.  Similar to my PAC paper intro or Estelle's

Laser driven plasma waves offer orders of magnitude increases in accelerating
gradient over standard accelerating structures \cite{TajimaPRL79} (which are limited by electrical
breakdown), thus holding the promise of much shorter particle
accelerators \cite{EsareyRMP09}. High quality electron beams of energy up-to 1 GeV 
have been produced in just a few centimeters \cite{GeddesNature04,ManglesNature04,FaureNature04,LeemansNature06}, 
with 10 GeV stages being planned as modules of a high energy collider \cite{SchroederAAC08}.

As the laser propagates through a plasma, it displaces electrons while ions remain essentially static, 
creating a pocket of positive charges that the displaced electrons rush to fill. The 
resulting coherent periodic motion of the electrons oscillating around their original position 
creates a wake with periodic structure following the laser. The alternate concentration of positive and negative charges in 
the wake creates very intense electric fields. An electron (or positron) beam injected with 
the right phase can be accelerated by those fields to high energy in a much shorter distance than is 
possible in conventional particle accelerators.
The efficiency and quality of the acceleration is governed by several factors which require precise three-dimensional 
shaping of the plasma column, as well as the laser and particle beams, and understanding of their evolution. 

Computer simulations have had a profound impact on the design and understanding of past and present 
experiments  \cite{GeddesSciDAC09}, with  
accurate modeling of the wake formation and beam
acceleration requiring fully kinetic methods (usually Particle-In-Cell) with large computational resources
due to the wide range of space and time scales involved \cite{GeddesJP08,HuangSciDAC09}. For
example, modeling 10 GeV stages for the LOASIS (LBNL) BELLA proposal
\cite{BELLA} demanded as many as 5,000 processor hours for a one-dimension simulation on a
NERSC supercomputer \cite{BruhwilerAAC08}. Various reduced models have been 
developed to allow multidimensional simulations at manageable computational costs: 
fluid approximation \cite{ShadwickPoP09}, quasistatic approximation \cite{SpranglePRL90,Quickpic,FengJCP09}, 
laser envelope models \cite{Quickpic}, scaled parameters \cite{CormierAAC08,GeddesPAC09}. However, the 
various approximations that they require result in a narrower range of applicability. 
As a result, even using several models concurrently does not usually provide 
a complete description. For example, scaled simulations of 10 GeV LPA stages do 
not capture correctly some essential transverse physics, e.g. the laser and beam 
betatron motion, which can lead to inaccurate beam emittance (a measure of the beam quality). 
An envelope description can 
capture these effects correctly at full scale for the early propagation through the plasma 
but can fail as the laser spectrum broadens due to energy depletion as it propagates 
further in the plasma.

An alternative approach allows for orders of magnitude speedup 
%which does not require any approximation 
%beyond the standard discretization of the Vlasov-Maxwell system of equations, %performed in Particle-In-Cell simulations, 
of simulations, whether at full or reduced scale, via the proper choice of a reference frame 
moving near the speed of light in the direction of the laser \cite{VayPRL07}. It does so  
without alteration to the fundamental equations of particle motion 
or electrodynamics, 
provided that the high-frequency part of the light emitted counter to the direction 
of propagation of the beam can be neglected.
This approach exploits the properties of space and 
time dilation and contraction associated with the Lorentz transformation. As shown 
in \cite{VayPRL07}, the ratio of longest to shortest space
and time scales of a system of two or more components crossing at relativistic
velocities is not invariant under such a transformation (a laser crossing a 
plasma is just such a relativistic crossing).
Since the number of computer operations (e.g., time steps),
for simulations based on formulations from first principles, is proportional
to the ratio of the longest to shortest time scale of interest, it follows that
such simulations will eventually have different computer runtimes, yet
equivalent accuracy, depending solely upon the choice of frame of
reference. 

%For a laser wakefield stage, the number of time steps scales with 
%the time it takes to the laser and/or the accelerated beam to cross the plasma, divided by the 
%shortest time scale of interest which is the laser period. To first approximation, i.e. assuming that the 
%laser propagates through the plasma at the speed of light in vacuum, this is proportional to the 
%ratio of the plasma length divided by the laser wavelength. In a reference frame moving at a fraction $\beta$
%of the speed of light, the length of the laser is multiplied by $\left(1+\beta\right)\gamma$ with respect to the 
%length in the laboratory frame, while the plasma is contracted by $\gamma$, where $\gamma=1/\sqrt{1-\beta^2}$ 
%is the relativistic factor of the moving frame. As a result, the ratio of the plasma length to the laser 
%wavelength is now reduced by $1/\left(1+\beta\right)\gamma^2$, hence the number of time steps. In reality, the 
%laser group velocity in the plasma is less than the velocity of light in vacuum, and  the laser dilation effect 
%vanishes until it actually reverses as the velocity of the frame of reference approaches, and 
%then passes above the group velocity of the laser in the plasma. For a 10 GeV stage as planed for the 
%BELLA project, the group velocity of the laser in the plasma corresponds to a relativistic factor above 
%$100$, translating into potential speedup above $100^2=10,000$ for calculations in a boosted frame 
%at $\gamma$ of $100$ or more, as compared to a calculation in the laboratory frame. 

The procedure appears straightforward: identify the frame of reference which will minimize the range of space 
and/or time scales and perform the calculation in this frame. However, several practical complications arise. 
First, the input and output data are usually known from, or compared to, experimental data. Thus, calculating in 
a frame other than the laboratory entails transformations of the data between the calculation frame and the laboratory 
frame.
Second, while the fundamental equations of electrodynamics and particle motion are written in a covariant form, 
the numerical algorithms that are derived from them may not retain this property, and calculations in frames 
moving at different velocities may not be successfully conducted with the use of the exact same algorithms. 
For example, it was shown 
in \cite{VayPOP08} that calculating the propagation of ultra-relativistic charged particle beams in an 
accelerator using standard Particle-In-Cell techniques lead to large numerical errors, which were fixed by 
developing a new particle pusher.
%   It was demonstrated  
%that the errors originated in the structure of the widely used Boris pusher, and were eliminated by using 
%a novel particle pusher. The new solver was combined in a Particle-In-Cell code with a specifically developed 
%field solver retaining electrostatic, magnetostatic and longitudinal inductive fields, leading to the demonstration of 
%three orders of magnitude speed-up between a calculation using a reference frame moving near the speed of 
%light and a calculation using the laboratory frame \cite{VayPRL07,VayPOP08}. 
The modeling of a laser plasma accelerator (LPA) stage in a boosted frame 
involves the fully electromagnetic modeling of a plasma propagating at near the speed of light, for which Numerical Cerenkov 
\cite{BorisJCP73,HaberICNSP73} is a potential issue.
Third, electromagnetic calculations that include wave propagation will include waves propagating forward 
and backward in any direction. 
%A change of frame of reference will result, via the Lorentz transformation, in dilation of the waves propagating 
%in the direction of the frame of reference, but contraction of the ones propagating in the opposite direction. 
%Depending upon the physics of interest to the simulator, the choice of an optimum frame may thus involve 
%selecting the waves that are deemed of most importance. Such a choice is not necessary in the modeling 
%of a free electron laser, where an electron beam 
%propagates through a succession of inversely polarized magnet dipoles resulting in transverse oscillations of 
%the electrons, and emission of coherent light. Viewed in the beam frame, the electrons emit identically in the 
%forward and backward direction. In the laboratory frame, the forward propagating wave is shortened while the 
%backward wave lengthens. As a result, any frame between the beam frame and the laboratory frame will 
%allow resolution of the backward emitted wave, provided that the forward emitted wave is resolved. As 
%shown in \cite{FawleyAAC08}, the beam frame is optimal and results in orders of magnitude saving over the 
%laboratory frame. While this analysis holds for the description of the waves emitted by the electron or 
%positron beam in a LPA stage, it does not for the ones emitted by the plasma. 
%To first approximation, the 
%waves emitted by the plasma in the laboratory frame have identical wavelength in the forward and backward directions.  
For a frame of reference moving in the direction of the accelerated beam (or equivalently the wake of the laser), 
waves emitted by the plasma in the forward direction expand 
while the ones emitted in the backward direction contract, following the properties of the Lorentz transformation. 
If one is to resolve both forward and backward propagating 
waves emitted from the plasma, there is no gain in selecting a frame different from the laboratory frame. However, 
the physics of interest for a laser wakefield is the laser driving the wake, the wake, and the accelerated beam. 
Backscatter is weak in the short-pulse regime, and does not 
interact as strongly with the beam as do the forward propagating waves 
which stay in phase for a long period. It is thus often assumed that the backward propagating waves 
can be neglected in the modeling of LPA stages. The accuracy  of this assumption is shown by 
comparison between explicit codes which include both forward and backward waves and envelope or quasistatic codes which neglect backward waves 
\cite{GeddesJP08,GeddesPAC09,CowanAAC08}.

After the idea and basic scaling for performing simulations of LPA in a Lorentz boosted frame were published in \cite{VayPRL07}, 
there have been several reports of the application of the technique to various regimes of LPA 
\cite{BruhwilerAAC08,VayPAC09,MartinsPAC09,VaySciDAC09,HuangSciDAC09,VayDPF09,MartinsCPC10}. Speedups varying between 
several to a few thousands were reported with various levels of accuracy in agreement between simulations performed in a Lorentz 
boosted frames and in a laboratory frame. High-frequency instabilities were reported to develop in 2D or 3D calculations, that were limiting 
the velocity of the boosted frame and thus the attainable speedup \cite{BruhwilerPC08,VayDPF09,MartinsCPC10}.

In this paper, we present numerical techniques that were implemented in the Particle-In-Cell code Warp \cite{Warp} for mitigating 
the numerical Cerenkov instability, including a solver with tunable coefficients, and show that these techniques are effective 
for suppressing the high frequency instability observed in boosted frame simulations. A detailed study of the application of 
these techniques to the simulations of downscaled LPA stages reveals that choosing the frame of the wakefield as the 
reference frame allows for more aggressive application of the standard techniques mitigating numerical Cerenkov, than 
is possible in laboratory frame simulations. It is shown that the instability that develops with high-boost frames is well 
controlled, allowing for the first time 2D and 3D simulations of LPA 
in the wakefield frame, for 100 GeV and 1 TeV class stages, achieving the maximum theoretical speedups 
of over $10^5$ and $10^6$ respectively. 

This paper is organized as follows. The theoretical speedup expected for performing the modeling of 
a LPA stage in a boosted frame is derived in Section 2. Section 3 addresses the issue of input and output of data 
in a boosted frame. High frequency instability issues and remedies are presented in Section 4. 
These techniques enable accurate modeling of 0.1 GeV-1 TeV LPA stages. Stage modeling results are presented in
 section 5, and observed speedup is contrasted to the theoretical speedup of section 2.

\section{Theoretical speedup dependency with the frame boost}
The obtainable speedup is derived as an extension of the formula that was derived in \cite{VayPRL07}, taking in addition 
into account the group velocity of the laser as it traverses the plasma. In \cite{VayPRL07}, the laser was assumed to propagate at the velocity of light in vacuum during the entire process, which is a good approximation when the relativistic factor of the frame boost $\gamma$ is small compared to the relativistic factor of the laser wake $\gamma_w$ in the plasma. The expression is generalized here to higher values of $\gamma$, for which the actual group velocity of the wake in the plasma must be taken into account. We shall show that for a 10 GeV class LPA stage, the 
maximum attainable speedup is above four orders of magnitude.  

Assuming that the simulation box is a 
fixed number of plasma periods long, which implies the use (which is standard) of a moving window following 
the wake and accelerated beam, the speedup is given by the ratio of the time taken by the laser pulse and the plasma to cross each other, divided by the shortest time scale of interest, that is the laser period. Assuming for simplicity that the wake propagates at the group velocity of plane waves in a uniform plasma of density $n_e$, the group velocity of the wake is given by

\begin{equation}
v_w/c=\beta_w=\left(1+\frac{\omega_p^2}{\omega^2}\right)^{-1/2}
\end{equation}

where $\omega_p=\sqrt{(n_e e^2)/(\epsilon_0 m_e)}$ is the plasma frequency, $\omega=2\pi c/\lambda$ is the laser frequency, $\epsilon_0$ is the permittivity of vacuum, $c$ is the speed of light in vacuum, and $e$ and $m_e$ are respectively the charge and mass of the electron.

In the simulations presented herein, the runs are stopped when the last electron beam macro-particle exits the plasma, and a measure of the total time of the simulation is given by

\begin{equation}
T=\frac{L+\eta \lambda_p}{v_w-v_p}
\end{equation}

where $\lambda_p\approx 2\pi c/\omega_p$ is the wake wavelength, $L$ is the plasma length, $v_w$ and $v_p=\beta_p c$ are respectively the velocity of the wake and of the plasma relative to the frame of reference, and $\eta$ is an adjustable parameter for taking into account the fraction of the wake which exited the plasma at the end of the simulation.
The numerical cost $R_t$ scales as the ratio of the total time to the shortest timescale of interest, which is the inverse of the laser frequency, and is thus given by

\begin{equation}
R_t=\frac{T c}{\lambda}=\frac{\left(L+\eta \lambda_p\right)}{\left(\beta_w-\beta_p\right) \lambda}
\end{equation}

In the laboratory, $v_p=0$ and the expression simplifies to 

\begin{equation}
R_{lab}=\frac{T c}{\lambda}=\frac{\left(L+\eta \lambda_p\right)}{\beta_w \lambda}
\end{equation}

In a frame moving at $\beta c$, the quantities become
\begin{eqnarray}
\lambda_p^*&=&\lambda_p/\left[\gamma \left(1-\beta_w \beta\right)\right] \\
L^*&=&L/\gamma \\
\lambda^*&=& \gamma\left(1+\beta\right) \lambda\\
\beta_w^*&=&\left(\beta_w-\beta\right)/\left(1-\beta_w\beta\right) \\
v_p^*&=&-\beta c \\
T^*&=&\frac{L^*+\eta \lambda_p^*}{v_w^*-v_p^*} \\
R_t^*&=&\frac{T^* c}{\lambda^*} = \frac{\left(L^*+\eta \lambda_p^*\right)}{\left(\beta_w^*+\beta\right) \lambda^*}
\end{eqnarray}
where $\gamma=1/\sqrt{1-\beta^2}$.

The expected speedup from performing the simulation in a boosted frame is given by the ratio of $R_{lab}$ and $R_t^*$

\begin{equation}
S=\frac{R_{lab}}{R_t^*}=\frac{\left(1+\beta\right)\left(L+\eta \lambda_p\right)}{\left(1-\beta\beta_w\right)L+\eta \lambda_p}
\label{Eq_scaling1d0}
\end{equation}

Assuming that $\gamma<<\gamma_w$, and that $\beta_w\approx1$ (which is a valid approximation for most practical cases of interest), this expression is consistent with the expression derived in \cite{VayPRL07} for the LPA case which states that $R_t^*=\alpha R_t/\left(1+\beta\right)$ with $\alpha=\left(1-\beta+l/L\right)/\left(1+l/L\right)$, where $l$ is the laser length which is generally proportional to $\eta \lambda_p$, and $S=R_t/R_T^*$.

The linear theory predicts that for the intense lasers (a$\gtrsim$1) typically used for acceleration, the laser depletes its energy over approximately the same length $L_d=\lambda_p^3/2\lambda^2$ over which the particles dephase from the wake \cite{TajimaPRL79}. Acceleration is compromised beyond $L_d$ and in practice, the plasma length is proportional to the dephasing length, i.e. $L= \xi L_d$. In most cases, $\gamma_w^2>>1$, which allows the approximations $\beta_w\approx1-\lambda^2/2\lambda_p^2$, and $L=\xi \lambda_p^3/2\lambda^2\approx \xi \gamma_w^2 \lambda_p/2>>\eta \lambda_p$, so that Eq.(\ref{Eq_scaling1d0}) becomes

\begin{equation}
%S\approx\frac{\left(1+\beta\right)\left(\xi\lambda_p^2+2\eta \lambda^2\right)}{\left(1-\beta\beta_w\right)\xi\lambda_p^2+2\eta \lambda^2}
S=\left(1+\beta\right)^2\gamma^2\frac{\xi\gamma_w^2}{\xi\gamma_w^2+\left(1+\beta\right)\gamma^2\left(\xi\beta/2+2\eta\right)}
\label{Eq_scaling1d}
\end{equation}

For low values of $\gamma$, i.e. when $\gamma<<\gamma_w$, Eq.(\ref{Eq_scaling1d}) reduces to

\begin{equation}
S_{\gamma<<\gamma_w}=\left(1+\beta\right)^2\gamma^2
\label{Eq_scaling1d_simpl2}
\end{equation}

Conversely, if $\gamma\rightarrow\infty$, Eq.(\ref{Eq_scaling1d}) becomes

\begin{equation}
S_{\gamma\rightarrow\infty}=\frac{4}{1+4\eta/\xi}\gamma_w^2
\label{Eq_scaling_gamma_inf}
\end{equation}

Finally, in the frame of the wake, i.e. when $\gamma=\gamma_w$, assuming that $\beta_w\approx1$, Eq.(\ref{Eq_scaling1d}) gives

\begin{equation}
S_{\gamma=\gamma_w}\approx\frac{2}{1+2\eta/\xi}\gamma_w^2
\label{Eq_scaling_gamma_wake}
\end{equation}

Since $\eta$ and $\xi$ are of order unity, and the practical regimes of most interest satisfy $\gamma_w^2>>1$, the speedup that is obtained by using the frame of the wake will be near the maximum obtainable value given by Eq.(\ref{Eq_scaling_gamma_inf}).

Note that without the use of a moving window, the relativistic effects that are at play in the time domain would also be at play in the spatial domain, as shown in \cite{VayPRL07}, and the $\gamma^2$ scaling would transform to $\gamma^4$. In the frame of the wake, there is no need of the moving window, thus 
simplifying the procedure, while in a frame traveling faster than the wake in the laboratory, a moving window propagating in the backward direction is needed. However, the scaling shows that there would be very little gain in doing the latter.
%\begin{equation}
%S=\frac{2\left(1+\beta_w\right)L}{\left(1-\beta_w^2\right)L+\eta \lambda_p\left(1+\beta_w\right)}
%\end{equation}

%\begin{equation}
%S=\frac{2\gamma_w^2\left(1+\beta_w\right)L}{L+\eta \lambda_p\left(1+\beta_w\right)\gamma_w^2}
%\end{equation}

\subsection{Estimated speedup for 0.1-100 GeV stages}
Formula (\ref{Eq_scaling1d}) is used to estimate the speedup for the calculations of 100 MeV, 1 GeV, 10 GeV and 100 GeV class stages, assuming 
a laser wavelength $\lambda=0.8\mu m$. Using parameters and scaling laws from \cite{CormierAAC08}, the corresponding 
initial plasma densities $n_e$ are respectively  $10^{19}$cc, $10^{18}$cc, $10^{17}$cc and $10^{16}$cc, while the plasma lengths $L$ are 1.5 mm, 4.74 cm, 1.5 m, and 47.4 m, with $\xi\approx1.63$. 
For these values, the wake wavelengths $\lambda_p$ are respectively $10.6\mu m$, $33.4\mu m$,  $106.\mu m$, $334.\mu m$, and relativistic factors $\gamma_w$ are $13.2$, $41.7$, $132$ and $417.$ 
In the simulations presented in this paper, the beam is injected near the end of the wake period (first "bucket"). In first approximation, the beam has propagated through about half a wake period to reach full acceleration, and we set $\eta\approx0.5$. For a beam injected into the $n^{th}$ bucket, $\eta$ would be set to $n-1/2$. If positrons were considered, they would be injected half a wake period ahead of the location of the electrons injection position for a given period, and one would have $\eta=n-1$. For the parameters considered here, $L\approx \lambda_p/\gamma_w^2$, and (\ref{Eq_scaling_gamma_inf}) gives $S_{\gamma\rightarrow\infty}\approx 2\gamma_w^2$.

\begin{figure}[htb]
    \centering
    \includegraphics*[width=90mm, trim=0.in 0.8in 0.in 0.in]{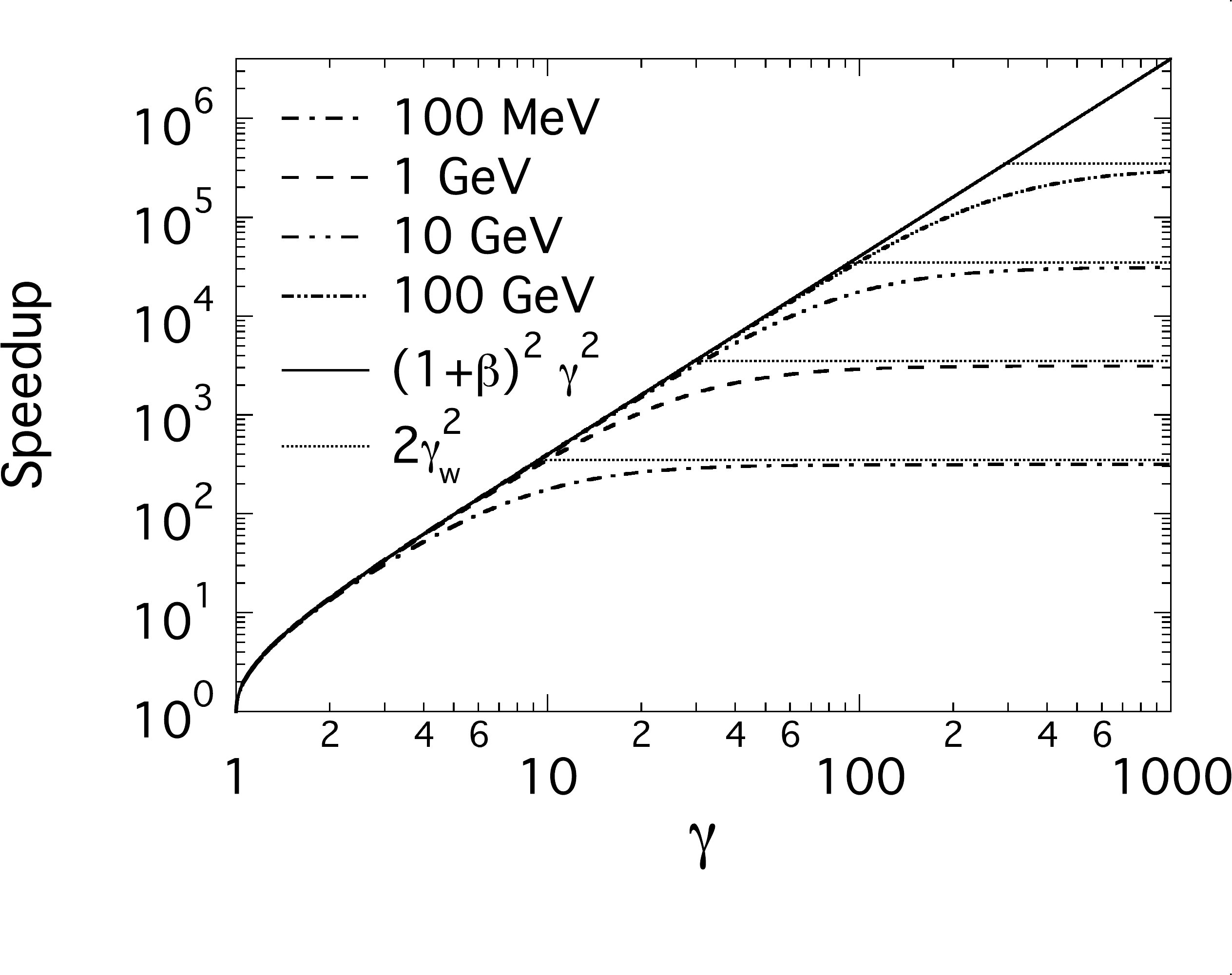}
    \caption{Speedup versus relativistic factor of the boosted frame from Eq.(\ref{Eq_scaling1d})  for 100 MeV - 100 GeV LPA class stages.}
   \label{Scaling1d}
\end{figure}

The speedup versus the relativistic factor of the boosted frame $\gamma$ is plotted in Fig. \ref{Scaling1d}. As expected, for low values of $\gamma$, the speedup scales as (\ref{Eq_scaling1d_simpl2}), and asymptotes to a value slightly lower than $2\gamma_w^2$ for large values of $\gamma$. It is of interest to note that the qualitative behavior is identical to the one obtained in \cite{VayPRL07} (see Fig. 1 and accompanying analysis) in the analysis of the crossing of two rigid identical beams, confirming the generality of the generic analysis presented in \cite{VayPRL07}.  
For a 100 GeV class stage, the maximum estimated speedup is as large as 300,000.

\section{Input and output to and from a boosted frame simulation} 
%So far, it has been common practice to perform simulations in the laboratory frame, for direct comparison with experimental results, or in another frame (beam frame, center of mass frame, etc.) which offers an advantage of symmetry, simplification, or other convenience, in comparing the results to those of analytical theory or experiment. However, the analysis that was provided in \cite{VayPRL07} shows that the frame that will minimize the computational requirements may not be any of the above. In this case, one may need to apply Lorentz transformations between the frame of calculation and the frame in which input data are known and/or the frame in which the output data are desired. Because of the relativity of simultaneity that is inherent to the Lorentz transformation, this requires a process that goes beyond a mere algebraic manipulation of data. 

This section describes the procedures that have been implemented in the Particle-In-Cell framework Warp \cite{Warp} to handle the input and output of data between the frame of calculation and the laboratory frame. Simultaneity of events between two frames is valid only for a plane that is perpendicular to the relative motion of the frame. As a result, the input/output processes involve the input of data (particles or fields) through a plane, as well as output through a series of planes, all of which are perpendicular to the direction of the relative velocity between the frame of calculation and the other frame of choice. 

\subsection{Input}
\subsubsection{Particles}
Particles are launched through a plane using a technique which applies to many calculations in a boosted frame, including LPA, and is illustrated using the case of a positively charged particle beam propagating through a background of cold electrons in an assumed continuous transverse focusing system, leading to a growing transverse instability  \cite{VayPRL07}. In the laboratory frame, the electron background is initially at rest and a moving window is used to follow the beam progression. Traditionally, the beam macroparticles are initialized all at once in the window, while background electron macroparticles are created continuously in front of the beam on a plane that is perpendicular to the beam velocity. In a frame moving at some fraction of the beam velocity in the laboratory frame, the beam initial conditions at a given time in the calculation frame are generally unknown and one must initialize the beam differently. However, it can be taken advantage of the fact that the beam initial conditions are often known for a given plane in the laboratory, either directly, or via simple calculation or projection from the conditions at a given time. Given the position and velocity $\{x,y,z,v_x,v_y,v_z\}$ for each beam macroparticle at time $t=0$ for a beam moving at the average velocity $v_b=\beta_b c$ (where $c$ is the speed of light) in the laboratory, and using the standard synchronization ($z=z'=0$ at $t=t'=0$) between the laboratory and the calculation frames, the procedure for transforming the beam quantities for injection in a boosted frame moving at velocity $\beta c$ in the laboratory is as follows (the superscript $'$ relates to quantities known in the boosted frame while the superscript $^*$ relates to quantities that are know at a given longitudinal position $z^*$ but different times of arrival):

\begin{enumerate}
\item project positions at $z^*=0$ assuming ballistic propagation
\begin{eqnarray} 
    t^* &=& \left(z-\bar{z}\right)/v_z \label{Eq:t*}\\
    x^* &=& x-v_x t^* \label{Eq:x*}\\
    y^* &=& y-v_y t^* \label{Eq:y*}\\
    z^* &=& 0 \label{Eq:z*}
\end{eqnarray}
the velocity components being left unchanged, 
\item apply Lorentz transformation from laboratory frame to boosted frame
\begin{eqnarray} 
    t'^* &=& -\gamma t^* \label{Eq:tp*}\\
    x'^* &=& x^* \label{Eq:xp*}\\
    y'^* &=& y^* \label{Eq:yp*}\\
    z'^* &=& \gamma\beta c t^* \label{Eq:zp*}\\
    v'^*_x&=&\frac{v_x^*}{\gamma\left(1-\beta \beta_b\right)} \label{Eq:vxp*}\\
    v'^*_y&=&\frac{v_y^*}{\gamma\left(1-\beta \beta_b\right)} \label{Eq:vyp*}\\
    v'^*_z&=&\frac{v_z^*-\beta c}{1-\beta \beta_b} \label{Eq:vzp*}
\end{eqnarray}
where $\gamma=1/\sqrt{1-\beta^2}$. With the knowledge of the time at which each beam macroparticle crosses the plane into consideration, one can inject each beam macroparticle in the simulation at the appropriate location and time. 

\item synchronize macroparticles in boosted frame, obtaining their positions at a fixed $t'(=0)$ which is before any particle is injected
\begin{eqnarray} 
    z' &=& z'^*-\bar{v}'^*_z t'^* \label{Eq:zp}
\end{eqnarray}
    This additional step is needed for setting the electrostatic or electromagnetic fields at the plane of injection. In a Particle-In-Cell code, the three-dimensional fields are calculated by solving the Maxwell equations (or static approximation like Poisson, Darwin or other \cite{VayPOP08}) on a grid on which the source term is obtained from the macroparticles distribution. This requires generation of a three-dimensional representation of the beam distribution of macroparticles at a given time before they cross the injection plane at $z'^*$. This is accomplished by expanding the beam distribution longitudinally such that all macroparticles (so far known at different times of arrival at the injection plane) are synchronized to the same time in the boosted frame. To keep the beam shape constant, the particles are "frozen" until they cross that plane: the three velocity components and the two position components perpendicular to the boosted frame velocity are fixed, while the remaining position component is advanced at the average beam velocity. As particles cross the plane of injection, they become regular "active" particles with full 6-D dynamics.

\end{enumerate}

\begin{figure}[htb]
   \centering
   \includegraphics*[width=120mm]{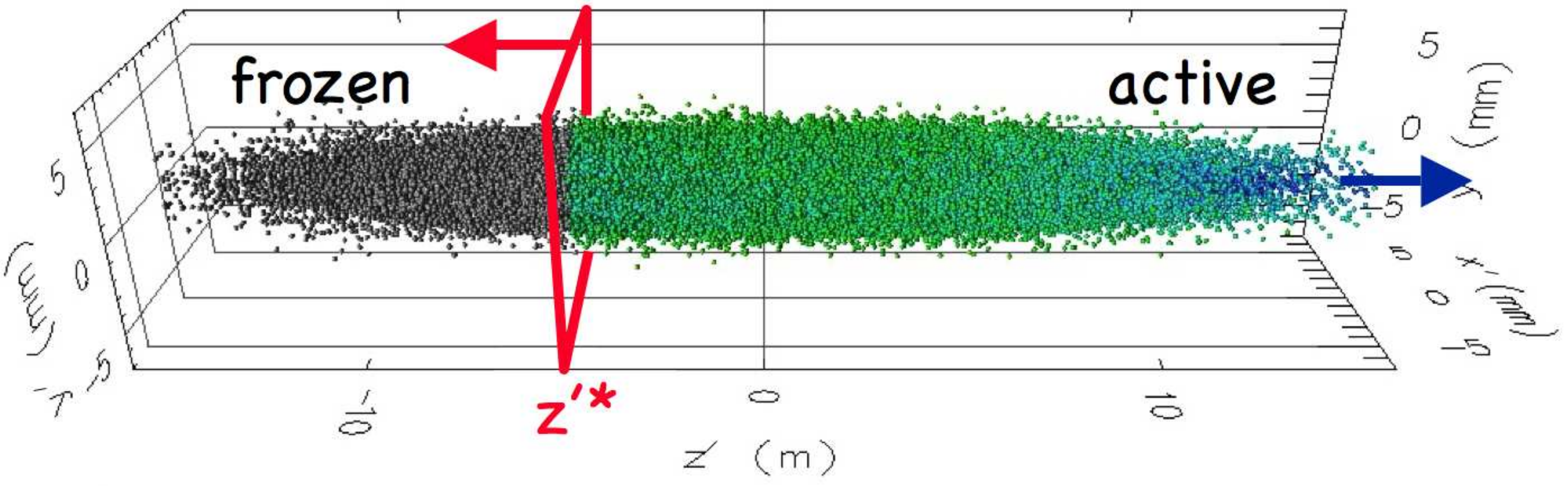}
   \includegraphics*[width=120mm]{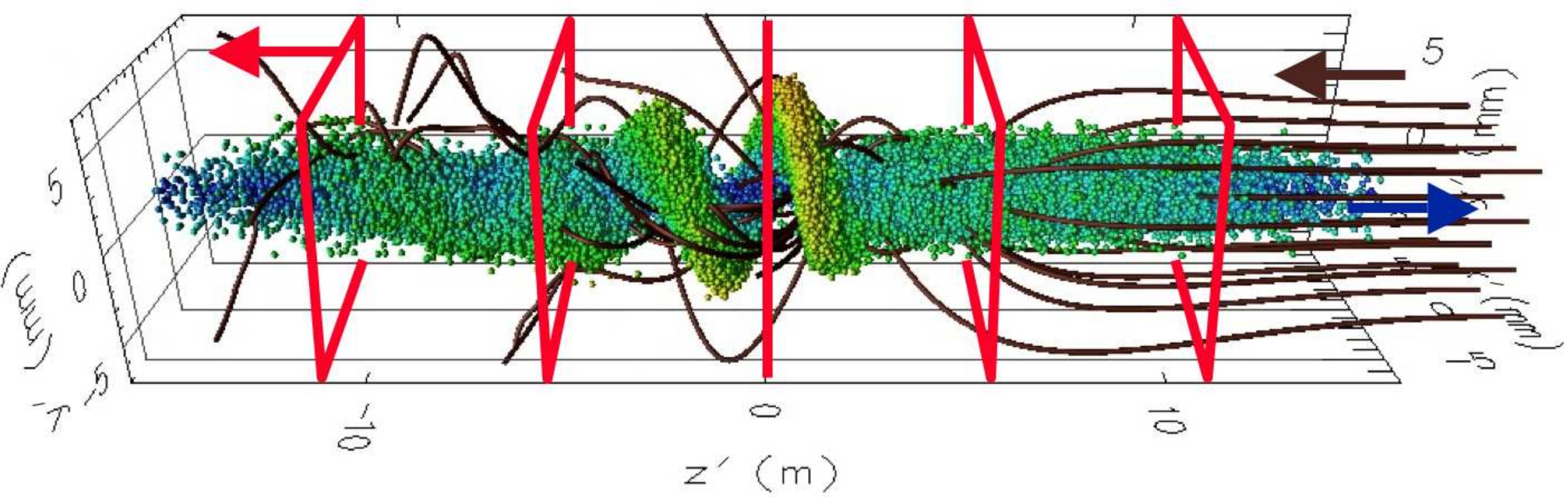}
   \caption{(top) Snapshot of a particle beam ``frozen" (grey spheres) and ``active" (colored spheres) macroparticles traversing the injection plane (red rectangle). (bottom) Snapshot of the beam macroparticles (colored spheres) passing through the background of electrons (dark brown streamlines) and the diagnostic stations (red rectangles). The electrons,  the injection plane and the diagnostic stations are fixed in the laboratory plane, and are thus counterpropagating to the beam in a boosted frame. }
   \label{Fig_inputoutput}
\end{figure}

Figure \ref{Fig_inputoutput} (top) shows a snapshot of a beam that has passed partly through the injection plane. As the frozen beam macroparticles pass through the injection plane (which moves opposite to the beam in the boosted frame), they are converted to ``active" macroparticles. The charge or current density is accumulated from the active and the frozen particles, thus ensuring that the fields at the plane of injection are consistent. 

\subsubsection{Laser}

Similarly to the particle beam, the laser is injected through a plane perpendicular to the axis of propagation of the laser (by default $z$). 
The electric field $E_\perp$ that is to be emitted is given by the formula
\begin{equation}
E_\perp\left(x,y,t\right)=E_0 f\left(x,y,t\right) \sin\left[\omega t+\phi\left(x,y,\omega\right)\right] 
\end{equation}
where $E_0$ is the amplitude of the laser electric field, $f\left(x,y,t\right)$ is the laser envelope, $\omega$ is the laser frequency, $\phi\left(x,y,\omega\right)$ is a phase function to account for focusing, defocusing or injection at an angle, and $t$ is time. By default, the laser envelope is a three dimensional gaussian of the form 
\begin{equation}
 f\left(x,y,t\right)=e^{-\left(x^2/2 \sigma_x^2+y^2/2 \sigma_y^2+c^2t^2/2 \sigma_z^2\right)}
 \end{equation}
 where $\sigma_x$, $\sigma_y$ and $\sigma_z$ are the dimensions of the laser pulse; or it can be defined arbitrarily by the user at runtime.
If $\phi\left(x,y,\omega\right)=1$, the laser is injected at a waist and parallel to the axis $z$. 

If, for convenience, the injection plane is moving at constant velocity $\beta_s c$, the formula is modified to take the Doppler effect on frequency and amplitude into account and becomes
\begin{equation}
E_\perp\left(x,y,t\right)=\left(1-\beta_s\right)E_0 f\left(x,y,t\right) \sin\left[\left(1-\beta_s\right)\omega t+\phi\left(x,y,\omega\right)\right].
\end{equation}

The injection of a laser of frequency $\omega$ is considered for a simulation using a boosted frame moving at $\beta c$ with respect to the laboratory. Assuming that the laser is injected at a plane that is fixed in the laboratory, and thus moving at $\beta_s=-\beta$ in the boosted frame, the injection in the boosted frame is given by
\begin{eqnarray}
E_\perp\left(x',y',t'\right)&=&\left(1-\beta_s\right)E'_0 f\left(x',y',t'\right) \sin\left[\left(1-\beta_s\right)\omega' t'+\phi\left(x',y',\omega'\right)\right]\\
&=&\left(E_0/\gamma\right) f\left(x',y',t'\right) \sin\left[\omega t'/\gamma+\phi\left(x',y',\omega'\right)\right]
\end{eqnarray}
since $E'_0/E_0=\omega'/\omega=1/\left(1+\beta\right)\gamma$.

The electric field is then converted into currents that get injected via two dual 2-D arrays of macro-particles, with one positive and one negative macro-particle per cell in the plane of injection, whose weights and motion are governed by $E_\perp\left(x',y',t'\right)$. Injecting using these dual arrays of macroparticles offer the advantages of automatically including the longitudinal component which arise from emitting into a boosted frame, and to verify the discrete Gauss law thanks to the use of the Esirkepov current deposition scheme \cite{EsirkepovCPC01}.

The technique implemented in Warp presents several advantage over other procedures that have been proposed elsewhere \cite{BruhwilerAAC08,MartinsCPC10}. In \cite{MartinsCPC10}, the laser beam is initialized entirely in the computational box, leading to larger boxes transversely in a boosted frame, as the Rayleigh length of the laser shortens and the overall laser pulse radius rises, eventually offsetting the benefits of the boosted frame. The transverse broadening of the box is avoided in \cite{BruhwilerAAC08} at the cost of a more complicated injection scheme, requiring to launch the laser from all but one faces of the simulation box. The method presented here avoids the caveat of the broadening and retains simplicity with a standard injection technique through one plane.

\subsection{Output}
Some quantities, e.g. charge, are Lorentz invariant, while others, like dimensions perpendicular to the boost velocity, are the same in the laboratory frame. Those quantities are thus readily available from standard diagnostics in the boosted frame calculations. Quantities which do not fall in this category are recorded at a number of regularly spaced ``stations", immobile in the laboratory frame, at a succession of time intervals to record data history, or averaged over time. A visual example is given on Fig.  \ref{Fig_inputoutput} (bottom). Since the space-time locations of the diagnostic grids in the laboratory frame generally do not coincide with the space-time positions of the macroparticles and grid nodes used for the calculation in a boosted frame, some interpolation is performed at runtime during the data collection process. As a complement or an alternative, selected particle or field quantities are dumped at regular interval for post-processing. The choice of the methods depends on the requirements of the diagnostics and particular implementations.

%\clearpage
\section{High frequency instability and Numerical Cerenkov}
\begin{figure}[htb]
   \centering
       \includegraphics*[width=140mm]{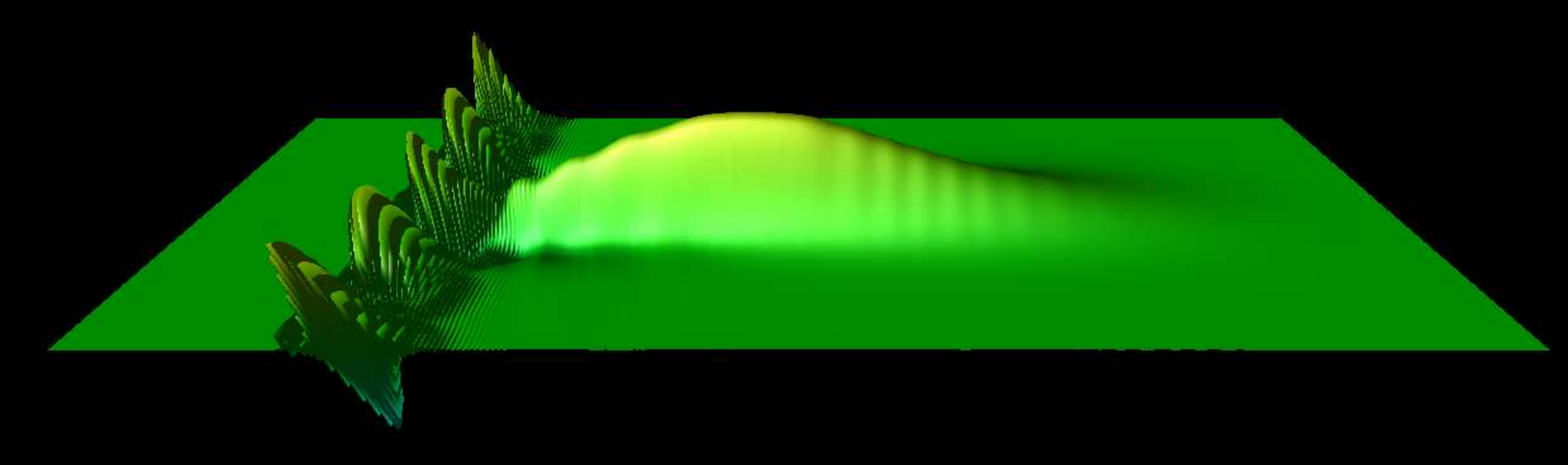} 
%       \includegraphics*[width=140mm]{viewinsta2.pdf} 
%      \includegraphics*[0,0][500,200]{viewinsta.pdf} 
 %  \caption{(left) Average beam energy versus longitudinal position in the laboratory frame from calculations in a frame moving at $\gamma_f=1$, $2$, $5$ and $10$; (right) CPU time recorded as the beam crosses successive stations in the laboratory frame.}
   \caption{Snapshot of a surface plot of the longitudinal field from a 2-1/2D simulation of a full scale 10GeV LPA in a boosted frame at $\gamma=130$ (elevation is proportional to the magnitude of the electric field). The laser is propagating from left to right and the plasma from right to left. A fast growing short wavelength instability is developing at the front of the plasma.}
   \label{Fig_viewinsta}
\end{figure}

As reported in \cite{VayDPF09} and \cite{MartinsCPC10}, for simulations using a boosted frame at $\gamma\geq 10-20$ (depending on parameters), a fast growing short wavelength instability was observed developing at the front of the plasma (see Fig. \ref{Fig_viewinsta}). The presence and growth rate of the instability was observed to be very sensitive to the resolution (slower growth rate at higher resolution), choice of field solver, and to the amount of damping of high frequencies and smoothing of short wavelengths. The instability is always propagating at some angle from the longitudinal axis, and is observed in 2D and 3D runs but was never observed in any of the 1D runs performed by the authors. When modeling an LPA setup in a relativistically boosted frame, the background plasma is traveling near the speed of light and it has been conjectured \cite{MartinsCPC10} that he observed instability might be caused by numerical Cerenkov. We investigate in this paper whether the instability that is observed in boosted frame simulations of LPA is indeed of numerical Cerenkov type and if the cures aimed at mitigating numerical Cerenkov are effective.

Due to spatial and time discretization of the Maxwell equations, numerical light waves may travel faster or slower on the computational grid than the actual speed of light in vacuum $c$, with the magnitude of the effect being larger at short wavelength, where discretization errors are the largest. When the numerical speed is lower than $c$, it is possible for fast macro-particles to travel faster than the wave modes, leading to numerical Cerenkov effects that may result in instabilities \cite{BorisJCP73,HaberICNSP73,GodfreyJCP74,GodfreyJCP75,GreenwoodJCP04}. The effect was studied analytically and numerically in detail for one-dimensional systems in \cite{GodfreyJCP74,GodfreyJCP75}. Several solutions were proposed: smoothing the current deposited by the macro-particles \cite{BorisJCP73,GodfreyJCP74}, damping the electromagnetic field \cite{GreenwoodJCP04,GodfreyICNSP80,FriedmanJCP90}, solving the Maxwell equations in Fourier space \cite{HaberICNSP73}, or using a field solver with a larger stencil to provide lower numerical dispersion  \cite{GreenwoodJCP04}. 

Several of the abovementioned techniques to mitigate numerical Cerenkov and high frequency errors have been implemented in Warp. All the simulations presented in this paper employed cubic splines for current deposition and electromagnetic force gathering between the macro-particles and the grid \cite{AbeJCP86}, whose beneficial effects on standard LPA PIC simulations have been demonstrated in \cite{CormierPRE08}. In addition, a Maxwell solver with tunable coefficients was implemented, as well as a damping scheme, and filtering of the deposited current and gathered electromagnetic fields, which are described in this section. The use of Fourier based Maxwell solvers is not considered in this paper.

%\clearpage
\subsection{Wideband lowpass digital filtering}
\begin{figure}[htb]
   \centering
   \includegraphics*[width=85mm]{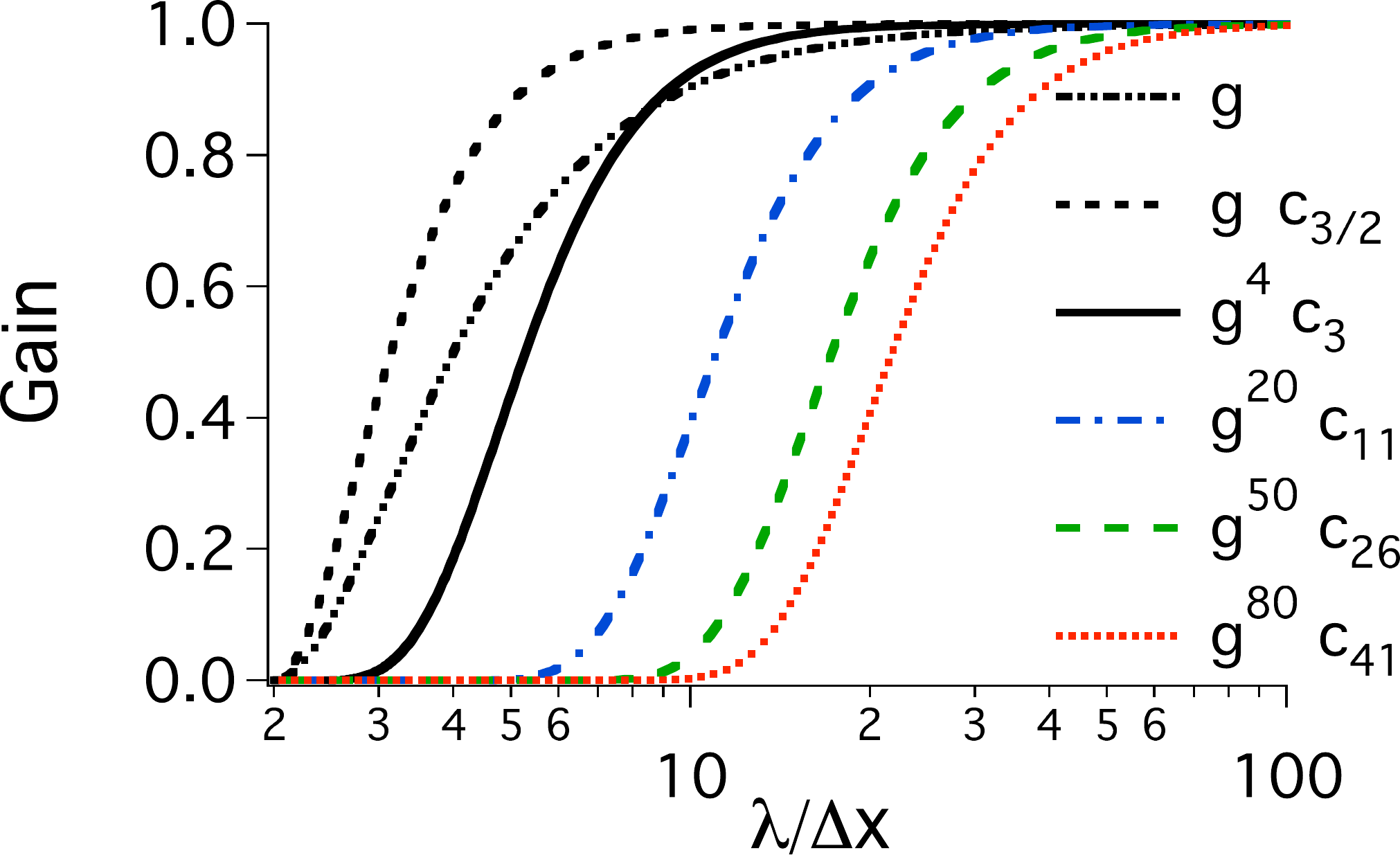}
   \caption{Gain versus wavelength for the bilinear filter without compensation ($g=g(1/2,k)$), with compensation ($g\cdot c_{3/2}=g(1/2,k)\cdot g(3/2,k)$), and n-pass bilinear filters with compensation ($g^n\cdot c_{\alpha_c}=g(1/2,k)^n\cdot g(\alpha_c,k)$) for $n=4$, $20$, $50$ and $80$.}
   \label{Fig_bilinear}
\end{figure}

It is common practice to apply digital filtering to the charge or current density in Particle-In-Cell simulations, for smoothing or compensation purpose \cite{BirdsallLangdon}. The most commonly used filter is the three points filter                                                                                                           
\begin{equation}
\phi_j^f=\alpha \phi_j+\left(1-\alpha\right)\frac{\phi_{j-1}+\phi_{j+1}}{2}
\label{Eq:3ptfilter}
\end{equation}
where $\phi^f$ is the filtered quantity. This filter is called a binomial filter when $\alpha=0.5$. Assuming $\phi=e^{j k x}$ and $\phi^f=g\left(\alpha,k\right) e^{j k x}$, where $g$ is the filter gain, which is function of the filtering coefficient $\alpha$ and the wavenumber $k$, we find from (\ref{Eq:3ptfilter}) that
\begin{eqnarray}
g\left(\alpha,k\right) & = & \alpha+\left(1-\alpha\right)\cos\left(k \delta x\right)\\
& \approx & 1-\left(1-\alpha\right) \frac{\left(k \delta x\right)^2}{2}+O\left(k^4\right)
\end{eqnarray}
For $n$ successive applications of filters of coefficients $\alpha_1$...$\alpha_n$, the total attenuation $G$ is given by
\begin{eqnarray}
G & = & \prod_{i=1}^n g\left(\alpha_i,k\right)\\
& \approx & 1-\left(n-\sum_{i=1}^n\alpha_i\right) \frac{\left(k \delta x\right)^2}{2}+O\left(k^4\right)
\end{eqnarray}
If $\alpha_n=n-\sum_{i=1}^{n-1}\alpha_i$ then $G \approx 1+O\left(k^4\right)$, providing a sharper cutoff in $k$ space. Such step is called a {\it compensation} step \cite{BirdsallLangdon}. For the bilinear filter ($\alpha=1/2$), the compensation factor is $\alpha_c=2-1/2=3/2$. For a succession of $n$ applications of the bilinear factor, it is $\alpha_c=n/2+1$. The gain versus wavelength is plotted in Fig. \ref{Fig_bilinear} for the bilinear filter without compensation ($G=g(1/2,k)$), with compensation ($G=g(1/2,k)\cdot g(3/2,k)$), and four n-pass bilinear filters with compensation ($G=g(1/2,k)^n\cdot g(3/2,k)$)  for $n=4$, $20$, $50$ and $80$.

The bilinear filter provides complete suppression of the signal at the grid Nyquist wavelength (twice the grid cell size). Suppression of the signal at integers of the Nyquist wavelength can be obtained by using a stride $s$ in the filter
\begin{equation}
\phi_j^f=\alpha \phi_j+\left(1-\alpha\right)\frac{\phi_{j-s}+\phi_{j+s}}{2}
\label{Eq:3ptfilter_stride}
\end{equation}
for which the gain is given by
\begin{eqnarray}
g\left(s,\alpha,k\right) & = & \alpha+\left(1-\alpha\right)\cos\left(s k \delta x\right)\\
& \approx & 1-\left(1-\alpha\right) \frac{\left(s k \delta x\right)^2}{2}+O\left(k^4\right)
\end{eqnarray}
\begin{figure}[htb]
   \centering
   \includegraphics*[width=65mm]{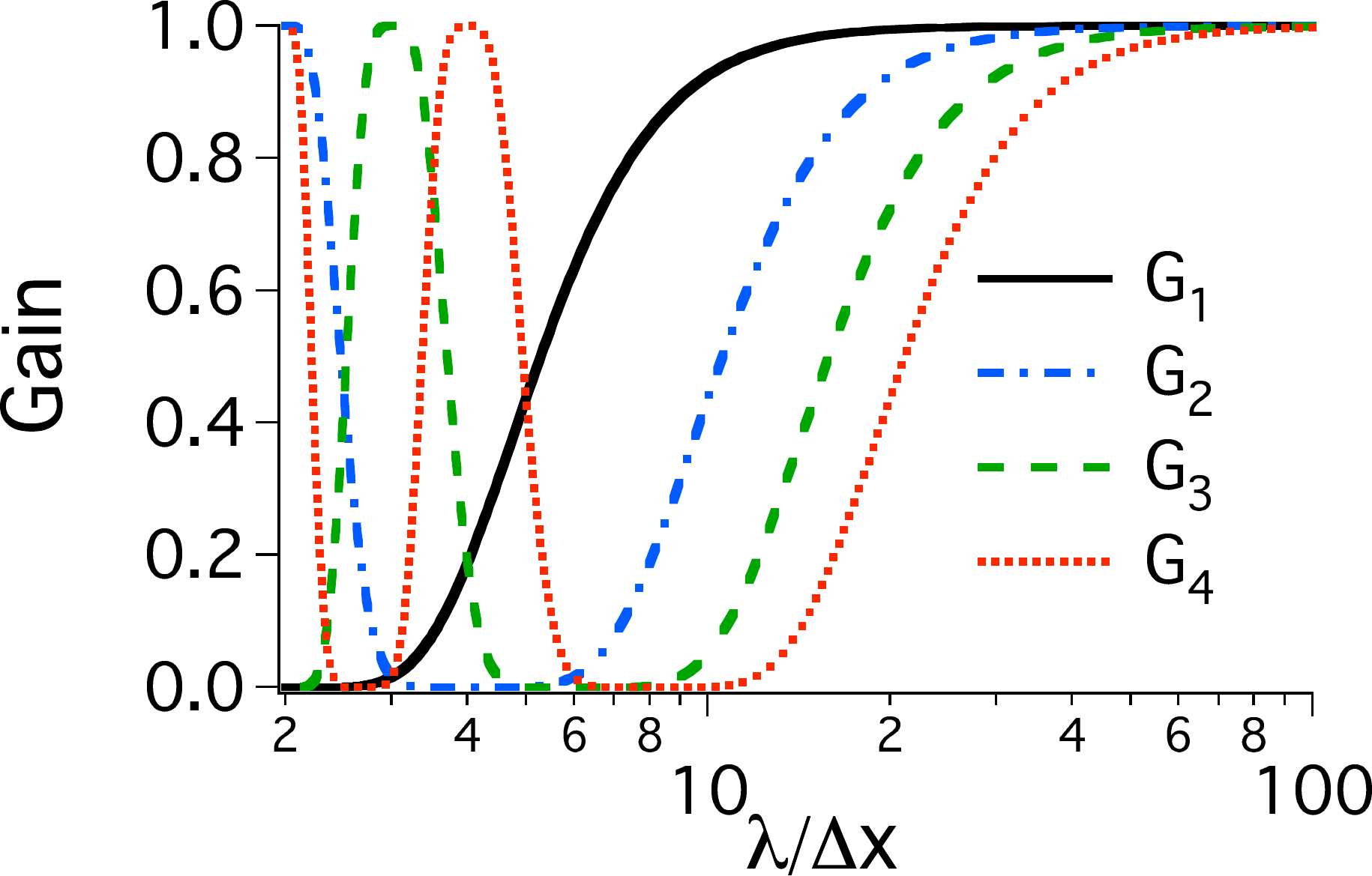}
   \includegraphics*[width=65mm]{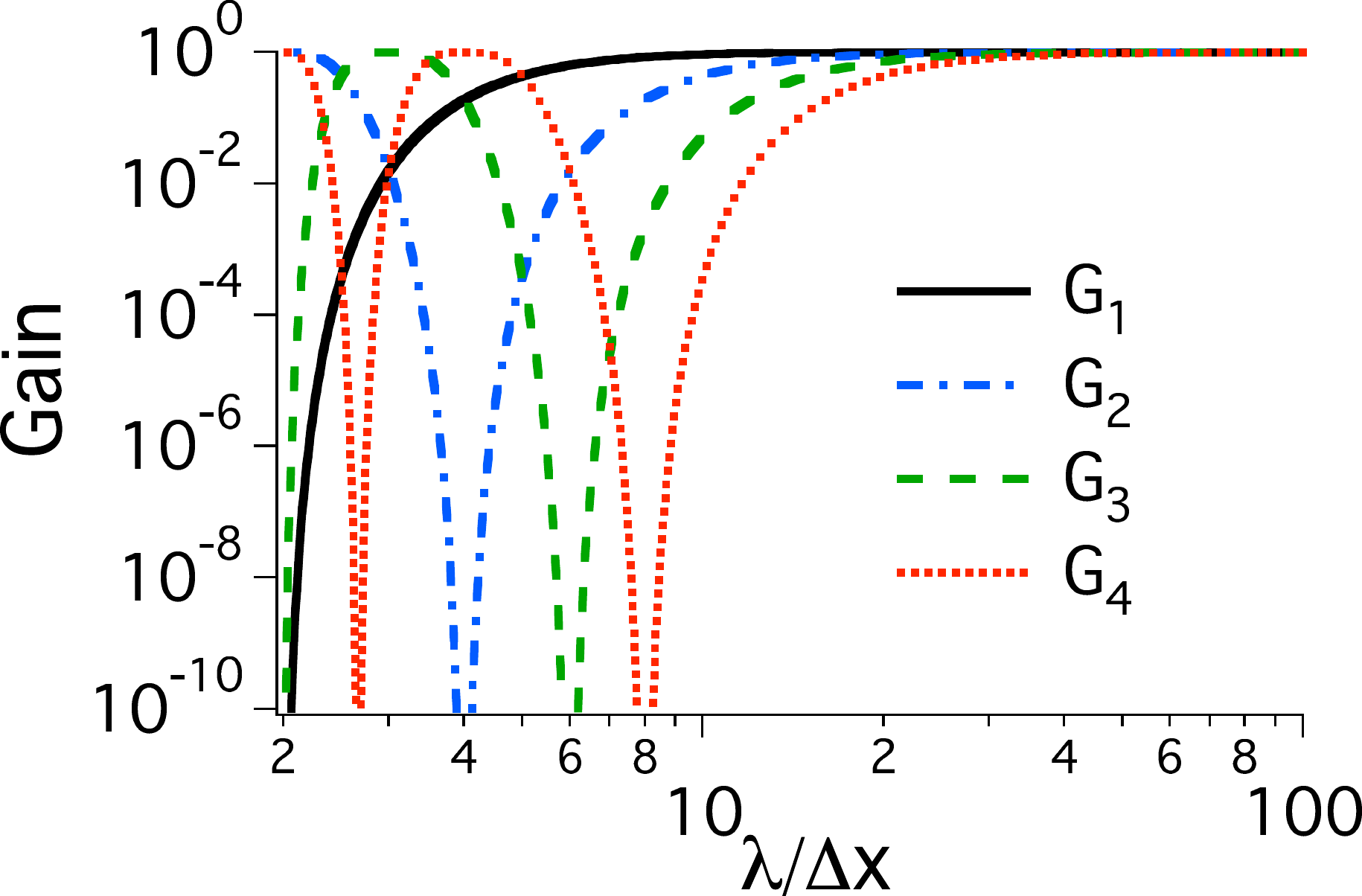}
   \includegraphics*[width=65mm]{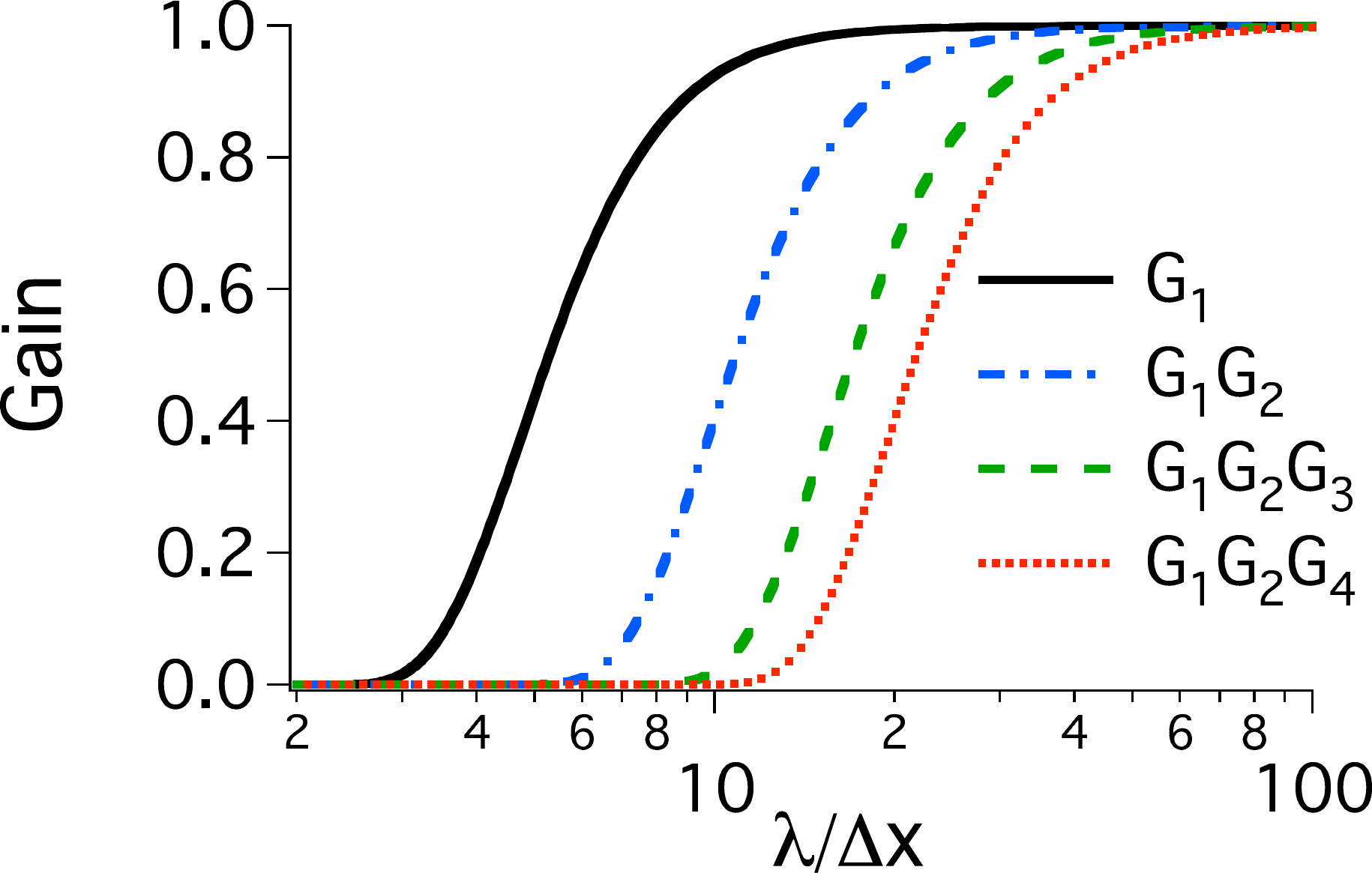}
   \includegraphics*[width=65mm]{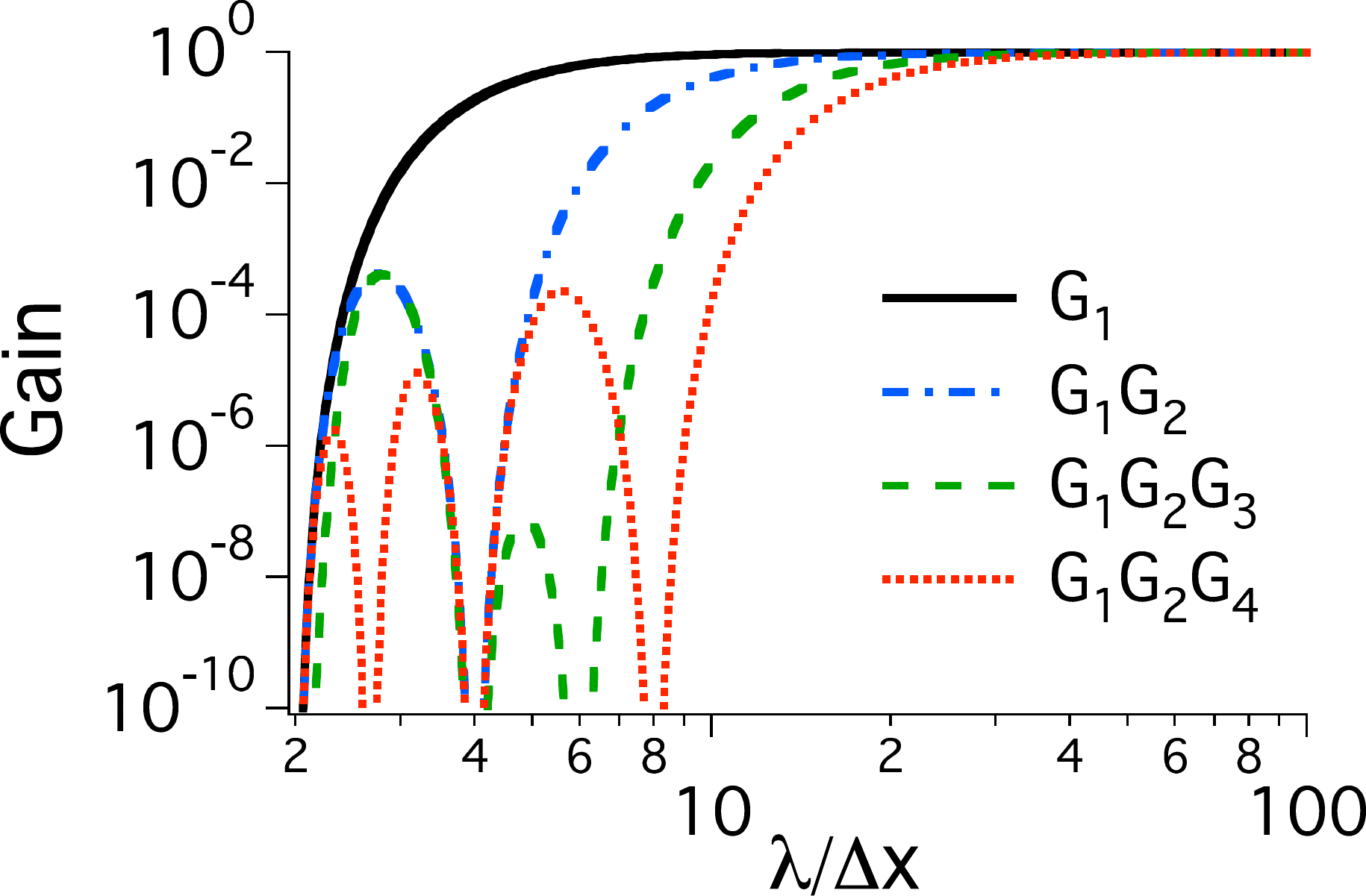}
   \caption{ (top) gain for four passes bilinear filters with compensation ($G_s=g(s,1/2,k)^4\cdot g(s,3/2,k)$)  for strides s=1 to 4 linear with (left) linear ordinate (right) logarithmic ordinate; (bottom) gain for four low pass filters combining the $G_1$ to $G_4$ filters with (left) linear ordinate (right) logarithmic ordinate.}
   \label{Fig_gg}
\end{figure}
The gain is plotted in Fig. \ref{Fig_gg} (top) for four passes bilinear filters with compensation ($G=g(s,1/2,k)^4\cdot g(s,3/2,k)$)  for strides s=1 to 4. For a given stride, the gain is given by the gain of the bilinear filter shifted in k space, with the pole $g=0$ shifted from $\lambda=2/\delta x$ to $\lambda=2s/\delta x$, with additional poles, as given by
\begin{equation}
s k \delta x = \arccos\left(\frac{\alpha}{\alpha-1}\right) \pmod {2\pi}
\end{equation}
The resulting filter is pass band between the poles, but since the poles are spread at different integer values in k space, a wide band low pass filter can be constructed by combining filters at different strides. Examples are given in Fig. \ref{Fig_gg} (bottom) for combinations of the filter with stride 1 to 4.  

\begin{figure}[htb]
   \centering
   \includegraphics*[width=65mm]{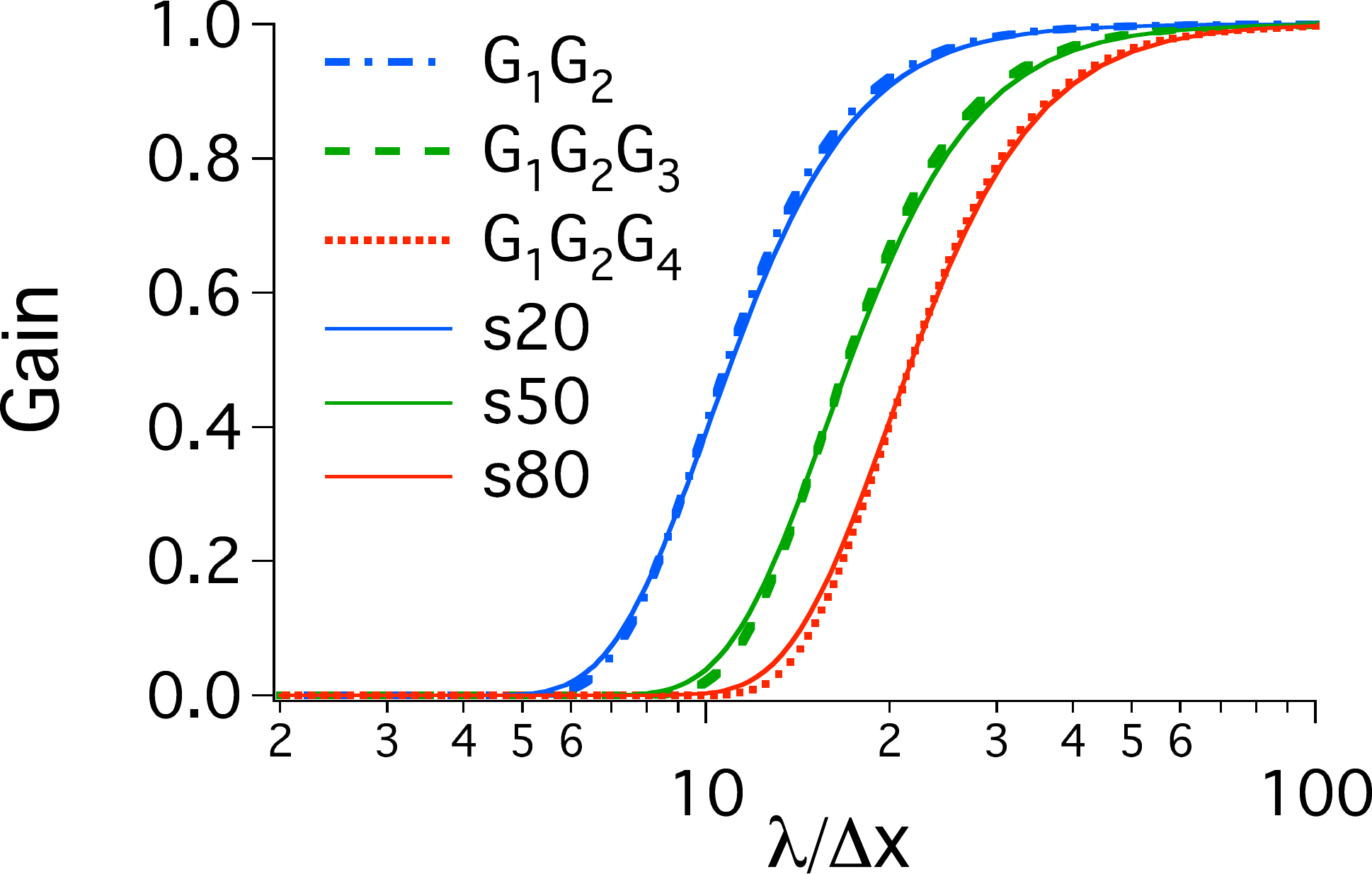}
   \includegraphics*[width=65mm]{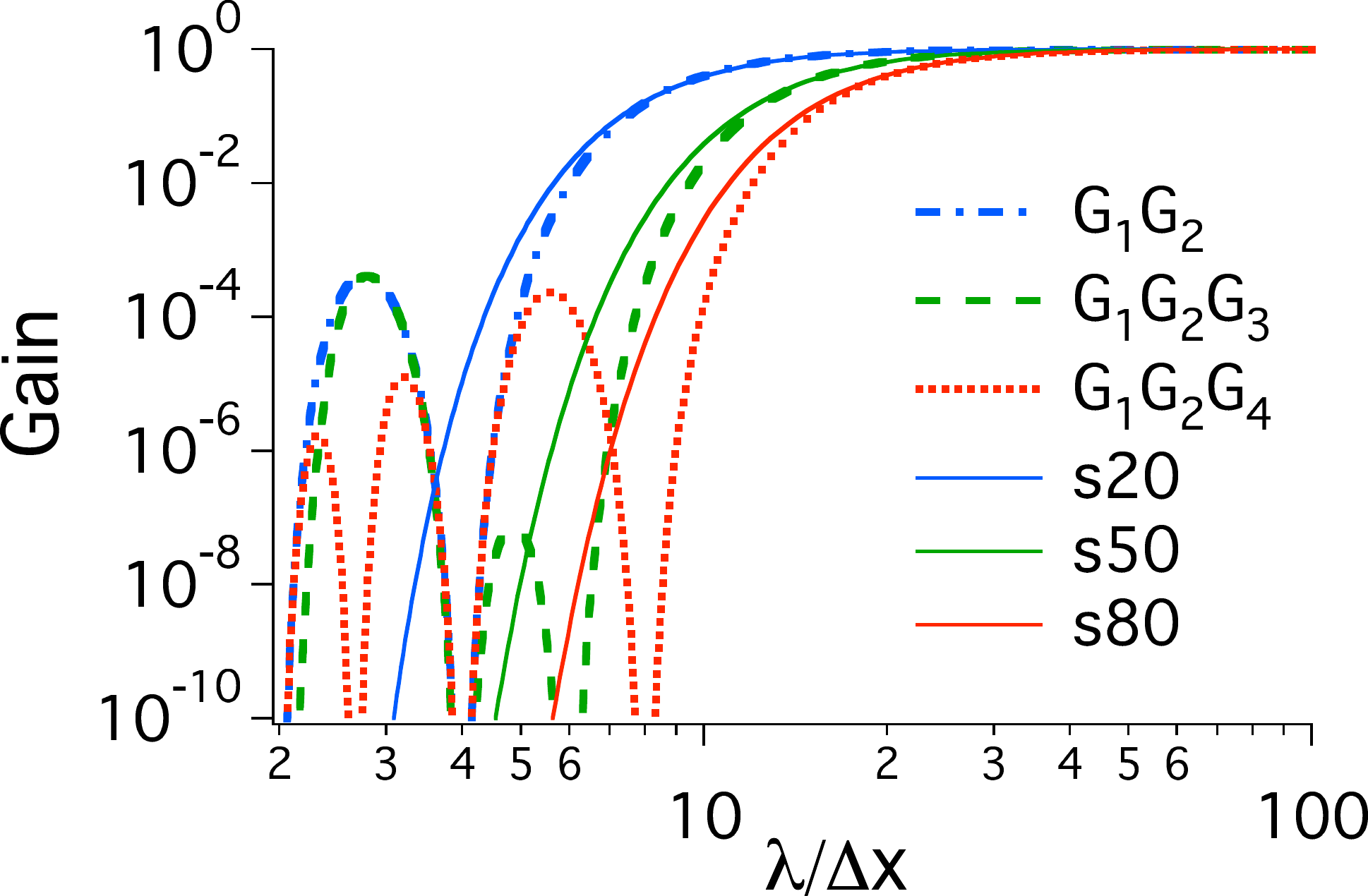}
   \caption{Comparison between filters with stride and filter s20-80 with (left) linear ordinate (right) logarithmic ordinate.}
   \label{Fig_ggcomp}
\end{figure}

The combined filters with strides 2, 3 and 4 have nearly equivalent fall-offs in gain (in linear scale) to the 20, 50 and 80 passes of the bilinear filter (see Fig. \ref{Fig_ggcomp}). Yet, the filters with stride need respectively 10, 15 and 15 passes of a three-point filter while the n-pass bilinear filer need respectively 21, 51 and 81 passes, giving gains of respectively 2.1, 3.4 and 5.4 in number of operations in favor of the filters with stride.
%Although such filters can be combined into one pass filter with large stencil, applying a large stencil is more challenging for parallel calculation if the size of the stencil is to surpass the size of the grid on a processor, resulting in a much more complicated message passing procedure than the standard message exchanges between neighboring zones. 
%Nonetheless, for efficiency, consecutive steps can be combined into one, to a degree that is dictated by the implementation and the machine hardware used for the run.

\clearpage
\subsection{Tunable solver} 
\label{Sec_lowdisp}
%(note that Cole stencil categorized as a (3, 3, 1, 0) stencil based scheme in Finkelstein/Kastner  classification)

In \cite{ColeIEEE1997} and \cite{ColeIEEE2002}, Cole introduced an implementation of the  source-free Maxwell's wave equations for narrow-band applications based on non-standard finite-differences (NSFD). In \cite{KarkICAP06}, Karkkainen et al adapted it for wideband applications. At the Courant limit for the time step and for a given set of parameters, the stencil proposed in \cite{KarkICAP06} has no numerical dispersion along the principal axes, provided that the cell size is the same along each dimension (i.e. cubic cells in 3D). The solver from \cite{KarkICAP06} was modified to be consistent with the Particle-In-Cell methodology and implemented in the code Warp, with the ability given to the user of setting the solver adjustable coefficients, providing tunability of the numerical properties of the solver to better fit the requirements of a particular application.

The "Cole-Karkkainnen"'s solver \cite{KarkICAP06} uses a non-standard finite difference formulation (extended stencil) of the Maxwell-Ampere equation. For implementation into a Particle-In-Cell code, the formulation must introduce the source term into Cole-Karkkainen's source free formulation in a consistent manner. However, modifying the NSFD formulation of the Maxwell-Ampere equation so that it includes the source term in a way that is consistent with the current deposition scheme is challenging. To circumvent this problem, Warp implementation departs from Karkkainen's by applying the enlarged stencil on the Maxwell-Faraday equations, which does not contain any source term but is formally equivalent to the source-free Maxwell-Ampere equation. Consequently, in Warp's implementation, the discretized Maxwell-Ampere equation is the same as in the Yee scheme, and the discretized Maxwell's equations read:
\begin{eqnarray}
\Delta_t \mathbf{B} & = & -\nabla^* \times \mathbf{E} \label{Eq:Faraday}\\
\Delta_t \mathbf{E} & = & c^2 \nabla \times \mathbf{B} - \frac{\mathbf{J}}{\epsilon_0} \label{Eq:Ampere}\\
\nabla \cdot \mathbf{E} & = & \frac{\rho}{\epsilon_0} \label{Eq:Gauss}\\
\nabla^* \cdot \mathbf{B} & = & 0 \label{Eq:divb}
\end{eqnarray}
where $\epsilon_0$ is the permittivity of vacuum, and Eq. \ref{Eq:Gauss} and \ref{Eq:divb} not being solved explicitly but verified via appropriate initial conditions and current deposition procedure. The differential operators are defined as 
\begin{eqnarray}
\nabla& = & \Delta_x \mathbf{\hat{x}} + \Delta_y \mathbf{\hat{y}} + \Delta_z \mathbf{\hat{z}} \\
\nabla^*& = & \Delta_x^* \mathbf{\hat{x}} + \Delta_y^* \mathbf{\hat{y}} + \Delta_z^* \mathbf{\hat{z}}, 
\end{eqnarray}
the finite differences and sums operators being
\begin{eqnarray}
\Delta_t G|^{n}_{i,j,k} & = & \frac{G|^{n+1/2}_{i,j,k}-G|^{n-1/2}_{i,j,k}}{\delta t} \\
\Delta_x G|^{n}_{i,j,k} & = & \frac{G|^{n}_{i+1/2,j,k}-G|^{n}_{i-1/2,j,k}}{\delta x} \\
\Delta^*_x & = & \left(\alpha+\beta S^1_x + \gamma S^2_x\right) \Delta_x
\end{eqnarray}
with
\begin{eqnarray}
S^1_x G|^{n}_{i,j,k} & = & G|^{n}_{i,j+1/2,k} + G|^{n}_{i,j-1/2,k}  \nonumber\\
& + & G|^{n}_{i,j,k+1/2}  + G|^{n}_{i,j,k-1/2}  \\
S^2_x G|^{n}_{i,j,k} & = & G|^{n}_{i,j+1/2,k+1/2} + G|^{n}_{i,j-1/2,k+1/2} \nonumber \\
& + & G|^{n}_{i,j+1/2,k-1/2} + G|^{n}_{i,j-1/2,k-1/2}
\end{eqnarray}

%\begin{equation}
%\begin{array}{ccc}
%\Delta_t H_x|^{n}_{i,j+1/2,k+1/2} & = & \Delta^*_z E_y|^{n}_{i,j+1/2,k+1/2} \\
%& - & \Delta^*_y E_z|^{n}_{i,j+1/2,k+1/2}
%\end{array}
%\end{equation}

The quantity $G$ is a sample vector component, $\delta t$ and $\delta x$ are respectively the time step and the grid cell size along $x$, while $\alpha$, $\beta$ and $\gamma$ are constant scalars verifying $\alpha+4\beta+4\gamma=1$. The operators along $y$ and $z$, i.e. $\Delta_y$, $\Delta_z$,  $\Delta^*_y$, $\Delta^*_z$,  $S^1_y$, $S^1_z$, $S^2_y$, and $S^2_z$, are obtained by circular permutation of the indices. 

In 2D, assuming the plane $(x,z)$, the enlarged finite operators simplify to
\begin{eqnarray}
\Delta^*_x & = & \left(\alpha+\beta S^1_x\right) \Delta_x\\
S^1_x G|^{n}_{i,j,k} & = & G|^{n}_{i,j+1/2,k} + G|^{n}_{i,j-1/2,k}.
\end{eqnarray}

An extension of this algorithm for non-cubic cells provided by Cowan in \cite{CowanICAP09} is not considered in this paper. However, all considerations given here for the solver implemented in Warp apply readily to the solver developed by Cowan.\\

\subsubsection{Numerical dispersion}
The dispersion relation of the solver is given by 
\begin{equation}
\left( \frac{ \sin\frac{\omega\delta t}{2}}{c\delta t}\right)^2 = C_x\left( \frac{ \sin\frac{k_x\delta x}{2}}{\delta x}\right)^2 +  C_y\left( \frac{ \sin\frac{k_y\delta y}{2}}{\delta y}\right)^2 +  C_z\left( \frac{ \sin\frac{k_z\delta z}{2}}{\delta z}\right)^2
\label{Eq:num_disp}
\end{equation}
with
\begin{eqnarray}
C_x &=&  \alpha+2\beta(c_y+c_z)+4\gamma c_y c_z \\
C_y &=&  \alpha+2\beta(c_z+c_x)+4\gamma c_z c_x \\
C_z &=&  \alpha+2\beta(c_x+c_y)+4\gamma c_x c_y
\label{Eq:coef_num_disp1}
\end{eqnarray}
and 
\begin{eqnarray}
c_x &=& \cos\left(k_x\delta_x\right) \\
c_y &=& \cos\left(k_y\delta_y\right) \\
c_z &=& \cos\left(k_z\delta_z\right)
\label{Eq:coef_num_disp2}
\end{eqnarray}

The Courant-Friedrichs-Lewy condition (CFL) is given by  

\begin{eqnarray}
c\delta t_c & \leq & \min[\delta x,\delta y, \delta z, \nonumber\\
&&1/\sqrt{\left(\alpha-4\gamma\right)\max\left[\kappa_x+\kappa_y,\kappa_x+\kappa_z,\kappa_y+\kappa_z\right]}, \nonumber \\
&&1/\sqrt{\left(\alpha-4\beta+4\gamma\right)\left(\kappa_x+\kappa_y+\kappa_z\right)}]
\label{Eq:cfl}
\end{eqnarray}

where $\kappa_x=1/\delta x^2$, $\kappa_y=1/\delta y^2$ and $\kappa_z=1/\delta z^2$.

Assuming cubic cells ($\delta x=\delta y=\delta z$), the coefficients given in \cite{KarkICAP06} ($\alpha=7/12$, $\beta=1/12$ and $\gamma=1/48$) allow $c \delta t=\delta x$, and thus no dispersion along the principal axes. 

It is of interest to note that (\ref{Eq:num_disp}) can be rewritten
\begin{equation}
\left( \frac{ \sin\frac{\omega\delta t}{2}}{c\delta t}\right)^2 = \left(s_x^2+s_y^2+s_z^2\right) + \beta'\left(s_x^2s_y^2+s_x^2s_z^2+s_y^2s_z^2\right) + \gamma'\left(s_x^2s_y^2s_z^2\right)
\label{Eq:num_disp_prime}
\end{equation}
with $s_x=\sin\left(k_x\delta_x/2\right)$, $s_y=\sin\left(k_y\delta_y/2\right)$, $s_z=\sin\left(k_z\delta_z/2\right)$,  $\beta' = -8\beta-16\gamma$ and $\gamma'=48\gamma$, for which the coefficients from \cite{KarkICAP06} take the nice values $\beta'=-1$ and $\gamma'=1$. 

\begin{table}[htdp]
\begin{center}
\begin{tabular}{|c|c|c|c|c|c|c|}
\hline
& Yee & CK & CK2 & CK3 &  CK4 & CK5\\
\hline
%$\beta'$  & $0$ & $-1$ & $-\frac{1}{2}$ & $-\frac{1}{2}$ & $-\frac{9}{10}$ \\
%$\gamma'$ & $0$ & 1 & $\frac{1}{2}$ & $0$ & $\frac{9}{10}$ \\
$\beta'$  & $0$ & $-1$ & $-1/2$ & $0$ & $-1/2$ & $-9/10$ \\
$\gamma'$ & $0$ & 1 & $1/2$ & $-1$ & $0$ & $9/10$ \\
\hline
%$\alpha$ & $1$ & $\frac{5}{12}$ & $\frac{19}{24}$ & $\frac{3}{4}$ & $\frac{5}{8}$ \\
%$\beta$  & $0$ & $\frac{1}{12}$ & $\frac{1}{24}$ & $\frac{1}{16}$ & $\frac{3}{40}$ \\
%$\gamma$ & $0$ & $\frac{1}{48}$ & $\frac{1}{96}$ & $0$ & $\frac{3}{160}$ \\
$\alpha$ & $1$ & $7/12$ & $19/24$ & $11/12$ & $3/4$ & $5/8$ \\
$\beta$  & $0$ & $1/12$ & $1/24$ & $1/24$ & $1/16$ & $3/40$ \\
$\gamma$ & $0$ & $1/48$ & $1/96$ & $-1/48$ & $0$ & $3/160$ \\
\hline
%$\frac{c \delta t}{\delta x}$ & $\frac{1}{\sqrt{3}}$ & $1$ & $\frac{1}{\sqrt{2}}$ & $\frac{\sqrt{2}}{\sqrt{3}}$ & $\frac{\sqrt{5}}{\sqrt{6}}$ \\
$c \delta t/\delta x$ & $1/\sqrt{3}$ & $1$ & $1/\sqrt{2}$ & $1/\sqrt{2}$ & $\sqrt{2}/\sqrt{3}$ & $\sqrt{5}/\sqrt{6}$ \\
\hline
\end{tabular}
\end{center}
\caption{List of coefficients }
\label{Table:CKcoefs}
\end{table}%

\begin{figure}[htb]
   \centering
{\small
 \begin{tabular}{@{}c@{}c@{}} % @{} removes extra space
%  \hline
  \hspace{0.cm} Yee & \hspace{0.cm} CK  \vspace{0.mm}\\
   \includegraphics*[width=60mm]{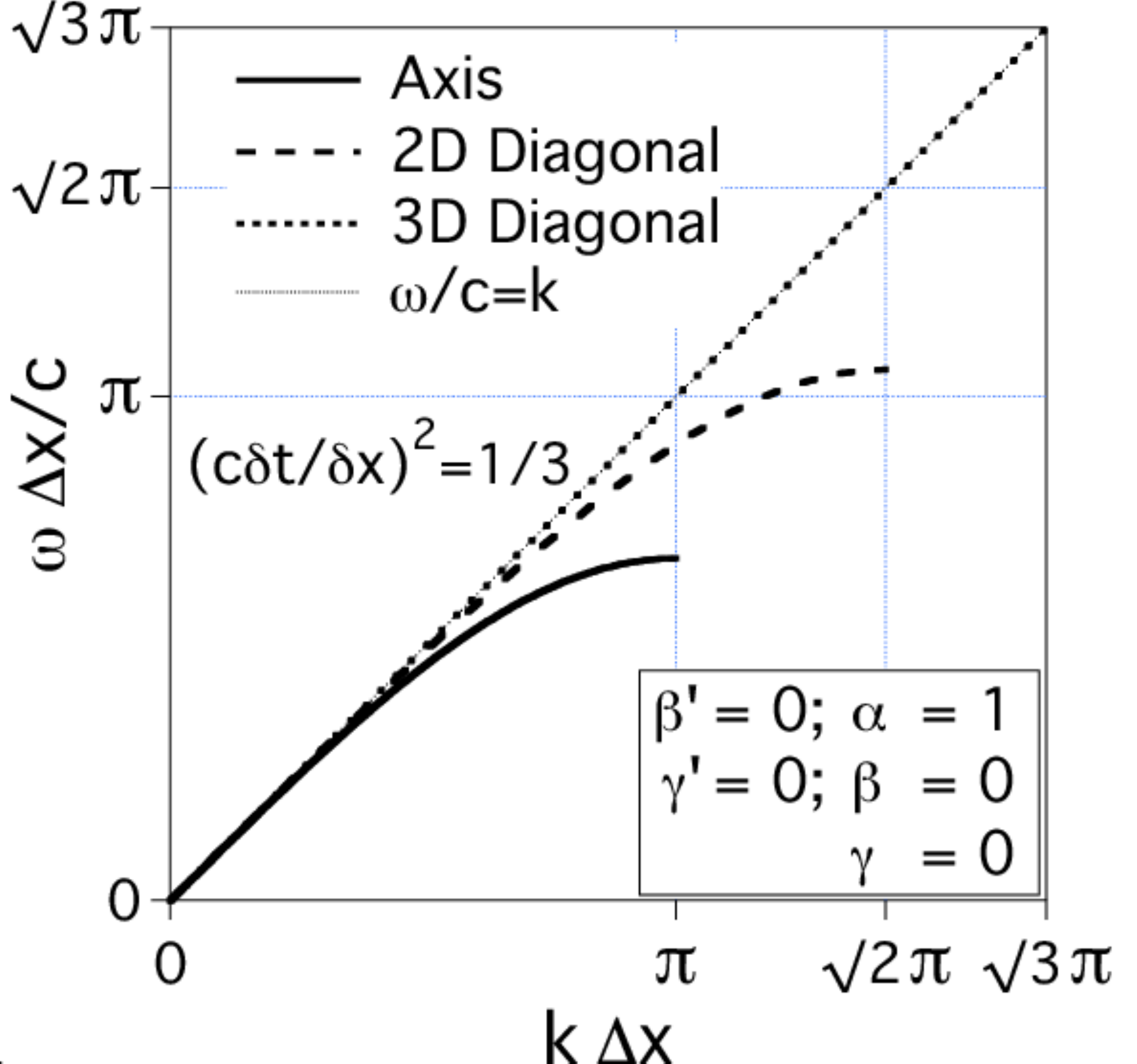}  \hspace{0.mm}&
   \includegraphics*[width=60mm]{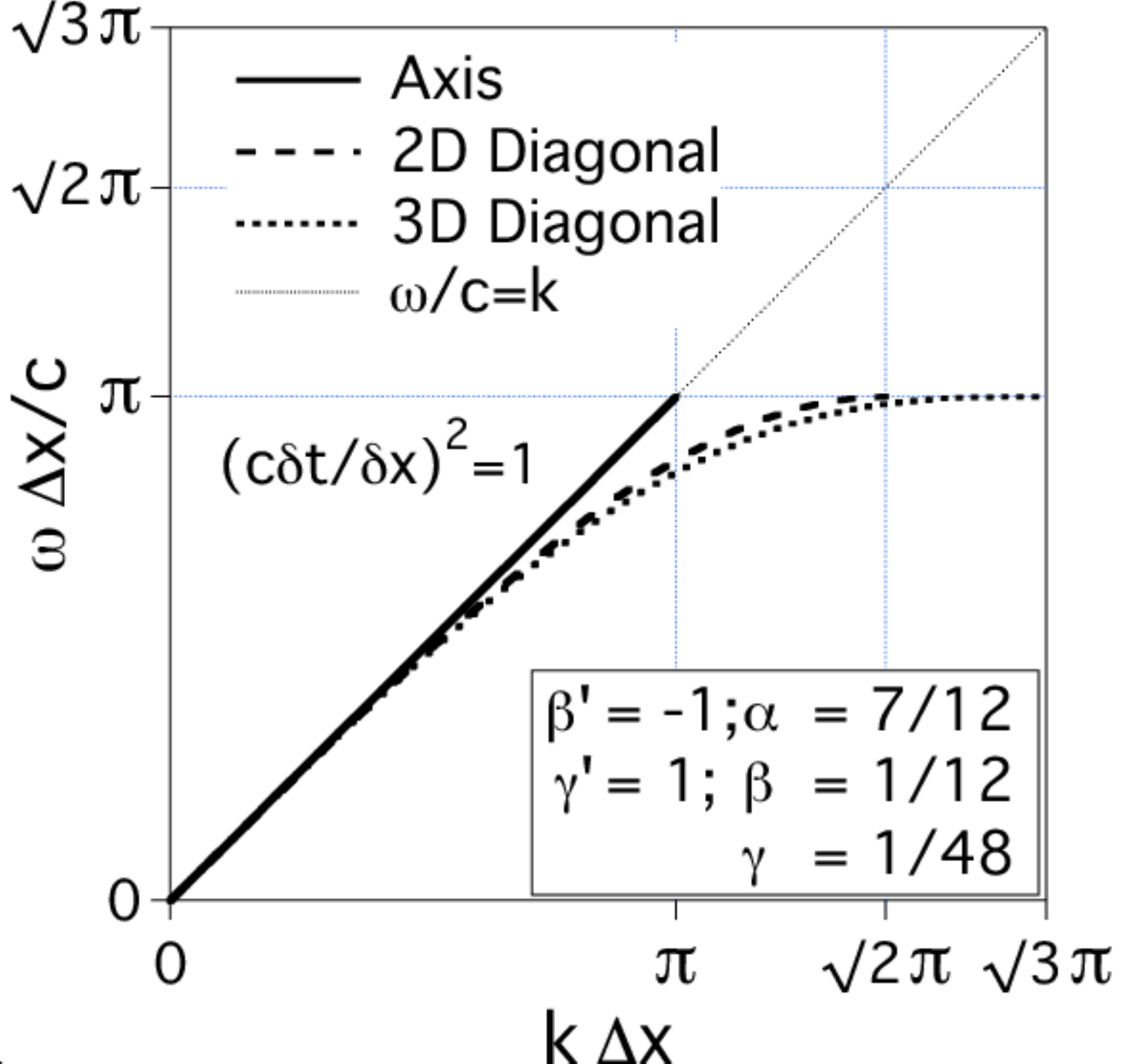}  \vspace{2mm}\\
  \hspace{0.cm} CK2& \hspace{0.cm} CK3\vspace{0.mm}\\
   \includegraphics*[width=60mm]{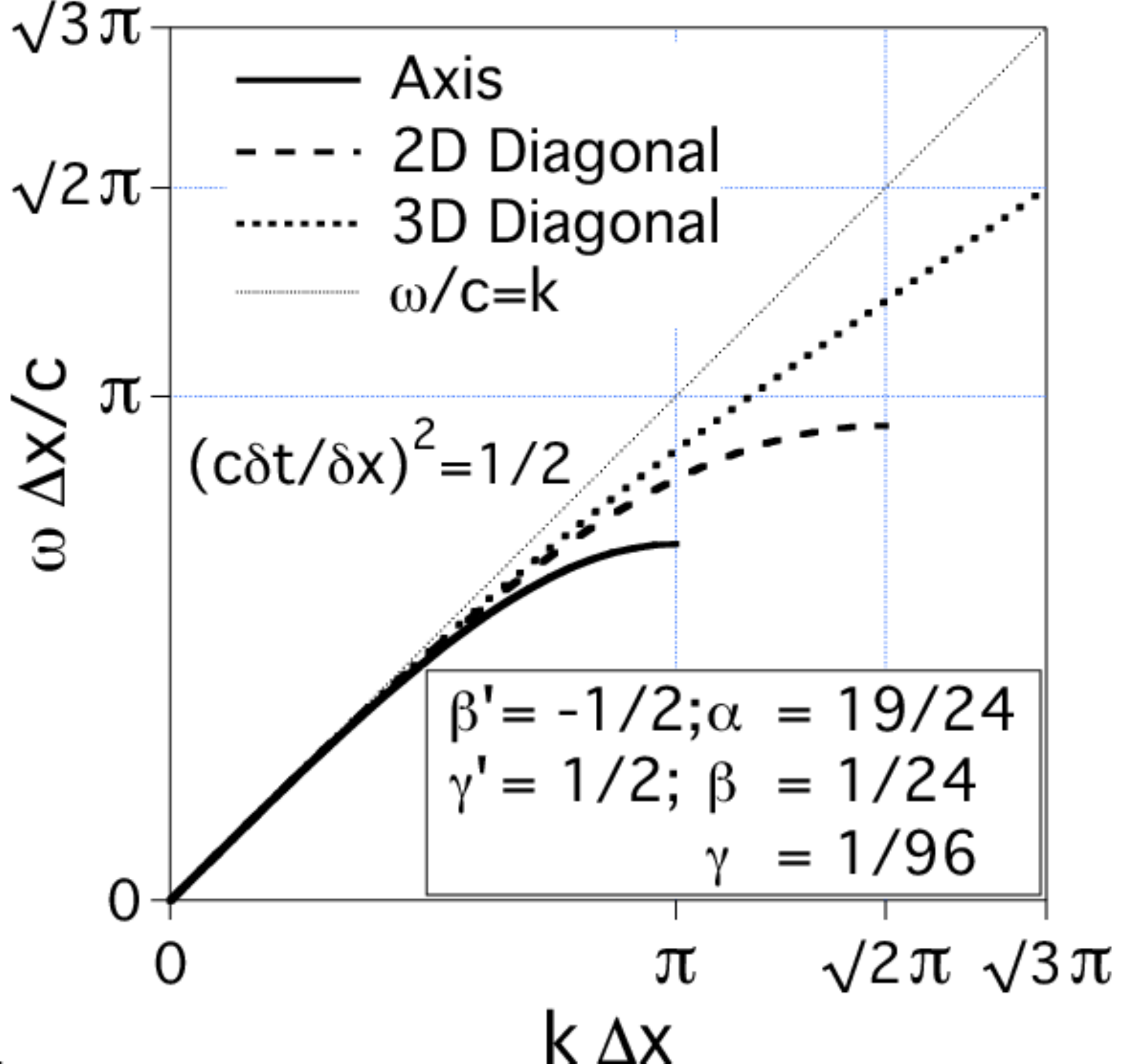} \hspace{0.mm}&
   \includegraphics*[width=60mm]{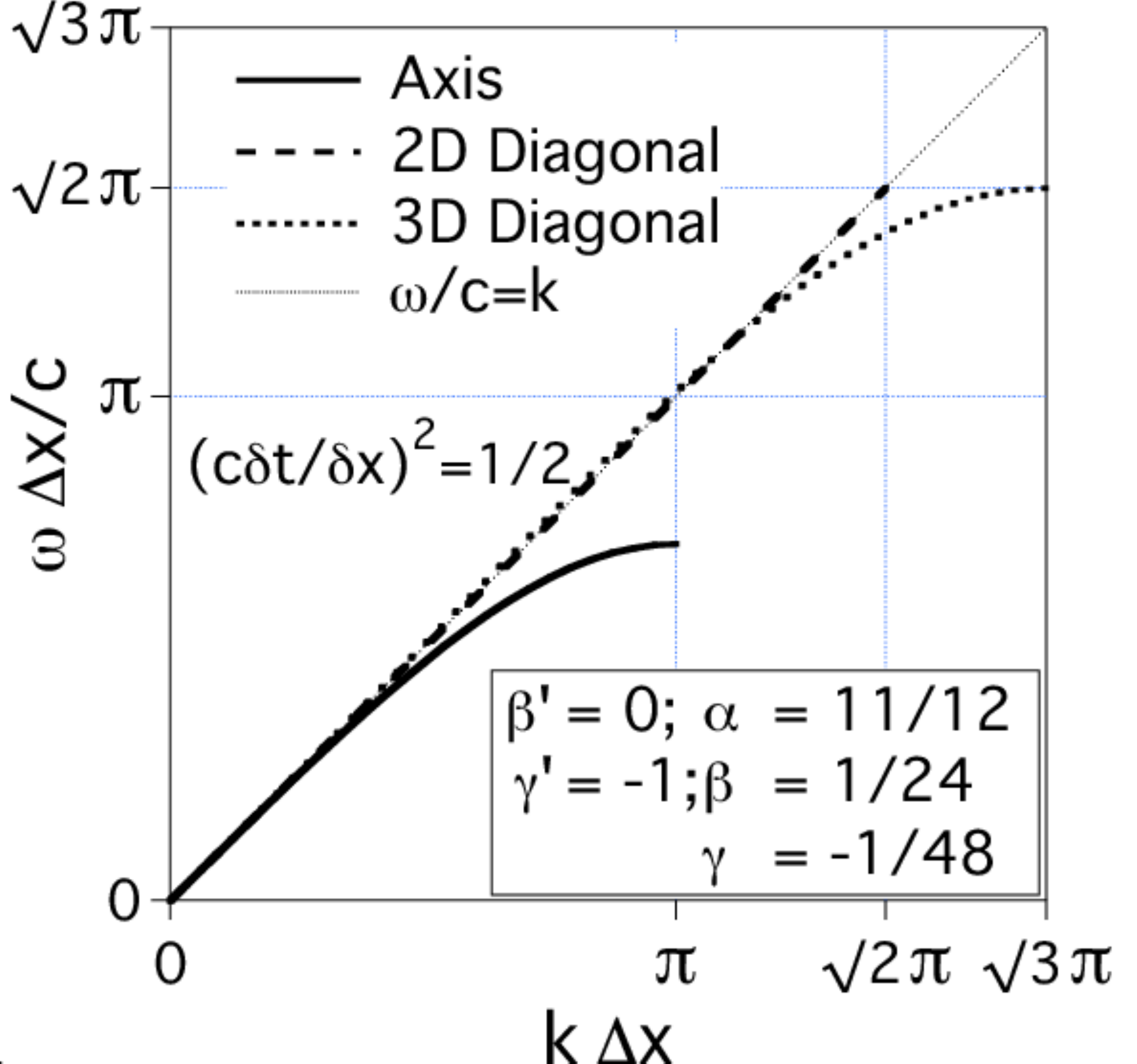} \vspace{2mm}\\
  \hspace{0.cm} CK4& \hspace{0.cm} CK5\vspace{0.mm}\\
   \includegraphics*[width=60mm]{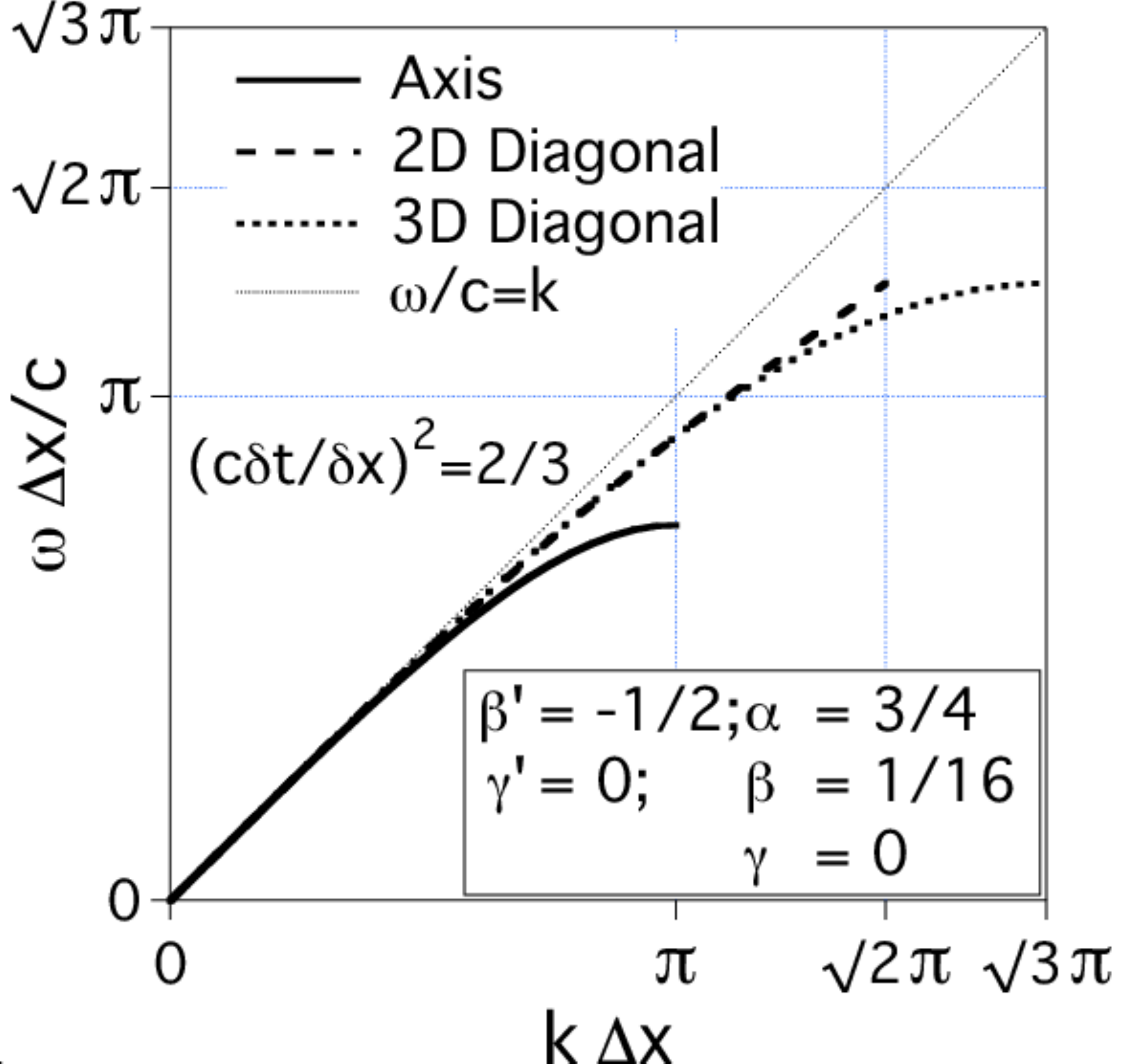}\hspace{0.mm}&
   \includegraphics*[width=60mm]{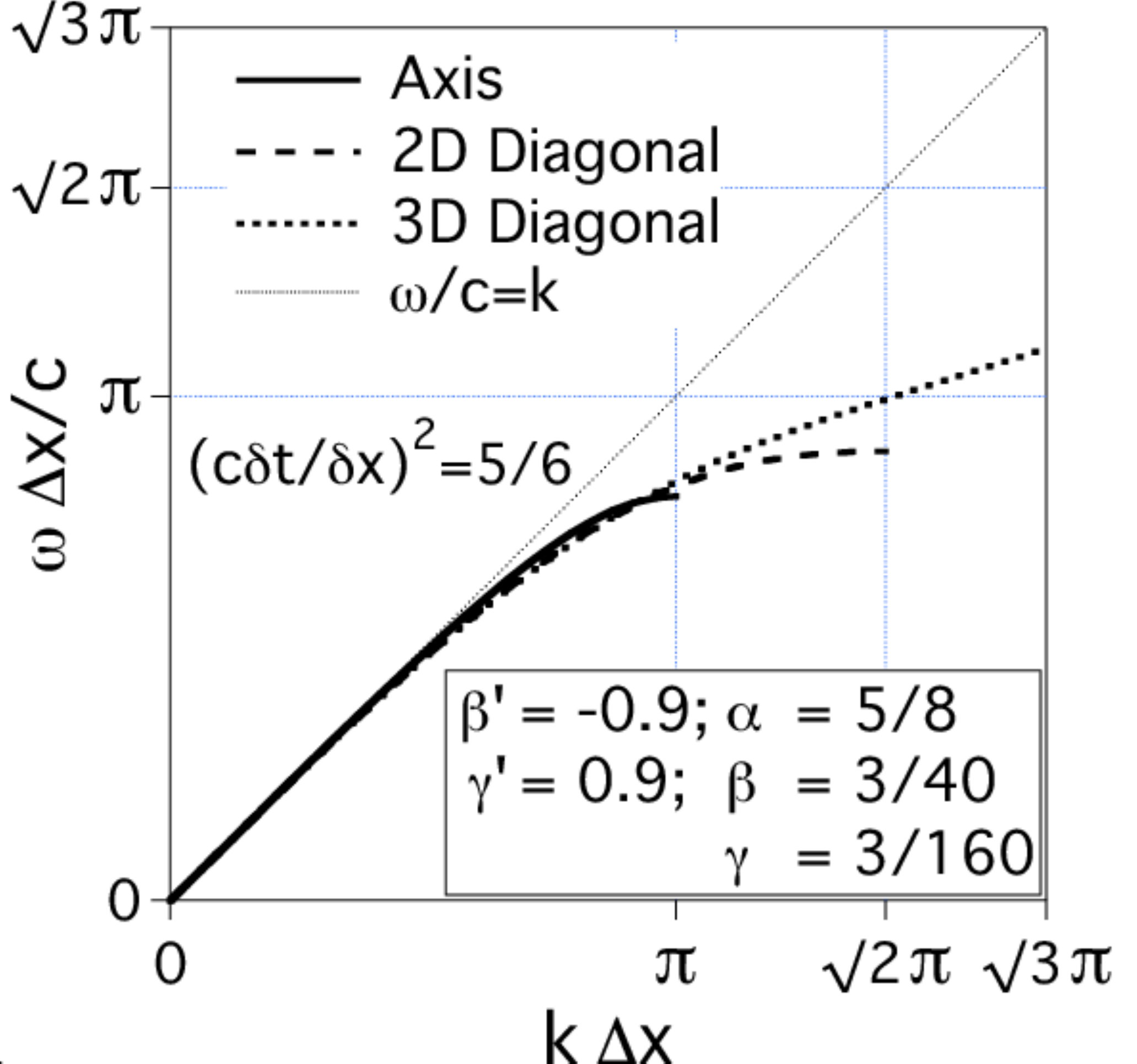}
\end{tabular}
}
   \caption{Numerical dispersion along the principal axis and diagonals for cubic cells ($\delta x=\delta y=\delta z$) at the Courant limit for the solver with adjustable numerical dispersion using the parameters from Table \ref{Table:CKcoefs}.}
   \label{Fig_dispersion}
\end{figure}

Sets of possible coefficients and the corresponding CFL condition, assuming cubic cells, are given in Table \ref{Table:CKcoefs}. The numerical dispersion using those coefficients are plotted in figure \ref{Fig_dispersion} along the principal axes and diagonals for cubic cells ($\delta x=\delta y=\delta z$) and contrasted with the one of the Yee solver (all taken at each solver's CFL time step limit). At the CFL limit, the Yee algorithm offers no numerical dispersion along the 3D diagonal, but relatively large numerical dispersion at the Nyquist frequency along the main axes. Conversely, the Cole-Karkkainen solver (CK) offers no numerical dispersion along the main axes but significant dispersion along the diagonals. The CK solver also allows larger time steps than the Yee solver by almost a factor of two in 3D. The solver labeled "CK2" offers numerical dispersion that is intermediate between the Yee solver and the CK solver along the main axes and the 3D diagonal, but slightly degraded along the 2D diagonal. Conversely, while solver CK3 also offers intermediate numerical dispersion along the main axes and the 3D diagonal, it offers no numerical dispersion along the 2D diagonal. Solver CK4 improves slightly the numerical dispersion along the main axes over CK2 and CK3 at the expense of the dispersion along the diagonals. Finally, CK5 offers the highest level of isotropy. The CFL time steps of solvers CK2, 3, 4 and 5 are intermediate between the Yee and the CK CFL time steps. This provides 
solvers with a range of numerical dispersion among which some may be more favorable with regard to the mitigation of numerical instabilities for a given application.

To reduce numerical dispersion to its lowest level, it is desirable to operate the CK solver as close as possible to the CFL limit $c\delta t=\delta x$. However, an instability (other than numerical Cerenkov) arises at the Nyquist frequency in such a case. The analysis is given in 1D in Appendix I, as well as its mitigation using digital filtering. Since for the CK solver, the CFL limit is independent of dimensionality, the analysis and mitigation apply readily to 2D and 3D simulations. 

For absorption of outgoing waves at the computational box boundaries, the extension of the solver to a Perfectly Matched Layer \cite{BerengerJCP96} is given in Appendix II.

\clearpage
\subsubsection{Current deposition and Gauss' Law}
In most applications, it is essential to prevent accumulations of errors to the discretized Gauss' Law. This is accomplished by providing a method for depositing the current from the particles to the grid which is compatible with the discretized Gauss' Law, or by providing a mechanism for "divergence cleaning" \cite{BirdsallLangdon,LangdonCPC92,MarderJCP87,VayPoP98}. For the former, schemes which allow a deposition of the current that is exact when combined with the Yee solver is given in \cite{VillasenorCPC92} for linear form factors and in \cite{EsirkepovCPC01} for higher order form factors. Since the discretized Gauss' Law and Maxwell-Faraday equation are the same in our implementation as in the Yee solver, charge conservation is readily verified using the current deposition procedures from \cite{VillasenorCPC92} and \cite{EsirkepovCPC01}, and this was verified numerically. Hence divergence cleaning is not necessary.

 \subsection{Friedman adjustable damping}
The tunable damping scheme developed by Friedman \cite{FriedmanJCP90} was shown to be the most potent practical method for mitigating the numerical Cerenkov instability in \cite{GreenwoodJCP04}, among the selected methods that were considered. It is readily applicable to the solver presented above by modifying (\ref{Eq:Faraday}) to
\begin{equation}
 \mathbf{B}^{n+3/2}  =  \mathbf{B}^{n+1/2} - \delta t\nabla^* \times \left[ \left(1+\frac{\theta}{4}\right)\mathbf{E}^{n+1} -  \frac{1}{2}\mathbf{E}^{n} +  \left(\frac{1}{2}-\frac{\theta}{4}\right) \bar{\mathbf{E}}^{n-1}\right] \label{Eq:Faraday_Friedman}\\
\end{equation}
with
\begin{equation}
\bar{\mathbf{E}}^{n-1}=\left(1-\frac{\theta}{2}\right)\mathbf{E}^{n}+\frac{\theta}{2}\bar{\mathbf{E}}^{n-2}
\end{equation}
where $0\leq\theta\leq 1$ is the damping factor. 
The numerical dispersion becomes
\begin{equation}
\left( \frac{ \sin\frac{\omega\delta t}{2}}{c\delta t}\right)^2 = F\Omega^2
\label{Eq:num_disp_Friedman}
\end{equation}
where
\begin{equation}
F=1-\frac{2\theta\sin^2\left(\omega\delta t/2\right)}{2e^{-i\omega\delta t}-\theta}
\end{equation}
and 
\begin{equation}
\Omega^2=\left[C_x\left( \frac{ \sin\frac{k_x\delta x}{2}}{\delta x}\right)^2 +  C_y\left( \frac{ \sin\frac{k_y\delta y}{2}}{\delta y}\right)^2 +  C_z\left( \frac{ \sin\frac{k_z\delta z}{2}}{\delta z}\right)^2\right]
\end{equation}
The CFL is given by 
\begin{equation}
c\delta t^*_c=c\delta t_c\sqrt{\frac{2+\theta}{2+3\theta}}
\end{equation}
where $\delta t_c$ is the critical time step of the numerical scheme without damping ($\theta =0$), as given by (\ref{Eq:cfl}).

\begin{figure}[htb]
   \centering
{\small
 \begin{tabular}{@{}c@{}c@{}} % @{} removes extra space
%  \hline
  \hspace{0.5cm} Yee-Friedman & \hspace{0.5cm} Cole-Karkkainen-Friedman  \vspace{2.mm}\\
   \includegraphics*[width=65mm]{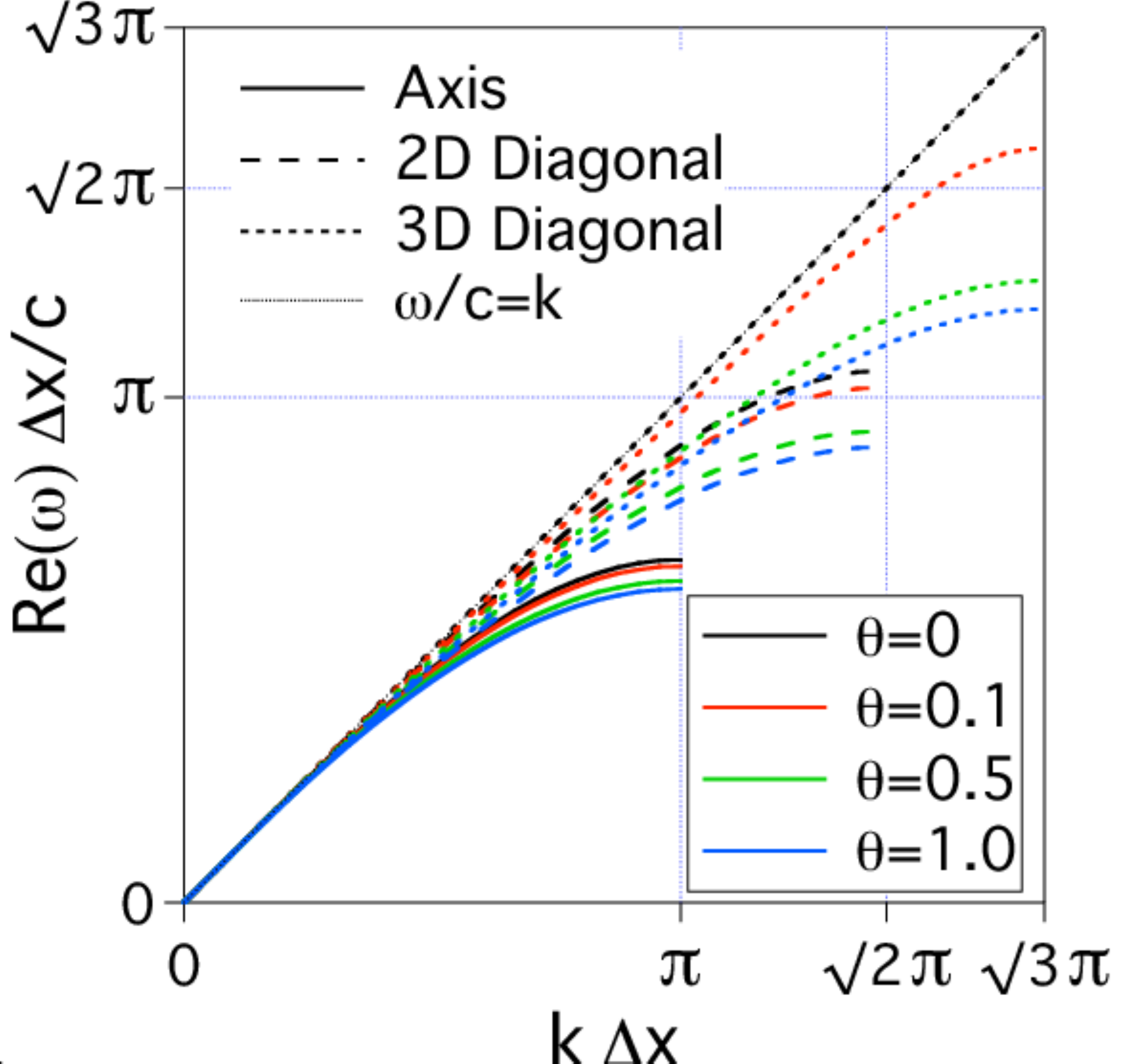}&
   \includegraphics*[width=65mm]{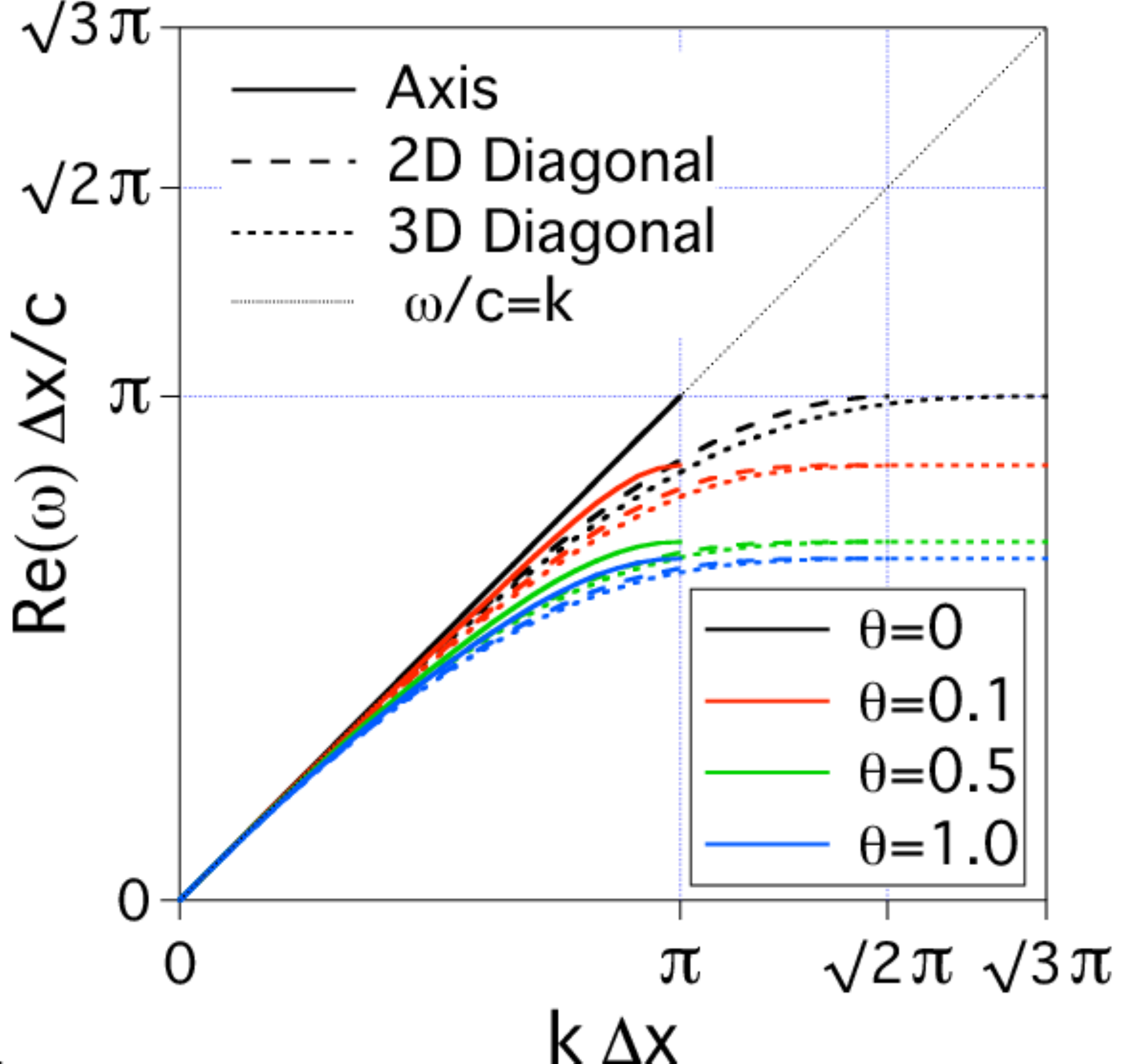} \\
   \includegraphics*[width=65mm]{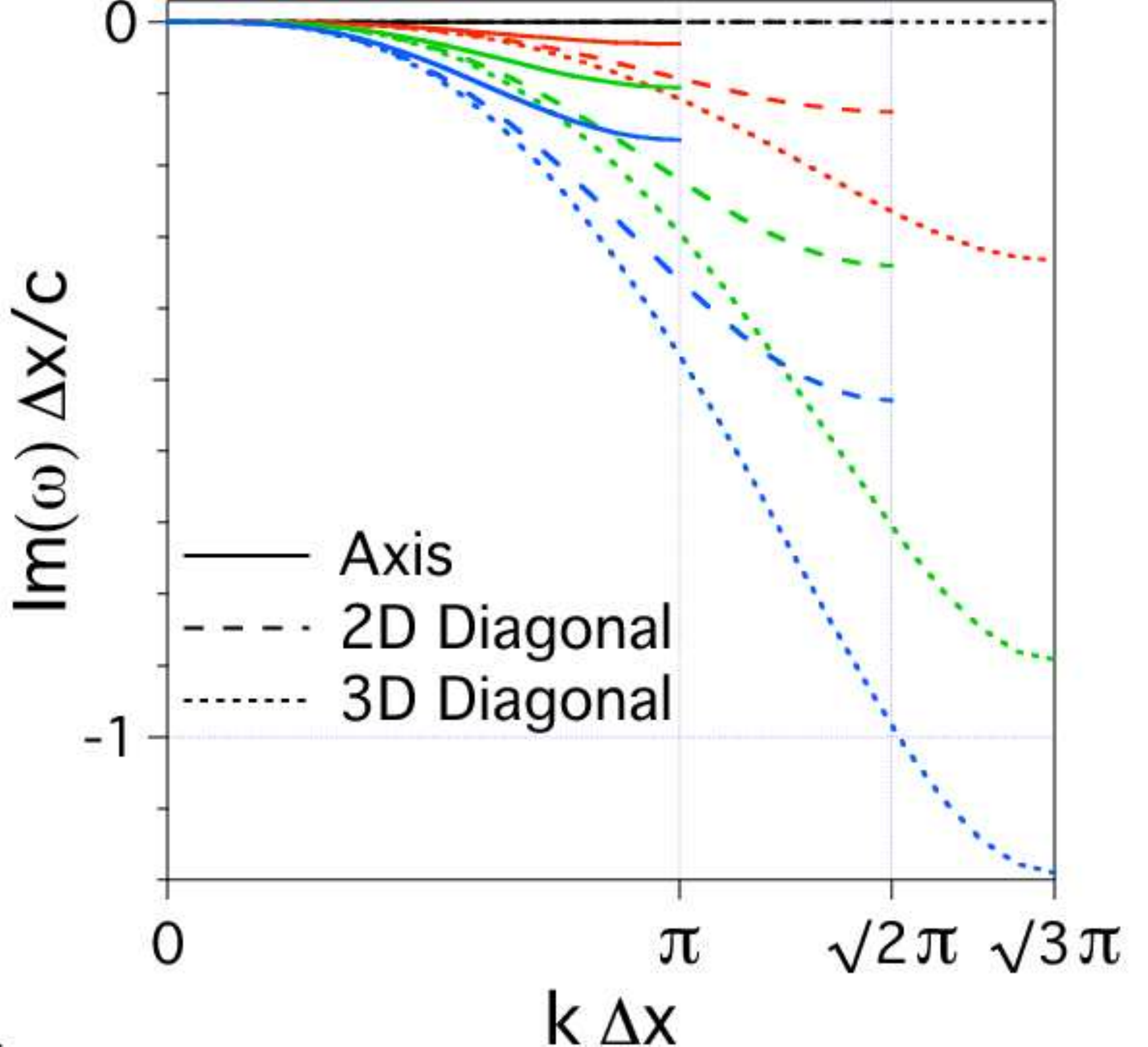} &
   \includegraphics*[width=65mm]{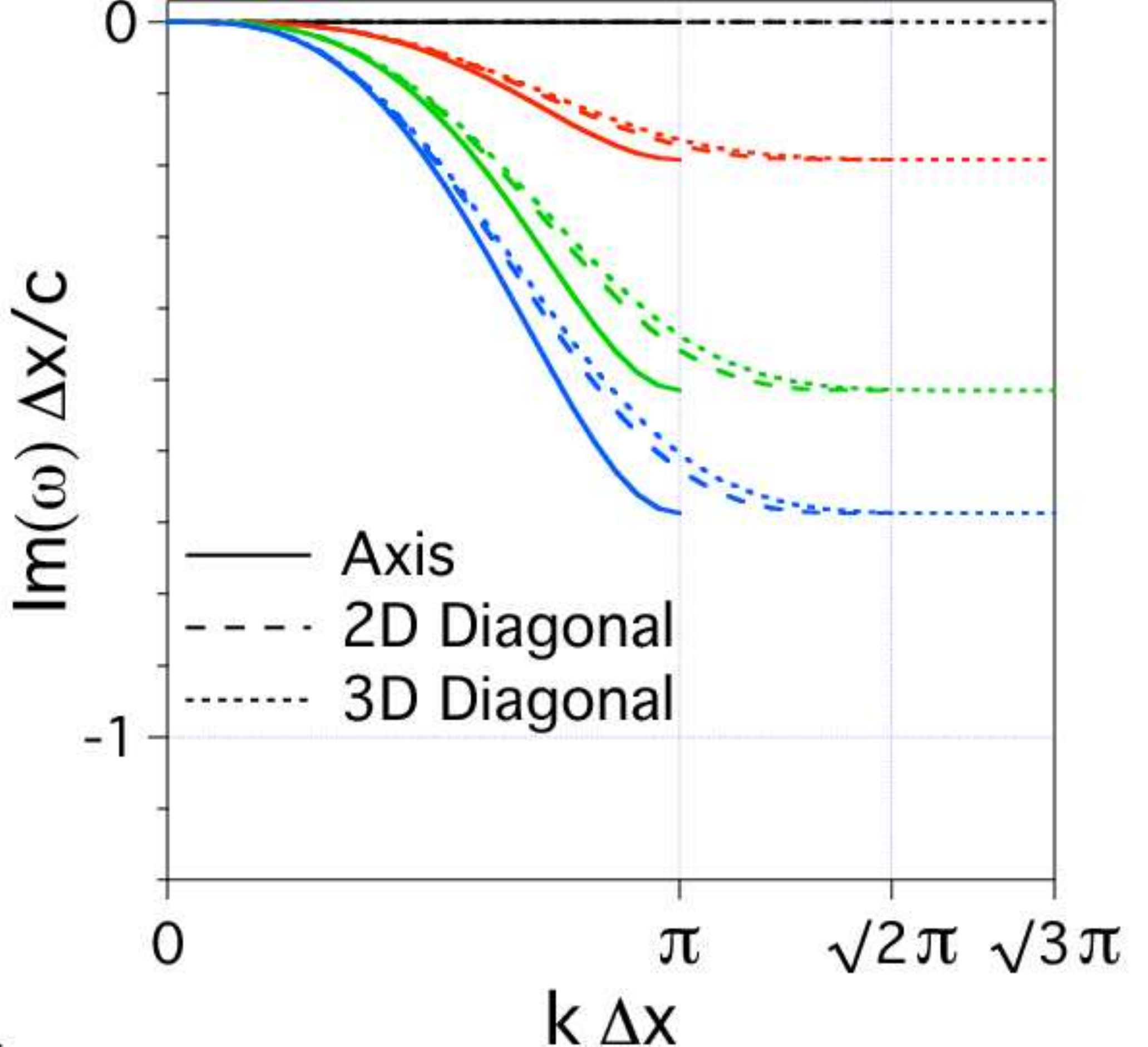} \\
\end{tabular}
}
   \caption{Numerical dispersion along the principal axis and diagonals for cubic cells ($\delta x=\delta y=\delta z$) at the Courant limit for: (left) the Yee-Friedman solver; (right) the Cole-Karkkainen-Friedman solver. The real part (phase) and the imaginary part (amplitude) are plotted respectively in the top and bottom rows.}
 \label{Fig_dispersion_friedman}
\end{figure}

The numerical dispersion of the Cole-Karkkainen-Friedman (CKF) solver (using the coefficients from the CK solver in Table \ref{Table:CKcoefs}) is plotted in figure \ref{Fig_dispersion_friedman} along the principal axis and diagonals for cubic cells ($\delta x=\delta y=\delta z$) and contrasted with the one of the Yee-Friedman (YF) solver (both taken at the Courant time step limit). 
The amount of phase error rises with the value of the damping parameter $\theta$ (partly due to the slightly more constraining limit on the critical time step). However, it was shown in \cite{GreenwoodJCP04} that the amount of damping provided by the YF solver was sufficient to counteract the slight degradation of numerical dispersion with raising $\theta$, reducing the numerical Cerenkov effects to an acceptable level for the problem that was considered. The damping is very isotropic with the CKF solver but not with the YF one. The YF implementation offers a higher level of damping of the shortest wavelengths along the 3D diagonals, while the CKF offers more damping along the axes, and the amount of damping along the 2D diagonals are similar. In summary, the YF implementation delivers respectively the highest/lowest level of damping in the direction of lowest/highest numerical dispersion, while the CKF implementation delivers a proportionally higher level of dispersion than the YF implementation along the direction of highest numerical dispersion. Thus it may be expected that the CKF implementation will be more efficient in reducing numerical Cerenkov effects. 

\clearpage
\section{Application to the modeling of laser wakefield acceleration}

This section presents applications of the methods to the modeling of 10 GeV LPA stages at full scale 
in 2-1/2D and 3D, which has not been done fully self-consistently with other methods. It has been shown that many parameters of high energy LPA stages can be accurately simulated at reduced cost by simulating stages of lower energy gain, with higher density and shorter acceleration distance, by scaling the physical quantities relative to the plasma wavelength, and this has been applied to design of 10 GeV LPA stages \cite{CormierAAC08, GeddesPAC09}. The number of oscillations of a mismatched laser pulse in the plasma channel however depends on stage energy and does not scale, though this effect is minimized for a channel guided stage as considered in \cite{CormierAAC08, GeddesPAC09}.  The number of betatron oscillations of the trapped electron bunch will also depend on the stage energy, and may affect quantities like the emittance of the beam.  For these reasons, and to prove validity of scaled designs of other parameters, it is necessary to perform full scale simulations, which is only possible by using reduced models or simulations in the boosted frame. 

As a benchmarking exercise, we first perform scaled simulations similar to the ones performed in \cite{CormierAAC08}, at a density of $n_e = 10^{19}$ cm$^{-3}$, using various values of the boosted frame relativistic factor $\gamma$ to show the accuracy and convergence of the technique.  These stages were shown to efficiently accelerate both electrons and positrons with low energy spread, and the scaled simulations predicted acceleration of hundreds of pC to 10 GeV energies using a 40 J laser. The accuracy of the technique 
is evaluated by modeling scaled stages \cite{CormierAAC08, GeddesPAC09} at 0.1 GeV, which 
allows for a detailed comparison of simulations using a reference frame ranging from the 
laboratory frame to the frame of the wake. Excellent agreement is obtained on wakefield histories on axis, beam average energy history and %
momentum spread at peak energy, with speedup over a hundred, in agreement with the theoretical estimates from Section 2. 
The downscaled simulations are also used for
an in-depth exploration of the effects of the methods presented in Sections 3 and 4, and evaluation of which 
techniques are required to permit maximum $\gamma$ boost while maintaining high accuracy. We then apply the boosted frame technique to provide full scale simulation of high efficiency quasilinear LPA stages at higher energy, verifying the scaling laws in the 10 GeV-1 TeV range. 

% and evaluating emittance effects not accessible to the scaled simulations to design stages suitable for a conceptual future LPA based collider \cite{SchroederAAC08}.

\subsection{Scaled 10 GeV stages}
The parameters were chosen to be close (though not identical) to the case where $k_pL=2$ in \cite{CormierAAC08} where $k_p$ is the plasma wavenumber and $L$ is the laser pulse length. 
In the cases considered in this paper, the beam is composed of test particles only, with the goal of testing the fidelity of the algorithm in modeling laser propagation and wake generation.  The results from simulations of LPA in a boosted frame where beam loading is present will be presented elsewhere. These
simulations are scaled replicas of $10$ GeV stages that would operate at actual
densities of $10^{17}$ cm$^{-3}$ \cite{CormierAAC08,GeddesPAC09} and allow
short run times to permit effective benchmarking between the algorithms. The main 
physical and numerical parameters of the simulation are given in Table \ref{Tablephyspar2d}. Unless noted otherwise, in all the simulations presented herein, the field is gathered from the grid onto the particles directly from the Yee mesh locations, i.e. using the 'energy conserving' procedure (see \cite{BirdsallLangdon}, chapter 10).

\begin{table}[htd]
\caption{List of parameters for scaled 10GeV class LPA stage simulation.}
\begin{center}
\begin{tabular}{lcc}
\hline
\hline
beam radius		 		&$R_b$		&$82.5$ nm\\
beam length		 		&$L_b$		&$85.$ nm\\
%beam peak density 			&$n_b$		&$10^{14}$ cm$^{-3}$\\
beam transverse profile             &                		&$\exp\left(-r^2/8 R_b^2\right)$\\
beam longitudinal profile           &                		&$\exp\left(-z^2/2 L_b^2\right)$\\
\hline
laser wavelength 			&$\lambda$	& $0.8$ $\mu$m\\
%laser length (HWHM)				&$L$		& $3.36$ $\mu$m\\
%laser size 					&$\sigma$	& $8.91$ $\mu$m\\
laser length (FWHM)			&$L$		& $10.08$ $\mu$m\\
normalized vector potential	&$a_0$		& $1$\\
%laser transverse profile		&$$			& gaussian\\
%laser transverse profile		&$$			& $\exp\left(-r^2/2\sigma^2\right)$\\
%laser longitudinal profile		&$$			&sinusoidal\\
laser longitudinal profile		&$$			& $\sin\left(\pi z/L\right)$\\
\hline
plasma density on axis		&$n_e$		&$10^{19}$ cm$^{-3}$\\
%plasma transverse profile		&			& parabolic well\\
plasma longitudinal profile	&			& flat\\
plasma length				&$L$& $1.5$ mm\\
%plasma radius				&& \\
plasma entrance ramp profile	&			& half sinus\\
plasma entrance ramp length	&			& $4$ $\mu$m\\
\hline
number of cells	 in x			& $N_x$		& 75\\
number of cells	 in z			& $N_z$		& 860 ($\gamma=13$)-1691 ($\gamma=1$)\\
cell size in x				& $\delta x$	& $0.65\mu m$\\
cell size in z				& $\delta z$	& $\lambda/32$\\
time step					& $\delta t$	& at CFL limit\\
particle deposition order		&			& cubic\\
\# of plasma particles/cell		&			& 1 macro-e$^-$+1 macro-p$^+$\\
\hline
\hline
\end{tabular}
\end{center}
\label{Tablephyspar2d}
\end{table}%

\begin{figure}[htb]
    \centering
    \includegraphics*[width=120mm]{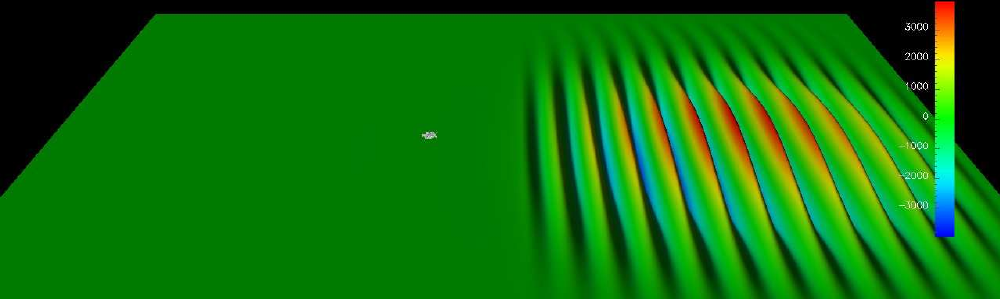} 
    \includegraphics*[width=120mm]{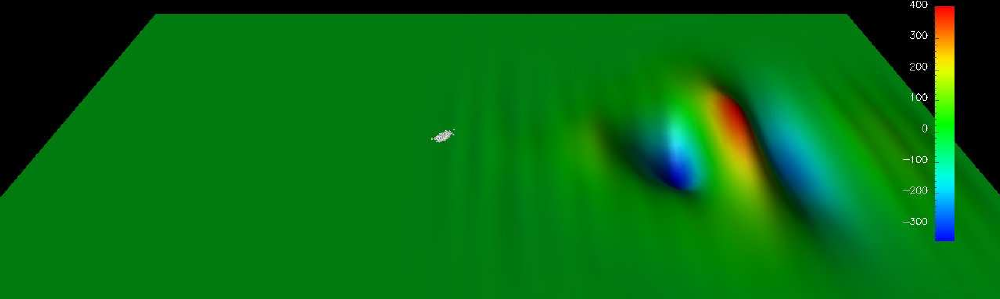}
    \caption{Colored surface rendering of the transverse electric field from a 2-1/2D Warp simulation of a laser wakefield acceleration stage in the laboratory frame (top) and a boosted frame at $\gamma=13$ (bottom),  with the beam (white) in its early phase of acceleration. The laser and the beam are propagating from left to right.}
    \label{Fig_surf2dey}
\end{figure}

\begin{figure}[htb]
    \centering
    \includegraphics*[width=120mm]{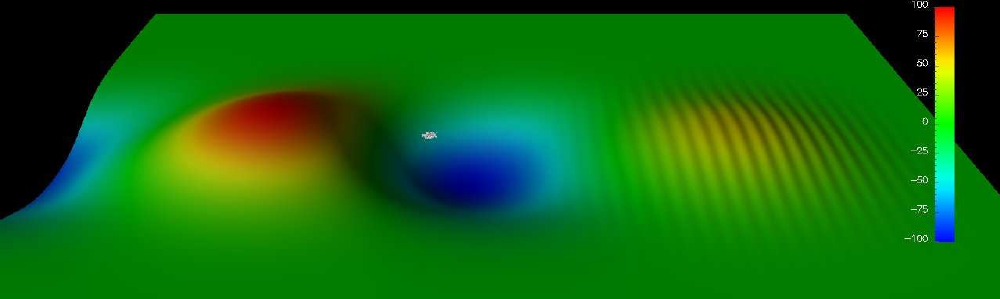} 
    \includegraphics*[width=120mm]{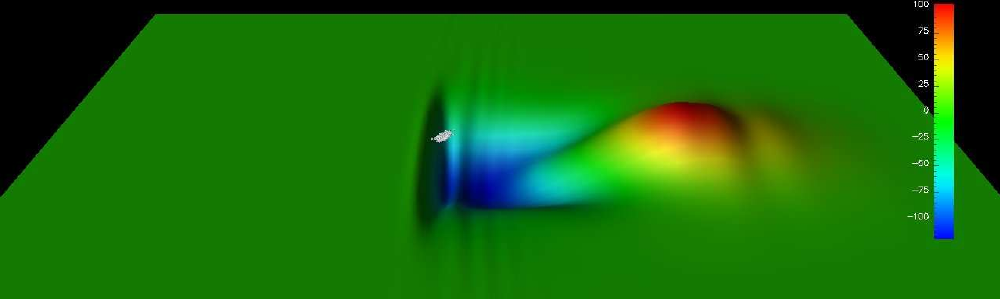}
    \caption{Colored surface rendering of the longitudinal electric field from a 2-1/2D Warp simulation of a laser wakefield acceleration stage in the laboratory frame (top) and a boosted frame at $\gamma=13$ (bottom),  with the beam (white) in its early phase of acceleration. The laser and the beam are propagating from left to right.}
    \label{Fig_surf2dez}
\end{figure}

\subsubsection{Using standard numerical techniques}
These runs were done using the standard Yee solver with no damping, and with the 4-pass stride-1 filter plus compensation, similarly to the simulations reported in \cite{CormierAAC08}. No signs of detrimental numerical instabilities were observed at the resolutions reported here with these settings. 

The approximate relativistic factor of the wake that is formed by the laser traveling in the plasma is given, according to linear theory, by $\gamma_w=2\pi c/\lambda \omega_p$ where $\omega_p=\sqrt{n_e e^2/\epsilon_0 m_e}$ is the electron plasma frequency. For the given parameters, $\gamma_w\approx 13.2$. Thus, Warp simulations were performed using reference frames moving between $\gamma=1$ (laboratory frame) and $13$.  
For a boosted frame associated with a value of $\gamma$ 
approaching $\gamma_w$ in the laboratory, the wake is expected to travel at low velocity in this boosted frame, and the physics to appear somewhat different from the one observed in the laboratory frame, in accordance to the properties of the Lorentz transformation. Figure \ref{Fig_surf2dey} and \ref{Fig_surf2dez} show surface renderings of the transverse and longitudinal electric fields respectively, as the beam enters its early stage of acceleration by the plasma wake, from a calculation in the laboratory frame and another in the frame at $\gamma=13$. The two snapshots offer strikingly different views of the same physical processes: in the laboratory frame, the wake is fully formed before the beam undergoes any significant acceleration and the imprint of the laser is clearly visible ahead of the wake; while in the boosted frame calculation, the beam is accelerated as the plasma wake develops, and the laser imprint is not visible on the snapshot. Close examination reveals that the short spatial variations which make the laser imprint in front of the wake are transformed into time variations in the boosted frame of $\gamma=13$. 

\begin{figure}[htb]
   {\small
   \begin{tabular}{@{}l@{}l@{}} % @{} removes extra space
     \hspace{3.5cm}2-1/2D & \hspace{3.5cm}3-D \vspace{3.mm}\\
     \includegraphics*[width=65mm]{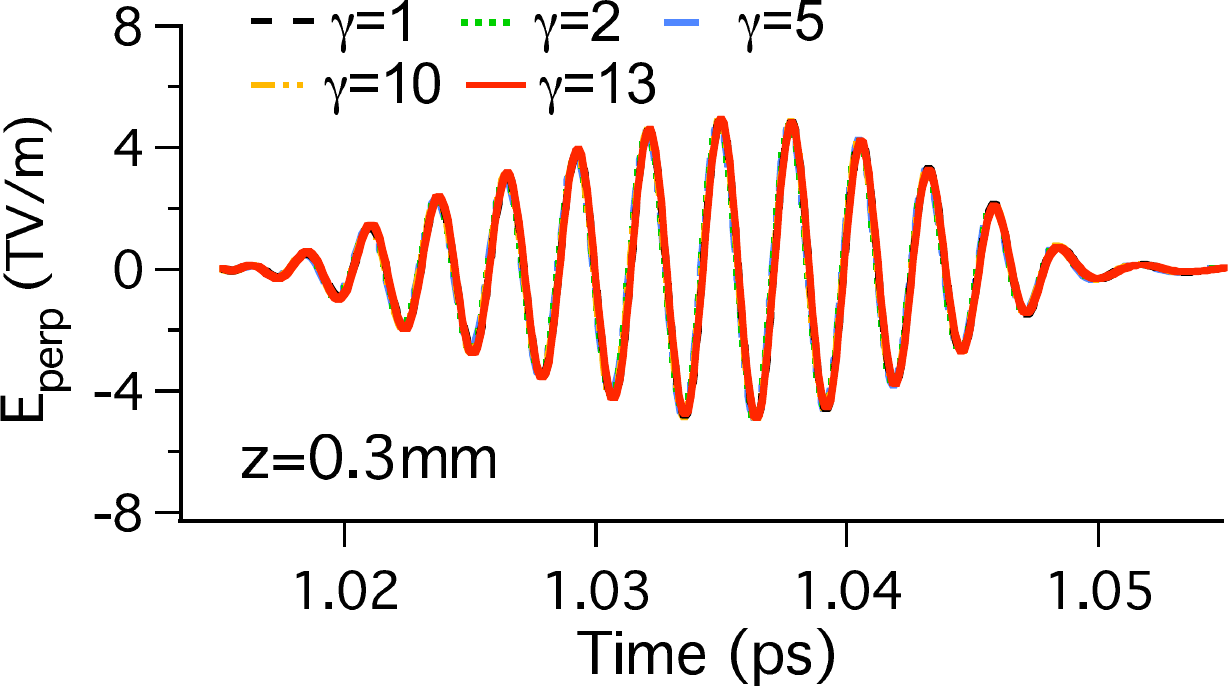}\hspace{5.mm}&
     \includegraphics*[width=65mm]{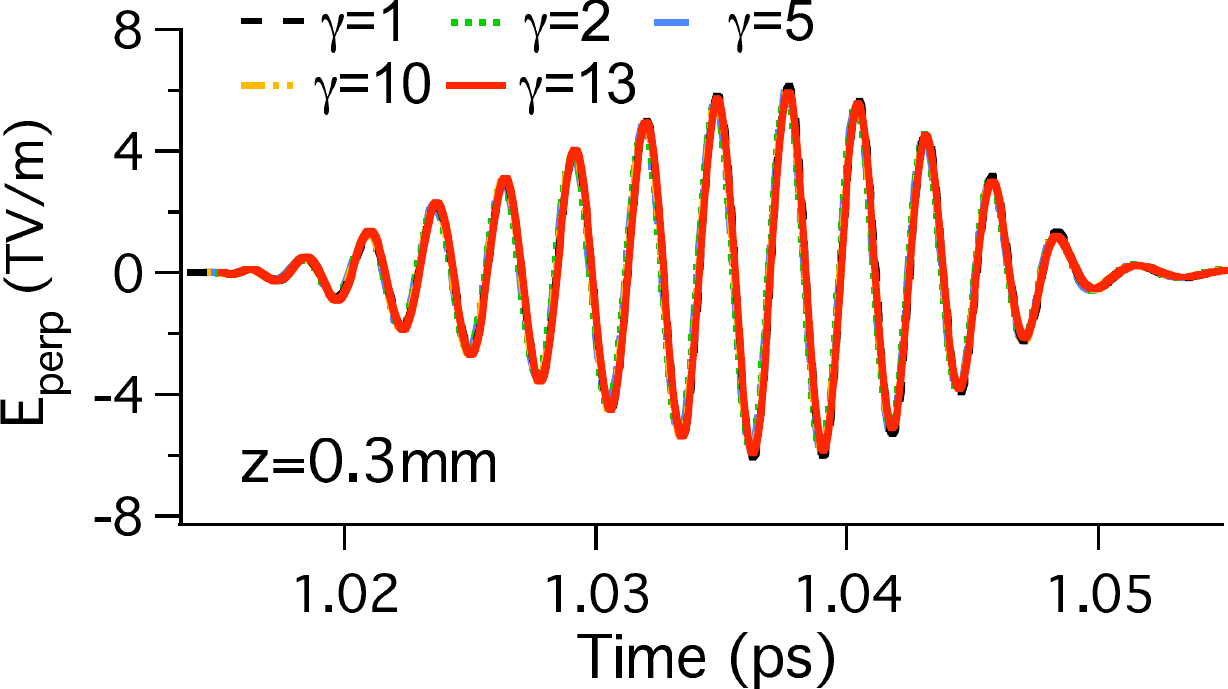} \\
     \includegraphics*[width=65mm]{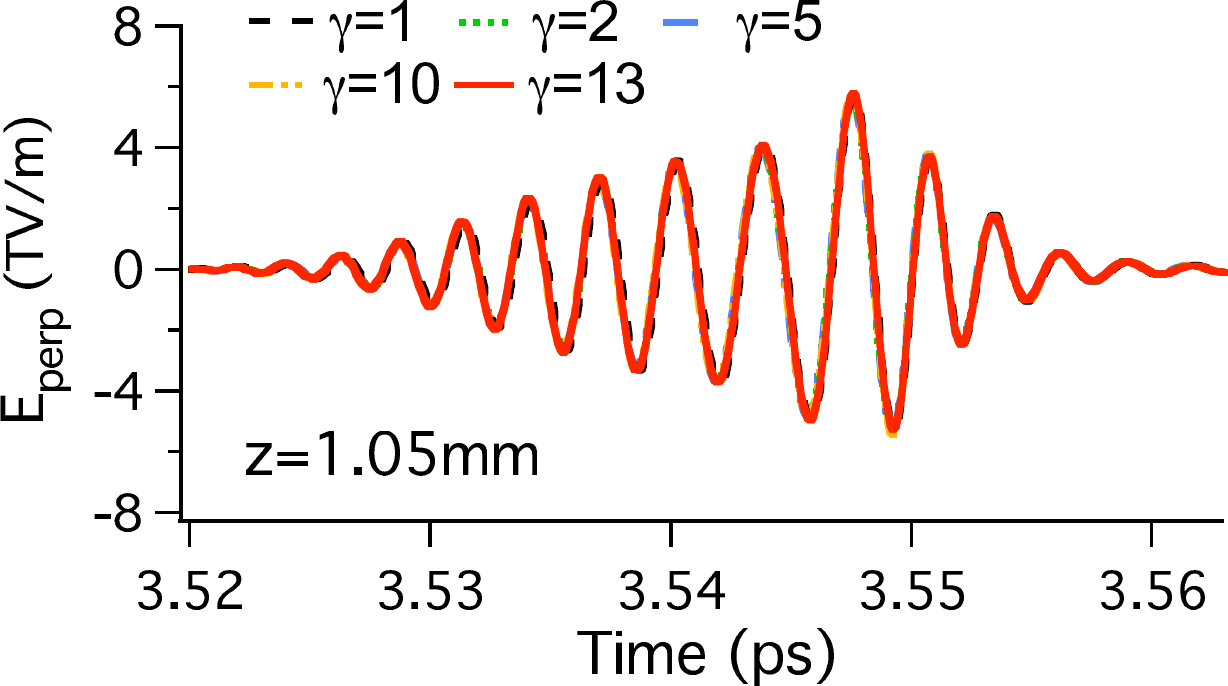}\hspace{5.mm} &
     \includegraphics*[width=65mm]{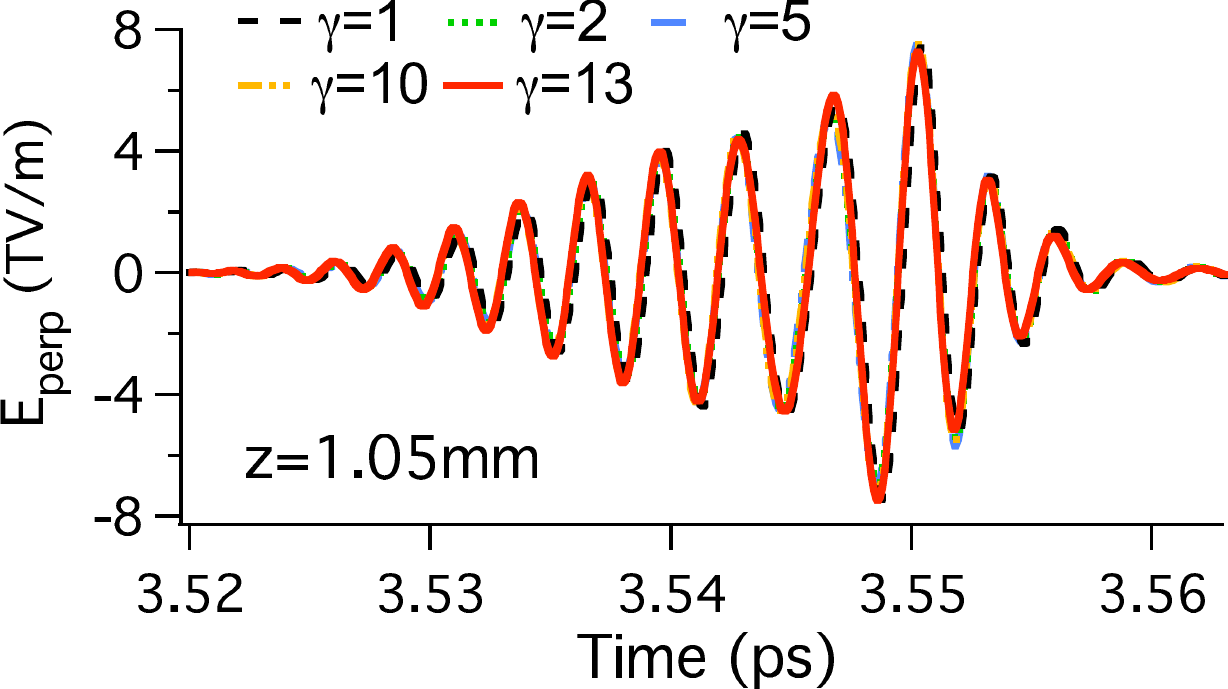} \\
   \end{tabular}}
   \caption{History of transverse electric field at the position $x=y=0$,  $z=0.3$ mm and $z=1.05$ mm (in the laboratory frame) from simulations in the laboratory frame ($\gamma=1$) and boosted frames at $\gamma=2$, $5$, $10$ and $13$.}
   \label{Fig_eystations2d}
\end{figure}

\begin{figure}[htb]
   {\small
   \begin{tabular}{@{}l@{}l@{}} % @{} removes extra space
     \hspace{3.5cm}2-1/2D & \hspace{3.5cm}3-D \vspace{3.mm}\\
     \includegraphics*[width=65mm]{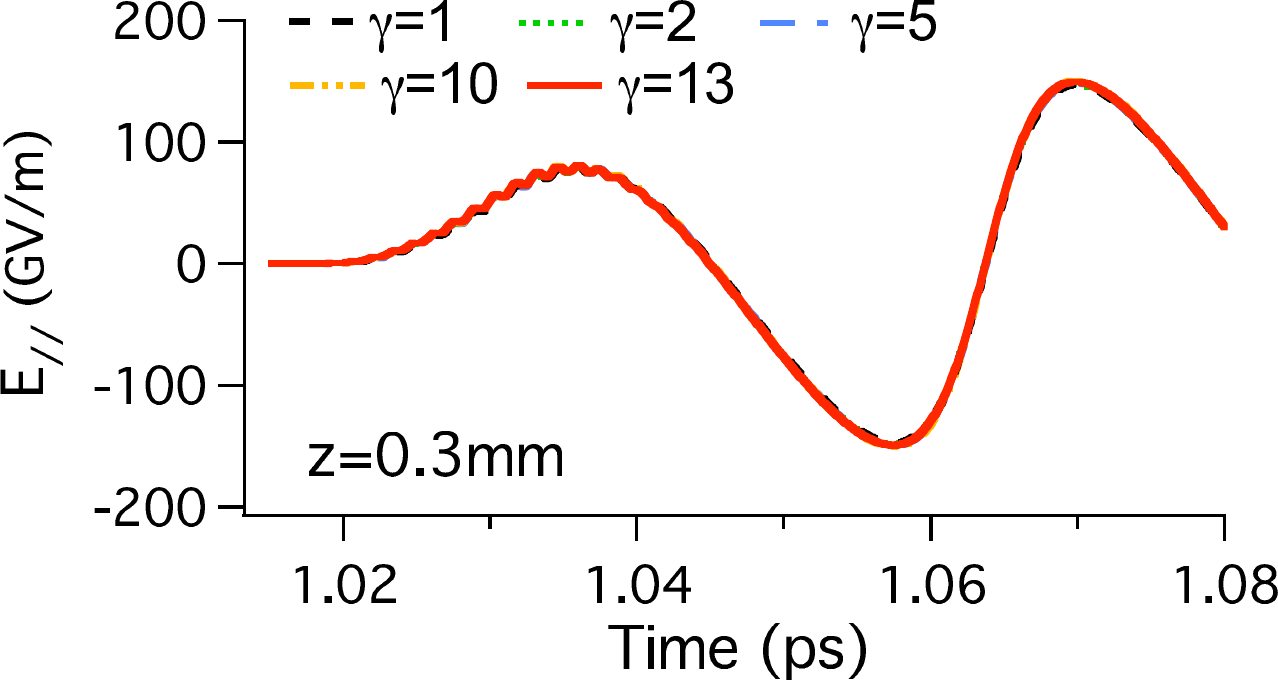}\hspace{5.mm}&
     \includegraphics*[width=65mm]{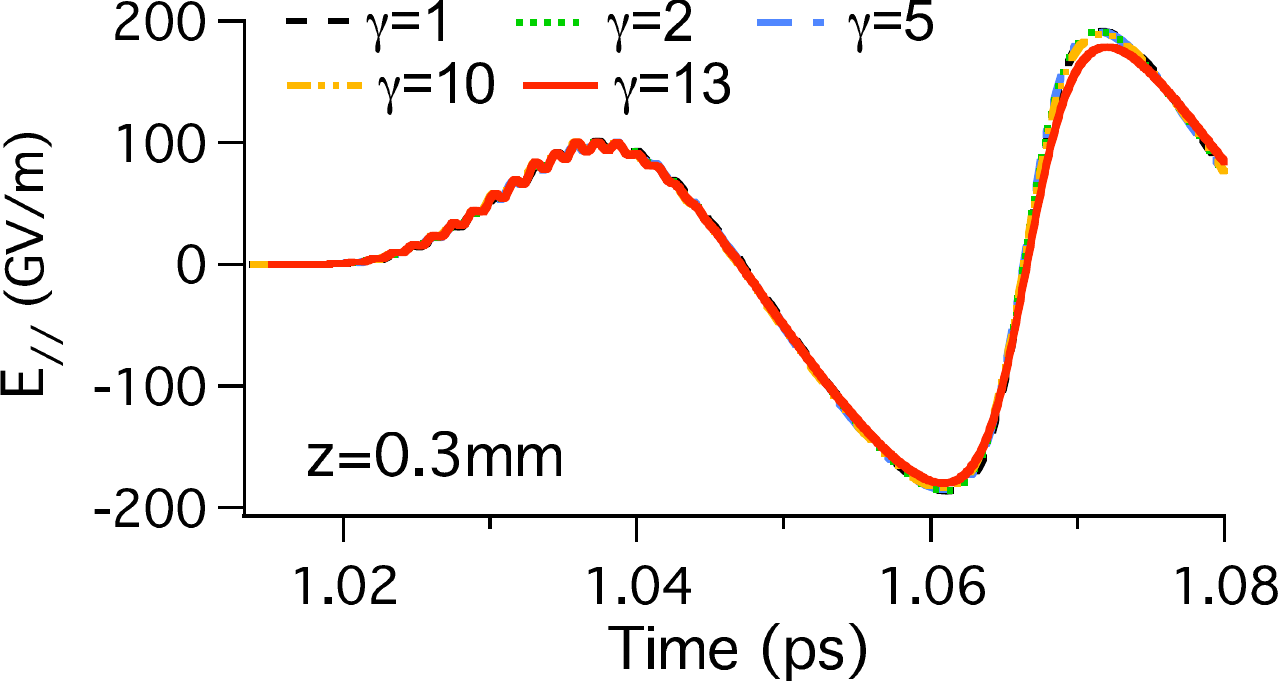} \\
     \includegraphics*[width=65mm]{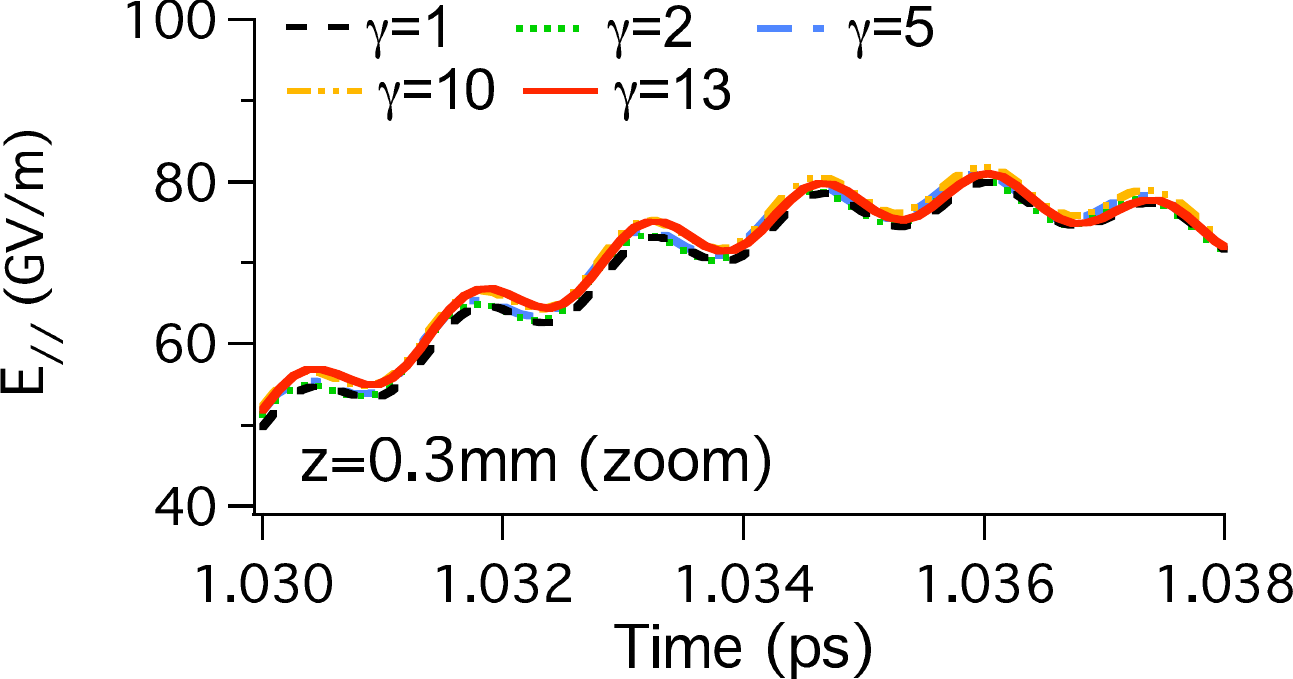}\hspace{5.mm} &
     \includegraphics*[width=65mm]{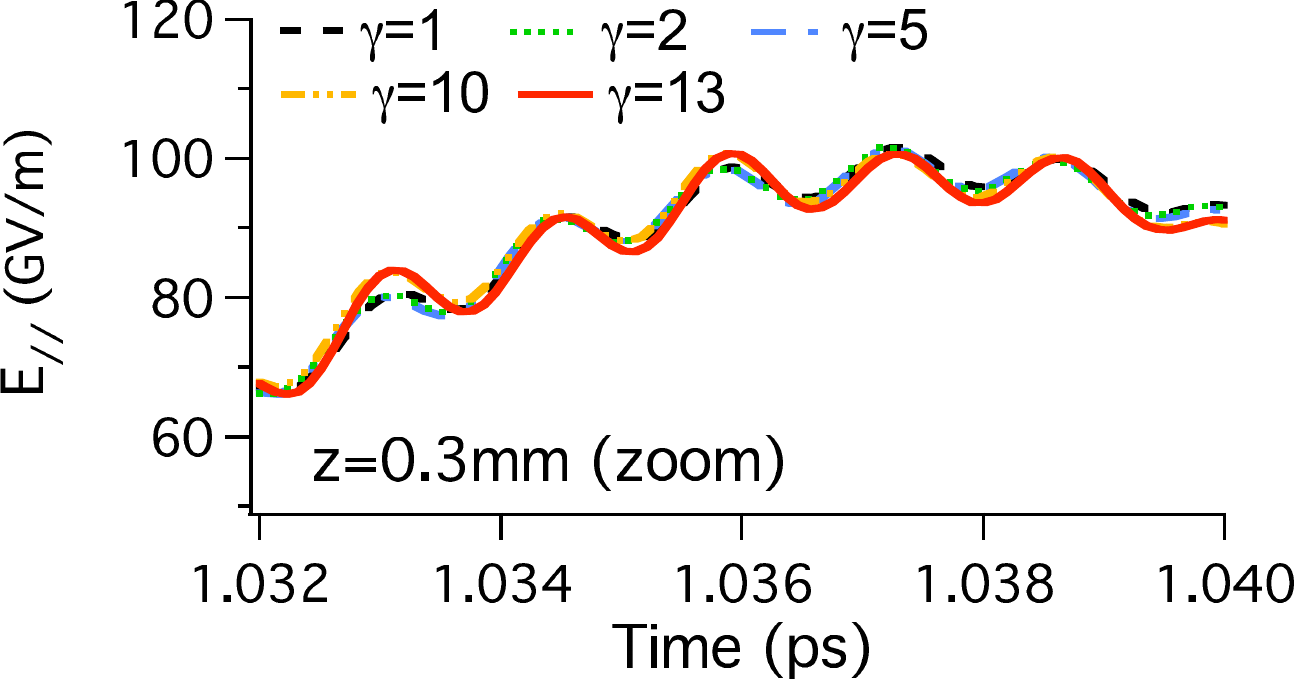}\\
     \includegraphics*[width=65mm]{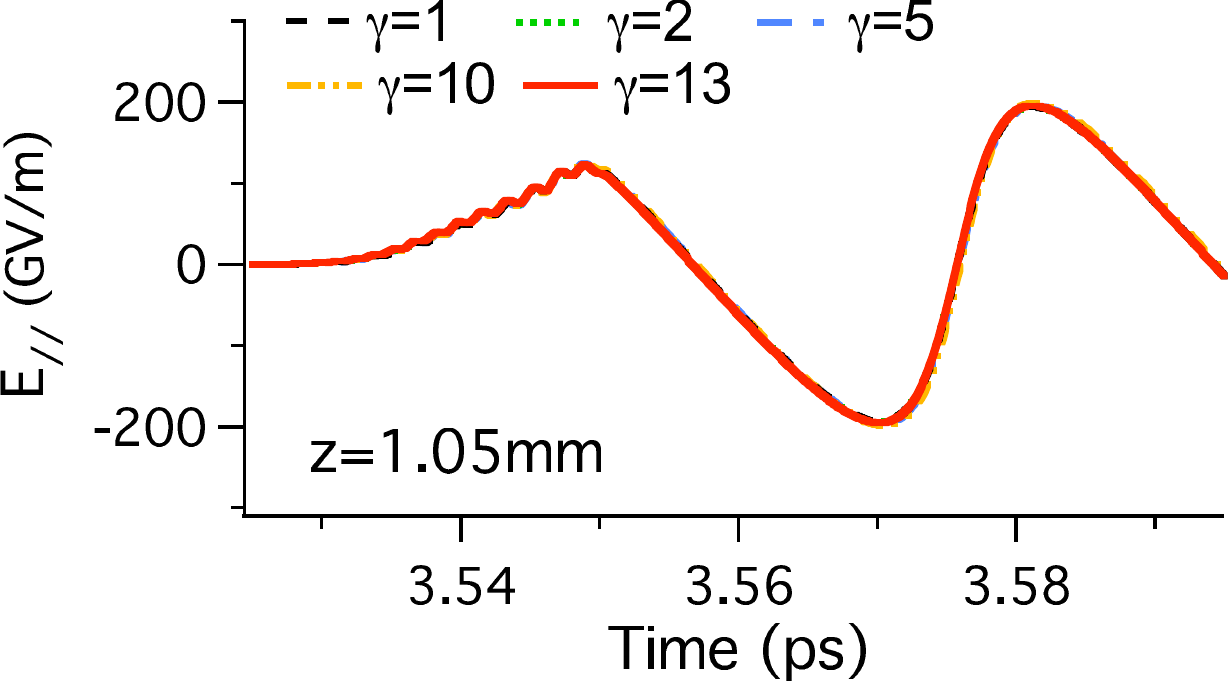}\hspace{5.mm} &
     \includegraphics*[width=65mm]{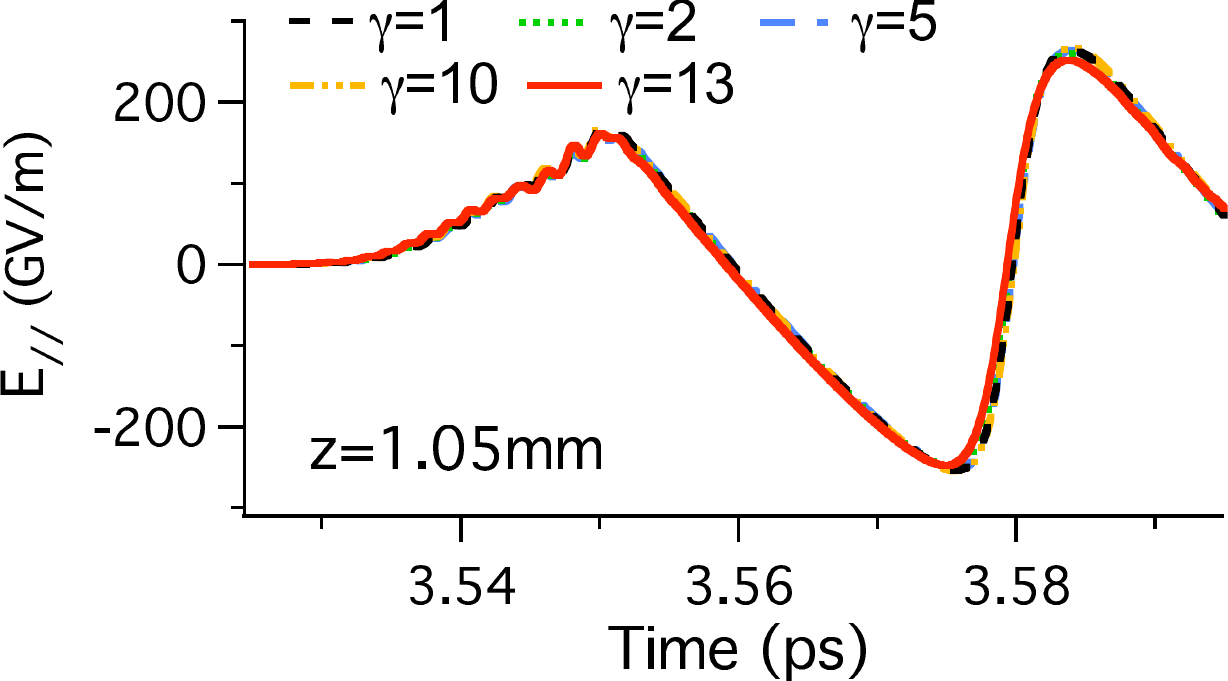} \\
   \end{tabular}}
   \caption{History of longitudinal electric field at the position $x=y=0$,  $z=0.3$ mm and $z=1.05$ mm (in the laboratory frame) from simulations in the laboratory frame ($\gamma=1$) and boosted frames at $\gamma=2$, $5$, $10$ and $13$.}
   \label{Fig_ezstations2d}
\end{figure}

\begin{figure}[htb]
   \centering
   {\small
   \begin{tabular}{@{}l@{}l@{}} % @{} removes extra space
   \hspace{3.5cm}2-1/2D & \hspace{3.5cm}3-D \vspace{3.mm}\\
   \includegraphics*[width=65mm]{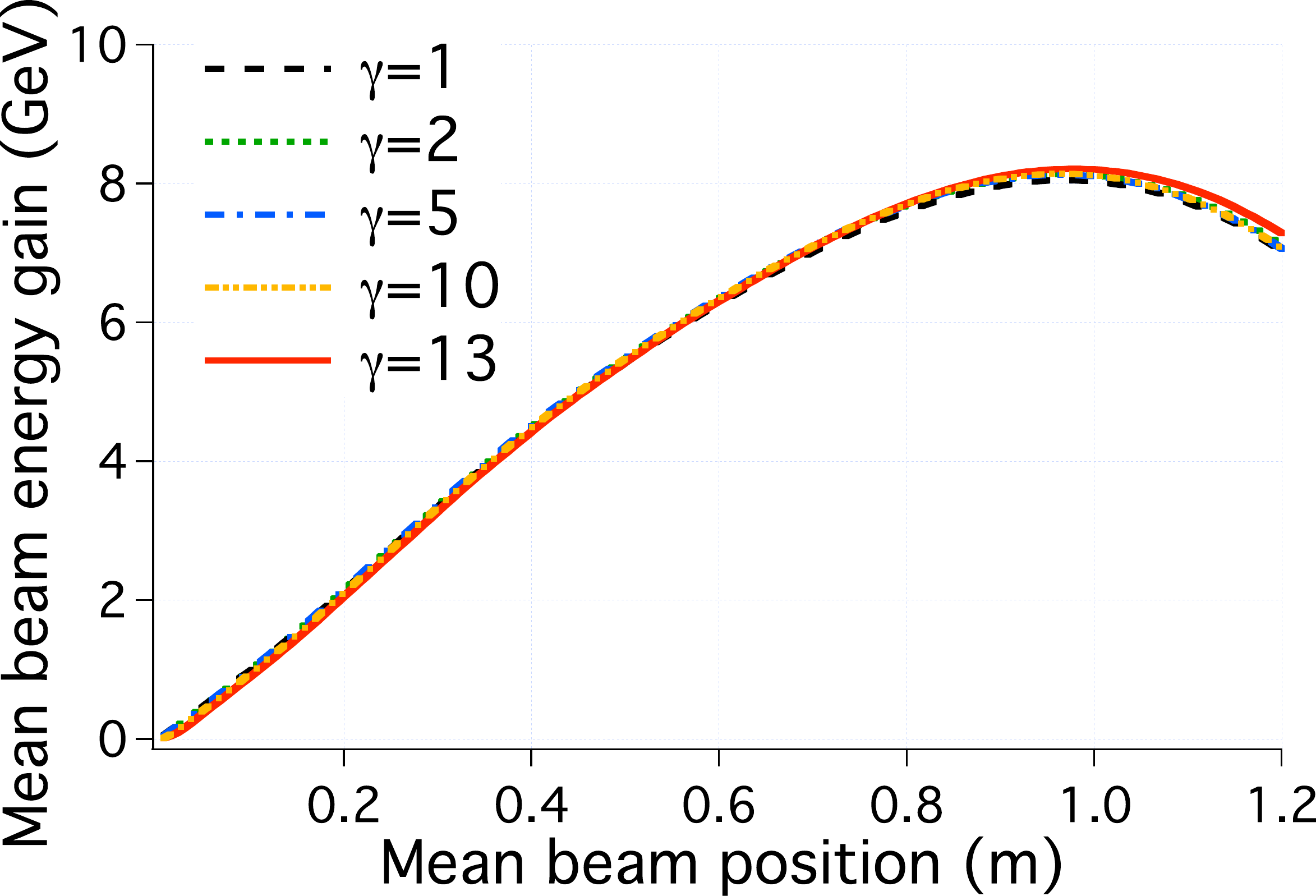}\hspace{5.mm}&
   \includegraphics*[width=65mm]{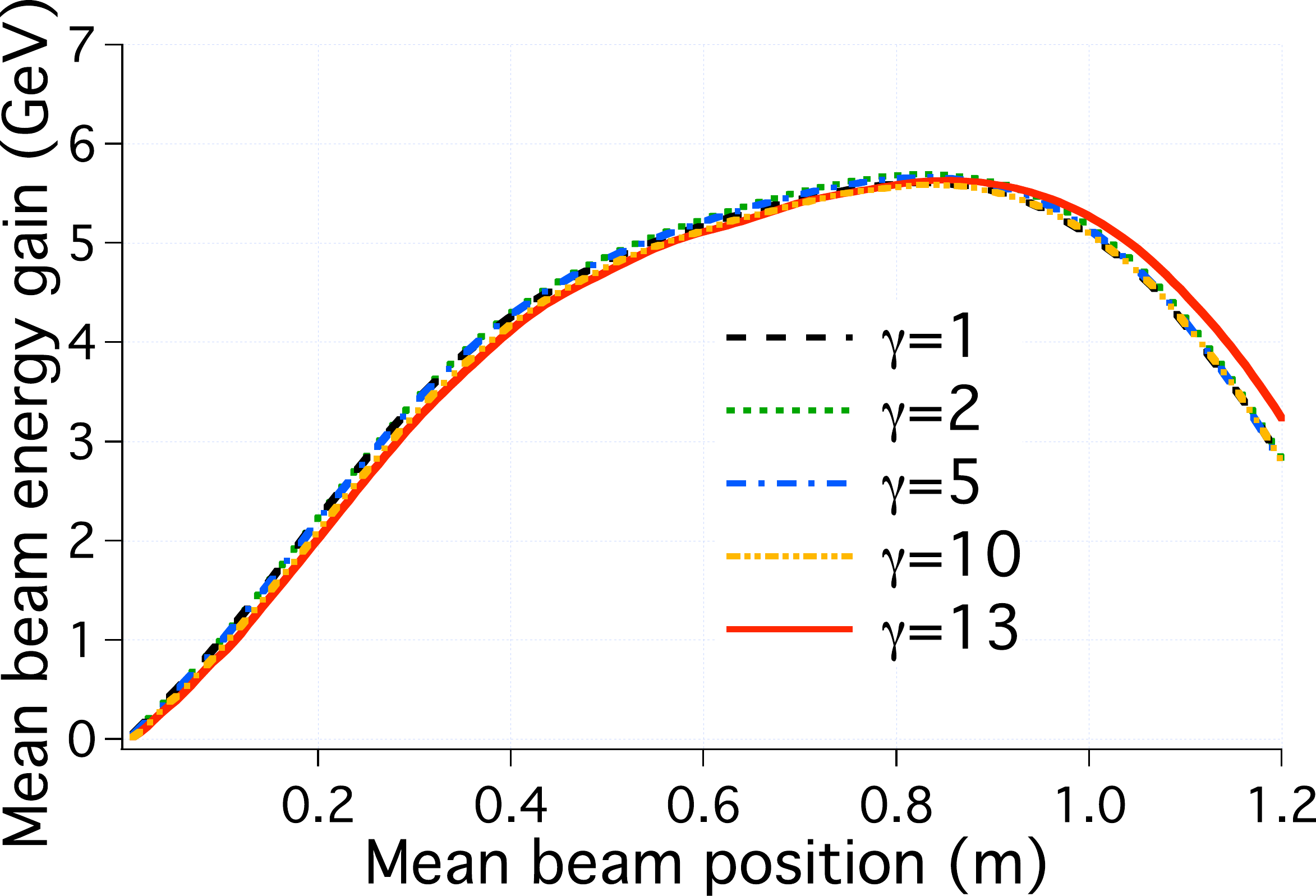}\\
   \includegraphics*[width=65mm]{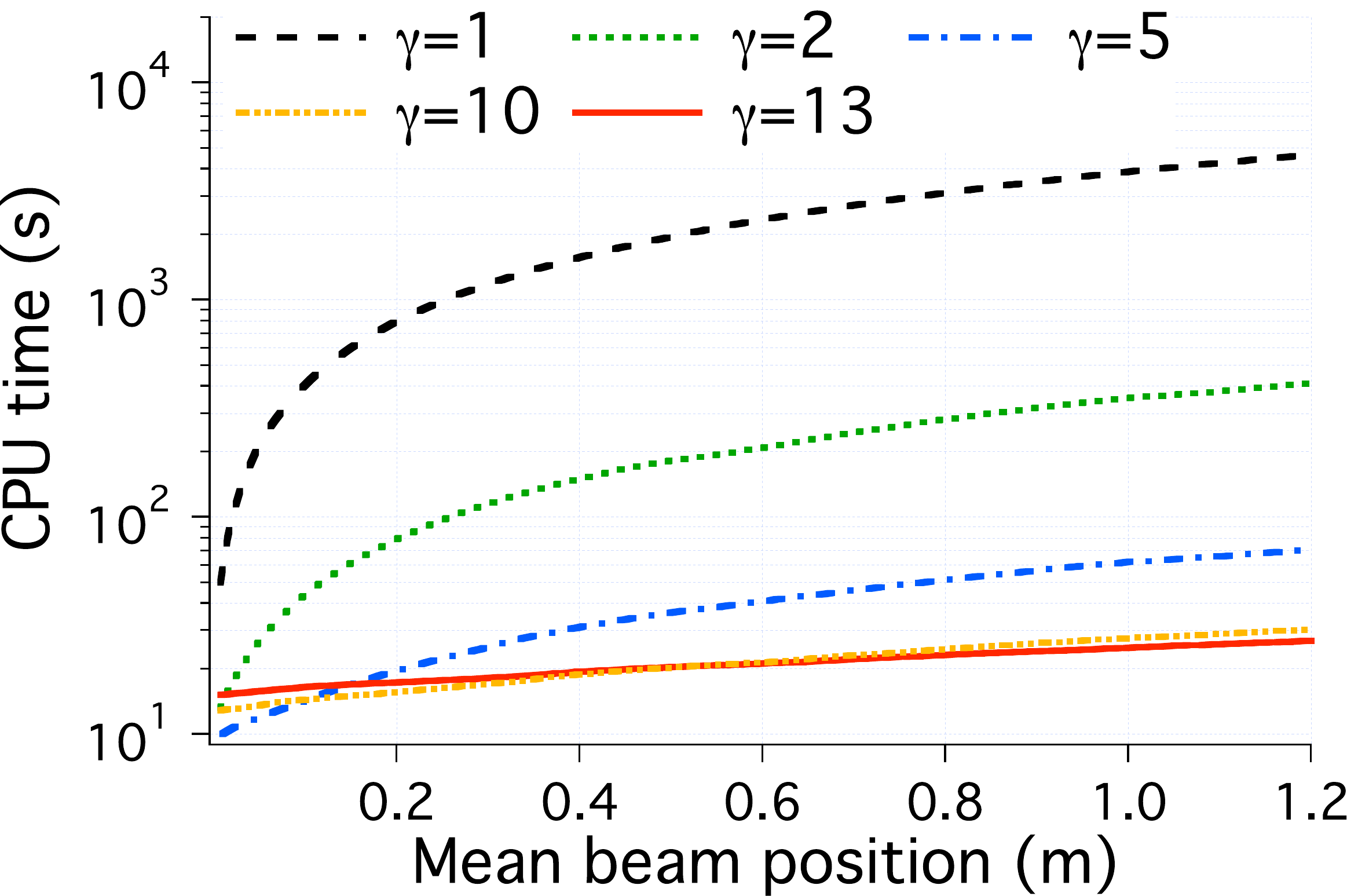}\hspace{5.mm}&
   \includegraphics*[width=65mm]{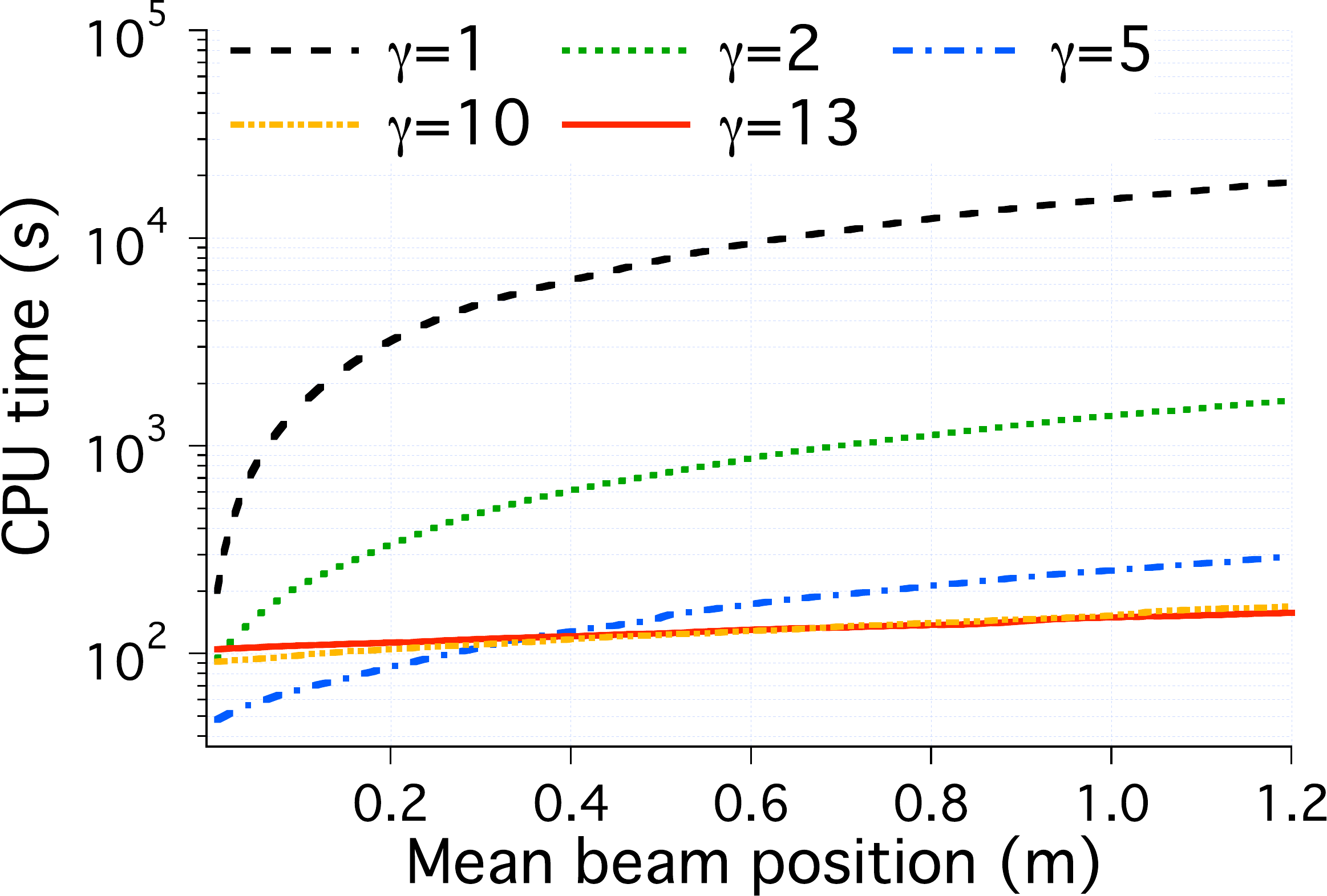}\\
   \includegraphics*[width=65mm]{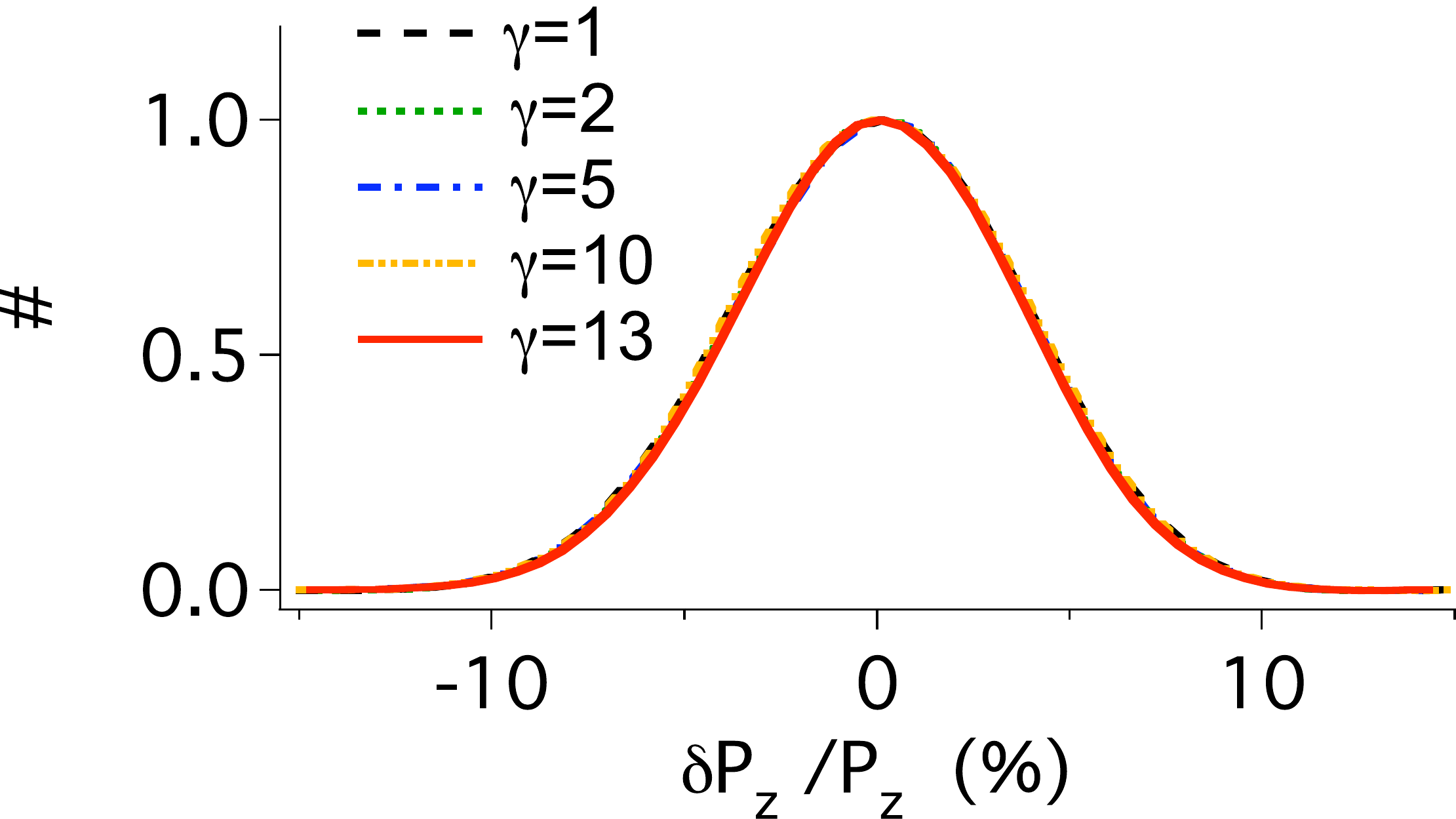}\hspace{5.mm}&
   \includegraphics*[width=65mm]{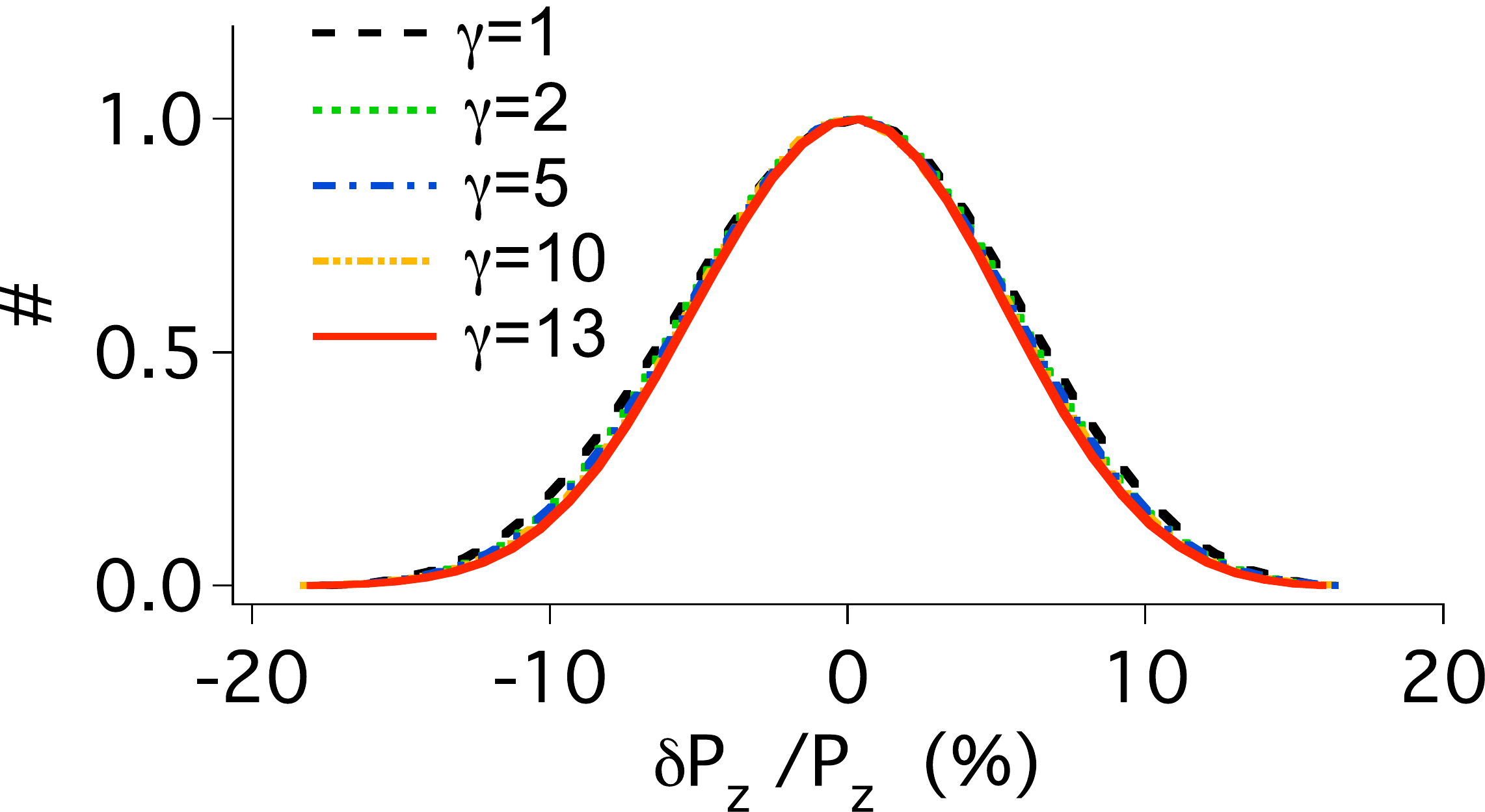}\\
   \end{tabular}}
   \caption{ (top) Average scaled beam energy gain and (middle) CPU time, versus longitudinal position in the laboratory frame from simulations; (bottom) distribution of relative longitudinal momentum dispersion at peak energy, in the laboratory frame ($\gamma=1$) and boosted frames at $\gamma=2$, $5$, $10$ and $13$.}
   \label{Fig_energytime2d}
\end{figure}

Histories of the perpendicular and longitudinal electric fields recorded at a number of stations at fixed locations in the laboratory offer direct comparison between the simulations in the laboratory frame ($\gamma=1$) and boosted frames at $\gamma=2$, $5$, $10$ and $13$. Figure \ref{Fig_eystations2d} and \ref{Fig_ezstations2d} show respectively the transverse and longitudinal electric fields collected at the positions $z=0.3$ mm and  $z=1.05$ mm (in the laboratory frame) on axis ($x=y=0$). The agreement is excellent and confirms that despite the apparent differences from snapshots taken from simulations in different reference frames, the same physics was recovered. This is further confirmed by the plot of the average scaled beam energy gain as a function of position in the laboratory frame, and of relative longitudinal momentum dispersion at peak energy (Fig. \ref{Fig_energytime2d}). The small differences observed on the mean beam energy histories and on the longitudinal momentum spread are attributed to a lack of convergence at the resolution that was chosen. The beam was launched with the same phase in the 2-1/2D and the 3D simulations, resulting in lower energy gain in 3D, due to proportionally larger laser depletion effects in 3D than in 2-1/2D. 

The CPU time recorded as a function of the average beam position in the laboratory frame  (Fig. \ref{Fig_energytime2d}-middle) indicates that the simulation in the frame of $\gamma=13$ took $\approx25$ s in 2-1/2D and $\approx150$ s in 3D versus $\approx5,000$ s in 2-1/2D and $\approx20,000$ s in 3D in the laboratory frame, demonstrating speedups of $\approx200$ in 2-1/2D and $\approx130$ in 3D, between calculations in a boosted frame at $\gamma=13$ and the laboratory frame.

All the simulations presented so far in this section were using the Yee solver, for which the Courant condition is given by $c\delta t < \left(1/\delta x^2+1/\delta z^2\right)^{-1/2}$ in 2D and $c\delta t < \left(1/\delta x^2+1/\delta y^2+1/\delta z^2\right)^{-1/2}$ in 3D where $\delta t$ is the time step and $\delta x$, $\delta y$ and $\delta z$ are the computational grid cell sizes in $x$, $y$ and $z$.  As $\gamma$ was varied, the transverse resolution was kept constant, while the longitudinal resolution was kept at a constant fraction of the incident laser wavelength $\delta z = \zeta\lambda$, such that in a boosted frame,  $\delta z^* = \zeta\lambda^* = \zeta \left(1+\beta\right)\gamma\lambda$. As a result, the speedup becomes, when using the Yee solver

\begin{equation}
S_{yee2D}=S \frac{\delta z\sqrt{1/\delta x^2+1/\delta z^2} } { \delta_z^*\sqrt{1/\delta x^2+1/\delta z^{*2} } }
\label{Eq_scaling2d}
\end{equation}

in 2D and 

\begin{equation}
S_{yee3D}=S \frac{\delta z\sqrt{1/\delta x^2+1/\delta y^2+1/\delta z^2}}{\delta_z^*\sqrt{1/\delta x^2+1/\delta y^2+1/\delta z^{*2}}}
\label{Eq_scaling3d}
\end{equation}

in 3D where $S$ is given by Eq. (\ref{Eq_scaling1d}). 

\begin{figure}[htb]
    \centering
    \includegraphics*[width=90mm, trim=0.in 0.6in 0.in 0.in]{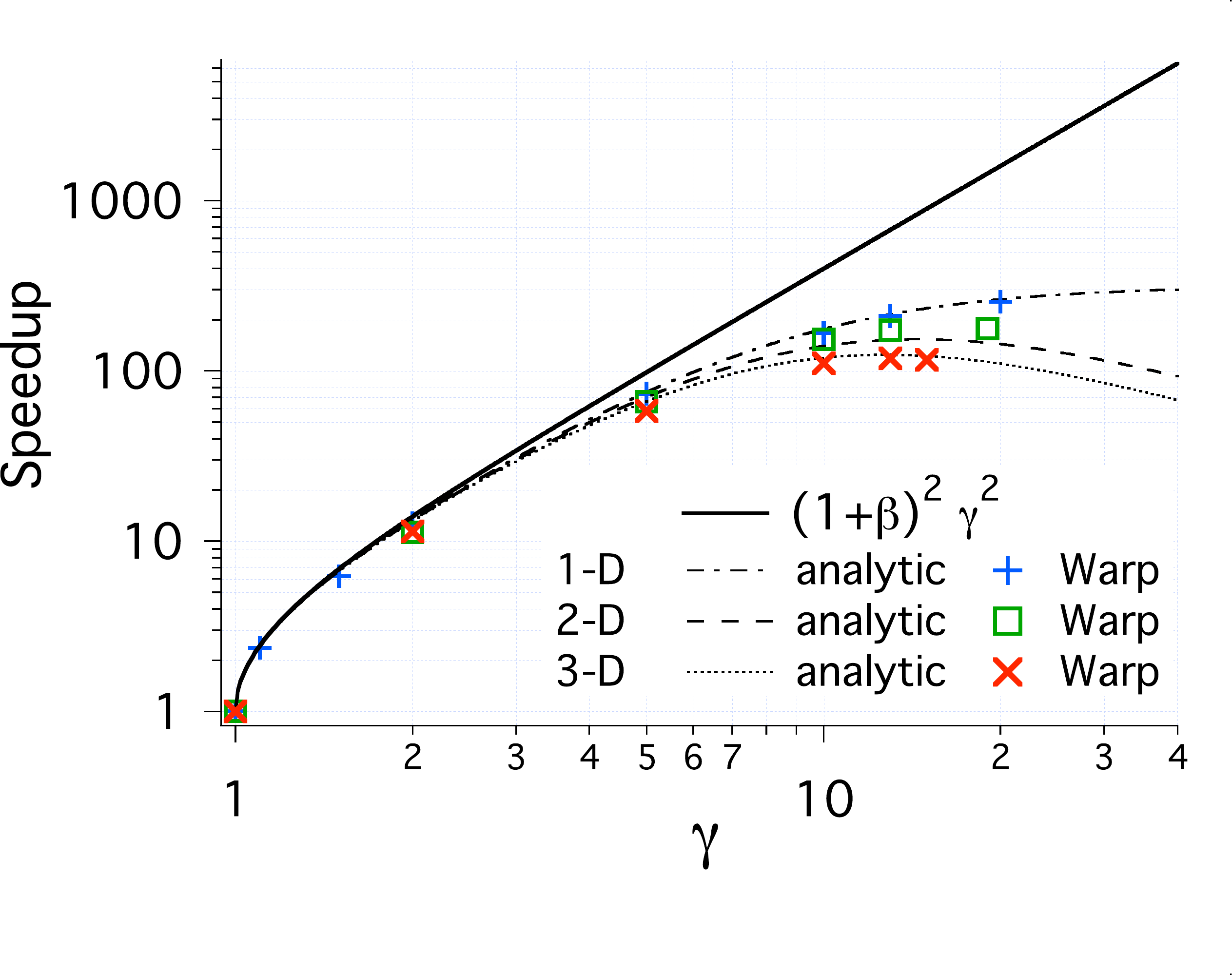}
    \caption{Speedup versus relativistic factor of the boosted frame from Eq. (\ref{Eq_scaling1d}), (\ref{Eq_scaling2d}), (\ref{Eq_scaling3d}), and Warp simulations.}
    \label{Fig_scaling123d}
\end{figure}

The speedup versus relativistic factor of the reference frame is plotted in Fig. \ref{Fig_scaling123d}, from (\ref{Eq_scaling1d}), (\ref{Eq_scaling2d}) and (\ref{Eq_scaling3d}), and contrasted with measured speedups from 1D, 2-1/2D and 3D Warp simulations, confirming the scaling obtained analytically.

\clearpage
\subsubsection{Effect of filtering, solver with adjustable dispersion and damping}
\begin{figure}[htb]
   \centering
   \includegraphics*[width=65mm]{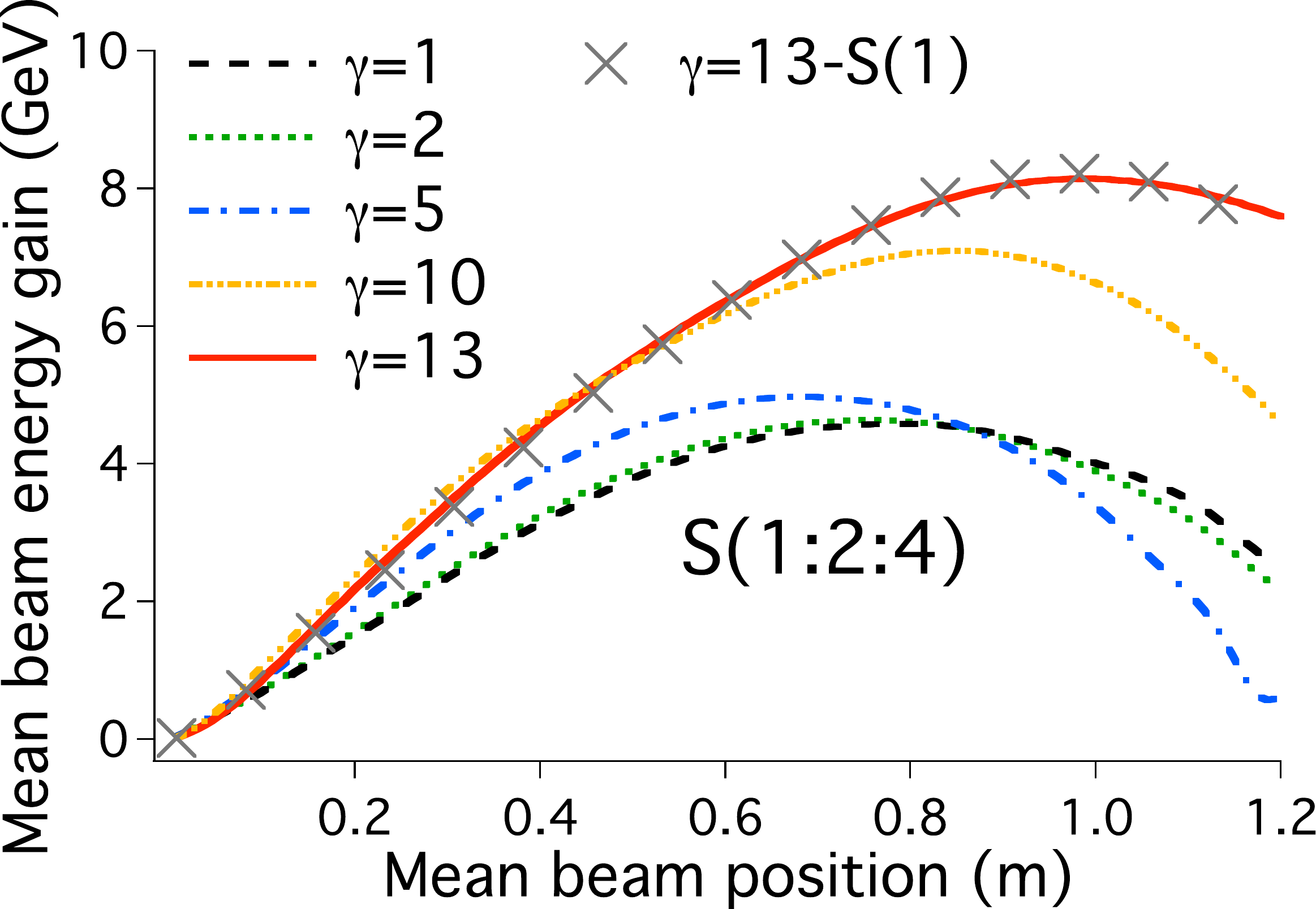}
   \includegraphics*[width=65mm]{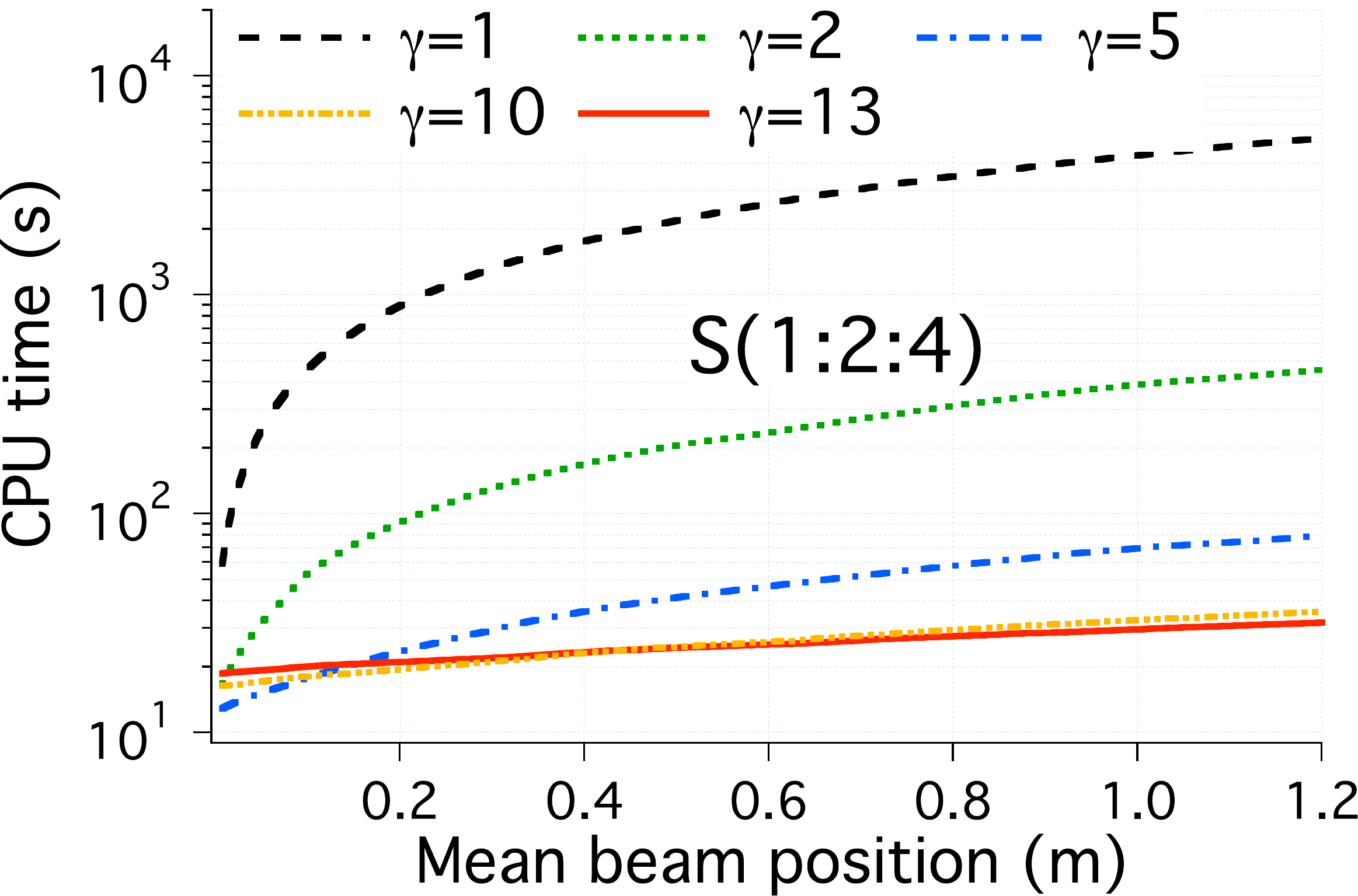}
   \caption{( (left) Average scaled beam energy gain and (right) CPU time, versus longitudinal position in the laboratory frame from simulations in the laboratory frame ($\gamma=1$) and boosted frames at $\gamma=2$, $5$, $10$ and $13$, using the Yee solver with digital filter S(1:2:4) (grey cross is reference from run with filter S(1)).}
   \label{Fig_energytime2dFilter}
\end{figure}

\begin{figure}[htb]
   \centering
   \includegraphics*[width=65mm]{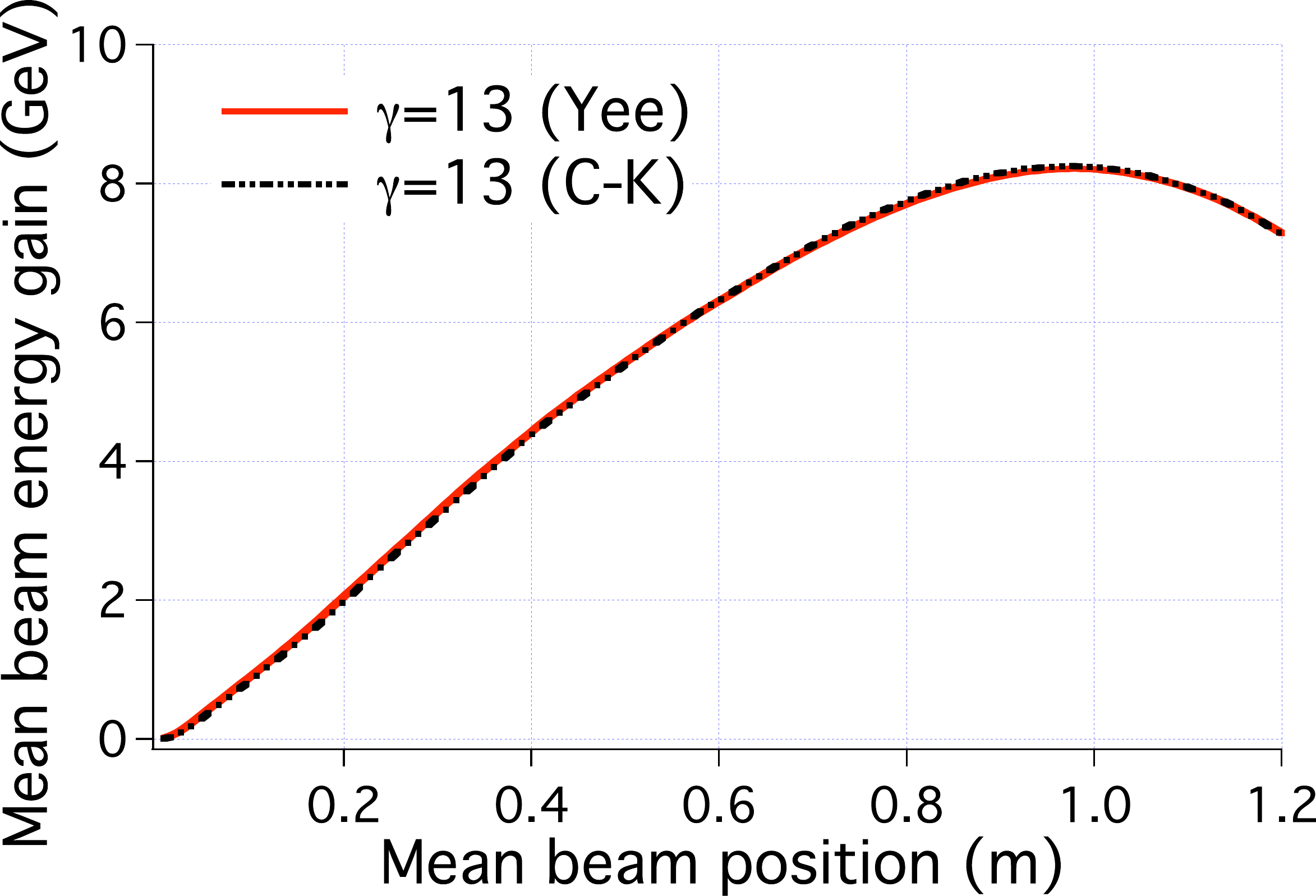}
   \includegraphics*[width=65mm]{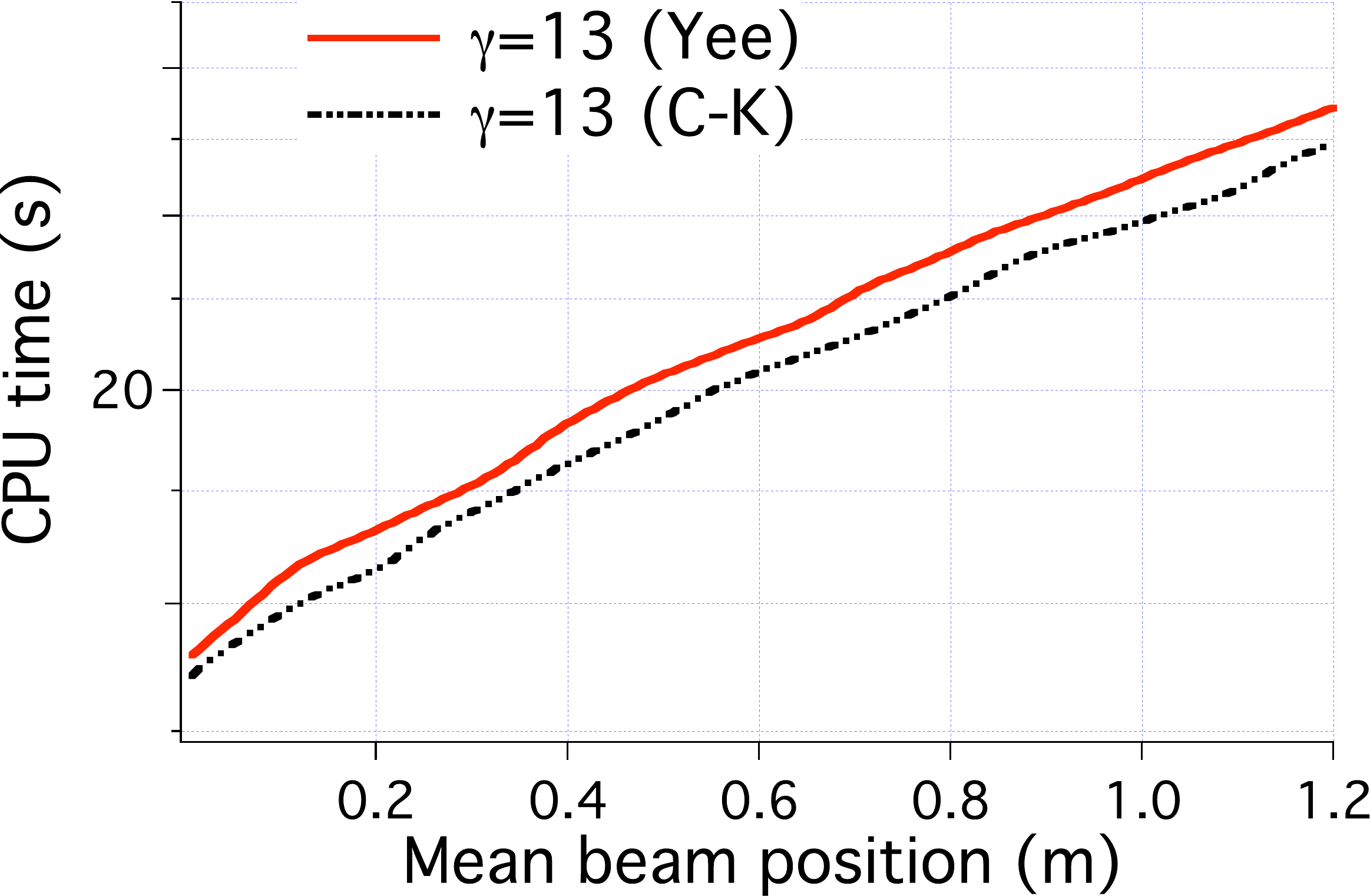}
   \caption{( (left) Average scaled beam energy gain and (right) CPU time, versus longitudinal position in the laboratory frame from simulations in the laboratory frame ($\gamma=1$) and boosted frames at $\gamma=2$, $5$, $10$ and $13$, using the Cole-Karkkainen solver with filter S(1) (red curve is reference from calculation with Yee solver and filter S(1)).}
   \label{Fig_energytime2dCK}
\end{figure}

\begin{figure}[htb]
   \centering
   \includegraphics*[width=65mm]{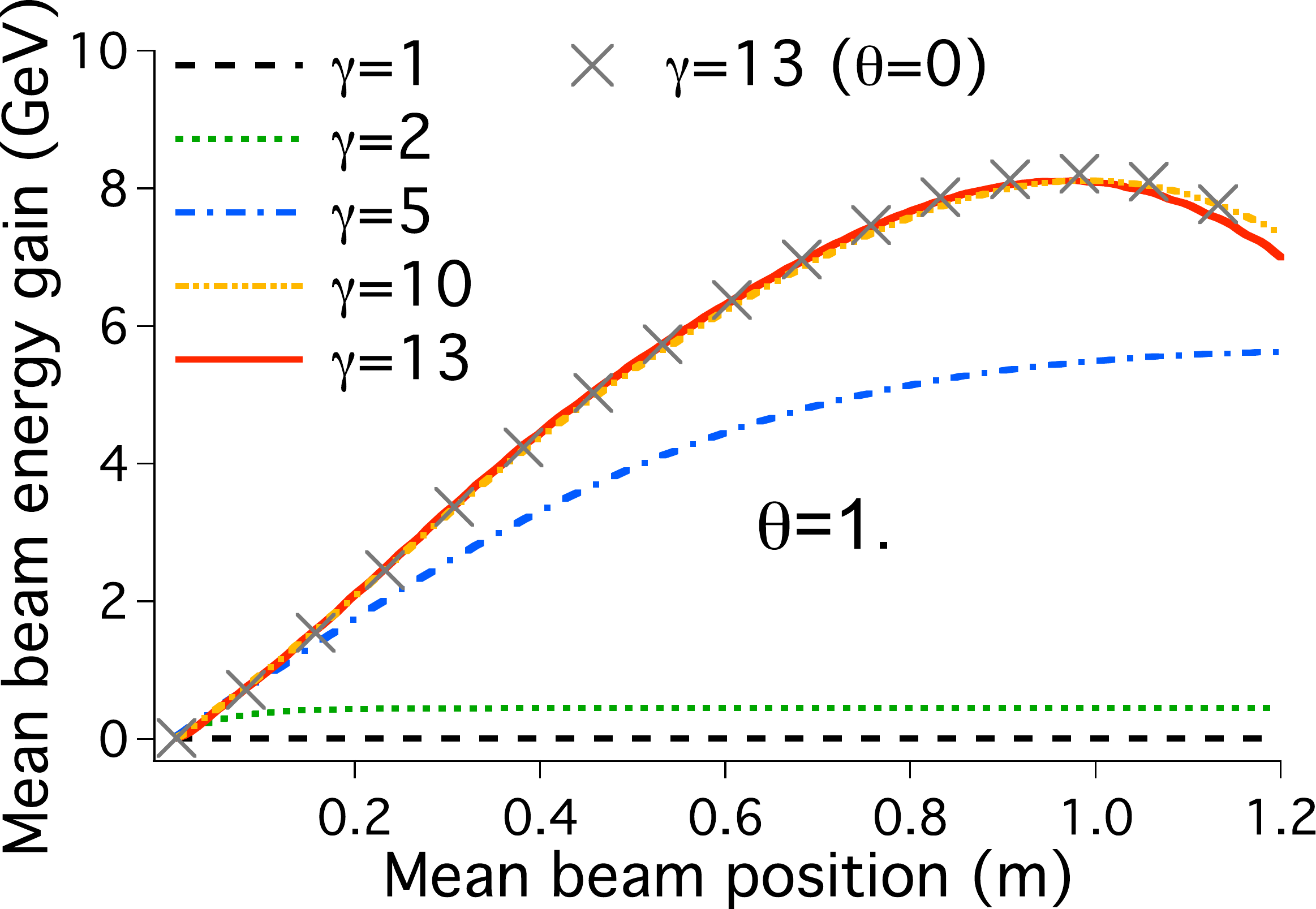}
   \includegraphics*[width=65mm]{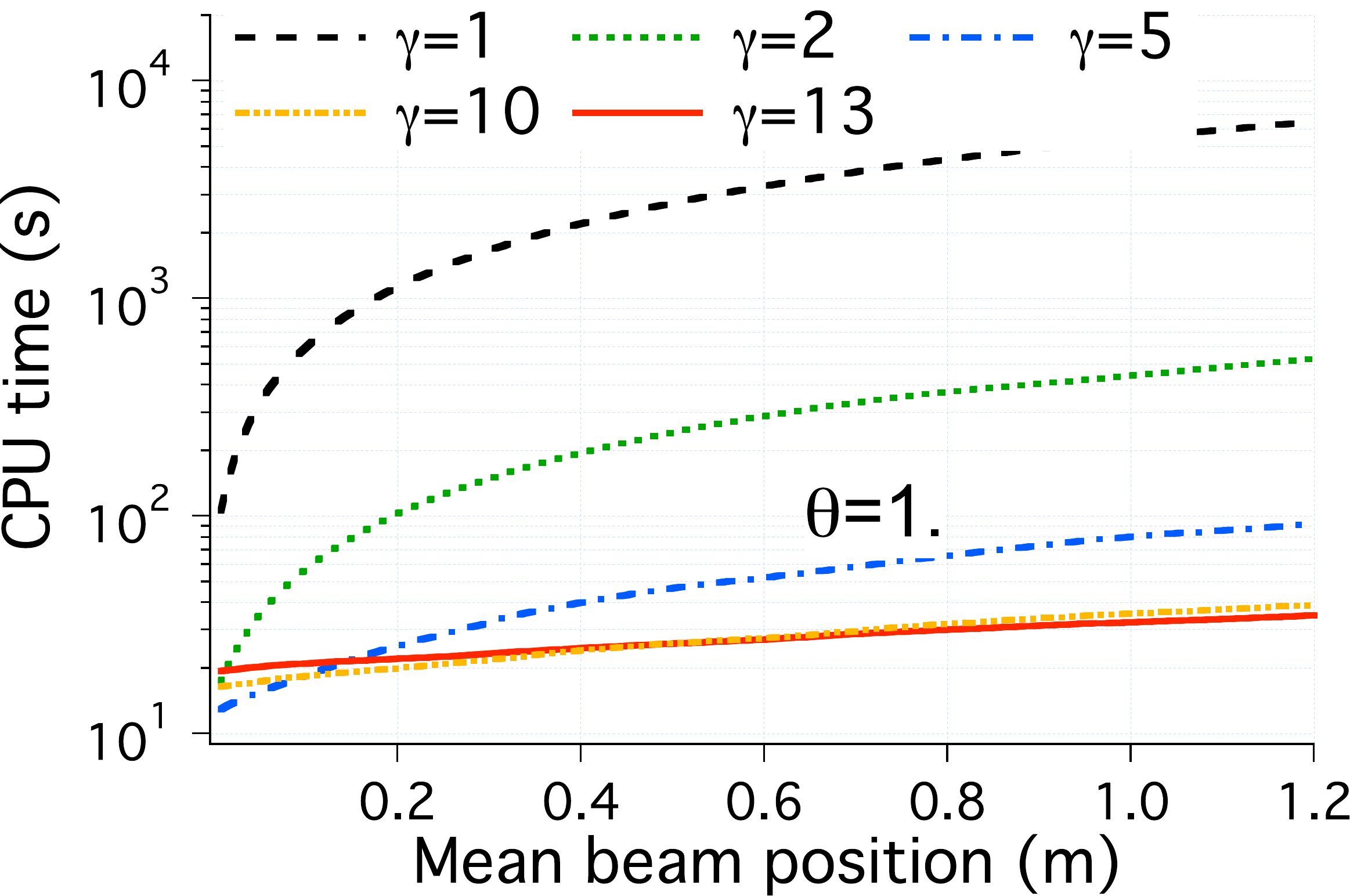}
   \caption{( (left) Average scaled beam energy gain and (right) CPU time, versus longitudinal position in the laboratory frame from simulations in the laboratory frame ($\gamma=1$) and boosted frames at $\gamma=2$, $5$, $10$ and $13$, using the Yee-Friedman solver with $\theta=1$ (grey cross is reference from run with no damping).}
   \label{Fig_energytime2dYF}
\end{figure}

The modeling of full scale stages, which allow for higher values of $\gamma$ for the reference frame, is more prone to the high frequency instability that was mentioned in a previous section, as we will show below. In anticipation of the application of the method presented above to mitigate the instability, simulations of the scaled stage were conducted using the Yee solver with digital filter S(1:2:4) as described above (Fig. \ref{Fig_energytime2dFilter}), the Cole-Karkkainen solver (Fig. \ref{Fig_energytime2dCK}) or the Yee-Friedman solver (Fig. \ref{Fig_energytime2dYF}). 
%The beam average energy and CPU timing histories are given in Fig. \ref{Fig_energytime2dFSCK}.
 %while the longitudinal electric field histories are given for the cases with filtering S(1:2:4) and damping in Fig. \ref{Fig_ezstations2dFSCK}. 

Smoothing with the wideband filter S(1:2:4) did not produce significant degradations for the calculation in the wake frame ($\gamma=13$) but did otherwise. The calculations with the Yee solver and the Cole-Karkkainen solver gave identical results, validating our implementation of the CK solver. Despite the more expensive stencil, the run with the CK solver was almost $40\%$ faster, due to a time step larger by $\sqrt{2}$.  Similarly to filtering, damping aggressively did not degrade the result in the range $10\leq\gamma\leq13$ but did significantly in the range $1\leq\gamma\leq5$. Comparing the timings with those of Fig. \ref{Fig_energytime2d} (middle-left) shows that the smoothing and the damping added less than a factor of two of total runtime to the simulations. 


%\begin{figure}[htb]
 %  \centering
 %  \includegraphics*[width=60mm]{ezstationsm1th1p0.pdf}
%   \includegraphics*[width=60mm]{ezstationsm1th1p0zoom.pdf}
%   \includegraphics*[width=60mm]{ezstationsm4.pdf}
%   \includegraphics*[width=60mm]{ezstationsm4zoom.pdf}
%   \caption{History of perpendicular electric field at the position $x=y=0$,  $z=0.3mm$ (in the laboratory frame) from simulations in the laboratory frame ($\gamma=1$) and boosted frames at $\gamma=2$, $5$, $10$ and $13$, using (top) the Yee-Friedman solver with $\theta=1$; (middle) the Yee solver with digital filter S(1:2:4).}
 %  \label{Fig_ezstations2dFSCK}
%\end{figure}

Those results lead to several observations: (i) while the grid dimensions  and number of cells were chosen such that square cells were obtained for $\gamma=13$, meaning a larger dispersion in the longitudinal direction with the Yee solver than with the Cole-Karkkainen solver, both gave the same result. This is significant since for simulations of LPA in the laboratory frame reported in the literature, the need to have nearly perfect numerical dispersion in the longitudinal direction imposes a constraint on the cell aspect ratio and thus on resolution \cite{TsungPoP2006,CowanPriv2010}. This constraint is removed when simulating in the frame of the wake ($\gamma=13\approx\gamma_w$); (ii) damping of high frequencies with the Yee-Friedman solver or wideband smoothing of short wavelength have a negligible effect on accuracy for simulations in the frame of the wake, but degrade the accuracy very significantly for slower moving reference frames. The dependency of the effect of damping and smoothing with $\gamma$ boost has two causes. First, simulations with a boost $\gamma\approx\gamma_w$ require fewer time steps than simulations using a lower value of $\gamma$. Thus, for a given value of the damping coefficient $\theta$, the integrated amount of damping will be lower for the simulations with $\gamma\approx\gamma_w$. Second, as mentioned above in the discussion of the surface renderings shown in Fig. \ref{Fig_surf2dey} and \ref{Fig_surf2dez}, a large fraction of the short wavelength content that is present in the simulations in the laboratory frame is transformed into time oscillations in simulations in the wake frame. Hence, filtering short wavelength has less effect on the physics when calculating in the wake frame than when calculating in the laboratory frame; (iii) the cost of using even the most aggressive damping or smoothing is low, especially considering that the simulations presented here were using only two plasma macro-particles per cell. 

In summary, calculating in a boosted frame near the frame following the wake ($\gamma\approx\gamma_w$) relaxes the constraint on the numerical dispersion in the direction of propagation of the laser (which is essential in simulations in the laboratory frame), and allows for more aggressive damping of high frequencies and smoothing of short wavelengths than is possible in standard laboratory frame calculations.

%\clearpage
\subsection{Full scale 10 GeV class stages}
As noted in \cite{BruhwilerAAC08}, full scale simulations using the laboratory frame of 10 GeV stages at plasma densities of  $10^{17}$ cm$^{-3}$ are not practical on present computers in 2D and 3D. At this density, the wake relativistic factor $\gamma_w\approx 132$, and 2-1/2D and 3D simulations were done in boosted frames up to $\gamma=130$. 

\begin{figure}[htb]
   \centering
 {\small
 \begin{tabular}{@{}c@{}c@{}} % @{} removes extra space
% \begin{tabular}{|@{}c@{}|@{}c@{}|} % @{} removes extra space
%  \hline
%  Yee & Cole-Karkkainen \\
%  \hline
%    \hspace{1.cm} Cole-Karkkainen-Friedman & \hspace{1.cm} Yee-Friedman  \vspace{2.mm}\\
    \includegraphics*[width=65mm]{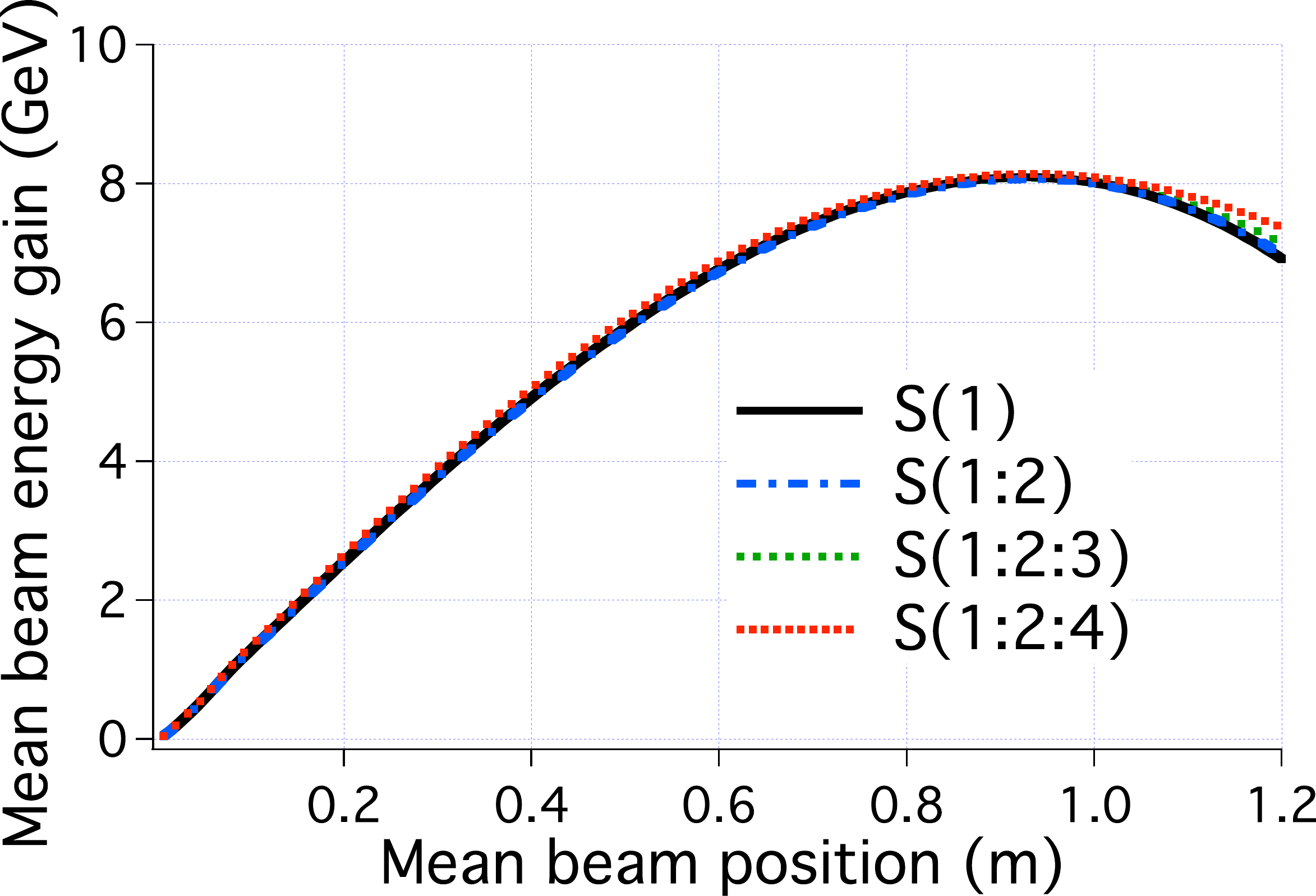} &
    \hspace{1mm}\includegraphics*[width=68mm]{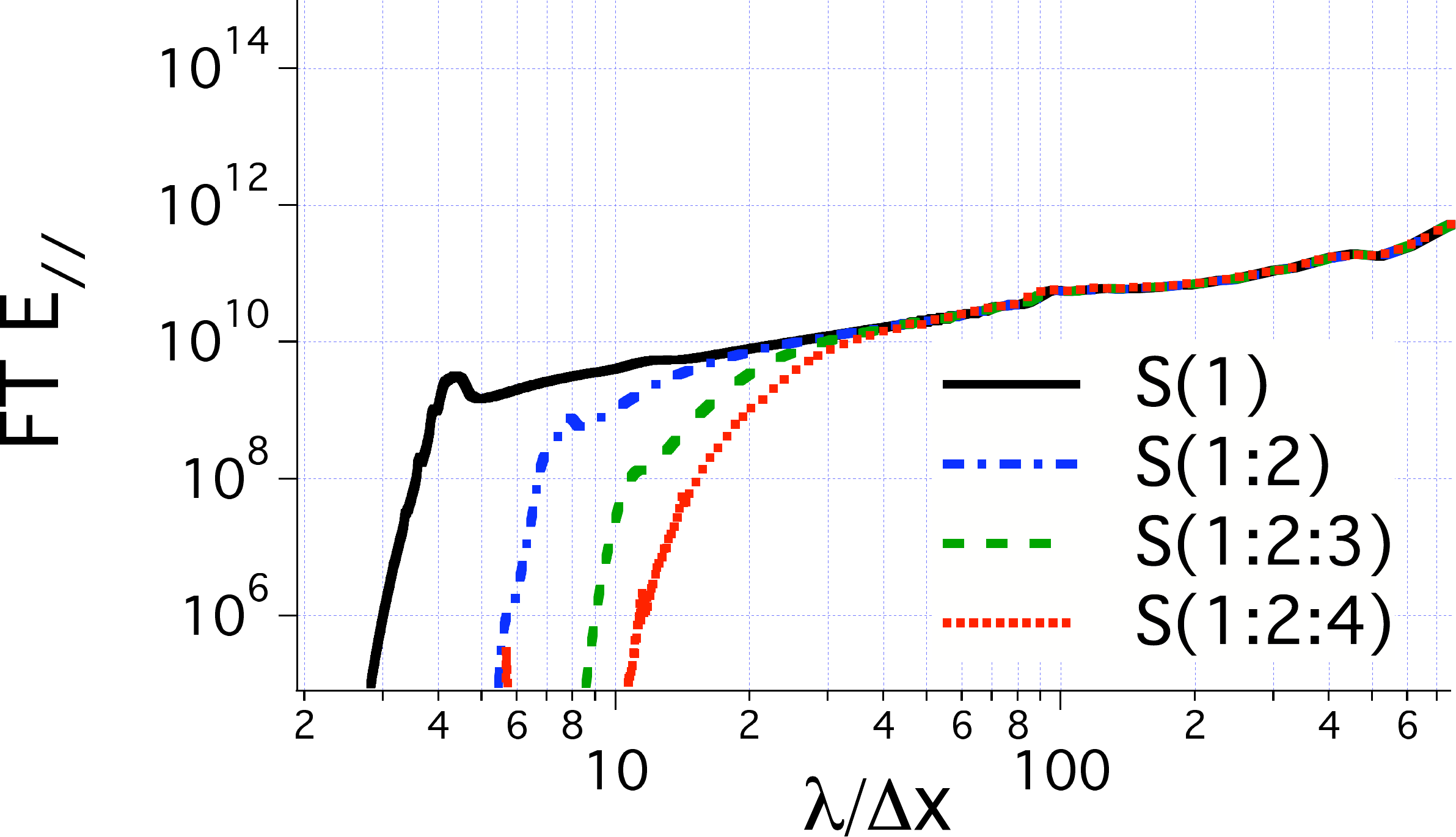} \\
 \end{tabular}
% $\theta=0.$\\
% \begin{tabular}{@{}c@{}c@{}} % @{} removes extra space
 %   \includegraphics*[width=65mm]{YFBhist0p1.pdf} &
  %  \hspace{1mm}\includegraphics*[width=68mm]{yftez0p1.pdf} 
%    \includegraphics*[width=65mm]{YFBhist0p5.pdf} &
%   \hspace{1mm}\includegraphics*[width=68mm]{yftez0p5.pdf} \\
%    \includegraphics*[width=65mm]{YFBhist1p0.pdf} &
%    \hspace{1mm}\includegraphics*[width=68mm]{yftez1p0.pdf} 
%\end{tabular}
% $\theta=0.1$\\}
}
   \caption{(left) Average beam energy gain versus longitudinal position (in the laboratory frame), (right) Fourier Transform of the longitudinal electric field at t=40 ps, averaged over whole domain, from 2D-1/2 simulations  of a full scale 10GeV LPA in a boosted frame at $\gamma=130$, using the Yee solver and various digital filter kernels. Square cells ($\delta x=\delta z=6.5\mu m$) and the CFL time step (c$\delta t/\delta z=1/\sqrt{2}$) were used.}
   \label{Fig_bhistfftez2dy}
\end{figure}

\subsubsection{Simulations in 2-1/2D}
\begin{figure}[htb]
   \centering
    \includegraphics*[width=65mm]{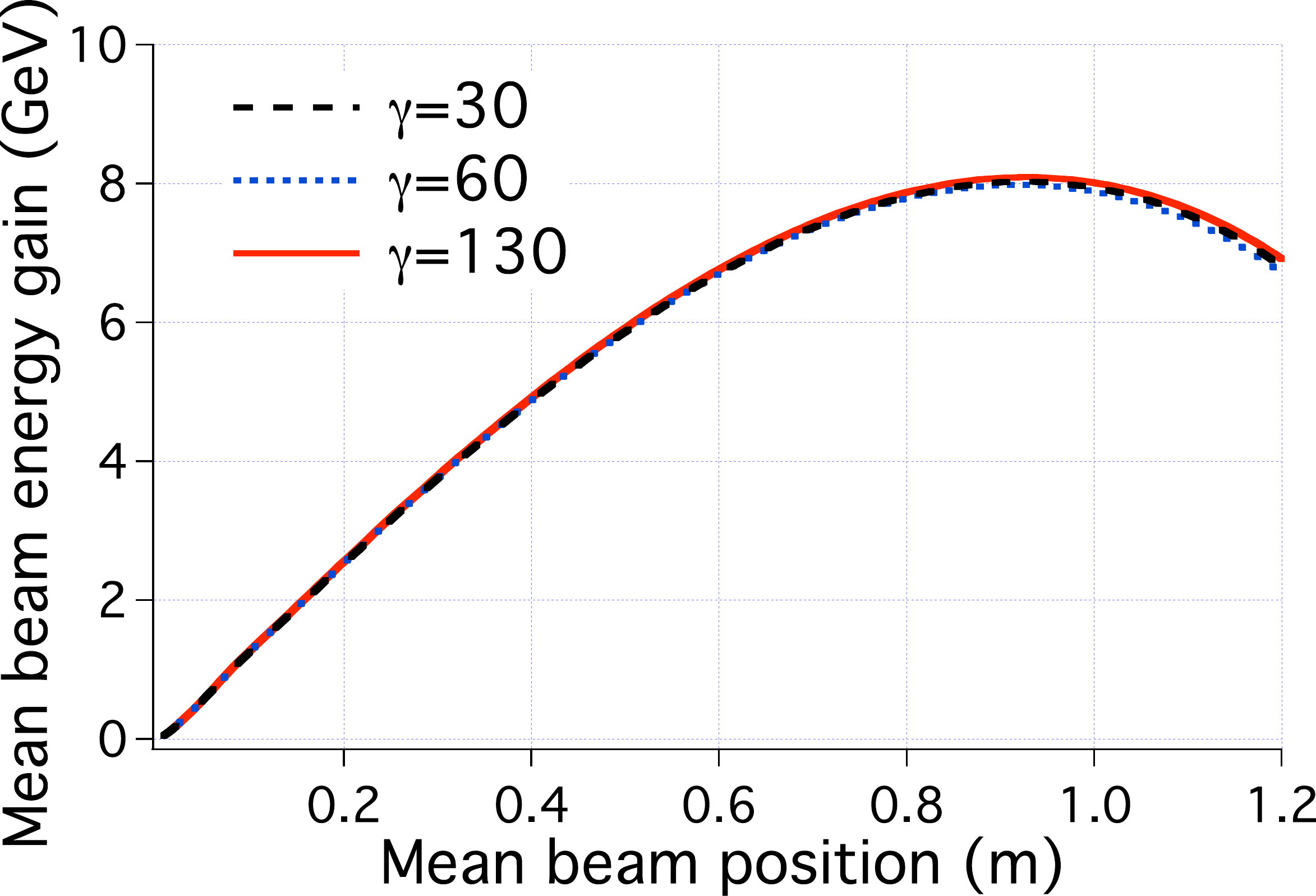} 
   \caption{Average beam energy gain versus longitudinal position (in the laboratory frame) from 2D-1/2 simulations  of a full scale 10GeV LPA in a boosted frame at $\gamma=30$, $60$ and $130$, using the Yee solver.}
   \label{Fig_bhistdf0p012d}
\end{figure}

Fig. \ref{Fig_bhistfftez2dy} shows the average beam energy gain versus longitudinal position and the averaged Fourier Transform of the longitudinal electric field taken at t=40 ps, from 2D-1/2 simulations of a full scale 10GeV LPA in a boosted frame at $\gamma=130$, using the Yee solver and various smoothing kernels.  Fig. \ref{Fig_bhistdf0p012d} shows the average beam energy gain versus longitudinal position from simulations in boosted frames at $\gamma=30$, $60$ and $130$. All runs gave the same beam energy history within a few percents, and no sign of instability is detected in the Fourier transform plot of the longitudinal electric field. The average energy gain peaks around 8 GeV, in agreement with the scaled simulations (see Fig. \ref{Fig_energytime2d}).

\subsubsection{Simulations in  3D}

\begin{figure}[htb]
   \centering
 {\small
 $\theta=0$
 \begin{tabular}{@{}c@{}c@{}} % @{} removes extra space
% \begin{tabular}{|@{}c@{}|@{}c@{}|} % @{} removes extra space
%  \hline
%  Yee & Cole-Karkkainen \\
%  \hline
    \includegraphics*[width=68mm]{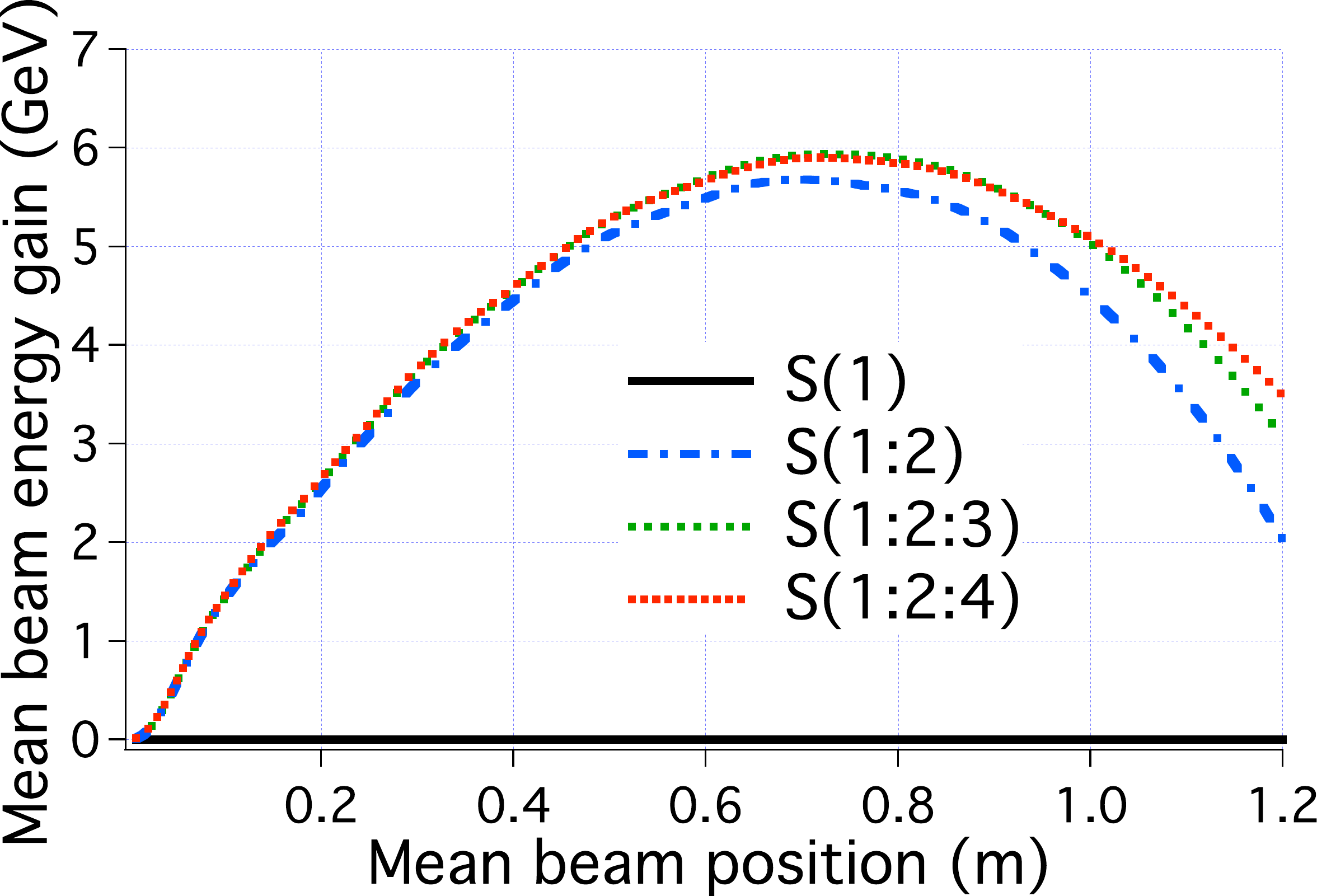} &
    \hspace{1mm}\includegraphics*[width=68mm]{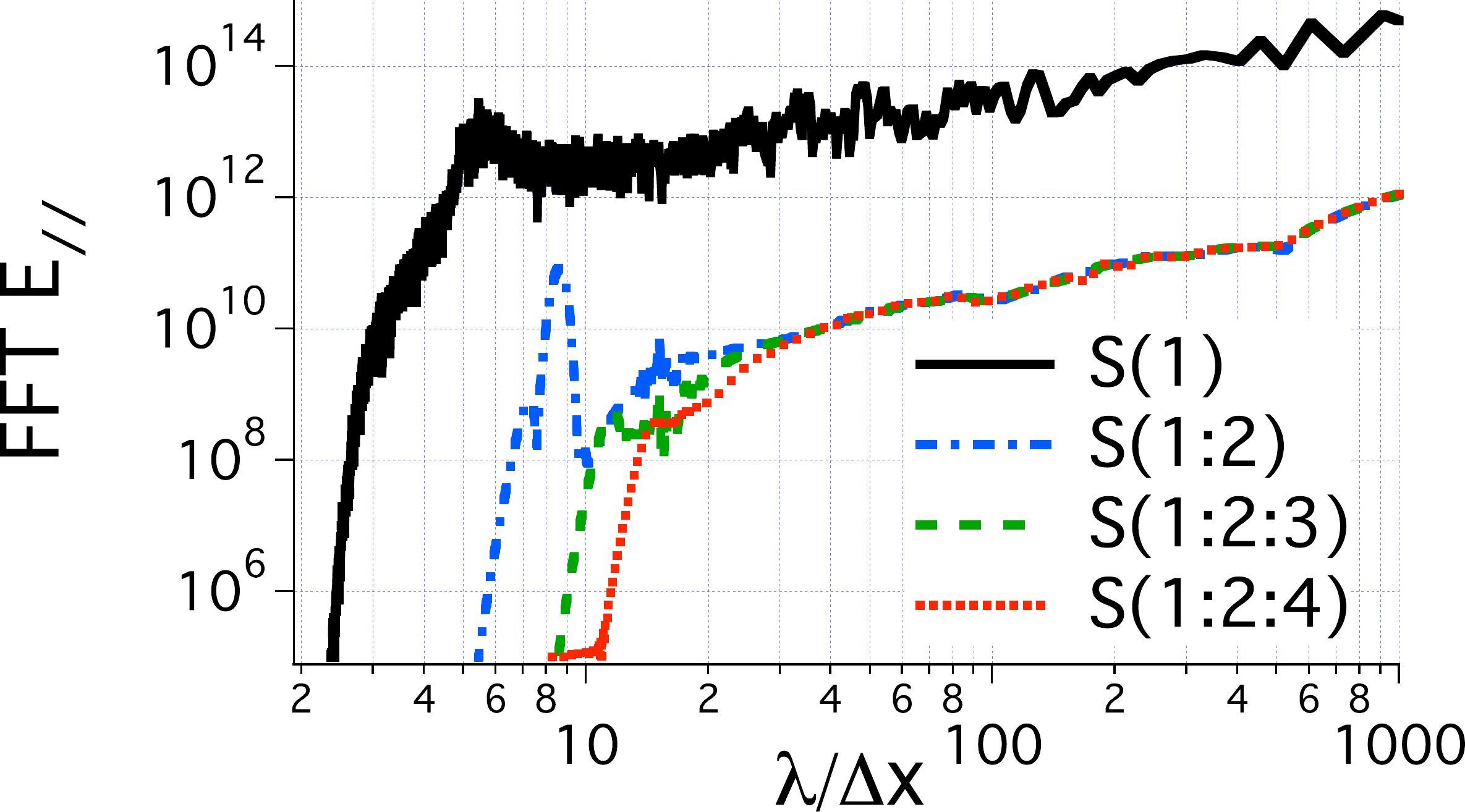} \\
\end{tabular}
 $\theta=0.1$
 \begin{tabular}{@{}c@{}c@{}} % @{} removes extra space
    \includegraphics*[width=68mm]{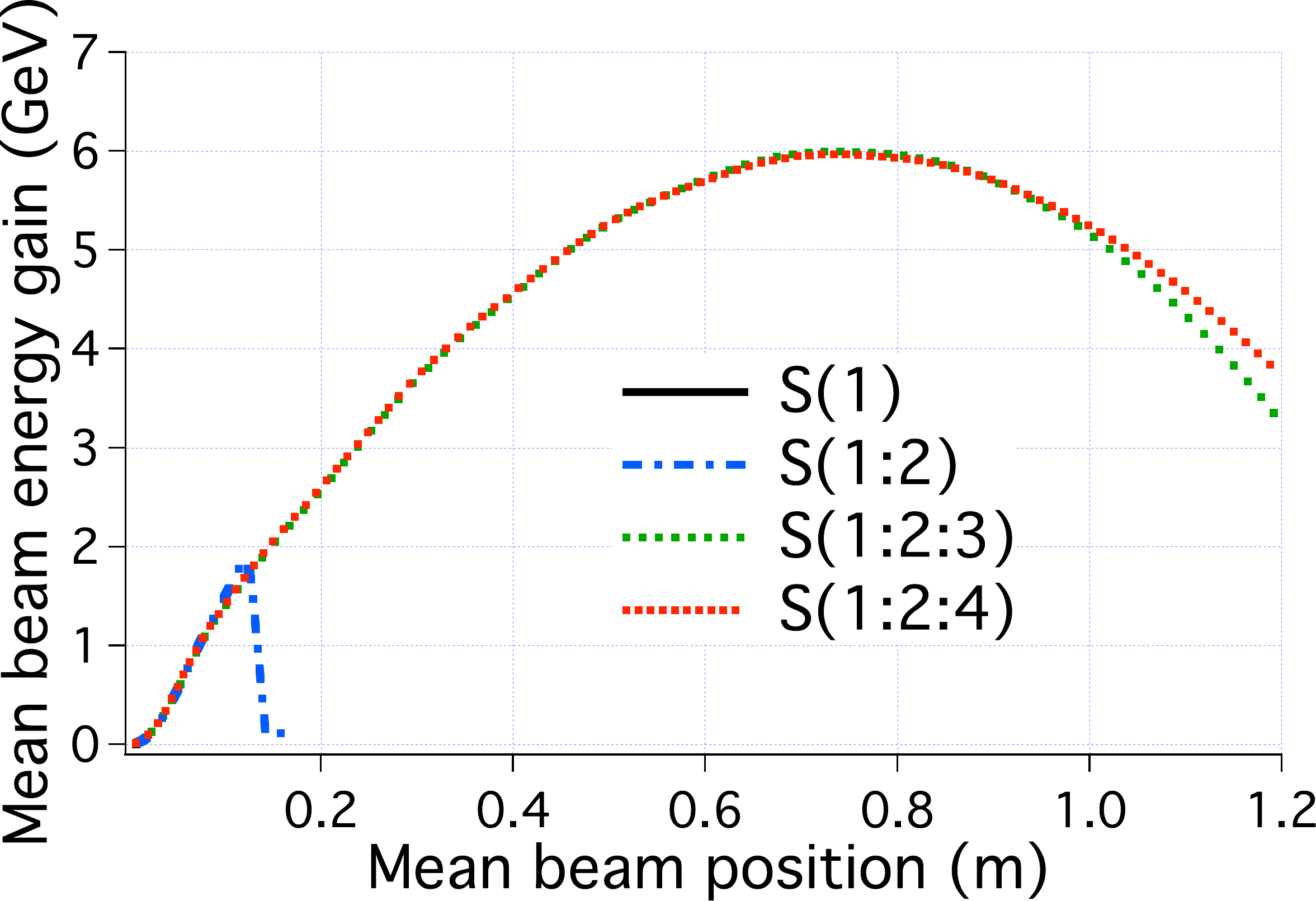} &
    \hspace{1mm}\includegraphics*[width=68mm]{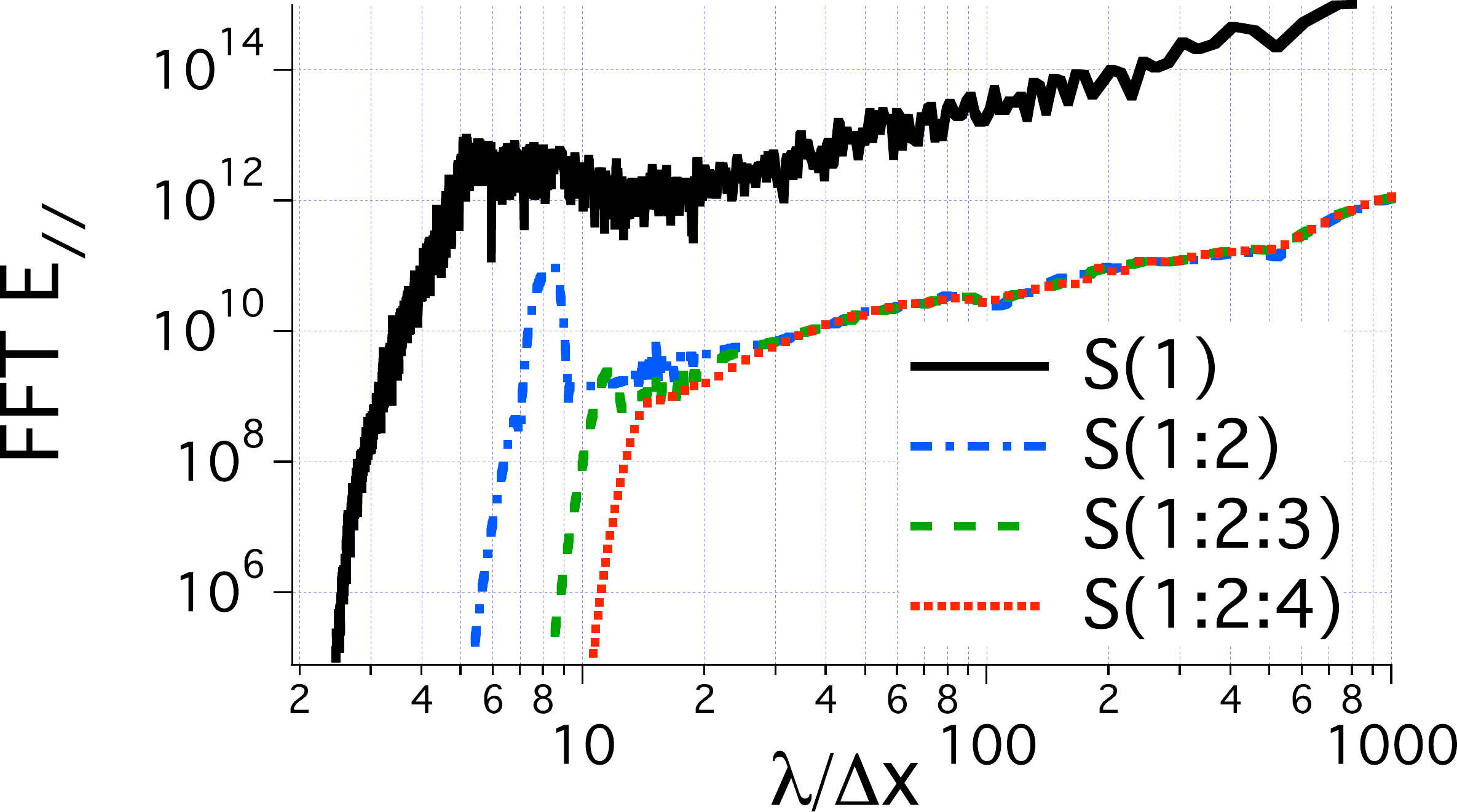} \\
\end{tabular}
 $\theta=0.5$
 \begin{tabular}{@{}c@{}c@{}} % @{} removes extra space
    \includegraphics*[width=68mm]{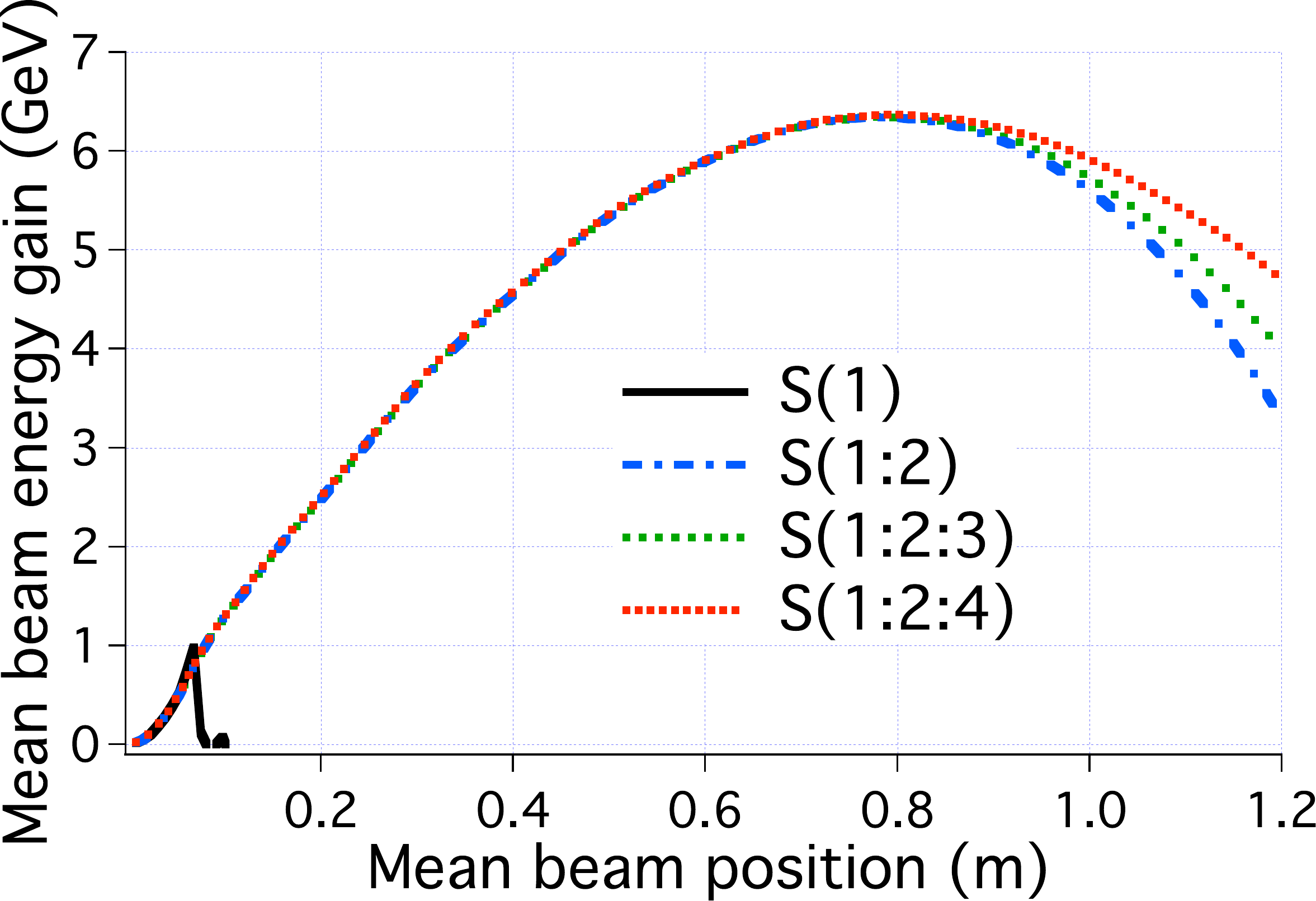} &
    \hspace{1mm}\includegraphics*[width=68mm]{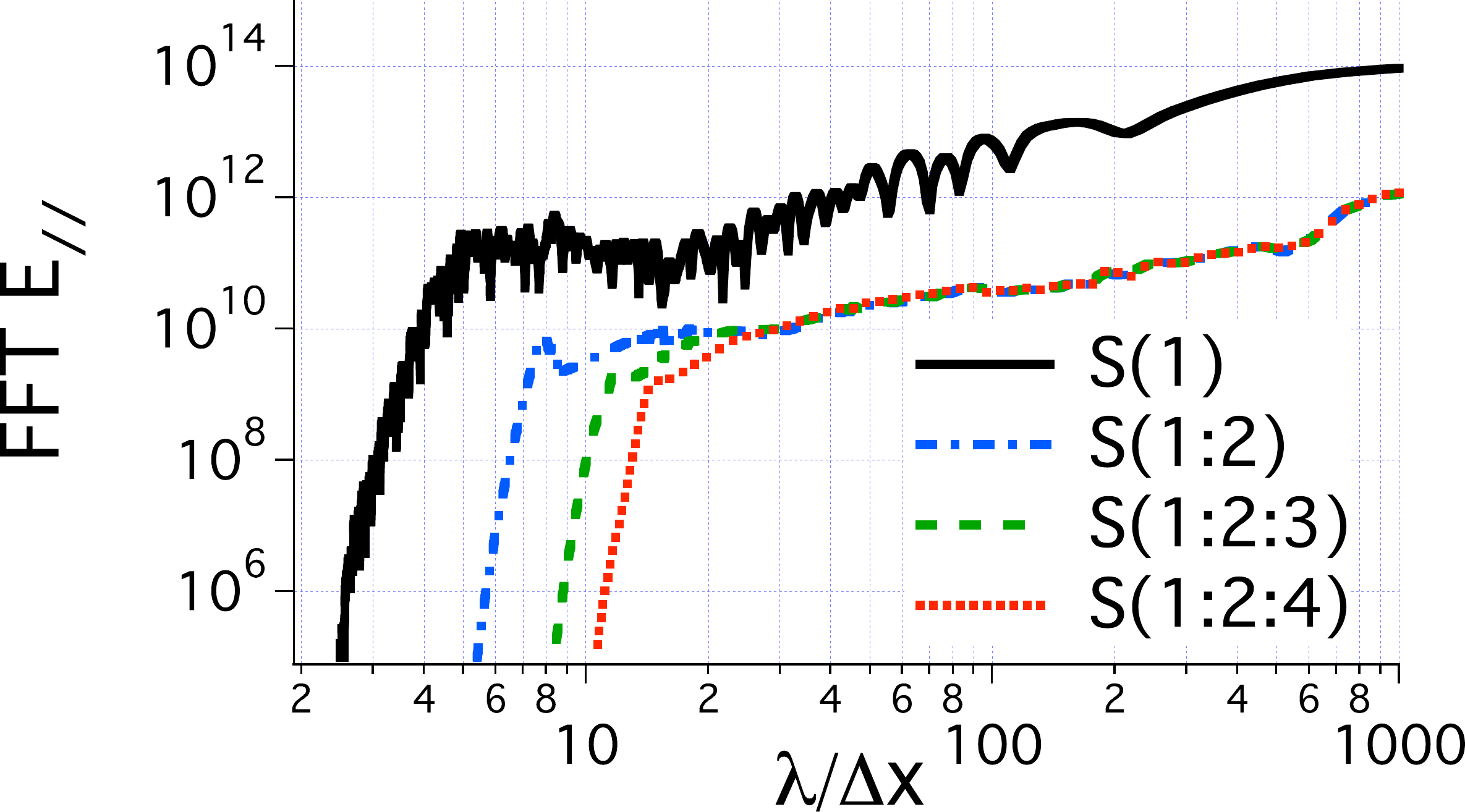} 
\end{tabular}}
   \caption{(left) Average beam energy gain versus longitudinal position (in the laboratory frame); (right) Fourier Transform of the longitudinal electric field at t=40 ps, averaged over  plane on axis perpendicular to laser polarization, from 3D simulations  of a full scale 10GeV LPA in a boosted frame at $\gamma=130$, using the Cole-Karkkainen-Friedman solver and various smoothing kernels, with (top) no numerical damping ($\theta=0$), (middle) damping with $\theta=0.1$ and (bottom) $\theta=0.5$.}
   \label{Fig_bhistfftez3dk}
\end{figure}

In 3D, all simulations at $\gamma=130$ using the Yee solver (using cubic cells and a time step at the CFL limit) developed the instability and loss of the beam, regardless of the amount of filtering or damping that has been tried. 
The failure of the 3D simulations using the Yee solver motivated use of the Cole-Karkkainen-Friedman (CKF)  solver, with various levels of filtering and damping. Data from 3D simulations using the CKF solver and various smoothing kernels are plotted in Fig. \ref{Fig_bhistfftez3dk}. Stability is attained when using a sufficient level of filtering. Damping is detrimental to stability at low levels ($\theta=0.1$) but is beneficial at a higher level ($\theta=0.5$). 

%though at the cost of a small loss of accuracy (compare peak energy between bottom and top plots in right column of Fig. \ref{Fig_bhistfftez3dk} ).

\begin{figure}[htb]
   \centering
 {\small
 \begin{tabular}{@{}c@{}c@{}} % @{} removes extra space
% \begin{tabular}{|@{}c@{}|@{}c@{}|} % @{} removes extra space
%  \hline
%  Yee & Cole-Karkkainen \\
%  \hline
%    \hspace{1.cm} Cole-Karkkainen-Friedman & \hspace{1.cm} Yee-Friedman  \vspace{2.mm}\\
    \includegraphics*[width=65mm]{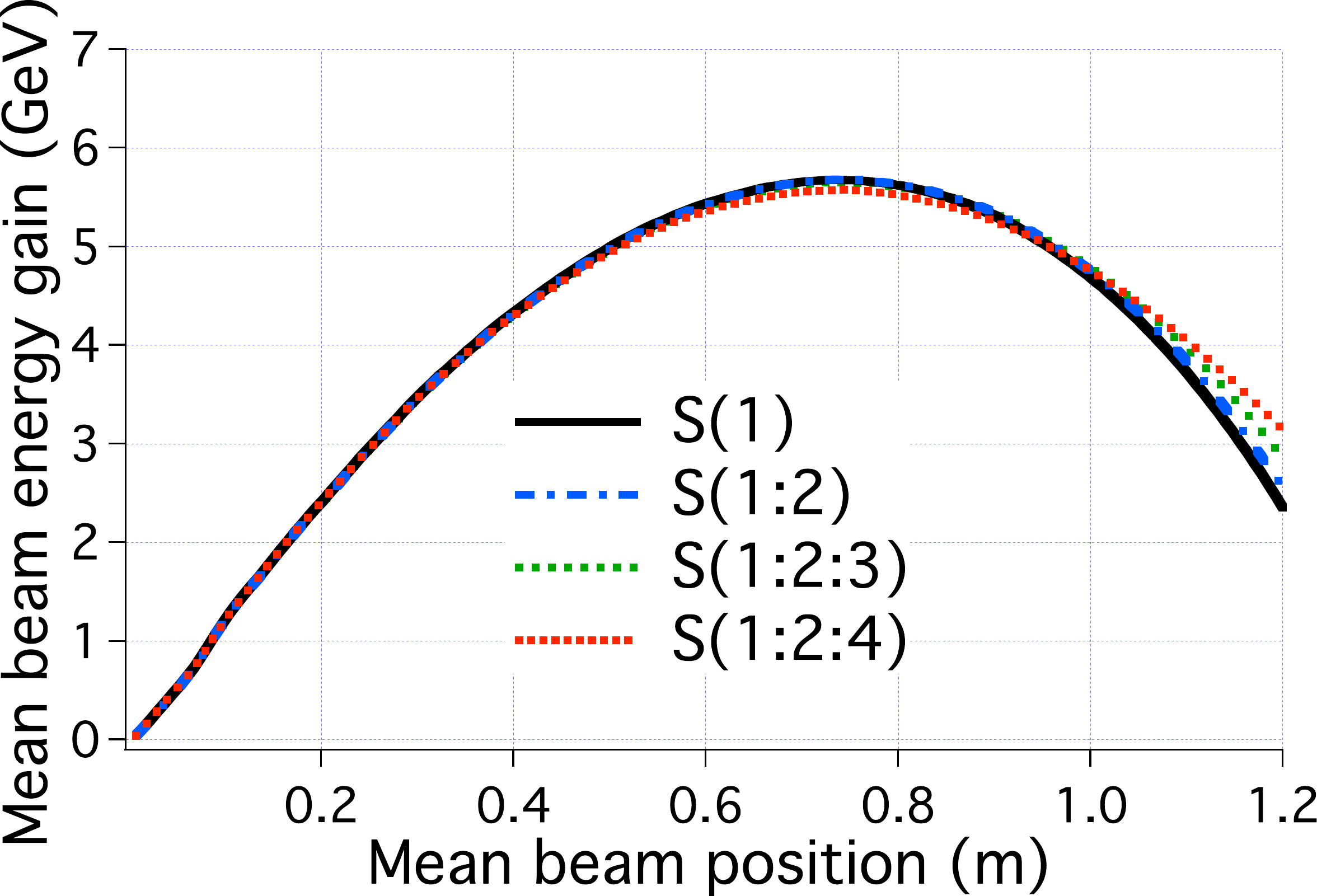} &
    \hspace{1mm}\includegraphics*[width=68mm]{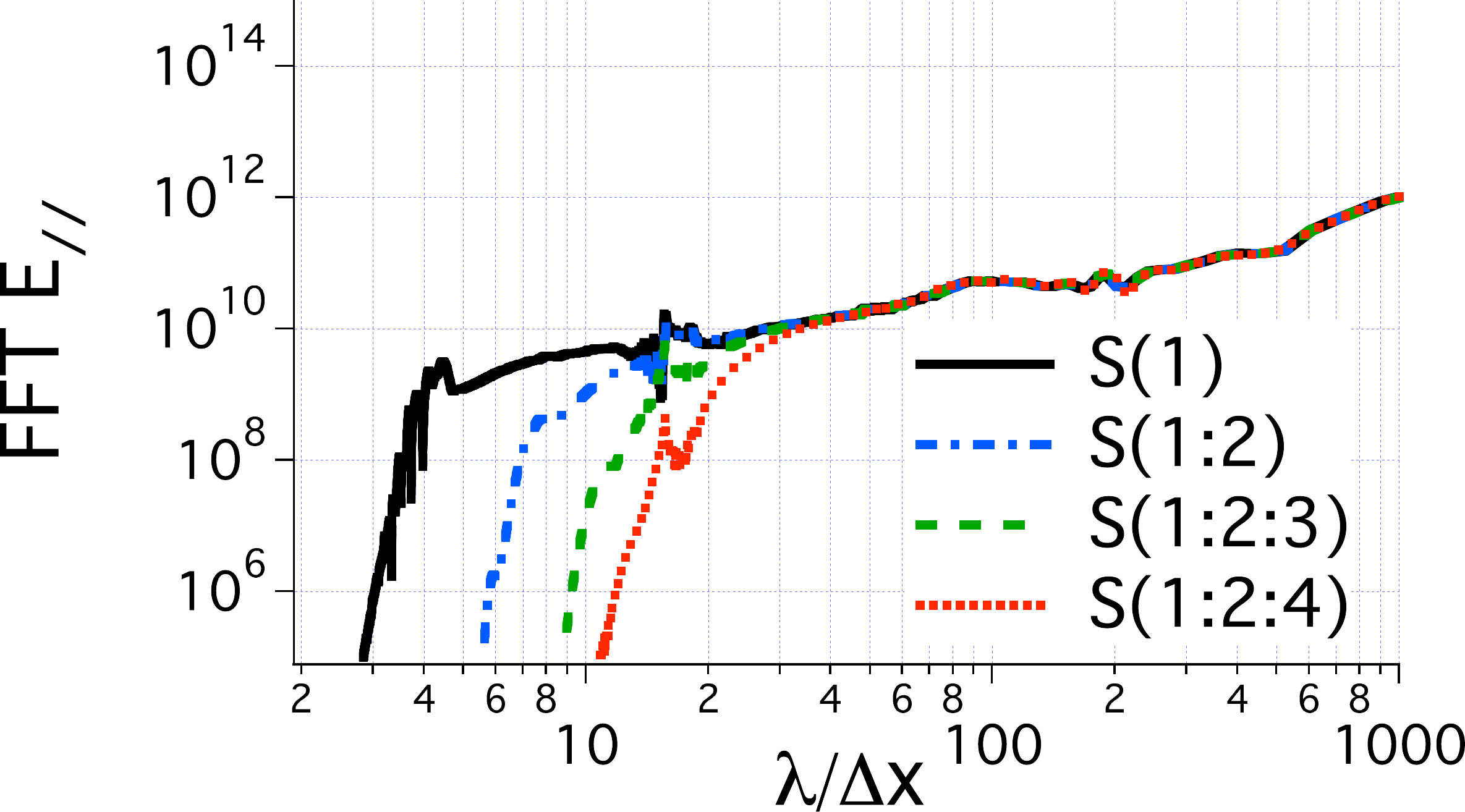} \\
 \end{tabular}
% $\theta=0.$\\
% \begin{tabular}{@{}c@{}c@{}} % @{} removes extra space
 %   \includegraphics*[width=65mm]{YFBhist0p1.pdf} &
  %  \hspace{1mm}\includegraphics*[width=68mm]{yftez0p1.pdf} 
%    \includegraphics*[width=65mm]{YFBhist0p5.pdf} &
%   \hspace{1mm}\includegraphics*[width=68mm]{yftez0p5.pdf} \\
%    \includegraphics*[width=65mm]{YFBhist1p0.pdf} &
%    \hspace{1mm}\includegraphics*[width=68mm]{yftez1p0.pdf} 
%\end{tabular}
% $\theta=0.1$\\}
}
   \caption{(left) Average beam energy gain versus longitudinal position (in the laboratory frame), (right) Fourier Transform of the longitudinal electric field at t=40 ps, averaged over whole domain, from 3D simulations  of a full scale 10GeV LPA in a boosted frame at $\gamma=130$, using the CK2 solver and various digital filter kernels.}
   \label{Fig_bhistfftez3dck2}
\end{figure}

\begin{figure}[htb]
   \centering
    \includegraphics*[width=65mm]{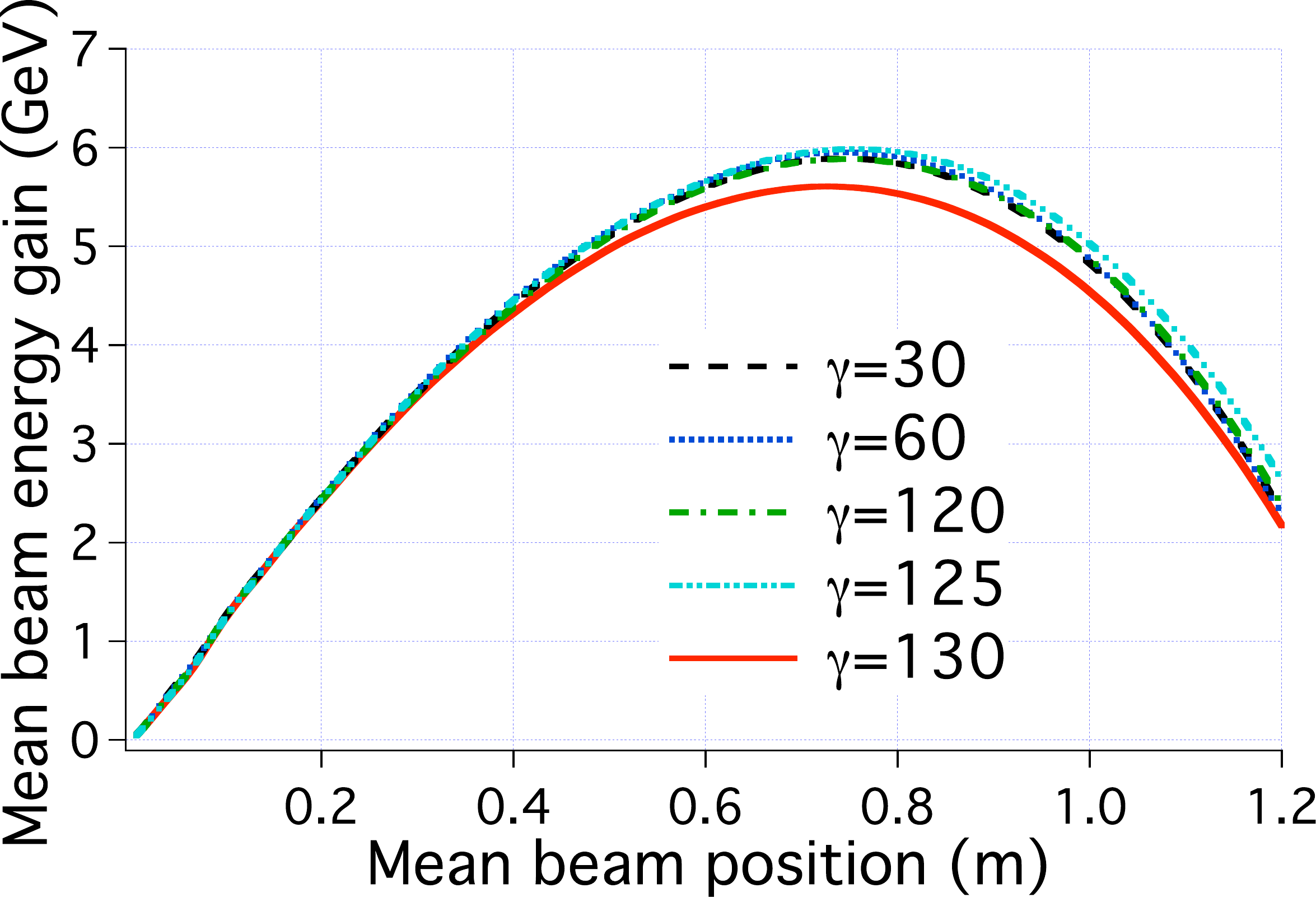}
   \caption{Average beam energy gain versus longitudinal position (in the laboratory frame) from 3D simulations  of a full scale 10GeV LPA in a boosted frame at $\gamma=30$, $60$, $120$, $125$ and $130$, using the Yee solver ($\gamma=30$ and $60$) and the CK2 solver ($\gamma=120-130$), with digital filter S(1) and with the time step set by c$\delta t/\delta z=1/\sqrt{2}$ for stability (see discussion below) .}
   \label{Fig_bhistfftez3dck2_gamma}
\end{figure}

Next, simulations using the solver coefficients CK2-5 from Table \ref{Table:CKcoefs} were performed, with the time step set at their respective CFL limit. The best results were obtained using solvers CK2 and CK3, while CK4 and CK5 did not offer substantial improvement over the CK solver. The results from the runs using CK2 and CK3 were nearly identical and hence only thoses from CK2 are reported in Fig. \ref{Fig_bhistfftez3dck2}, which show very consistent beam energy gain histories, and no sign of instability in the Fourier Transform plot of the longitudinal electric field at t=40 ps (closer inspection revealed that when using the lowest level of filtering S(1), a mild instability was developing but it was not affecting the average beam energy gain history). As shown on Fig. \ref{Fig_bhistfftez3dck2_gamma}, the results at $\gamma=30-125$ are in excellent agreement while the run at $\gamma=130$ predicts a slightly lower energy gain, all within 10 percent of the maximum energy gain predicted around 5.7 GeV by the scaled simulations shown on Fig. \ref{Fig_energytime2d} (top-right).

In summary, the full scale 6-7 GeV simulations using the frame of the wake performed in this subsection show: (i) 2-1/2D simulations using the Yee solver at the CFL limit (with square cells) were free of instability; %(ii) 3D simulations using the Yee solver at the CFL limit and 2-1/2 simulations at the 3D CFL limit developed strong instabilities at similar levels, which were not cured by damping or filtering; 
(ii) 3D simulations using the CK solver developed moderately strong instabilities that were mitigated using moderate to high levels of damping and/or filtering, the latter being the most effective; (iii) 3D simulations using the CK2 (or CK3) solver developed very mild instabilities that were mitigated with a low level of filtering.

%It is interesting to note that the solver coefficients (CK2 and CK3) which lead to the most stable simulations are not the one that give no dispersion along the grid axes (CK), but the ones that offer both an improved numerical dispersion along the main axes relative to the Yee solver and along the 3D diagonal relative to the CK solver. As noted above, the instability is observed to propagate at a small angle from the longitudinal axis and it makes sense that solvers that offer better numerical dispersion at an angle from the axis perform better at preventing numerical Cerenkov.

\clearpage
\subsection{Effects of numerical parameters on the observed instability}

\begin{figure}[htb]
   \centering
 {\small
 \begin{tabular}{@{}c@{}c@{}} % @{} removes extra space
% \begin{tabular}{|@{}c@{}|@{}c@{}|} % @{} removes extra space
%  \hline
%  Yee & Cole-Karkkainen \\
%  \hline
%    \includegraphics*[width=68mm]{BhistYth0p0.pdf} &
    \hspace{1mm}\includegraphics*[width=68mm]{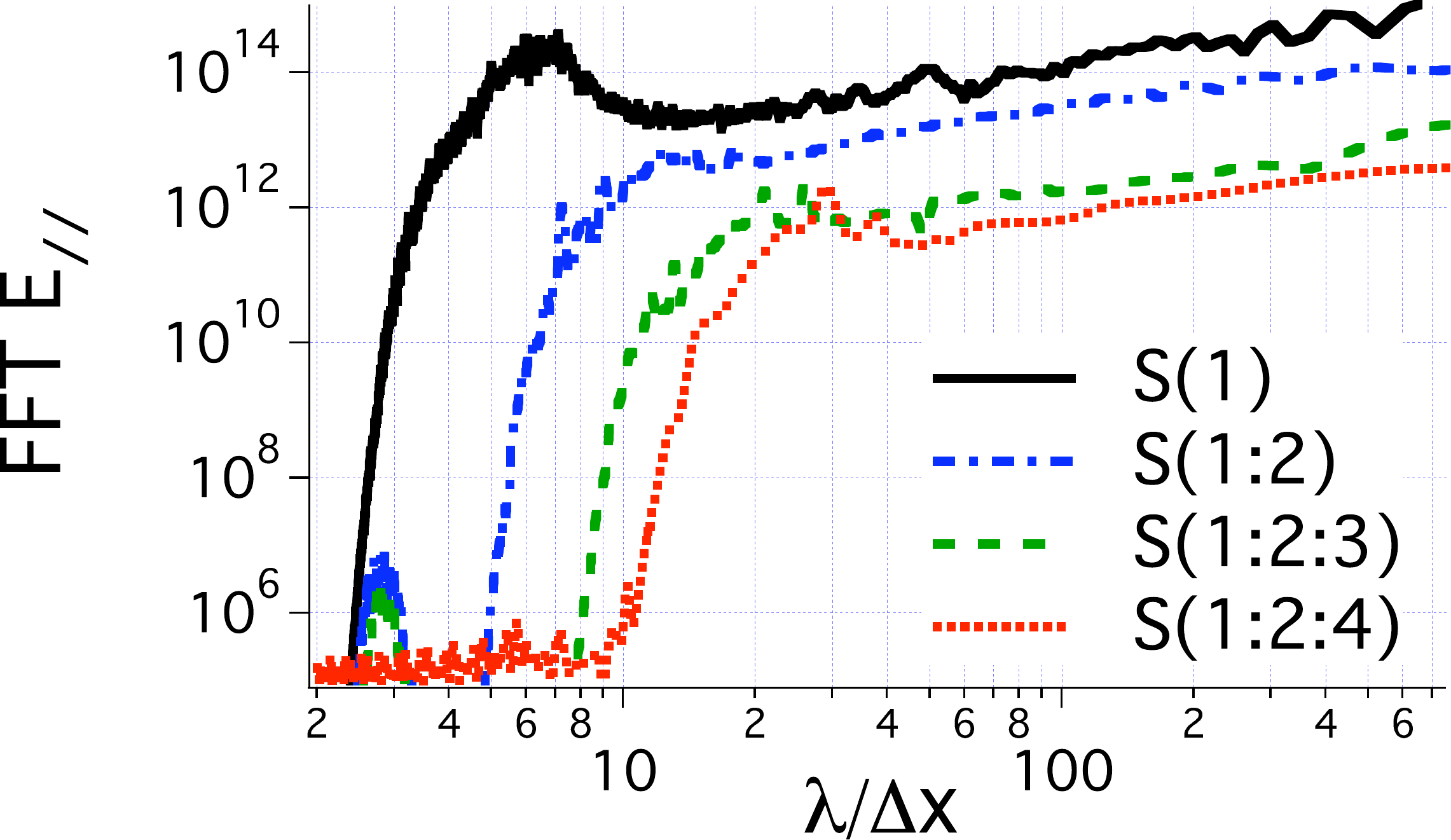} &
    \hspace{1mm}\includegraphics*[width=68mm]{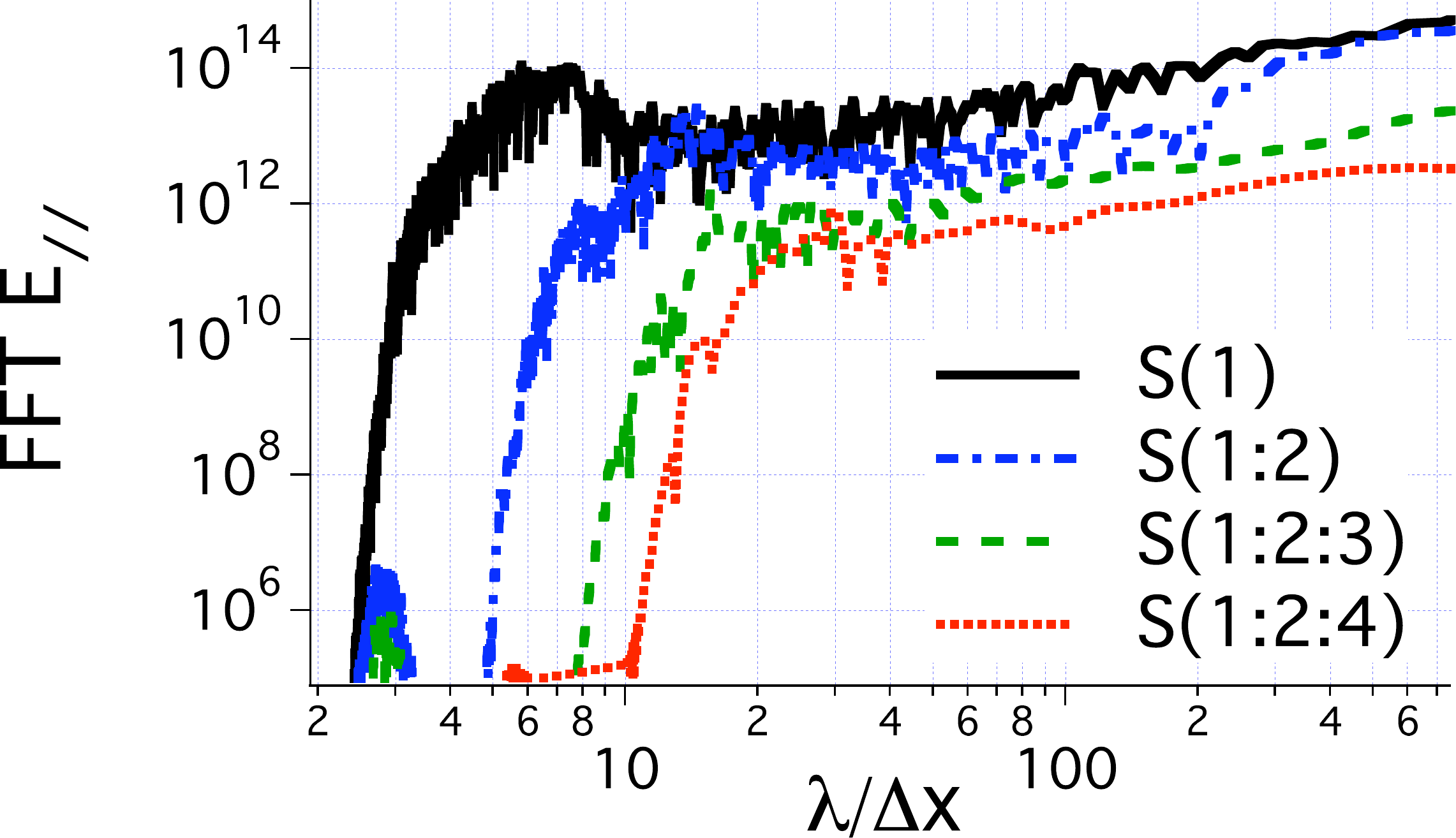} 
\end{tabular}}
   \caption{Fourier Transform of the longitudinal electric field at t=40 ps, averaged over plane on axis perpendicular to laser polarization, from (left) 3D and (right) 2D-1/2 simulations of a full scale 10GeV LPA in a boosted frame at $\gamma=130$, using the Yee solver and various smoothing kernels.  The same time step at the 3D CFL limit c$\delta t=\delta x/\sqrt{3}$ was used for both simulations.}
   \label{Fig_bhistfftezdt3dy}
\end{figure}
The Fourier transform of the longitudinal electric field averaged over the whole domain at t=40 ps, from 3D simulations using the Yee solver, is given in Fig. \ref{Fig_bhistfftezdt3dy} (left). It is contrasted to the same data taken from 2-1/2D simulations (right). Both simulations used the same time step at the 3D CFL limit c$\delta t=\delta z/\sqrt{3}$. The similarity of the two plots indicates that the degradation of the numerical dispersion that resulted from going from the 2D to the 3D CFL limit is the cause of the failure of the 3D runs using the Yee solver. Taking advantage of this observation, we study in this section the instability arising in 2-1/2D simulations using a time step at the 3D CFL limit. 

\subsubsection{Effects of spatial resolution}

\begin{figure}[htb]
   \centering
 {\small
 \begin{tabular}{@{}c@{}c@{}} % @{} removes extra space
   $E_{//}$ & FFT$ E_{//}$ \\
    \includegraphics*[width=65mm]{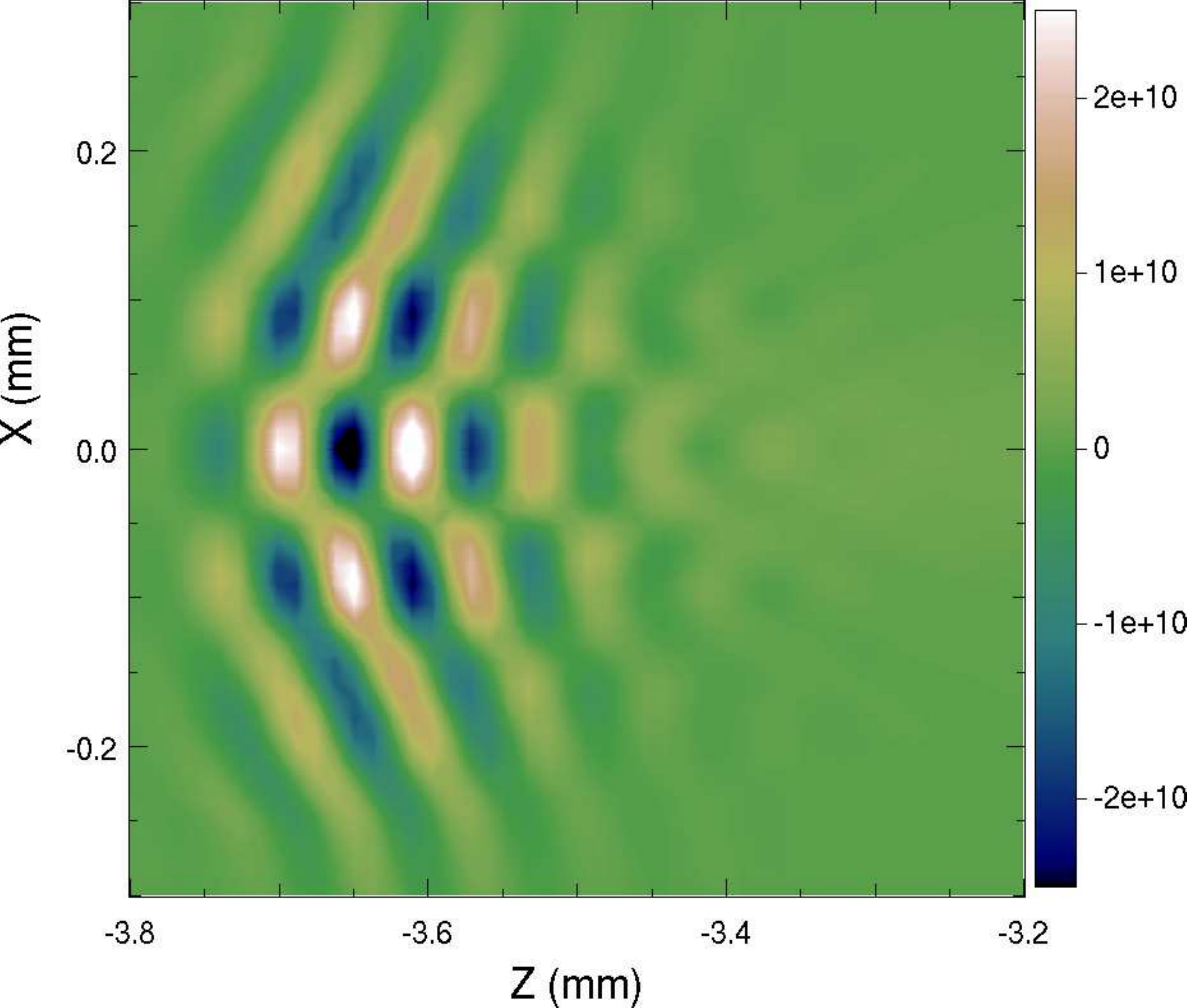} &
    \hspace{1mm}\includegraphics*[width=65mm]{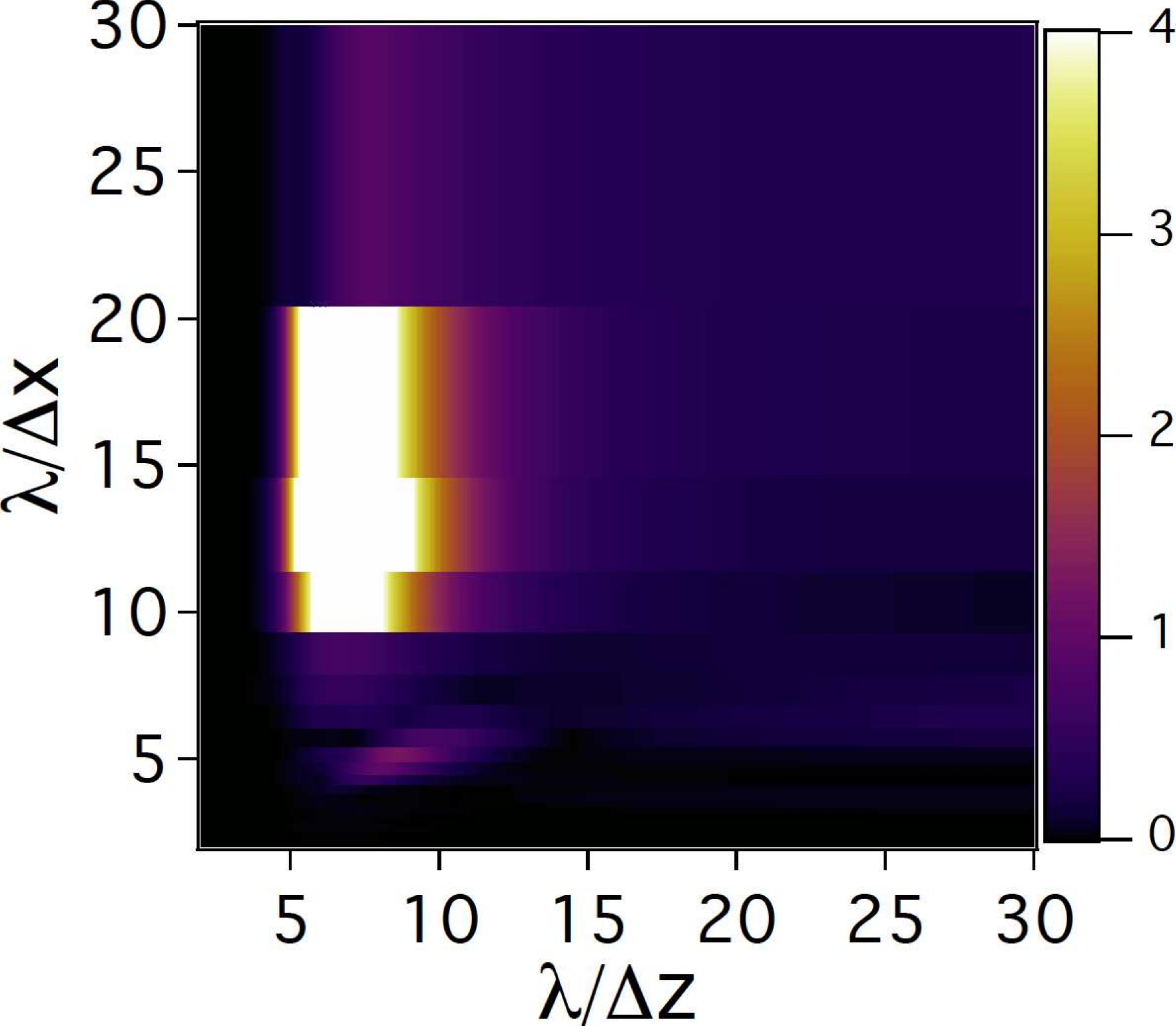} \\
    \includegraphics*[width=65mm]{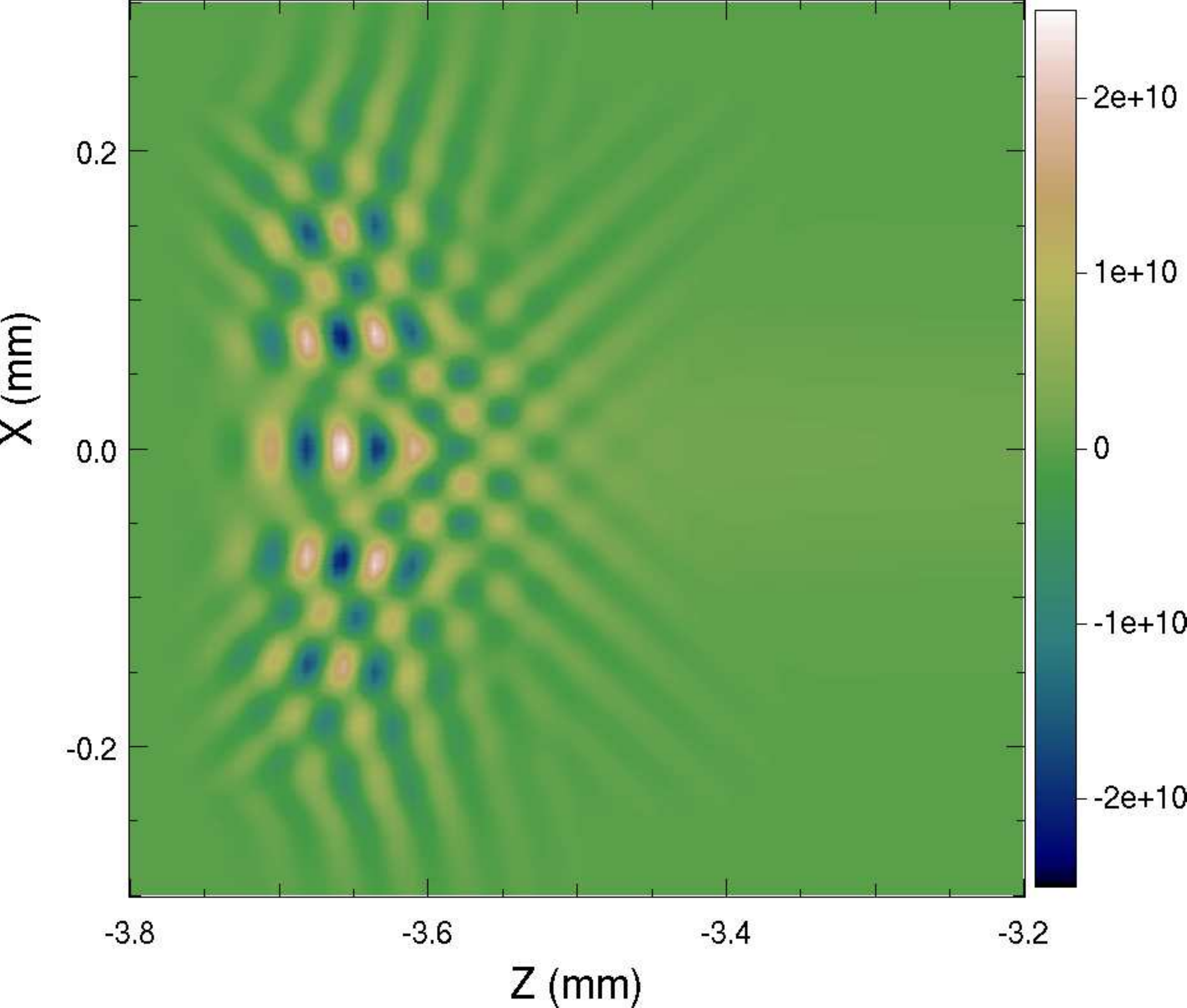} &
    \hspace{1mm}\includegraphics*[width=65mm]{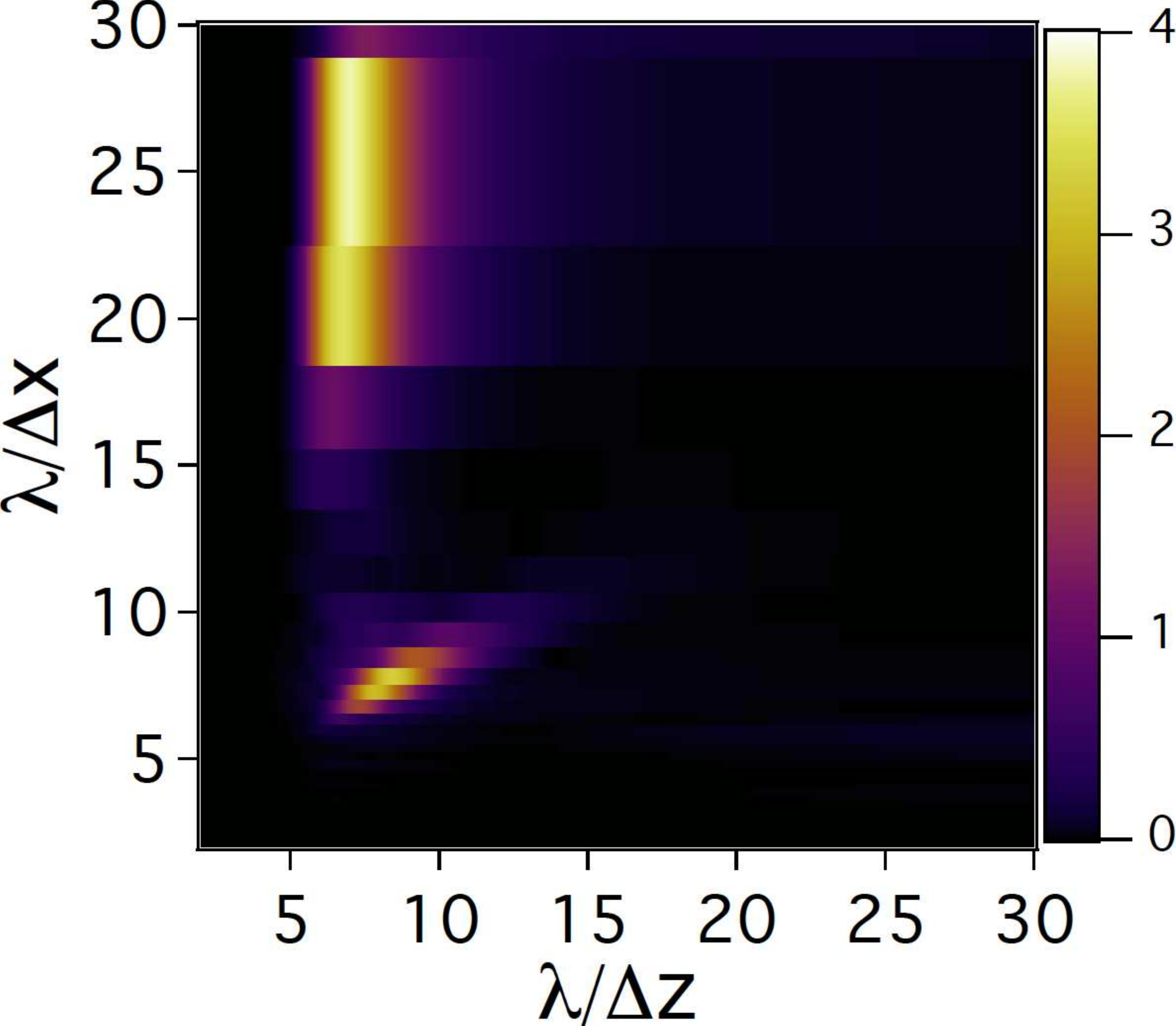} \\
    \includegraphics*[width=65mm]{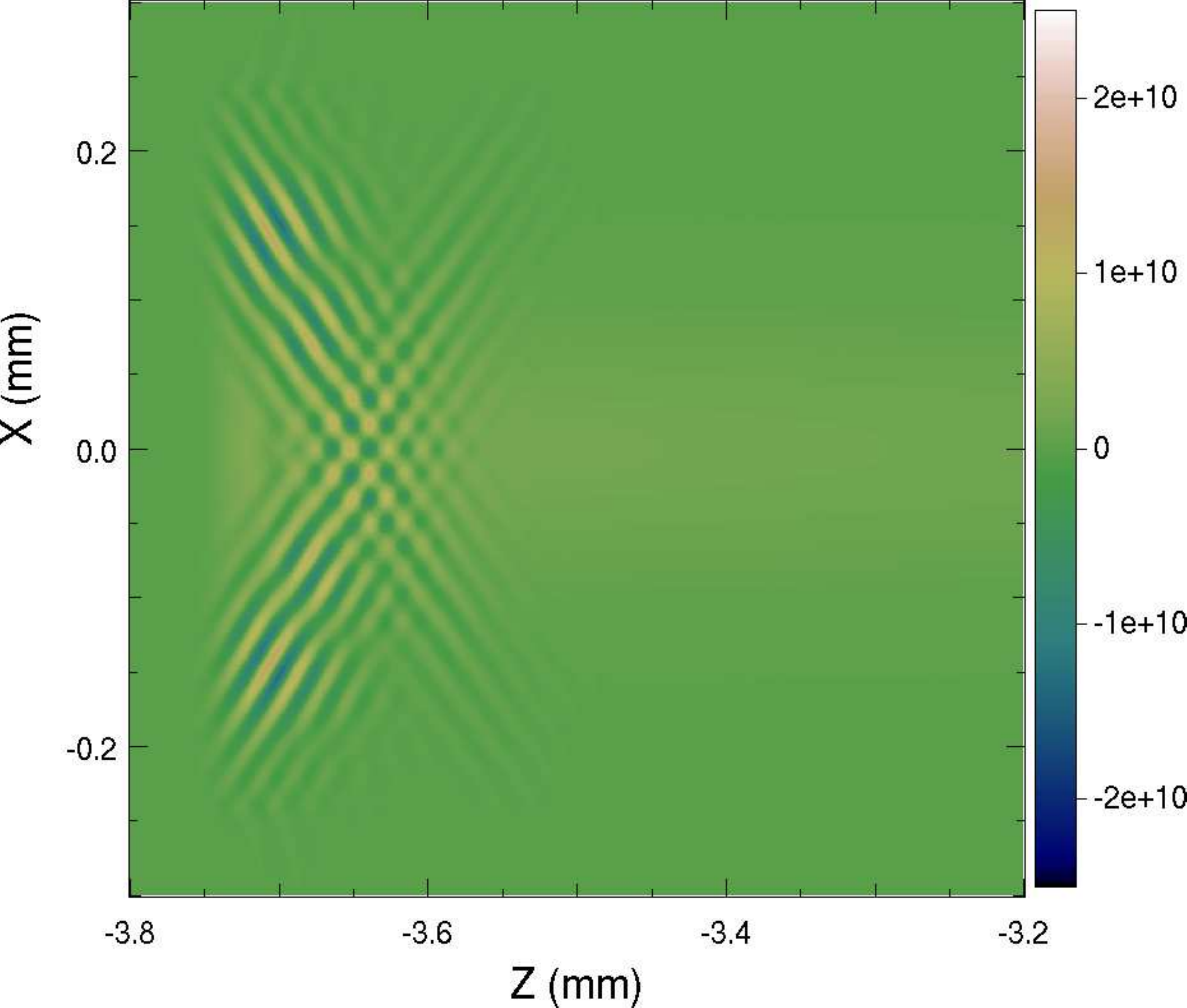} &
    \hspace{1mm}\includegraphics*[width=65mm]{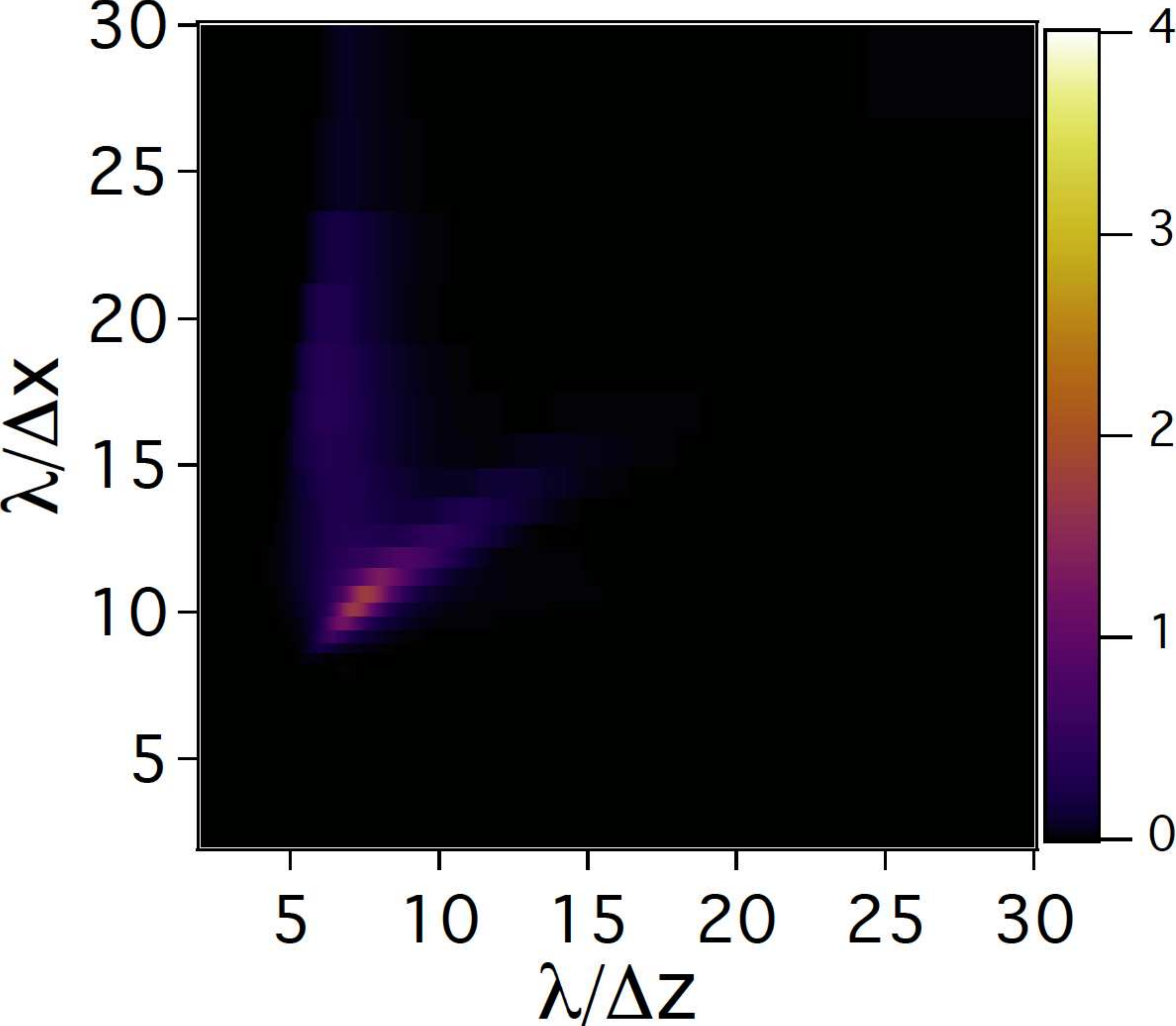} \\
 \end{tabular}
}
   \caption{(left) Snapshot of the longitudinal electric field ($E_{//}$) at the front of the plasma at $t=12.5$ ps; (right) Fourier Transform of the longitudinal electric field, from 2-1/2D simulations  of a full scale 10GeV LPA in a boosted frame at $\gamma=130$, using the Yee solver, for (top) $\delta x=\delta z=13\mu m$; (middle)  $\delta x=\delta z=6.5\mu m$; (bottom)  $\delta x=\delta z=3.25\mu m$. The time step at the 3D CFL limit c$\delta t=\delta z/\sqrt{3}$ was used for all three simulations. }
   \label{Fig_yeefft2dsqrt3}
\end{figure}

Snapshots of the longitudinal electric field at the front of the plasma taken at $t=12.5$ ps, and their corresponding Fourier transform, are given in Fig. \ref{Fig_yeefft2dsqrt3}, from 2-1/2D simulations using the Yee solver with the time step at the 3D CFL limit c$\delta t=\delta z/\sqrt{3}$. Three resolutions were considered: (a) $\delta_x=\delta_z=13\mu m$, (b)  $\delta_x=\delta_z=6.5\mu m$, and  (c)  $\delta_x=\delta_z=3.25\mu m$. The amplitude of the instability is roughly inversely proportional to the resolution. For this configuration, the instability exhibits two primary modes at various relative levels, both at a fixed number of grid cells in the longitudinal direction, but at a fixed absolute length in the transverse direction. This indicates that the transverse part of the modes is governed by the physical geometry of the problem while the longitudinal part is governed by numerical resolution. 

\begin{figure}[htb]
   \centering
 {\small
 \begin{tabular}{@{}c@{}c@{}} % @{} removes extra space
    \includegraphics*[width=70mm]{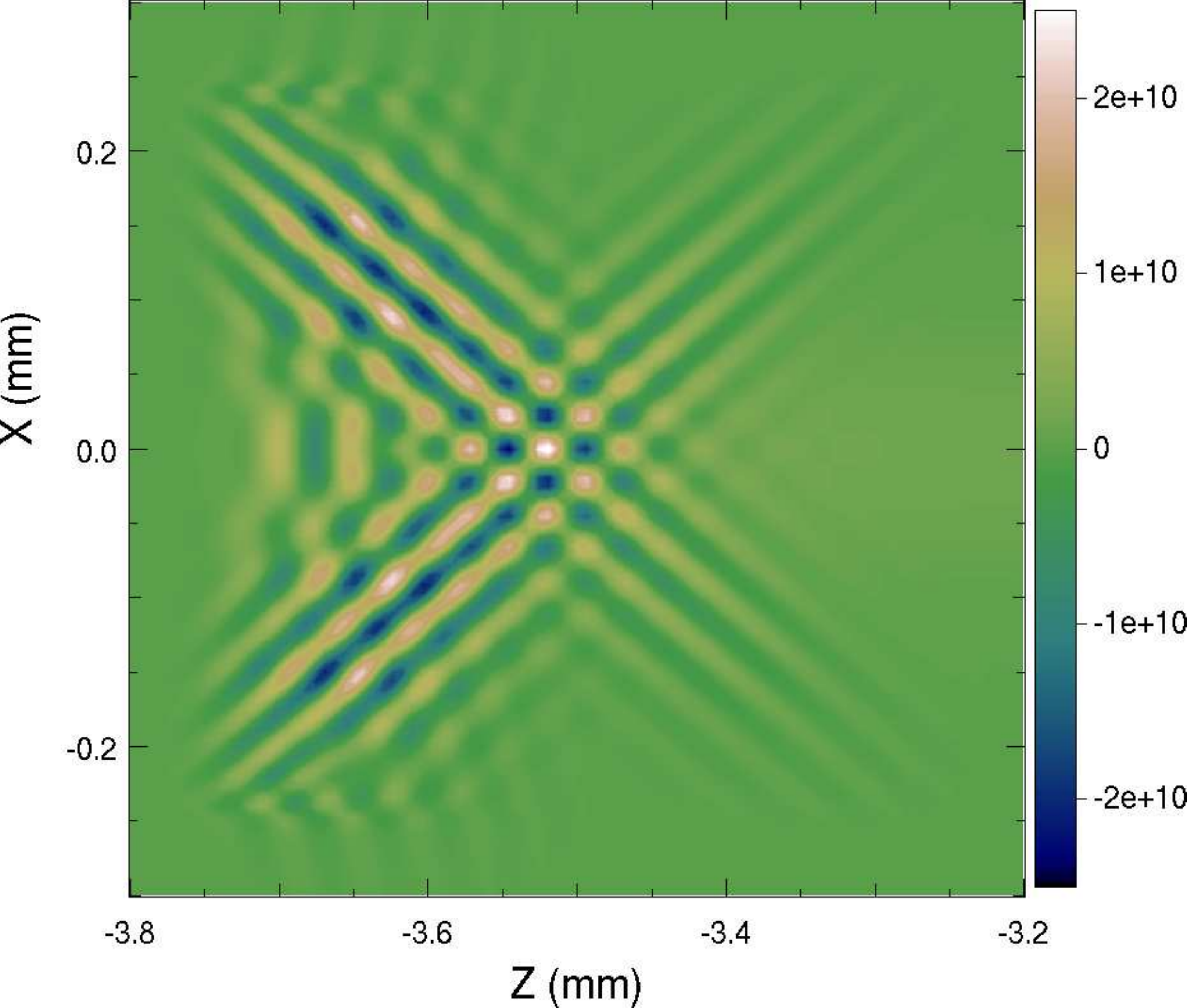} &
    \hspace{1mm}\includegraphics*[width=65mm]{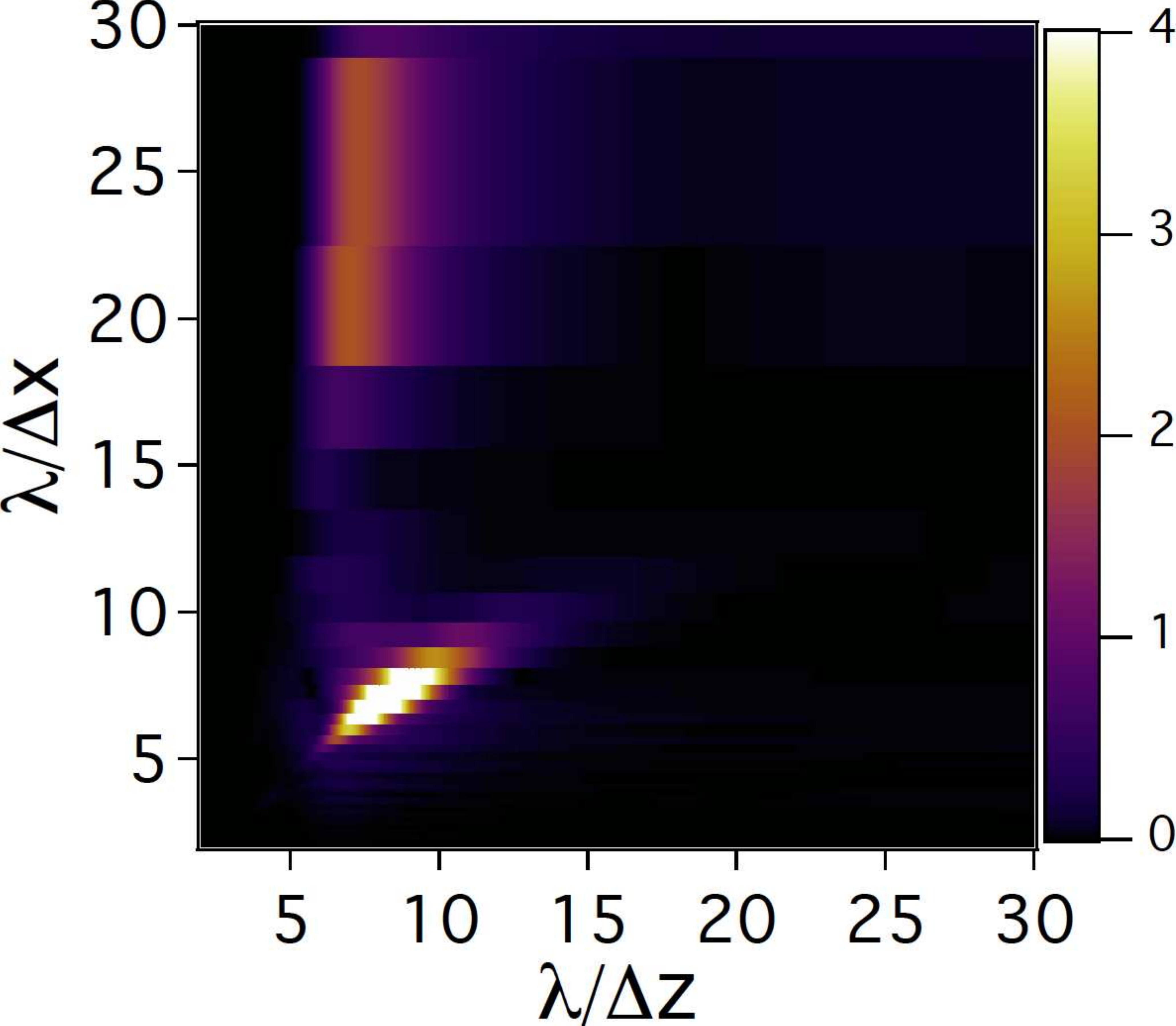} \\
 \end{tabular}
}
   \caption{(left) Snapshot of the longitudinal electric field ($E_{//}$) at the front of the plasma at $t=12.5$ ps; (right) Fourier Transform of the longitudinal electric field, from 2-1/2D simulations  of a full scale 10GeV LPA in a boosted frame at $\gamma=130$, with the CK solver, using $\delta x=\delta z=6.5\mu m$, and the time step at the 3D CFL limit c$\delta t=\delta x/\sqrt{3}$. }
   \label{Fig_fft2dsqrt3kark}
\end{figure}

Results from 2-1/2D simulation using the CK solver at the 3D CFL limit c$\delta t/\delta z=1/\sqrt{3}$ at the resolution $\delta_x=\delta_z=6.5\mu$ m are given in Fig. \ref{Fig_fft2dsqrt3kark}. The same two modes that were observed in the plots from the equivalent simulation using the Yee solver (see Fig. \ref{Fig_yeefft2dsqrt3}-middle), are present, and the overall amplitude of the instability is similar. These similarities on the details of the instability between the Yee and CK solvers indicate that the differences in numerical dispersion of the solvers do not constitute a key factor affecting the instability.

\subsubsection{Effects of time step}

\begin{figure}[htb]
   \centering
% \begin{tabular}{@{}c@{}c@{}} % @{} removes extra space
    \includegraphics*[width=68mm]{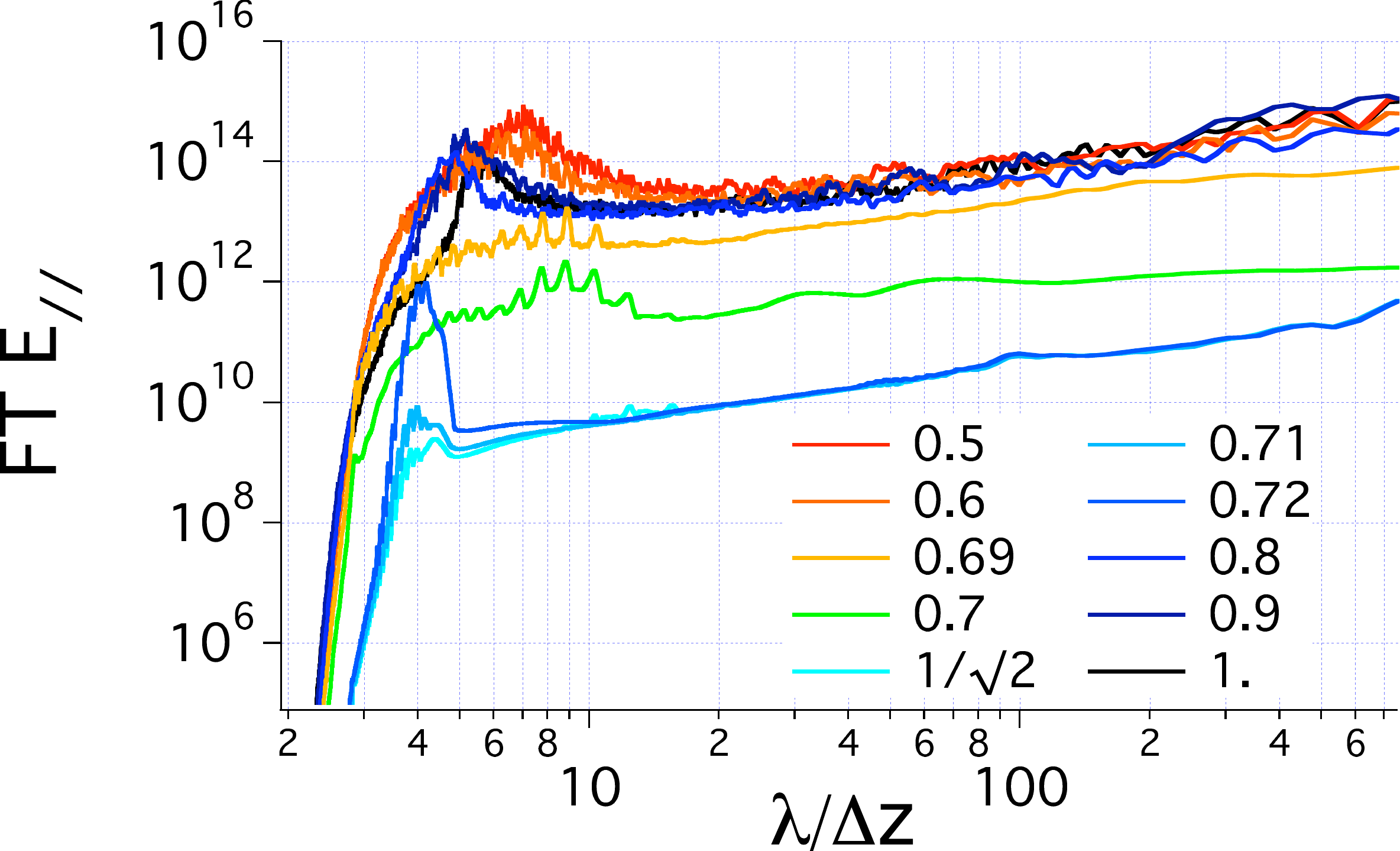} 
    \includegraphics*[width=60mm]{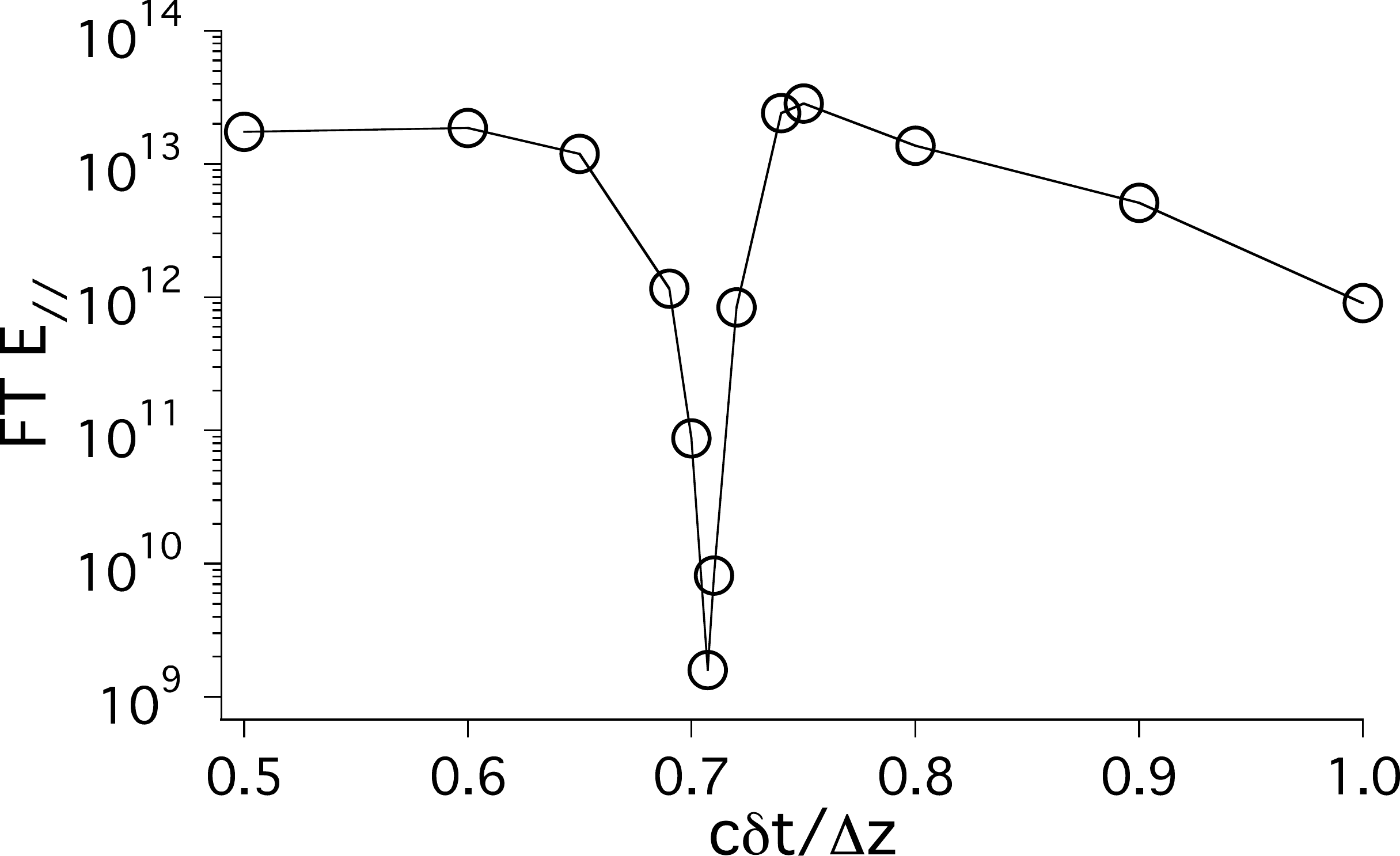} 
 %\end{tabular}
   \caption{Fourier Transform of the longitudinal electric field at t=40 ps, averaged over the whole domain, from 2-1/2D simulations  of a full scale 10GeV LPA in a boosted frame at $\gamma=130$, using the CK solver, for time steps between $c\delta t/\delta z=0.5$ and  $c\delta t/\delta z=1$, versus $\lambda/\delta z$ (left) and at $\lambda/\delta z=4$ (right).}
   \label{Fig_fft_scandt}
\end{figure}

It is striking that all the solvers that lead to the lowest levels of instability had the same CFL time step c$\delta t_{CFL}=\delta_z/\sqrt{2}$. For checking whether this is coincidental, simulations were performed using the CK solver, scanning the time step between c$\delta t/\delta z=0.5$ and  c$\delta t/\delta z=1$. The Fourier Transform of the longitudinal field averaged over the entire domain taken at $t=40$ ps, is given in Fig. \ref{Fig_fft_scandt}, exhibiting a sharp reduction of the instability level in a narrow band around c$\delta t=\delta_z/\sqrt{2}$. Since the numerical dispersion degrades in all directions when the time step diminishes, this indicates that the value of the time step value is of more importance than the numerical dispersion of the solver being used.

\begin{figure}[htb]
   \centering
 {\small
 \begin{tabular}{@{}c@{}c@{}} % @{} removes extra space
    \includegraphics*[width=65mm]{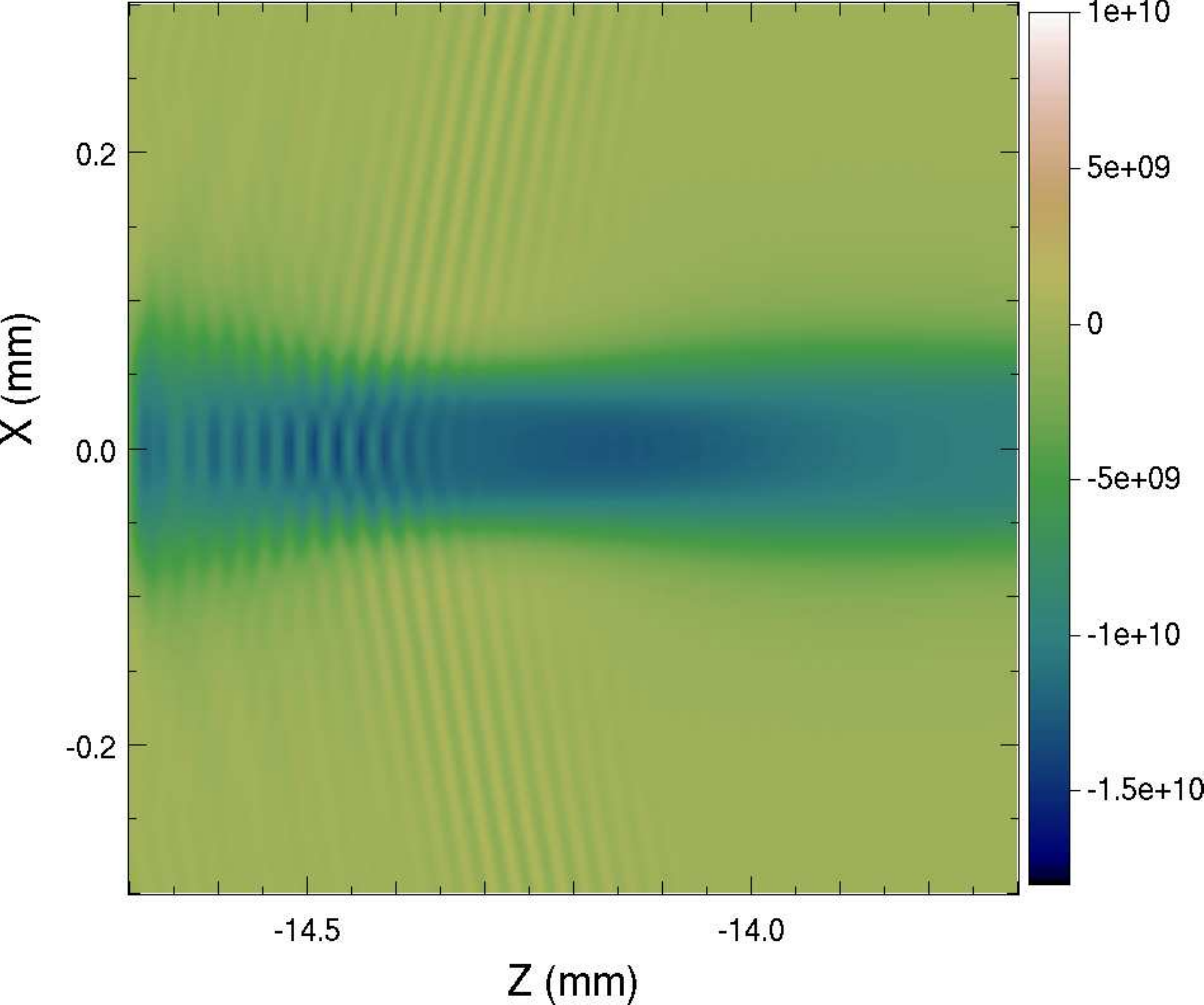} &
    \hspace{1mm}\includegraphics*[width=65mm]{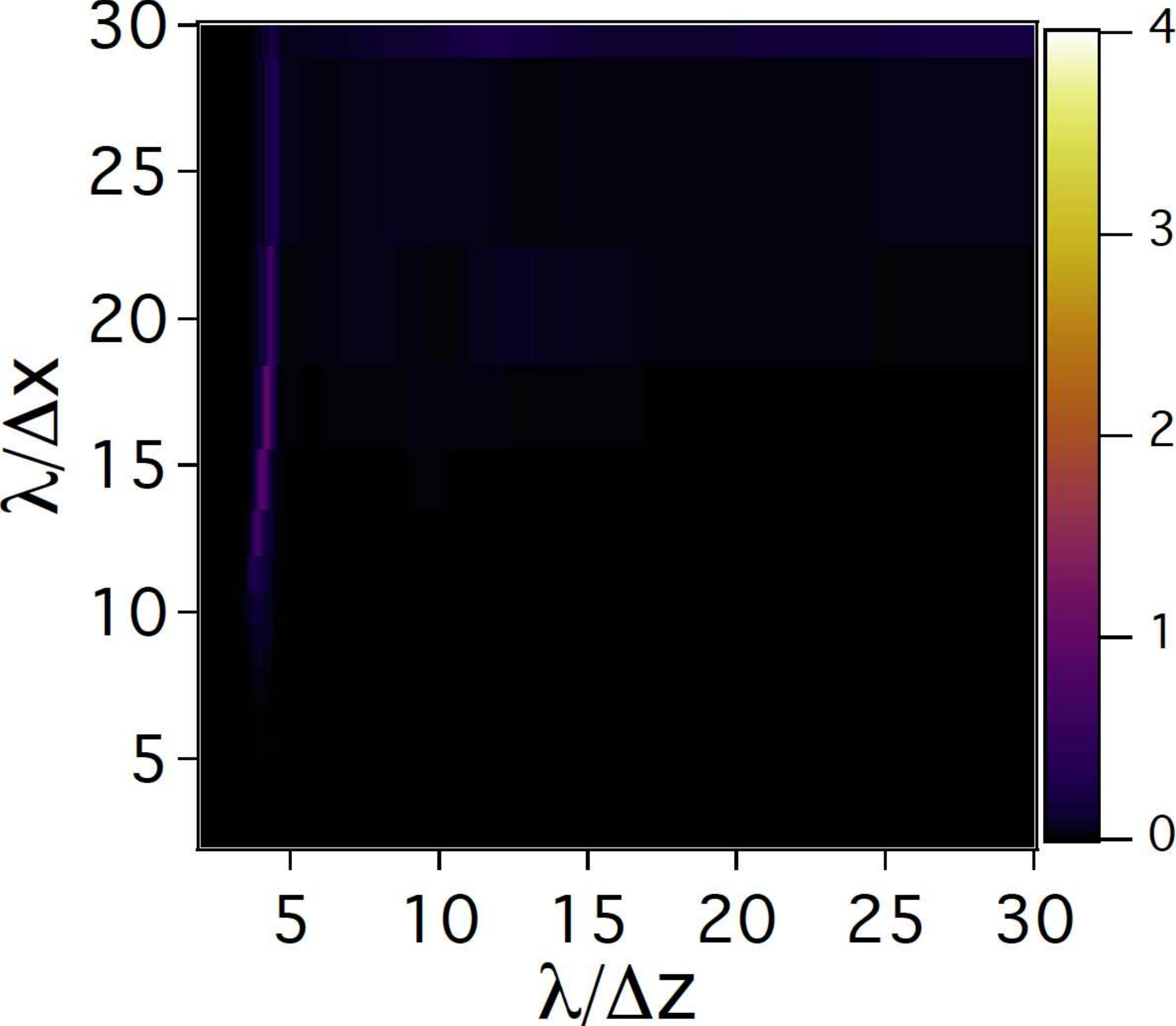} \\
    \includegraphics*[width=65mm]{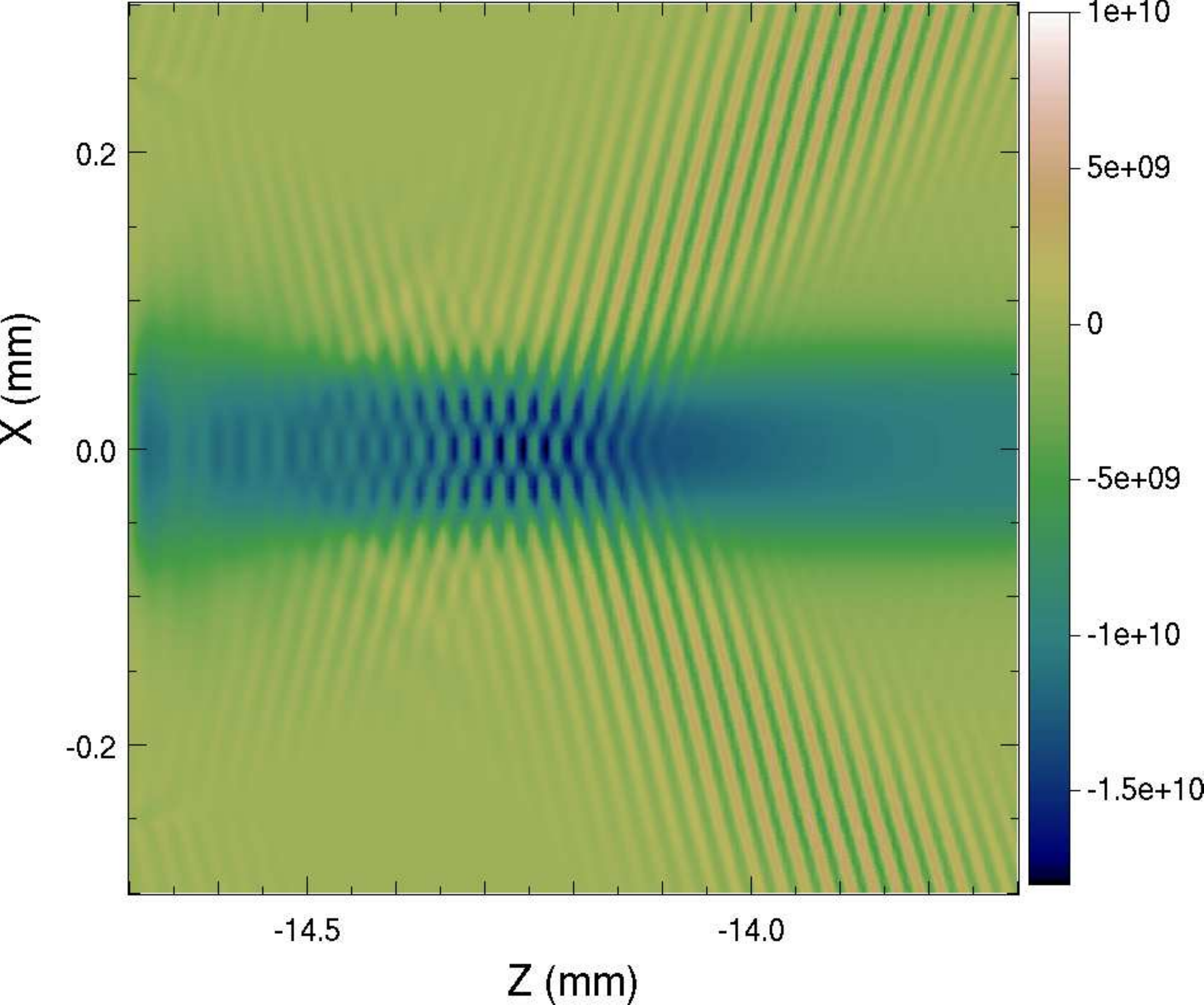} &
    \hspace{1mm}\includegraphics*[width=65mm]{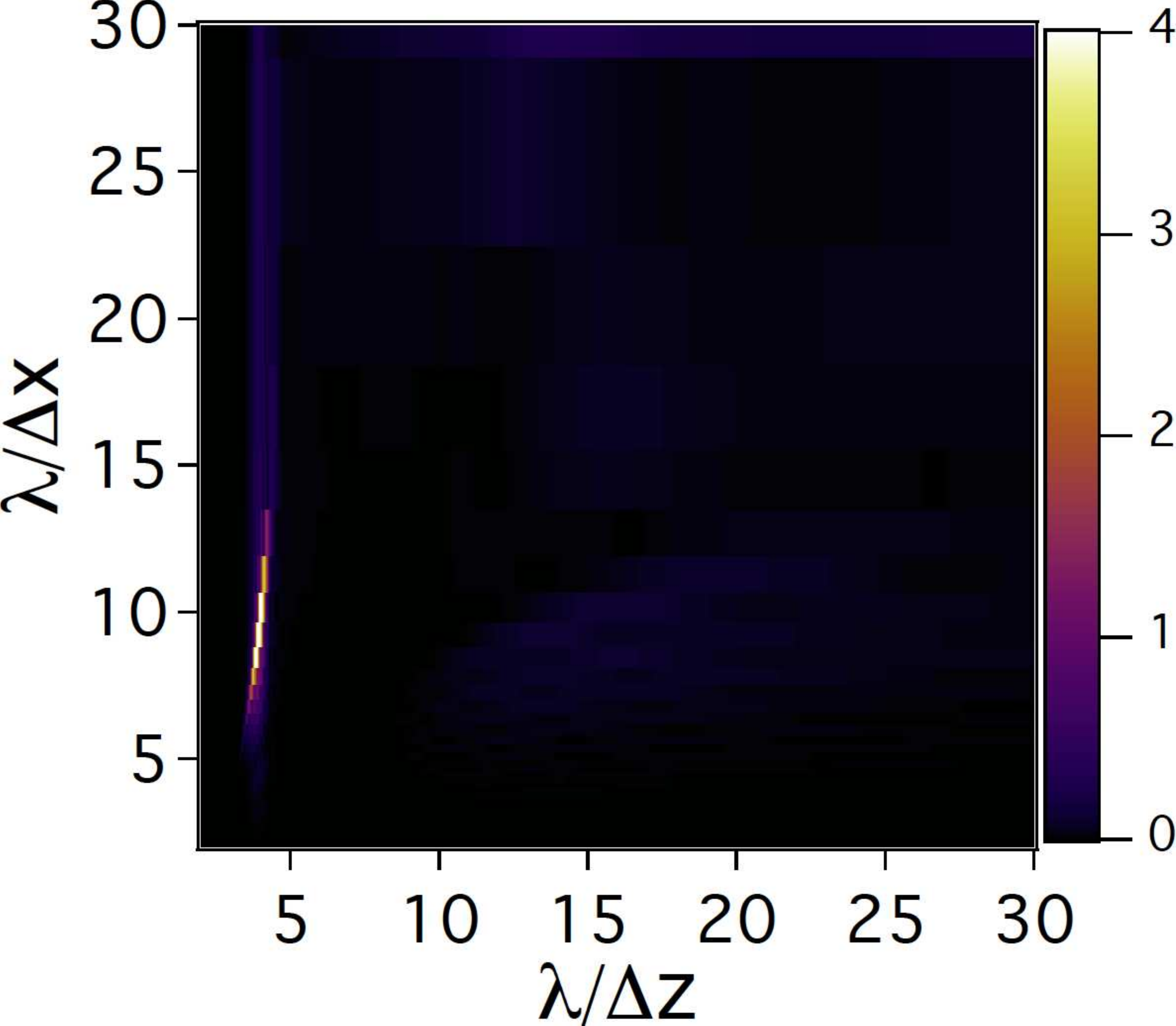} \\
 \end{tabular}
}
   \caption{(left) Snapshot of the longitudinal electric field ($E_{//}$) at the front of the plasma at $t=49$ ps; (right) Fourier Transform of the longitudinal electric field, from 2-1/2D simulations  of a full scale 10GeV LPA in a boosted frame at $\gamma=130$, using $\delta x=\delta z=6.5\mu m$, and the time step at the 2D CFL limit $c\delta t=\delta z/\sqrt{2}$, for (top) the Yee solver; (bottom) the CK solver.}
   \label{Fig_fft2dsqrt2}
\end{figure}

\begin{figure}[htb]
   \centering
 {\small
 \begin{tabular}{@{}c@{}c@{}} % @{} removes extra space
    \includegraphics*[width=65mm]{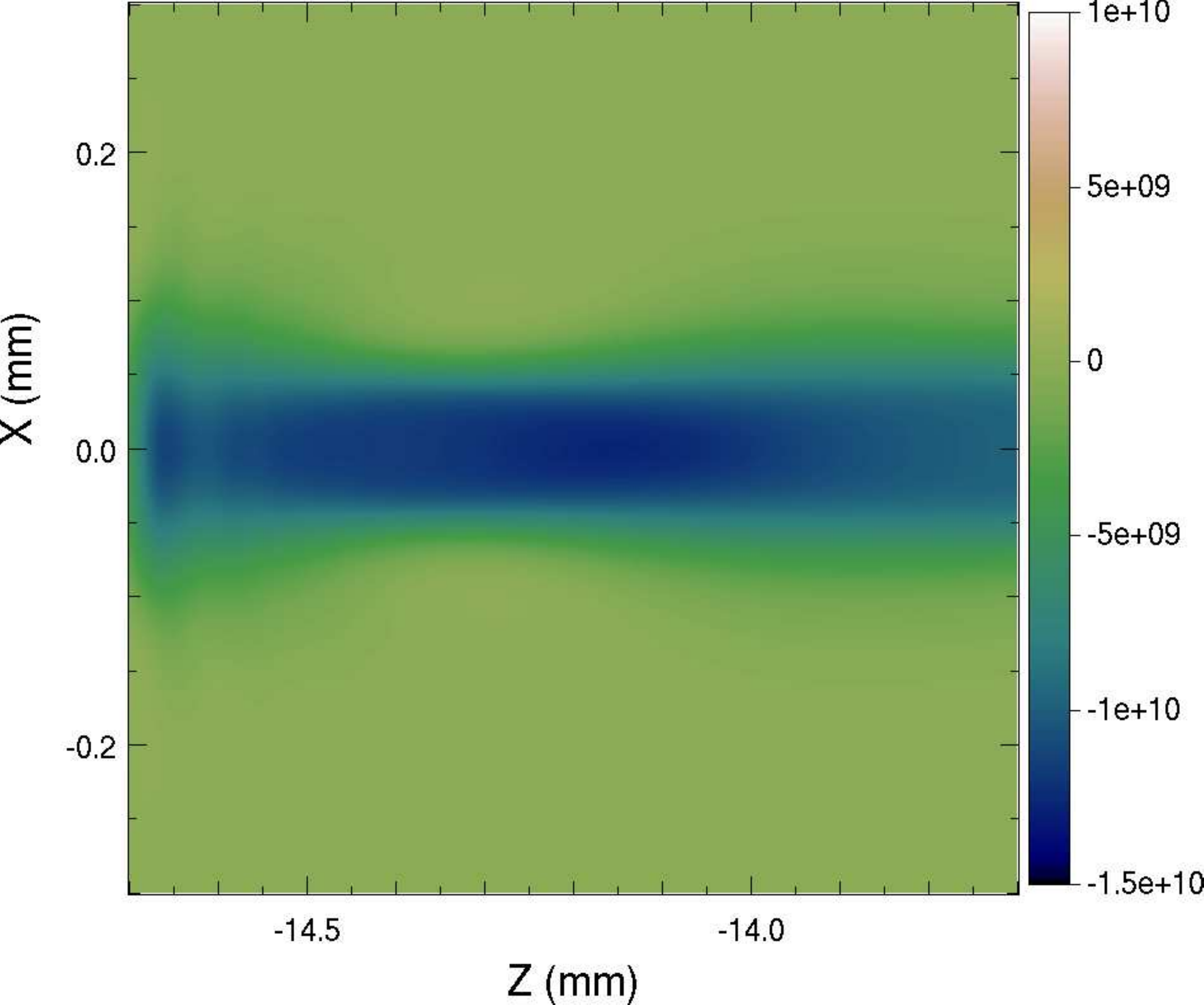} &
    \includegraphics*[width=65mm]{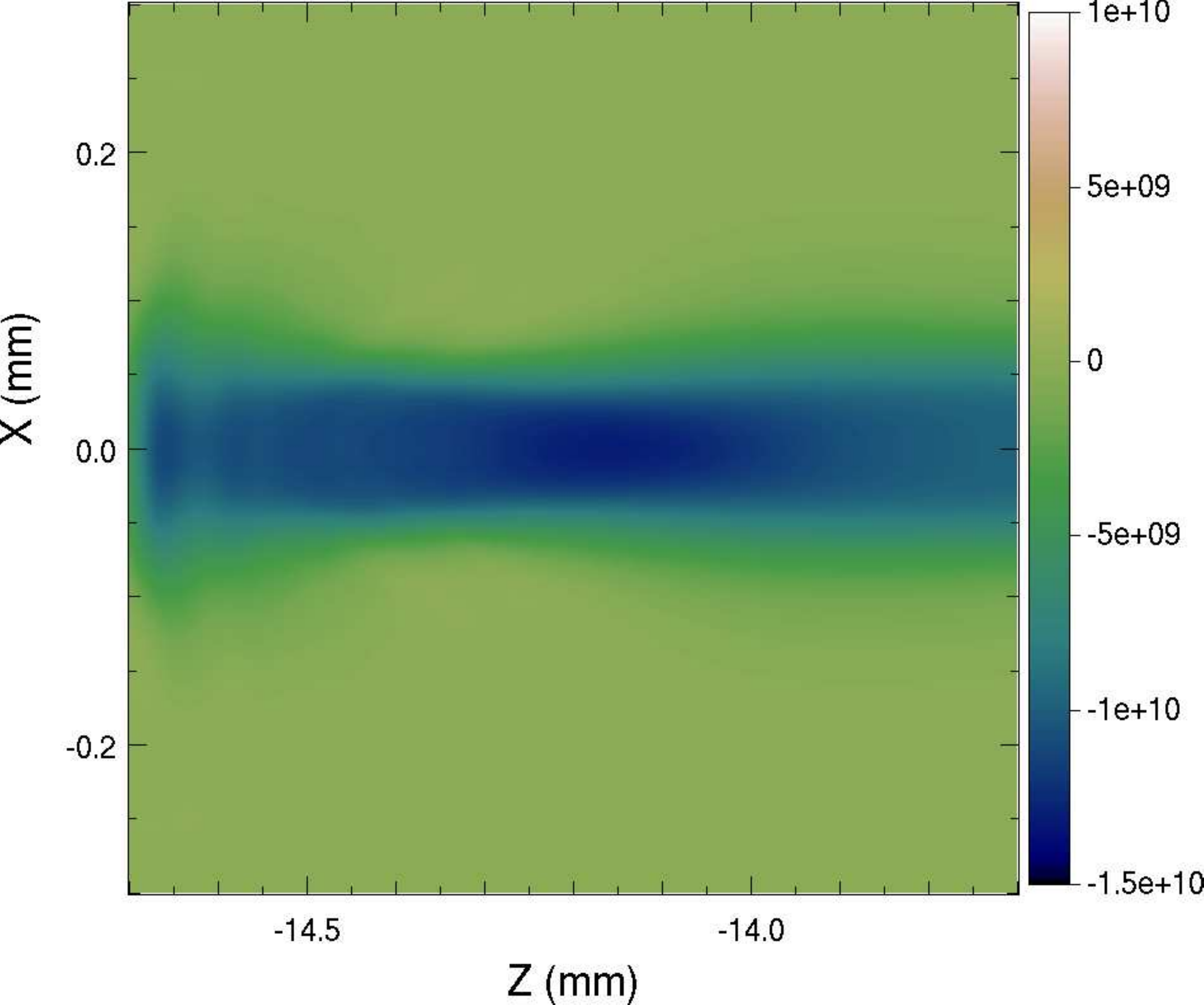} \\
 \end{tabular}
}
   \caption{Snapshot of the longitudinal electric field ($E_{//}$) at the front of the plasma at $t=49$ ps from 2-1/2D simulations of a full scale 10GeV LPA in a boosted frame at $\gamma=130$, using $\delta_x=\delta_z=6.5\mu m$, and the time step at the 2D CFL limit $c\delta t=\delta z/\sqrt{2}$, for (left) the Yee solver; (right) the CK solver. The filter S(1:2) was used to remove the instability that is visible in Fig. \ref{Fig_fft2dsqrt2}. The remaining feature is the wake.}
   \label{Fig_fft2dsqrt2_ksm2}
\end{figure}

Simulations using the Yee or the CK solver with the singular time step $c\delta t=\delta z/\sqrt{2}$ were performed and produced levels of instabilities that were much reduced (and delayed) compared to the 3D CFL time step (not shown). The snapshot of the electric field and its Fourier Transform taken at $t=49$ ps are given in Fig. \ref{Fig_fft2dsqrt2}. The Fourier spectrum is very similar in each case, although the instability is slightly stronger with the CK solver than with the Yee solver. In both cases, the instability is easily removed by using the S(1:2) filter (see Fig. \ref{Fig_fft2dsqrt2_ksm2}).

As mentioned in the previous section, the solvers CK, CK4 and CK5, which all have a CFL time step above the  singular time step c$\delta t=\delta z/\sqrt{2}$, produced significant levels of instability when run at their CFL limit. It was verified that using those solvers in 3D at the time step $c\delta t=\delta z/\sqrt{2}$ resulted in  greatly reduced levels of instability. It was also observed that running the Yee solver using non-cubic cells, e.g. with lower resolution transversely such as $\delta x=2\delta z$ at $\gamma=130$, or $\delta x=2.6\delta z$ at $\gamma=50$, produced the same pattern: a significant instability was present when using the CFL time step and was greatly reduced by using  $c\delta t=\delta z/\sqrt{2}$. Hence for the suppression of the instability, the choice of the solver seems to depends solely on whether its CFL condition allows stability at the special time step $c\delta t=\delta z/\sqrt{2}$ for a given grid cell aspect ratio, but not significantly on its numerical dispersion nor on the value of the grid cell aspect ratio.

\begin{figure}[htb]
   \centering
% \begin{tabular}{@{}c@{}c@{}} % @{} removes extra space
    \includegraphics*[width=68mm]{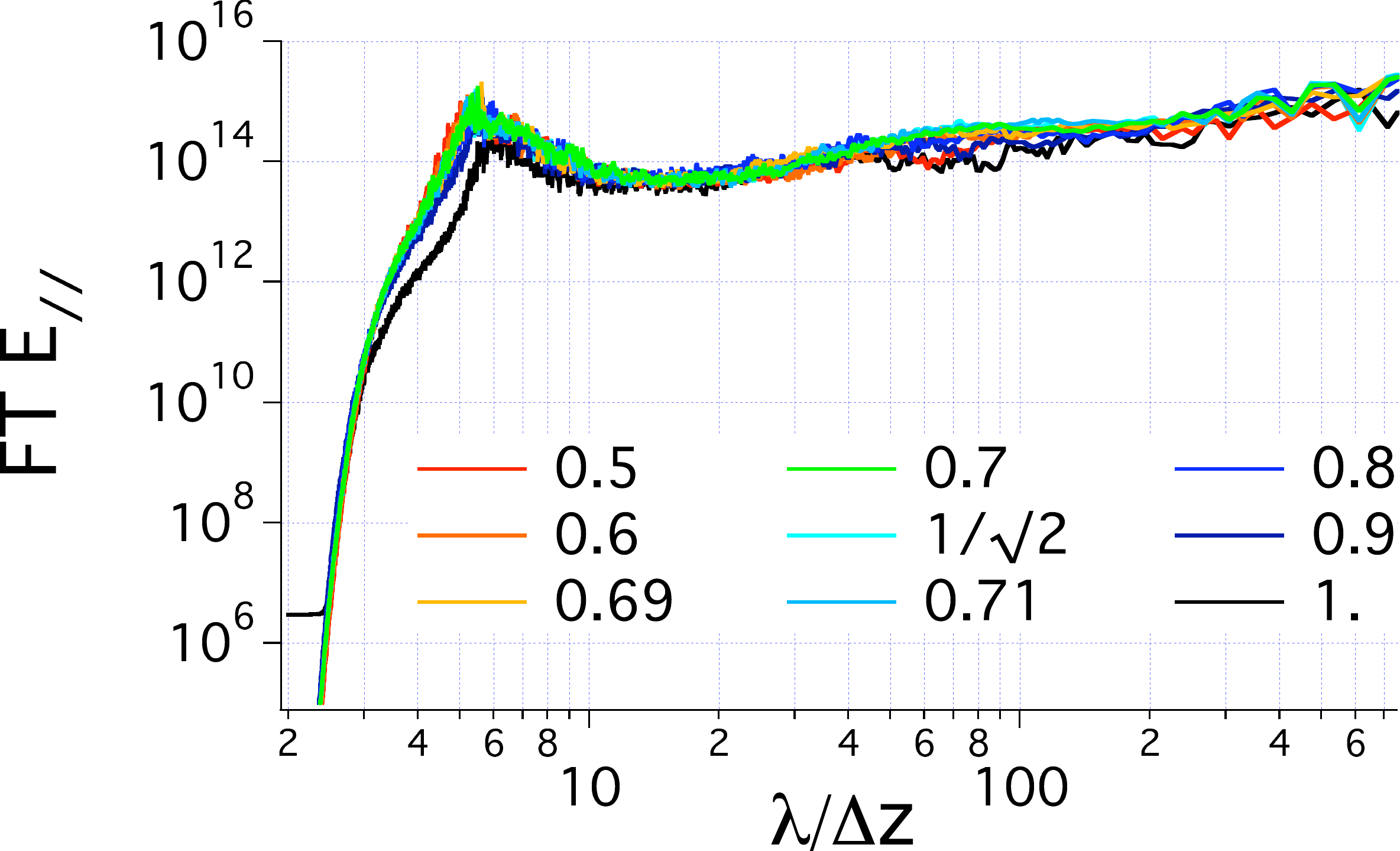} 
 %\end{tabular}
   \caption{Fourier Transform of the longitudinal electric field at t=40 ps, averaged over the whole domain, from 2-1/2D simulations  of a full scale 10GeV LPA in a boosted frame at $\gamma=130$, using the CK solver, for time steps between $c\delta t/\delta z=0.5$ and  $c\delta t/\delta z=1$, using a 'momentum conserving' field gathering scheme.}
   \label{Fig_fft_scandt_efetch1}
\end{figure}

\subsubsection{Effects of field gathering procedure}
The scan of time step was repeated using the 'momentum conserving' procedure \cite{BirdsallLangdon} , in which the field values are interpolated at the grid nodes before being gathered onto the particles. The result is given in Fig. \ref{Fig_fft_scandt_efetch1}. With the momentum conserving procedure, the level of instability is consistently high and independent of the time step. Since the numerical dispersion of the solver varies substantially with the time step, this result supports the conclusion that the instability may not be of numerical Cerenkov nature. The identification of the nature of the instability and the explanation of the singular time step $c\delta t_S$ call for a multidimensional (no instability was observed in 1D regardless of the field gathering method) analysis of the discretized Vlasov algorithm that was employed, which is left for future work.

The results that were obtained lead to the following conclusions: (i) the time step c$\delta t_S=\delta z/\sqrt{2}$ consistently produces the lowest levels of instability, regardless of dimensionality (2D vs 3D), the field solver being used, resolution, aspect ratio of cells (within the range of the finite number of cases that were experimented); (ii) the main advantage of the tunable field solver resides in allowing access to the singular time step $c\delta t_S$ rather than providing improved numerical dispersion, which consequently do not appear to be a primary driver of the instability; (iii) the instability is not completely removed at $c\delta t_S$ and filtering is still needed, albeit at lower levels; (iv) the field gathering procedure is key, as the existence of a singular time step at which the instability is greatly reduced is observed using an 'energy conserving' procedure, but not using a 'momentum conserving' procedure. These results indicate that the instability that is being observed may not be a type of numerical Cerenkov instability, as originally conjectured. 

\clearpage
\subsection{Full scale 100 GeV - 1 TeV class stages}

\begin{figure}[htb]
   \centering
 {\small
 \begin{tabular}{@{}c@{}c@{}} % @{} removes extra space
% \begin{tabular}{|@{}c@{}|@{}c@{}|} % @{} removes extra space
%  \hline
%  Yee & Cole-Karkkainen \\
%  \hline
%    \includegraphics*[width=68mm]{Ehist_100GeV2D_newphase.pdf} &
    \includegraphics*[width=75mm]{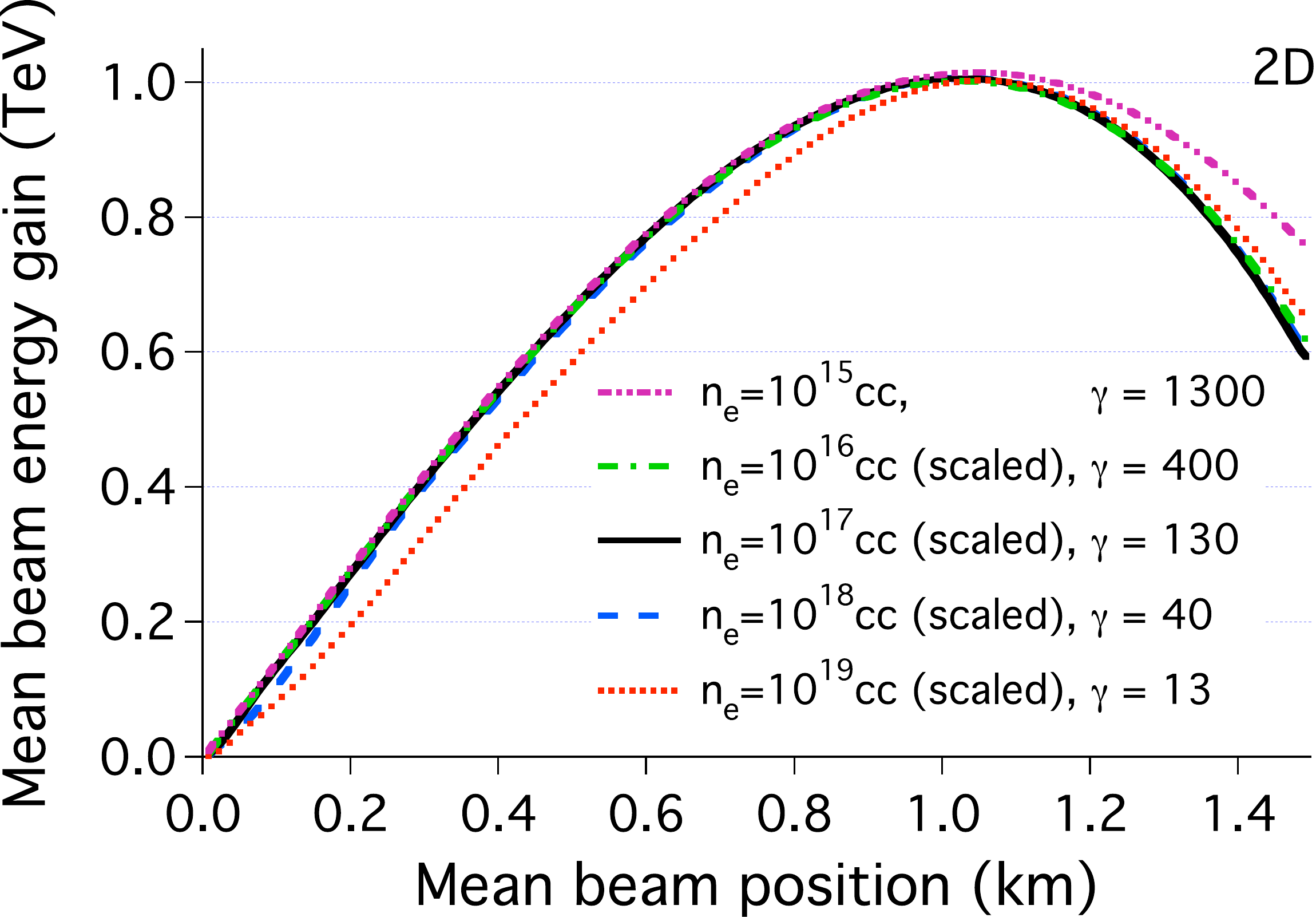} &
    \hspace{3mm}\includegraphics*[width=75mm]{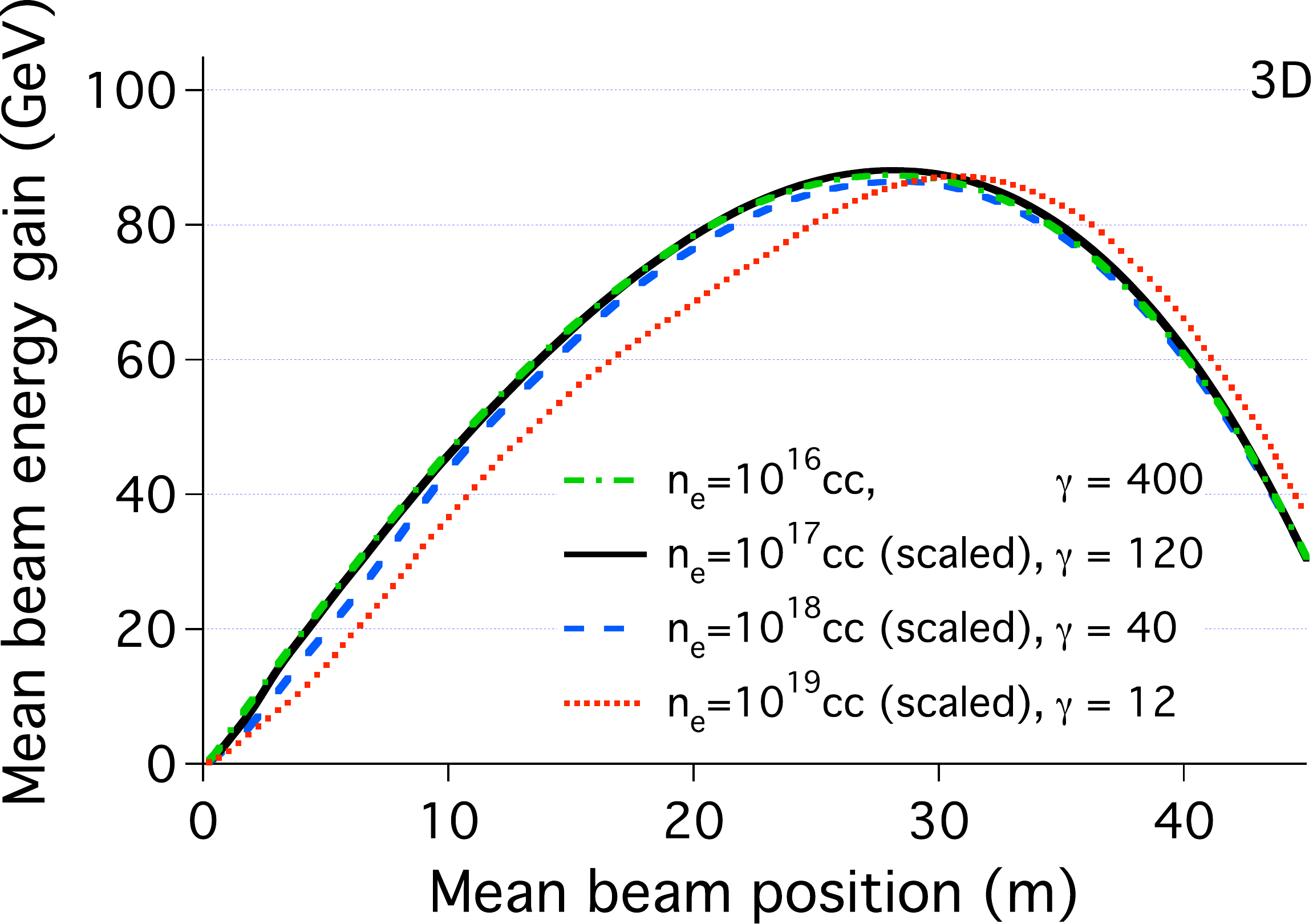} \\
\end{tabular}}
   \caption{Average beam energy gain versus longitudinal position (in the laboratory frame) for simulations at $n_e=10^{19}$ cc down to $10^{15}$ cc, using frames of reference between $\gamma=12$ and $\gamma=1300$, in 2-1/2D (left) and 3D (right).}
   \label{Fig_ehist100GeV}
\end{figure}

Using the knowledge acquired from the 10 GeV class study, simulations 
of stages in the range of 0.1 GeV-1 TeV were performed in 2-1/2D and in
the range of 0.1-100 GeV in 3D.  The plasma density $n_e$ scales inversely to the energy
gain, from $10^{19}$ cc down to $10^{15}$ cc in the 0.1 GeV-1 TeV range. These simulations used the parameters given in Table \ref{Tablephyspar2d} scaled
appropriately, and used the high speed of the boosted simulations to allow fast-turnaround improvement of the stage design  \cite{CormierAAC08,GeddesPAC09}. Scaled energy gain was increased by adjusting the  phase
of the beam injection behind the laser by $\sim12\%$ in 3D and 7$\%$ in
2D, with respect to the results presented in the preceding section. The 5$\%$
level difference between the 2D and 3D beam phases is likely due to small
differences in wake structure, laser depletion, and the small number of betatron oscillations of the laser. To minimize beam loss, the beam dimensions
were reduced by a factor of 3 in each dimension.   Simulations showing performance of this design in 2-1/2D
were performed using the Yee solver with filter S(1) for the 0.1-10 GeV runs,
S(1:2) for the 100 GeV and S(1:2:3) for the 1 TeV ones. The 3D simulations
were performed using the CK2 solver with filter S(1) for the 0.1-1 GeV runs,
and S(1:2) for the 10-100 GeV ones. The average beam energy gain history
is plotted in Fig. \ref{Fig_ehist100GeV}, scaling the 0.1-100 GeV runs to the 1 TeV range in
2-1/2D, and the 0.1-10 GeV runs to the 100 GeV range in 3D. The results
exhibit an excellent agreement on the peak scaled beam energy gain between
0.1-100 GeV runs, and on the scaled beam energy gain histories between the
1-100 GeV runs. A higher level of smoothing was needed for the 1TeV case,
explaining the deviation past 1 km. This deviation is of little importance in
practice, where one is mostly interested in the beam evolution up-to the peak energy point. The differences at $10^{19}$ on the scaled beam energy gain history can be attributed to the effects from having only a few laser oscillations per pulse.

Using (\ref{Eq_scaling1d}), the speedup of the full scale 100 GeV class run, which used a boosted frame of $\gamma=400$ as frame of reference, is estimated to be over 100,000, as compared to a run using the laboratory frame. Assuming the use of a few thousands of CPUs, a simulation that would require several decades to complete using standard PIC techniques in the laboratory frame, was completed in four hours using 2016 CPUs of the Cray system at NERSC. With the same analysis, the speedup of the 2-1/2D 1 TeV stage is estimated to be over a million.
%, which was verified by extrapolating runtimes from a few hundreds of time steps of a simulation in the laboratory frame.

\clearpage
\section{Conclusion and outlook}
The technique proposed in \cite{VayPRL07} was applied successfully to speedup by orders of magnitude calculations of laser-plasma accelerators from first principles. The theoretical speedup estimate from \cite{VayPRL07} was improved, while complications associated with the handling of input and output data between a boosted frame and the laboratory frame were discussed. Practical solutions were presented, including a technique for injecting the laser that is simpler and more efficient than methods proposed previously.

Control of an instability that was limiting the speedup of such calculations in previous work is demonstrated, via the use of a field solver with tunable coefficients and digital filtering. The tunable solver was shown to be compatible with existing "exact" current deposition techniques for conservation of Gauss Law, and accommodates Perfectly Matched Layers for efficient absorption of outgoing waves.

Extensive testing of the methods presented for numerical Cerenkov mitigation reveals that choosing the frame of the wake as the frame of reference allows for higher levels of filtering and damping than is possible in other frames with the same accuracy. It also revealed that there exists a singular time step for which the level of instability is minimal, independently of other numerical parameters, especially the numerical dispersion of the solver. This indicates that the observed instability may not be caused by numerical Cerenkov effects. Analysis of the nature of the instability is underway, but regardless of cause, the methods presented mitigate it effectively. The tunability of the field solver is key in providing stability in 3D at the singular time step, which is not attainable by the standard Yee solver.

The use of those techniques permitted the first calculations in the optimal frame of 10 GeV, 100 GeV and 1 TeV class stages, with speedups over 4, 5 and 6 orders of magnitude respectively over what would be required by "standard" laboratory frame calculations, which are impractical for such stages due to computational requirements. 

These results show that the technique can be applied to the modeling of 10 GeV stages, and future work will include the effects of beam loading, plasma density ramps, as well as particle trapping in the near future. Future work on the numerical methods include a comprehensive analysis of the instability and the existence of a singular time step under certain conditions, as well as the local application of filtering, smoothing and/or mesh refinement \cite{VayPoP04,VayCPC04} around the front of the plasma, where the instability develops. The latter is expected to provide mitigation of the instability while preserving accuracy in the core of the simulation.

\section{Appendix I: One dimensional analysis of the CK solver}
Although the most interesting applications of the CK solver require two or three dimensions, analysis of the method in one dimension reveals a potential issue when $c\delta t=\delta x$. In one dimension (choosing $x$), Equations (\ref{Eq:Faraday})-(\ref{Eq:Ampere}) reduce to
\begin{eqnarray}
B_y|_{i+1/2}^{n+1/2}&=&B_y|_{i+1/2}^{n-1/2}+\frac{\delta t}{\delta x}\left(E_z|_{i+1}^n-E_z|_i^n\right) \label{Eq:1db}\\
E_z|_i^{n+1}&=&E_z|_i^n+\frac{c^2\delta t}{\delta x}\left(B_y|_{i+1/2}^{n+1/2}-B_y|_{i-1/2}^{n+1/2}\right)-\frac{J_i^n}{\epsilon_0}\label{Eq:1de}
\end{eqnarray}

\begin{figure}[htb]
   \centering
{\small
 \begin{tabular}{@{}c@{}c@{}} % @{} removes extra space
%  \hline
  \hspace{0.7cm} Heaviside step excitation & \hspace{1.cm} oscillatory excitation  \vspace{2.mm}\\
   \includegraphics*[width=65mm]{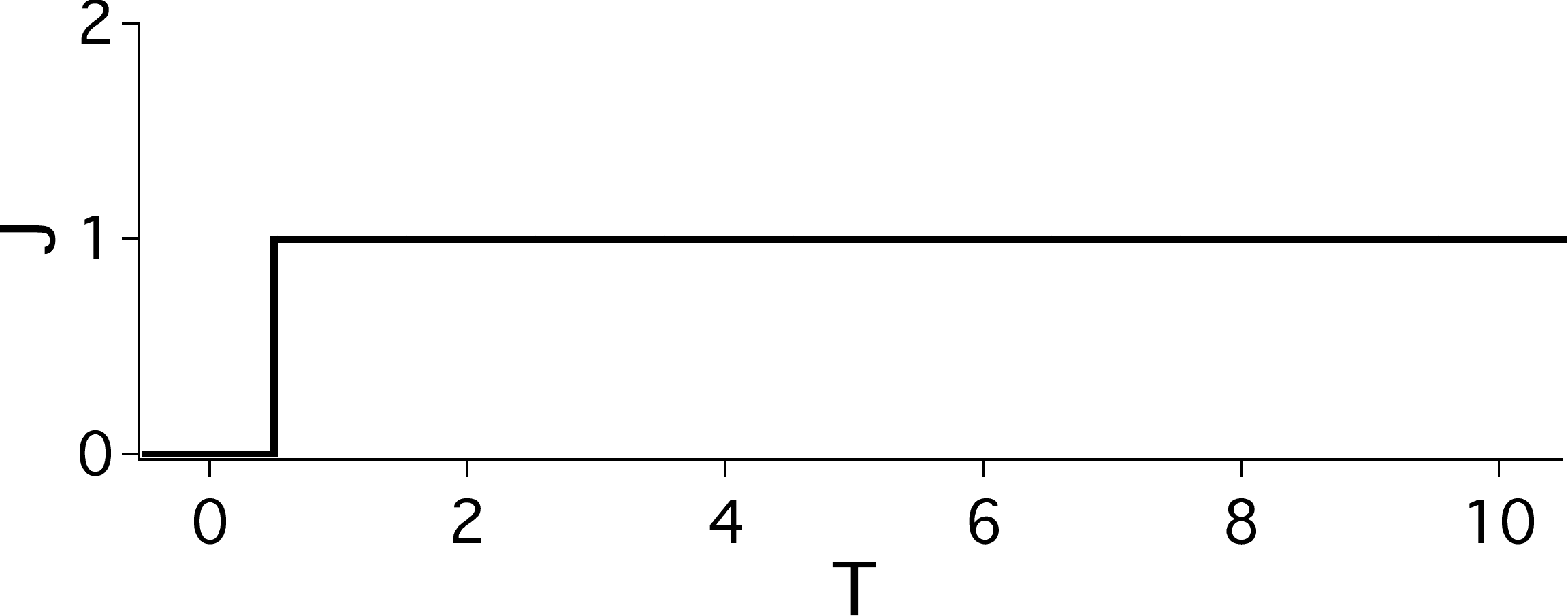} \hspace{1mm}&
   \includegraphics*[width=65mm]{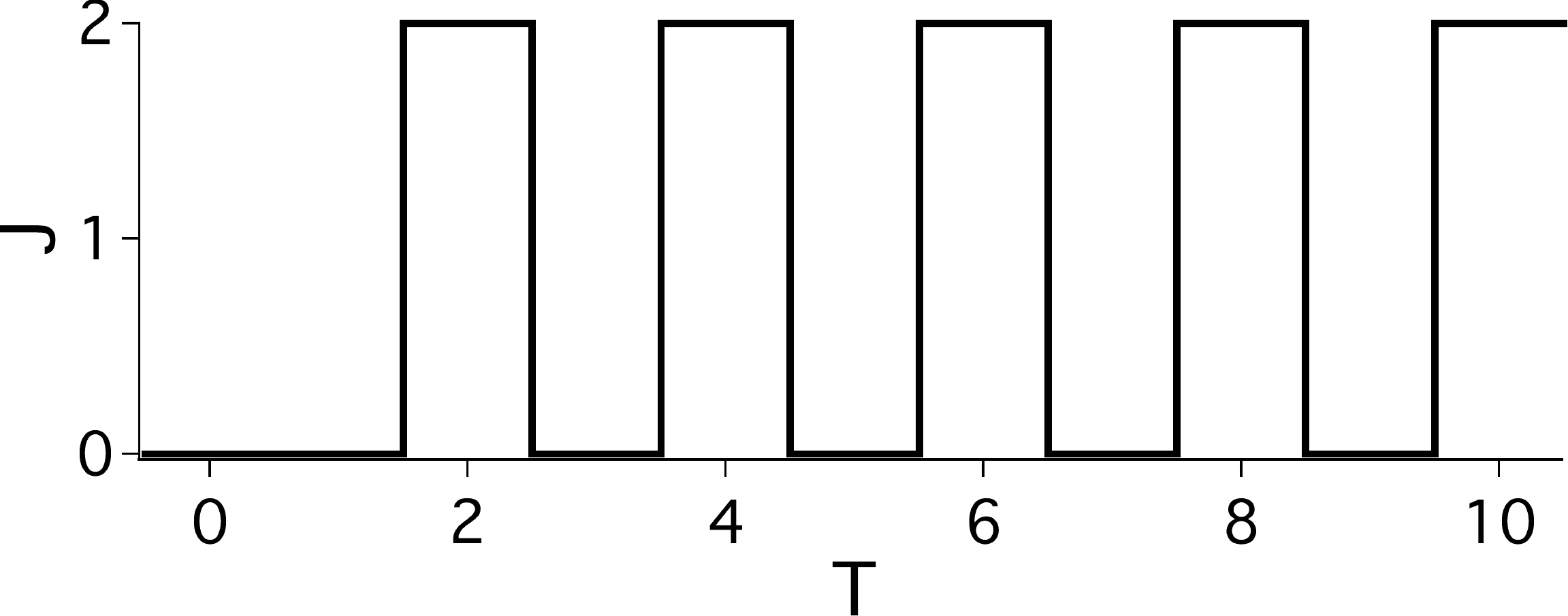}\\
   \includegraphics*[width=65mm]{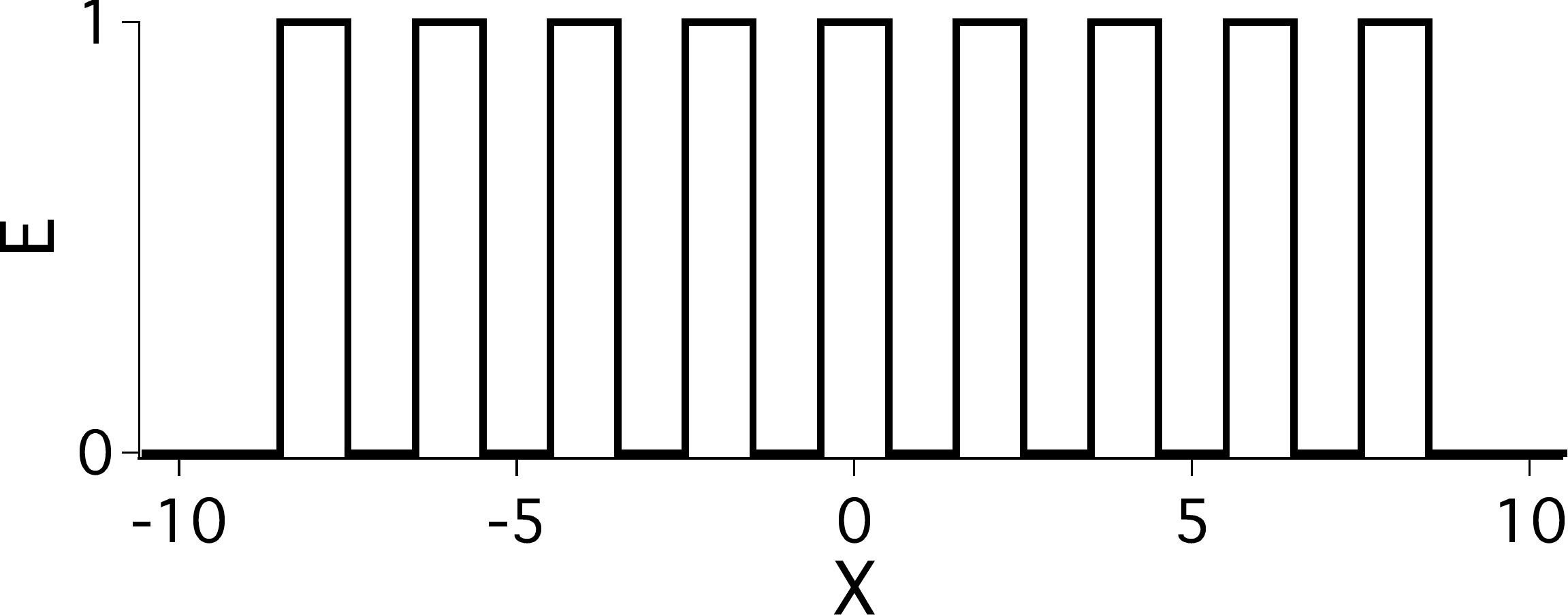}&
   \includegraphics*[width=65mm]{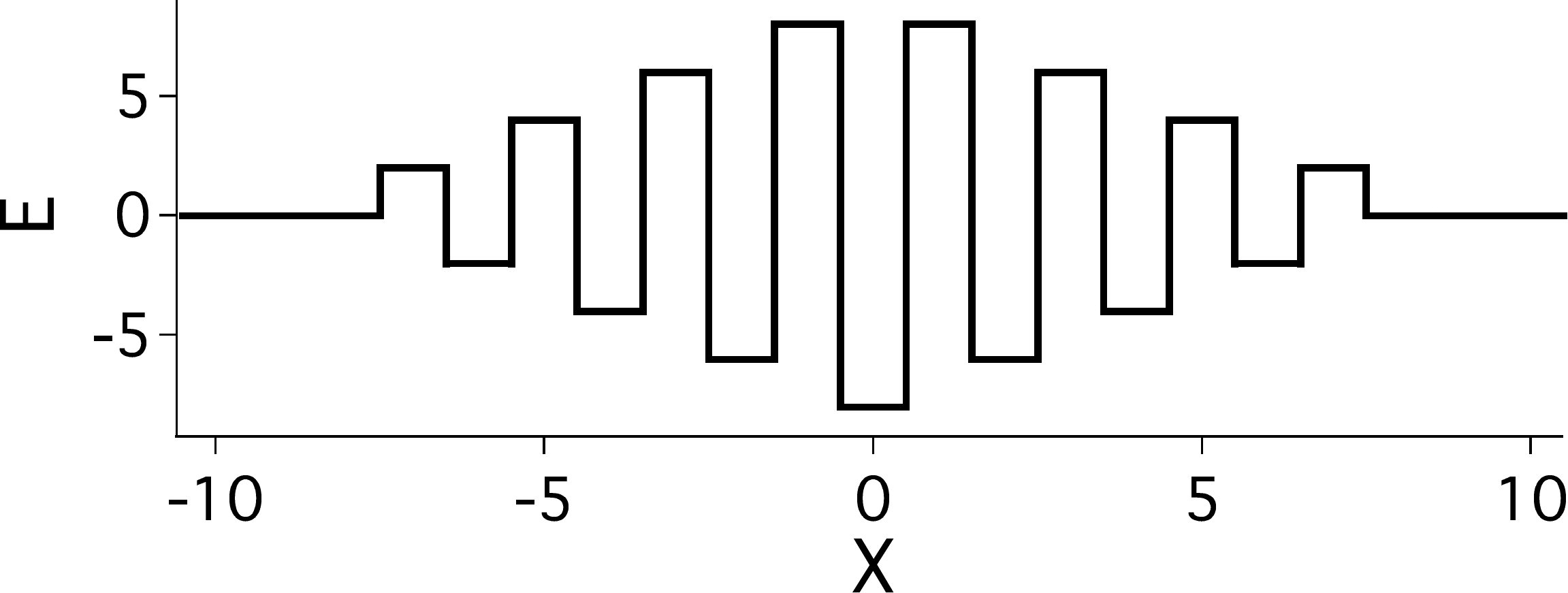}\\
   \includegraphics*[width=65mm]{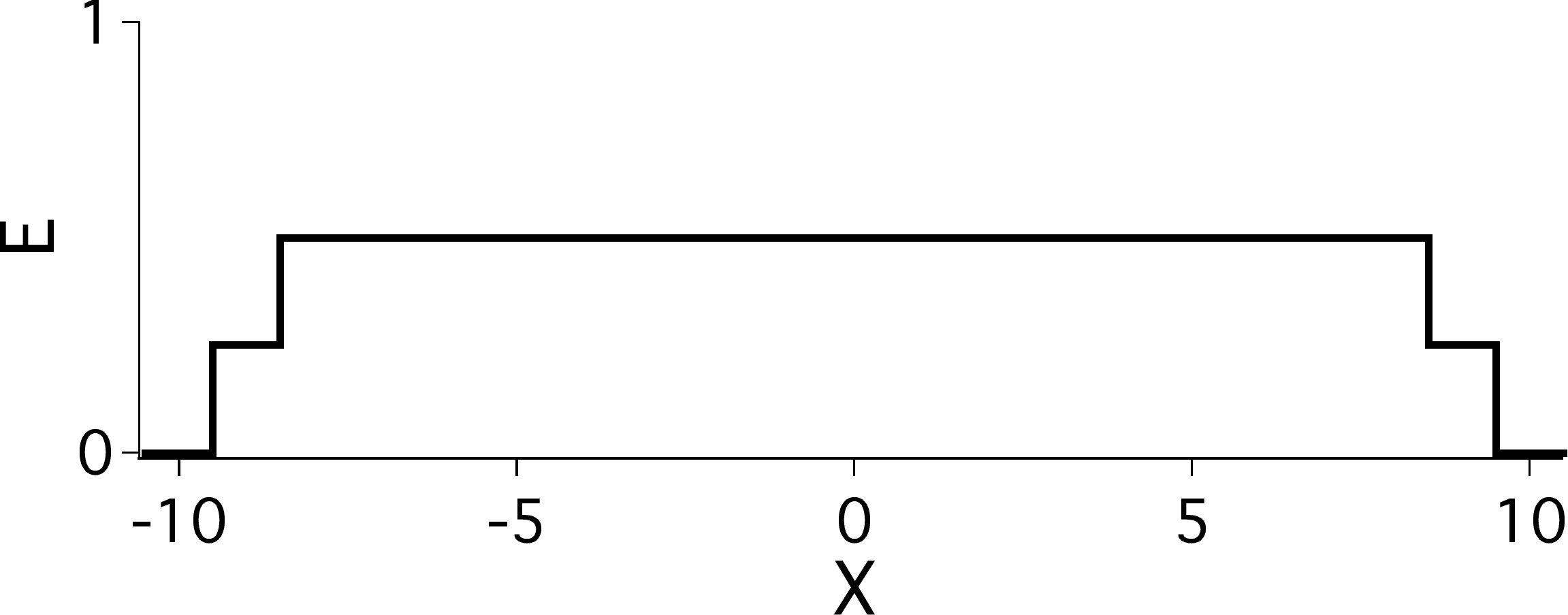}&
   \includegraphics*[width=65mm]{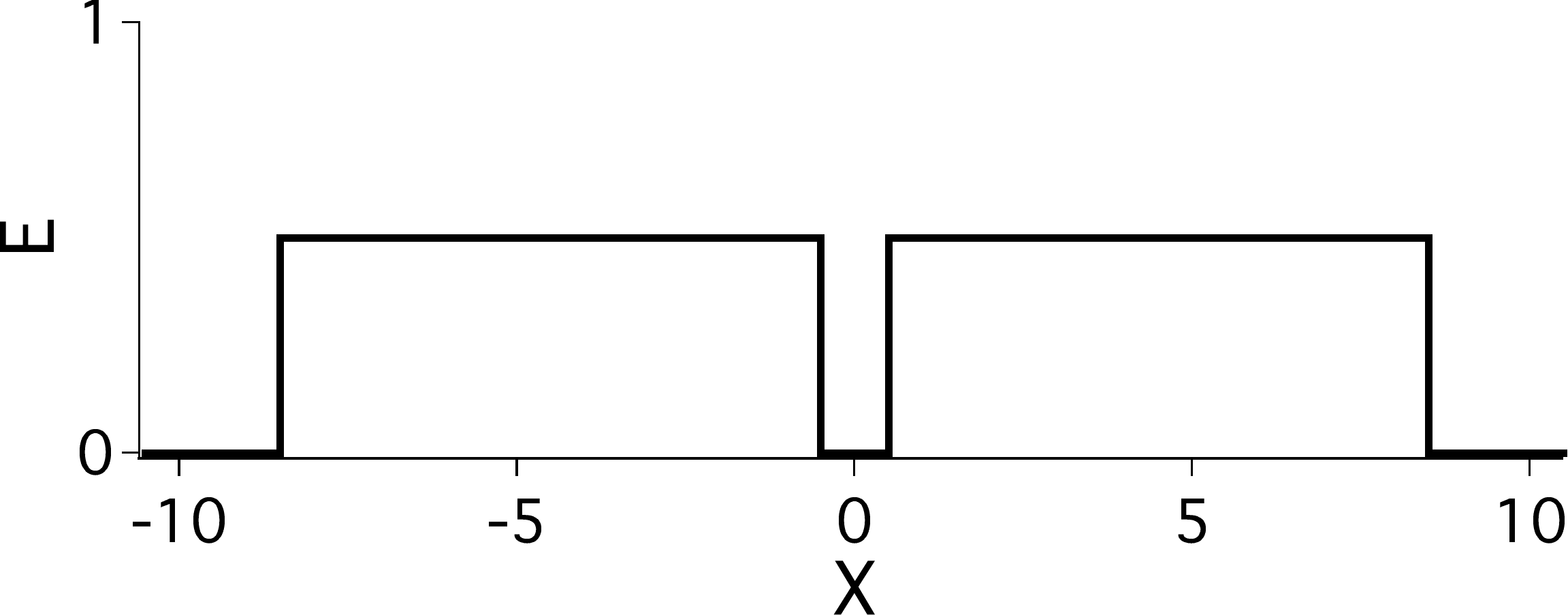}
\end{tabular}
}
   \caption{(top) time history (in time steps) of the current source for (left) a Heaviside step (right) a heaviside step modulated by a sinusoidal oscillation at the Nyquist frequency; (middle) response of the system of equations (\ref{Eq:1db})-(\ref{Eq:1de}) via a snapshot of the electric field after 10 time steps, without filtering of the source term; (bottom) response of the system of equations (\ref{Eq:1db})-(\ref{Eq:1de}) with application of bilinear digital filter of the source term in space. A time step of c$\delta t=\delta x$ was used in all runs and scaled constants $c=\epsilon_0=1$ were assumed.}
   \label{Fig_1d}
\end{figure}

Due to uniform time discretization and linearity, the response of the system (\ref{Eq:1db})-(\ref{Eq:1de}) to arbitrary distributions and evolutions of sources (i.e. macro-particles) can be written as the sum of its response to the excitation from a Heaviside function in time, at one location in the grid. Assuming a source term of the form $J|_i^n=H(t)$ where $H$ is the Heaviside function, and setting the time step at the Courant limit $c\delta t=\delta x$, the system (\ref{Eq:1db})-(\ref{Eq:1de}) produces a spurious "odd-even" oscillations at the Nyquist frequency, as shown in Fig. \ref{Fig_1d} (middle-left). If a sinusoidal signal oscillating at the Nyquist frequency is added to the source term, the amplitude of the spurious oscillation grows linearly with time, as  shown in Fig. \ref{Fig_1d} (middle-right). The spurious oscillation is effectively suppressed in both cases by the application of a "1-2-1" bilinear digital filter, as shown in Fig. \ref{Fig_1d} (bottom) . These types of filtering are of common use in Particle-In-Cell codes, often repeated a prescribed number of times and followed by a compensation stage to avoid excessive damping of long wavelengths \cite{BirdsallLangdon}. 

\begin{table}[htd]
\caption{List of parameters for scaled 10GeV class LPA stage simulation.}
\begin{center}
\begin{tabular}{lcc}
\hline
\hline
%beam radius		 		&$R_b$		&$82.5$ nm\\
beam length		 		&$L_b$		&$85$ nm\\
beam peak density 			&$n_b$		&$10^{14}$ cm$^{-3}$\\
%beam transverse profile             &                		&$\exp\left(-r^2/8 R_b^2\right)$\\
beam longitudinal profile           &                		&$\exp\left(-z^2/2 L_b^2\right)$\\
\hline
laser wavelength 			&$\lambda$	& $0.8$ $\mu$m\\
%laser length (HWHM)				&$L$		& $3.36$ $\mu$m\\
%laser size 					&$\sigma$	& $8.91$ $\mu$m\\
laser length (FWHM)			&$L$		& $10.08$ $\mu$m\\
normalized vector potential	&$a_0$		& $1$\\
%laser transverse profile		&$$			& gaussian\\
%laser transverse profile		&$$			& $\exp\left(-r^2/2\sigma^2\right)$\\
%laser longitudinal profile		&$$			&sinusoidal\\
laser longitudinal profile		&$$			& $\sin\left(\pi z/L\right)$\\
\hline
plasma density on axis		&$n_e$		&$10^{19}$ cm$^{-3}$\\
%plasma transverse profile		&			& parabolic well\\
plasma longitudinal profile	&			& flat\\
plasma length				&$L$& $1.5$ mm\\
%plasma radius				&& \\
plasma entrance ramp profile	&			& half sinus\\
plasma entrance ramp length	&			& $4$ $\mu$m\\
\hline
number of cells				& $N_z$		& 952\\
cell size					& $\delta z$	& $\lambda/24$\\
time step					& $\delta t$	& $\delta z/c$\\
particle deposition order		&			& cubic\\
\# of plasma particles/cell		&			& 10\\
\hline
\hline
\end{tabular}
\end{center}
\label{Tablephyspar}
\end{table}%

\begin{figure}[htb]
   \centering
{\small
 \begin{tabular}{@{}c@{}c@{}} % @{} removes extra space
%  \hline
  \hspace{1.cm} No filtering & \hspace{1.cm} bilinear filtering  \vspace{2.mm}\\
   \includegraphics*[width=65mm]{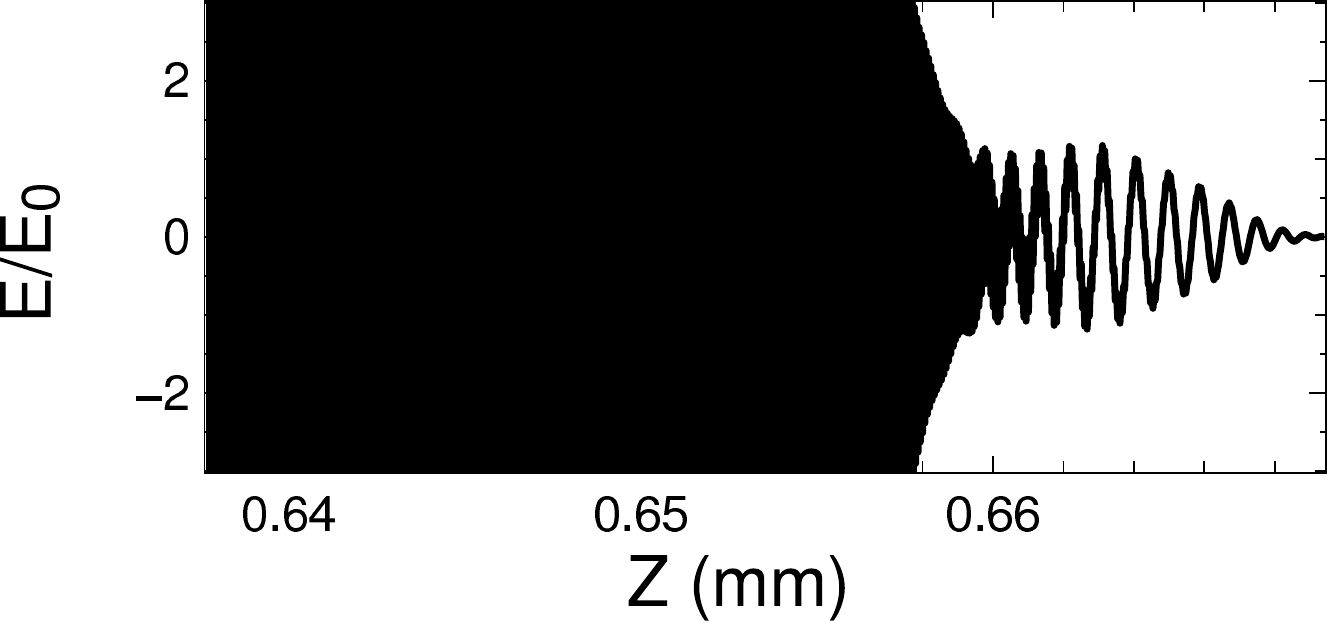} \hspace{2.mm}&
   \includegraphics*[width=65mm]{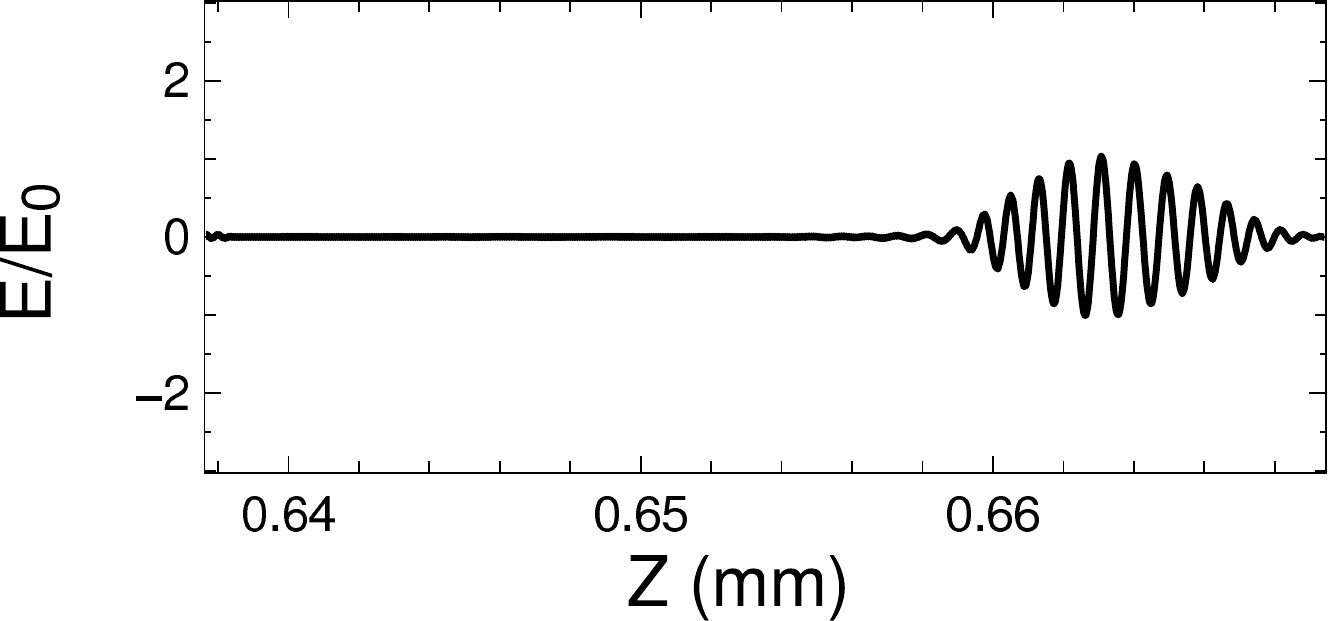}\\
   \includegraphics*[width=65mm]{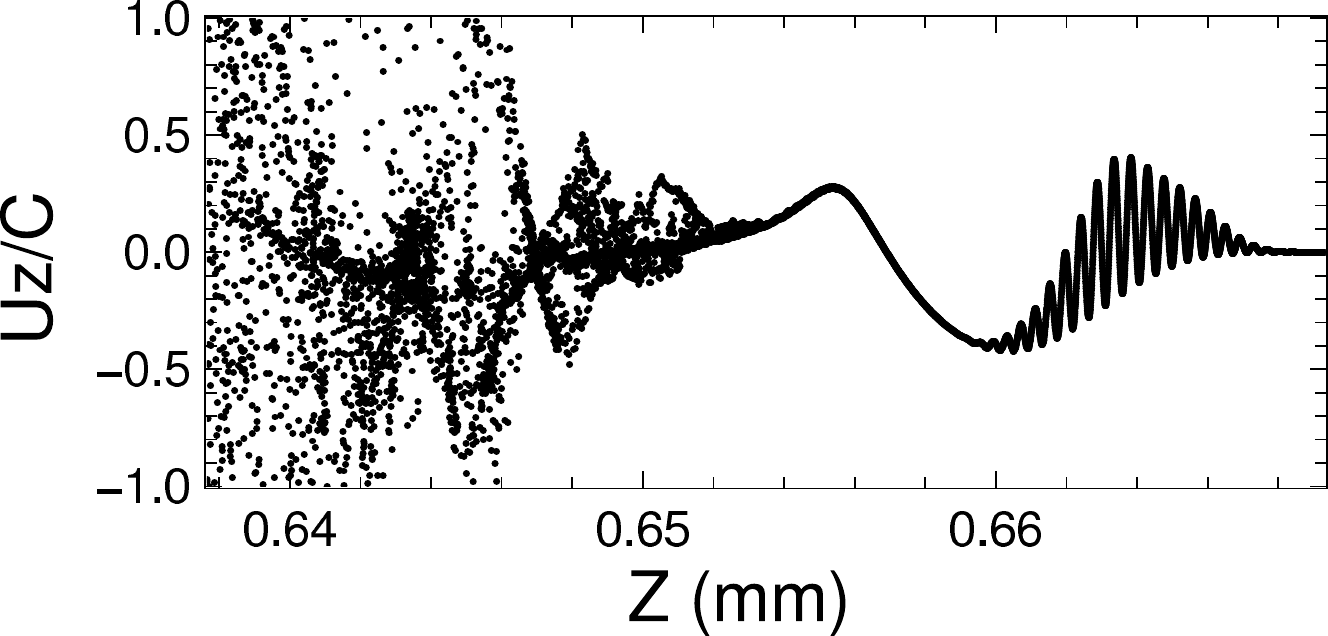} &
   \includegraphics*[width=65mm]{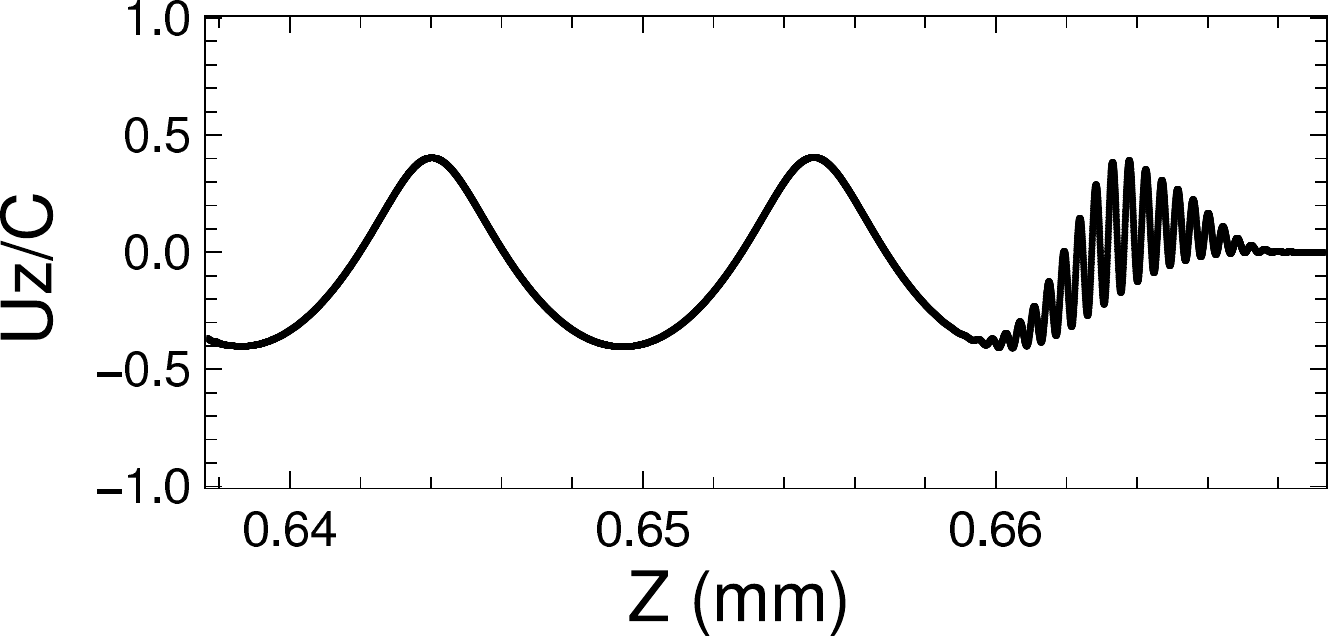}
 \end{tabular}}
   \caption{Snapshots from transverse electric field (normalized to maximum laser amplitude $E_0$) and plasma electrons longitudinal phase space projection, from a 1D simulation of a laser wakefield acceleration stage the CFL limits ($c\delta t=\delta x$) with (left) no filtering of current density; (right) application of a bilinear digital filter to the current density.}
   \label{Fig_1dlwfa}
\end{figure}

The impact of the spurious oscillations and the effectiveness of the bilinear filtering at suppressing it in actual simulations was tested on a 1D simulation of a scaled wakefield acceleration stage. The physical and numerical parameters of the simulation are given in table \ref{Tablephyspar}. Snapshots of the transverse electric field (aligned with the laser polarization) and the plasma electron phase space, taken once the laser has propagated about half way through the plasma (after $\sim$20,000 time steps) are given in Fig. \ref{Fig_1dlwfa}. Without filtering of the current density, an instability develops at the grid Nyquist frequency, severely disrupting the plasma wake, despite the fact that cubic splines were used to deposit current from macro-particles to the grid and gather the electromagnetic field from the grid to the macro-particles. One application of the bilinear filtering (without compensation) is sufficient to suppress the spurious instability and produce a steady and clean wake.
 
\section{Appendix II: Perfectly Matched Layer}
The split form of Perfectly Matched Layer (PML) \cite{BerengerJCP94} framework applies readily to Eqs (\ref{Eq:Faraday})-(\ref{Eq:Ampere}). The equations on the component along $z$ of the magnetic field are given by
\begin{eqnarray}
\left(\Delta_t + \sigma_x\right) B_{zx} & = & -\Delta^*_x E_y \\
\left(\Delta_t + \sigma_y\right) B_{zy} & = & \Delta^*_y E_x \\
\left(\Delta_t + \sigma_x\right) E_y & = & -c^2\Delta_x \left(B_{zx}+B_{zy}\right) \\
\left(\Delta_t + \sigma_y\right) E_x & = & c^2\Delta_y \left(B_{zx}+B_{zy}\right) 
\end{eqnarray}
where $\sigma_x$ and $\sigma_y$ are the absorbing layer coefficients along $x$ and $y$ respectively. The equations for the other components of the magnetic field and for the electric field are obtained similarly, applying the standard difference operator on the spatial derivatives of the electric field and the enlarged difference operator on the spatial derivatives of the magnetic field. The formula to update the fields is obtained by solving the finite-difference equations or by integrating over one time step, giving
\begin{eqnarray}
B_{zx}|^{n+1/2}_{i+1/2,j+1/2,k} & = & \xi_x B_{zx}|^{n-1/2}_{i+1/2,j+1/2,k} - \frac{1-\xi_x}{\sigma_x}\Delta^*_x E_y|^{n}_{i+1/2,j+1/2,k} \\
B_{zy}|^{n+1/2}_{i+1/2,j+1/2,k} & = & \xi_y B_{zy}|^{n-1/2}_{i+1/2,j+1/2,k} + \frac{1-\xi_y}{\sigma_y}\Delta^*_y E_x|^{n}_{i+1/2,j+1/2,k} \\
E_y|^{n+1}_{i,j+1/2,k}  & = & \xi_x E_y|^{n}_{i,j+1/2,k}-c^2\frac{1-\xi_x}{\sigma_x}\Delta_x \left(B_{zx}+B_{zy}\right)|^{n+1/2}_{i,j+1/2,k} \\
E_x|^{n+1}_{i+1/2,j,k}  & = & \xi_y E_x|^{n}_{i+1/2,j,k}+c^2\frac{1-\xi_y}{\sigma_y}\Delta_y \left(B_{zx}+B_{zy}\right)|^{n+1/2}_{i+1/2,j,k}
\label{Eq:PML_stencil}
\end{eqnarray}
where $\xi=\left(1-\sigma\delta t/2\right)/\left(1+\sigma\delta t/2\right)$ via direct solve, or $\xi=e^{-\sigma\delta t}$ via time integration (note that in our tests, both implementations gave nearly identical results). 
 
\begin{figure}[htb]
   \centering
 {\small
 \begin{tabular}{@{}c@{}c@{}} % @{} removes extra space
% \begin{tabular}{|@{}c@{}|@{}c@{}|} % @{} removes extra space
%  \hline
  Yee & Cole-Karkkainen \\
%  \hline
%  \begin{tabular}{@{\hspace{.5mm}}c@{\hspace{.5mm}}|@{\hspace{.5mm}}c@{\hspace{.5mm}}}
  \begin{tabular}{@{\hspace{.5mm}}c@{\hspace{.5mm}}@{\hspace{.5mm}}c@{\hspace{.5mm}}}
    $\sigma$ & $\sigma^*$\\
 %   \hline
    \includegraphics*[width=33mm]{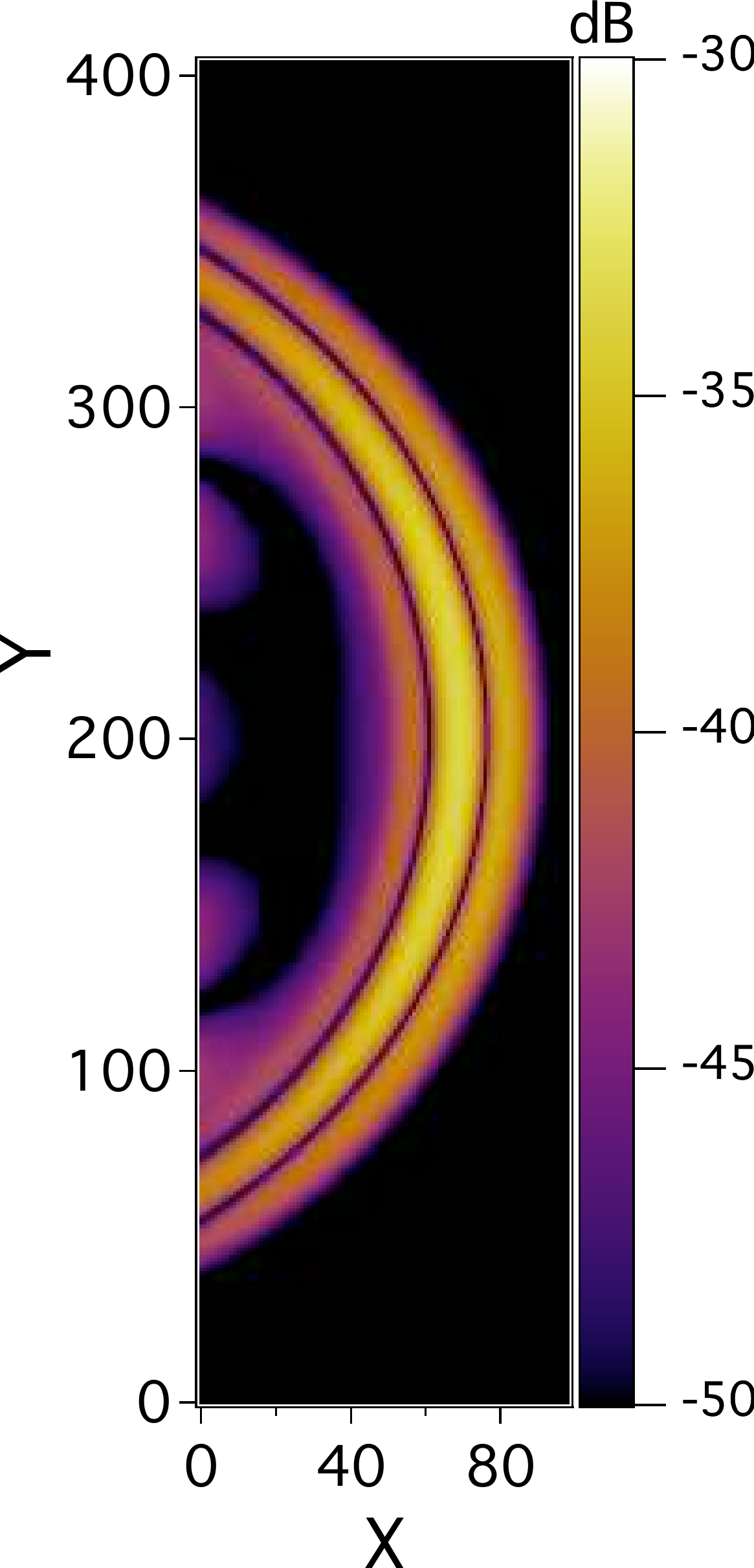} &
    \includegraphics*[width=33mm]{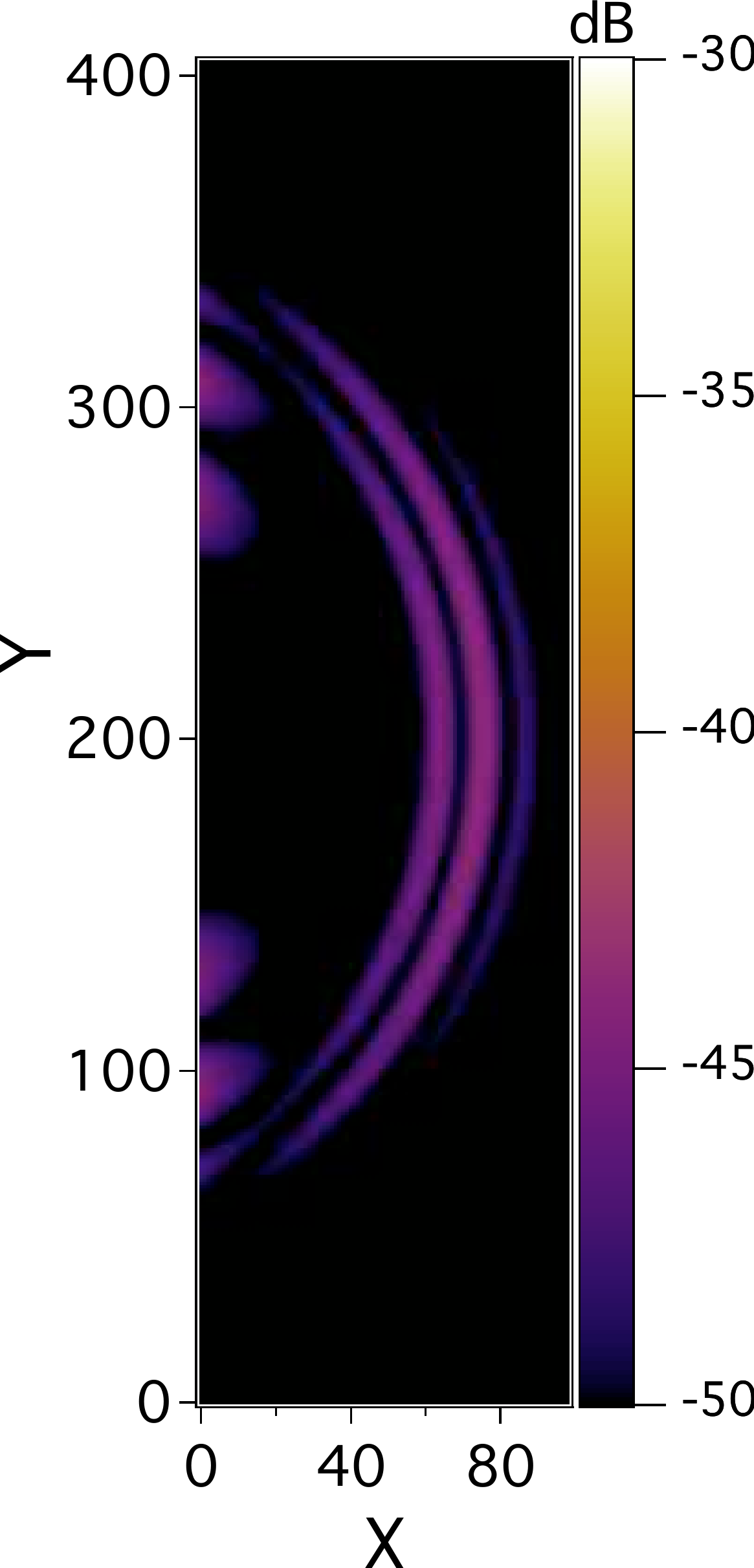}
  \end{tabular} &
%  \begin{tabular}{@{\hspace{.5mm}}c@{\hspace{.5mm}}|@{\hspace{.5mm}}c@{\hspace{.5mm}}}
  \begin{tabular}{@{\hspace{.5mm}}c@{\hspace{.5mm}}@{\hspace{.5mm}}c@{\hspace{.5mm}}}
    $\sigma$ & $\sigma^*$\\
%    \hline
    \includegraphics*[width=33mm]{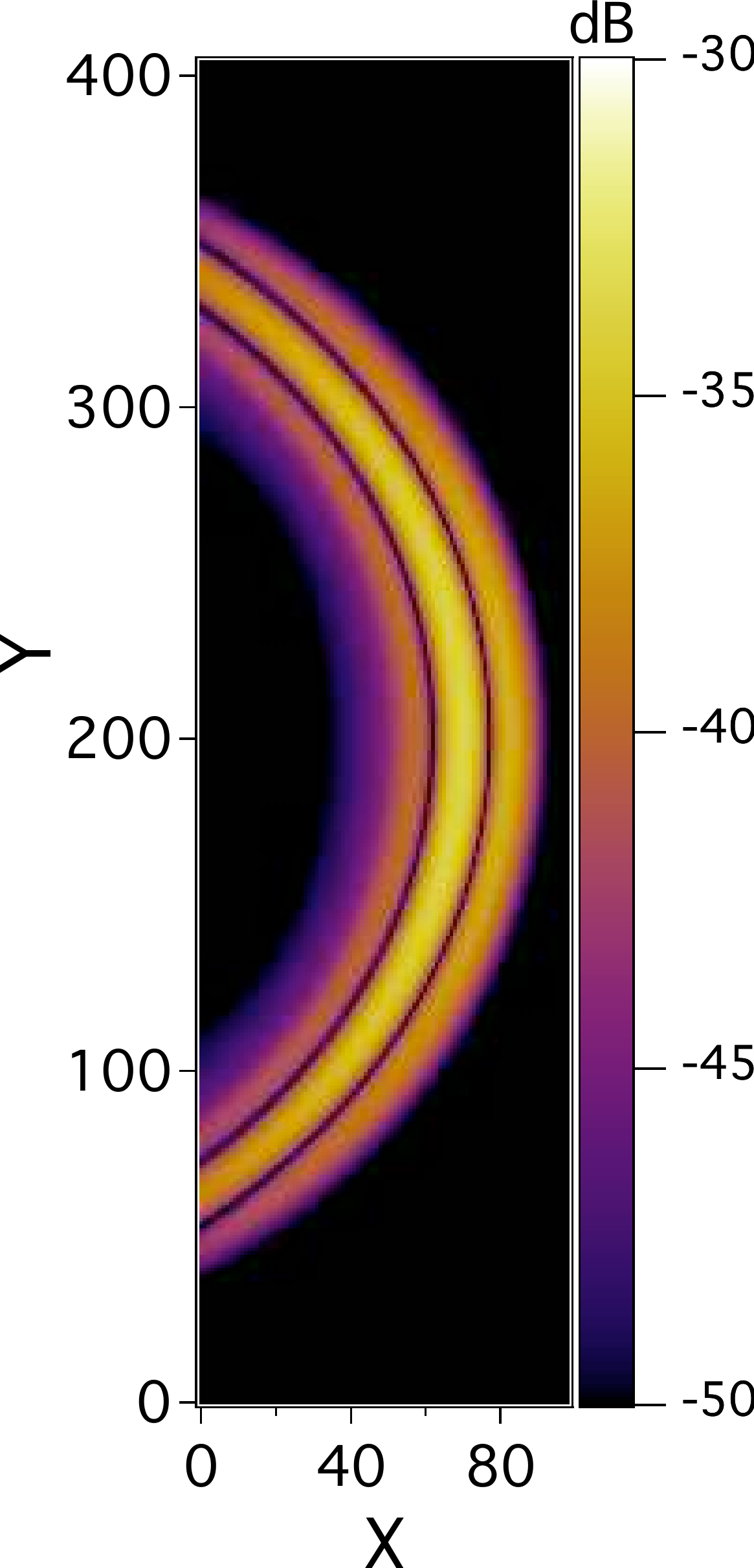} &
    \includegraphics*[width=33mm]{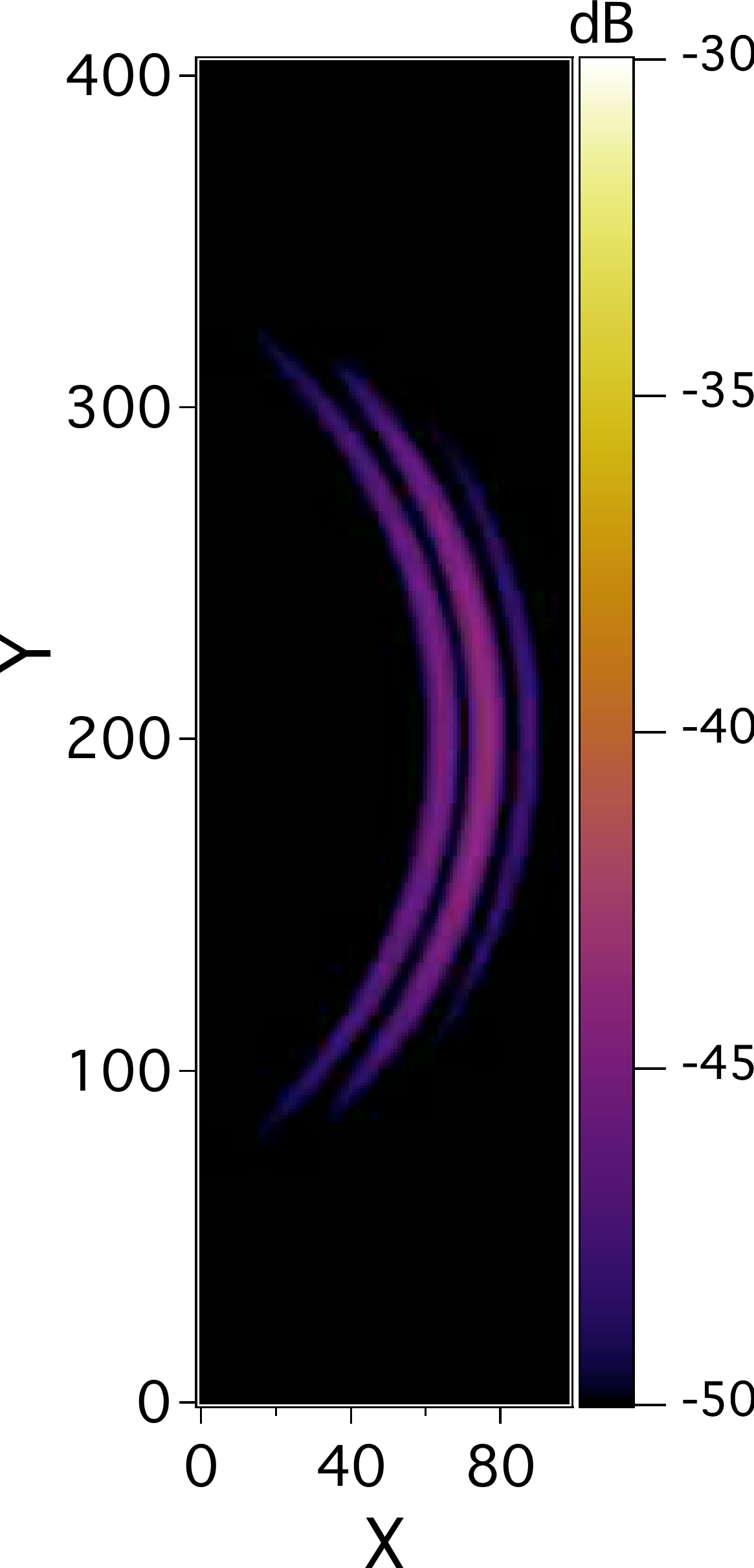}
  \end{tabular}\\
 %   \hline
\end{tabular}}
   \caption{Reflected signal (in dB) from a PML layer using the Yee or the Cole-Karkkainen solver. Each simulation was run for the time step set at the Courant limit.}
   \label{Fig_pml}
\end{figure}
 
The PML using the stencil given by (\ref{Eq:PML_stencil}) was tested and compared to the standard Yee implementation in 2D and 3D. Fig. \ref{Fig_pml} snapshots from 2D simulations of the reflected residue from a PML layer of a pulse with amplitude given by the Harris function $(10-15*\cos(2\pi c t/L)+6*\cos(4\pi c t/L)-cos(6\pi c t/L))/32$ where $t$ is time, $c$ is the speed of light and $L=50\delta x$ is the pulse length in cell size units. A grid of 400x400 cells was used with $\delta x=\delta y$. The absorbing layer was 8 cells deep and the dependency of the PML coefficients with the index position $i$ in the layer was $\sigma_i=\sigma_m\left(i\delta x/\Delta\right)^n$ with $\sigma_m=4/\delta x$, $\Delta=5\delta x$ and $n=2$. The alternative prescription for the coefficients given in \cite{VayJCP00,VayJCP02}, which reads $\sigma^*_i=\left(\xi_{i+1/2}-1/\xi_i\right)/\delta x$ with $\xi_i=e^{-\sigma_i\delta t}$ and $\sigma_i=\sigma_m\left(i\delta x/\Delta\right)^n$, was also tested. 

For the generic test case that has been considered, the new implementation exhibited a very low residue of reflections from the PML layer, which are qualitatively and quantitatively very similar to the residue obtained with a standard PML implementation. In agreement with results from \cite{VayJCP00,VayJCP02}, the use of the modified coefficients $\sigma^*$ led to an order of magnitude improvement over the use of the standard coefficients. 

The 3D tests gave similar absorption efficiency between the Yee and the new solver implementations of the PML, for all the CK solver coefficients given in Table \ref{Table:CKcoefs}.

It was shown in \cite{VayJCP00,VayJCP02} that the efficiency of the layer can be improved further for the standard PML by augmenting the equations with additional terms. However, a similar extension may not be readily available when using the Cole-Karkkainen stencil and is not considered here.

\section{Acknowledgments}

We are thankful to D. L. Bruhwiler, J. R. Cary, B. Cowan, E. Esarey, A. Friedman, C. Huang, S. F. Martins, W. B. Mori, B. A. Shadwick, and C. B. Schroeder for insightful discussions.

%\bibliographystyle{elsarticle-num}

\begin{thebibliography}{99}


\bibitem{VayPRL07}J.-L. Vay, {\it Phys. Rev. Lett.} {\bf 98} (2007) 130405. 

\bibitem{TajimaPRL79} T. Tajima, J. M. Dawson, {\it Phys. Rev. Lett.}, {\bf 43} (1979) 267.

\bibitem{EsareyRMP09} E. Esarey,  {\it et al.}, {\it Rev. Modern Phys.} {\bf 81}, 252 (2009) 1229.

\bibitem{GeddesNature04} C. G. R. Geddes, {\it et al.}, {\it Nature} {\bf 431}, 538 (2004).

\bibitem{ManglesNature04} S. P. D. Mangles, {\it et al.}, {\it Nature} {\bf 431}, 535 (2004).

\bibitem{FaureNature04} J. Faure, {\it et al.}, {\it Nature} {\bf 431}, 541 (2004).

\bibitem{LeemansNature06} W. P. Leemans, {\it et al.}, {\it Nature Physics} {\bf 2}, 696 (2006).

\bibitem{SchroederAAC08} C. B. Schroeder, {\it et al.} {\it Proc. 13th Advanced Accelerator Concepts Workshop}, Santa Cruz, CA (2008) 208.

\bibitem{GeddesSciDAC09} C. G. R. Geddes, {\it et al.}, "Laser Plasma Particle Accelerators: Large Fields for Smaller Facility Sources," {\it SciDAC Review} {\bf 13} (2009) 13.
%\bibitem{GeddesSciDAC09} C.G.R. Geddes, E. Cormier-Michel, E.H. Esarey, C.B. Schroeder, J-L. Vay, W.P. Leemans, and the LOASIS team, LBNL; D.L. Bruhwiler, J.R. Cary, B. Cowan, M. Durant, P. Hamill, P. Messmer, P. Mullowney, C. Nieter, K. Paul, S. Shasharina, S. Veitzer, and the VORPAL development team, Tech-X; G. Weber, O. Rubel, D. Ushizima, Prabhat, and E.W. Bethel, VACET; and J. Wu, SciDAC Scientific Data Management Center, "Laser Plasma Particle Accelerators: Large Fields for Smaller Facility Sources," SciDAC Review 13, pp. 13 (2009).

\bibitem{GeddesJP08} C. G. R. Geddes {\it et al.}, {\it J. Phys. Conf. Series V} {\bf 125} (2008) 12002/1-11.

\bibitem{HuangSciDAC09} C. Huang {\it et al.}, {\it J. Phys. Conf. Series} {\bf 180} (2009) 12005

\bibitem{BELLA} http://loasis.lbl.gov

\bibitem{BruhwilerAAC08} D. L Bruhwiler {\it et al.}, {\it Proc. 13th Advanced Accelerator Concepts Workshop}, Santa Cruz, CA (2008) 29.

\bibitem{ShadwickPoP09} B. A Shadwick, C. B. Schroeder, E. Esarey, {\it Phys. Plasmas} {\bf 16} (2009) 056704

\bibitem{SpranglePRL90} P. Sprangle, E. Esarey, and A. Ting, {\it Phys. Rev. Letters} {\bf 64} (1990) 2011-2014.

\bibitem{Quickpic} C. Huang {\it et al.}, {\it J. of Comput. Phys.} {\bf217} (2006) 658-679.

\bibitem{FengJCP09} B. Feng, C. Huang, V. Decyk, W.B. Mori, P. Muggli, T. Katsouleas, {\it J. Comput. Phys.} {\bf 228} (2009) 5340.

\bibitem{CormierAAC08} E. Cormier-Michel, {\it et al.} {\it Proc. 13th Advanced Accelerator Concepts Workshop}, Santa Cruz, CA (2008) 297.

\bibitem{GeddesPAC09} C. G. R. Geddes {\it et al.}, {\it Proc. Particle Accelerator Conference}, Vancouver, Canada (2009)  WE6RFP075.

\bibitem{VayPOP08}J.-L. Vay, {\it Phys. Plasmas}, {\bf 15} (2008) 056701. 

\bibitem{BorisJCP73}J. P. Boris, {\it J. Comput. Phys.} {\bf 12} (1973) 131-136.

\bibitem{HaberICNSP73} I. Haber, R. Lee, H. H. Klein, J. P. Boris, {\it Proc. Sixth Conf. Num. Sim. Plasmas}, Berkeley, CA (1973) 46-48.

\bibitem{CowanAAC08} B. Cowan,  {\it et al.} {\it Proc. 13th Advanced Accelerator Concepts Workshop}, Santa Cruz, CA (2008) 309.

\bibitem{VayPAC09} J.-L. Vay {\it et al.}, {\it Proc. Particle Accelerator Conference}, Vancouver, Canada (2009) TU1PBI04.

\bibitem{MartinsPAC09} S. F. Martins, {\it Proc. Particle Accelerator Conference}, Vancouver, Canada (2009) TH4GBC05.

\bibitem{VaySciDAC09} J.-L. Vay {\it et al.}, {\it J. Phys. Conf. Series} {\bf 180} (2009) 12006

\bibitem{VayDPF09}J. -L. Vay, W. M. Fawley, C. G. Geddes, E. Cormier-Michel, D. P. Grote, arXiv:0909.5603 (Sept. 2009)

\bibitem{MartinsCPC10} S. F. Martins, R. A. Fonseca, L. O. Silva, W. Lu, W. B. Mori, {\it Comput. Phys. Comm.} {\bf 182} (2010) 869-875.

%\bibitem{ecloud} {\it Proc. International Workshop on Electron-Cloud Effects}, Daegu, S. Korea (2007).

%\bibitem{fel} N. Kroll, P. Morton,  M. Rosenbluth, {\it IEEE J. Quantum Electron} {\bf QE-17} (1981) 1436.

%\bibitem{lwfa} T. Tajima and J. M. Dawson, {\it Phys. Rev. Lett.} {\bf 43} (1979) 267.

%\bibitem{BorisICNSP}J. P. Boris, {\it Proc. Fourth Conf. Num. Sim. Plasmas}, Naval Res. Lab., Wash., D. C. (1970) 3-67.

\bibitem{BruhwilerPC08} D. L. Bruhwiler, {\it Private Communication}.

\bibitem{Warp} D. P. Grote, A. Friedman, J.-L. Vay, I. Haber,  {\it AIP Conf. Proc.} {\bf 749} (2005) 55. 

\bibitem{EsirkepovCPC01} T. Esirkepov, {\it Comput. Phys. Comm.} {\bf 135} (2001) 144-53.

\bibitem{GodfreyJCP74}B. B. Godfrey, {\it J. Comput. Phys.} {\bf 15} (1974) 504-521.

\bibitem{GodfreyJCP75}B. B. Godfrey, {\it J. Comput. Phys.} {\bf 19} (1975) 58-76.

\bibitem{GreenwoodJCP04} A. D. Greenwood, K. L. Cartwright, J. W. Luginsland, E. A. Baca, {\it J. Comp. Phys.} {\bf 201} (2004) 665-684.

\bibitem{GodfreyICNSP80} B. B. Godfrey, {\it Proc. Ninth Conf. on Num. Sim. of Plasmas} (1980).

\bibitem{FriedmanJCP90} A. Friedman, {\it J. Comput. Phys.} {\bf 90} (1990) 292.

\bibitem{AbeJCP86} H. Abe, N. Sakairi, R. Itatani, H. Okuda, {\it J. Comput. Phys.} {\bf 63} (1986) 247-267.

\bibitem{CormierPRE08} E. Cormier-Michel, B. A. Shadwick, C. G. R. Geddes, E. Esarey, C. B. Schroeder, W. P. Leemans, {\it Phys. Rev. E} {\bf 78} (2008) 016404.

\bibitem{ColeIEEE1997} J. B. Cole, {\it IEEE Trans. Microw. Theory Tech.}, {\bf 45} (1997) 991Ð996.
% A high-accuracy realization of the Yee algorithm using non-standard finite differences

\bibitem{ColeIEEE2002} J. B. Cole, {\it IEEE Trans. Antennas Prop.}, {\bf 50} (2002) 1185Ð1191.
%A high-accuracy Yee algorithm based on nonstandard finite differences: new developments and verifications

\bibitem{KarkICAP06} M. Karkkainen, E. Gjonaj, T. Lau, T. Weiland, {\it Proc. International Computational Accelerator Physics Conference}, Chamonix, France (2006).

\bibitem{Yee} K. S. Yee, {\it IEEE Trans. Ant. Prop.} {\bf 14} (1966) 302-307

\bibitem{CowanICAP09} B. Cowan, {\it Proc. $10^{th}$ Internat. Comput. Accel. Phys. Conf.}, San Francisco, CA (2009).

\bibitem{BerengerJCP96} J.-P. B\'erenger, {\it J. Comput. Phys.} {\bf 114} (1994) 185.

\bibitem{TsungPoP2006} F. S. Tsung, {\it et al.}, {\it Phys. Plasmas} {\bf 13} (2006) 056708.

\bibitem{CowanPriv2010} B. Cowan, {\it Private communication}.

\bibitem{BirdsallLangdon} C. K. Birdsall and A. B. Langdon, Plasma Physics Via Computer Simulation (Adam-Hilger, 1991).

\bibitem{LangdonCPC92} A. B. Langdon, {\it Comput. Phys. Comm.} {\bf 70} (1992) 447. 

\bibitem{MarderJCP87} B. Marder, {\it J. Comput. Phys.} {\bf 68} (1987) 48.

\bibitem{VayPoP98} J.-L. Vay, C. Deutsch, {\it Phys. Plasmas} {\bf 5} (1998) 1190.

\bibitem{VillasenorCPC92} J. Villasenor, O. Buneman, {\it Comput. Phys. Comm.} {\bf 69} (1992) 306.

\bibitem{BerengerJCP94} J.-P. Berenger, {\it J. Comput. Phys.} {\bf 114} (1994) 185.
%A perfectly matched layer for the absorption of electromagnetic waves, 

\bibitem{VayJCP00} J.-L. Vay, {\it J. Comput. Phys.} {\bf 165} (2000) 511.

\bibitem{VayJCP02} J.-L. Vay, {\it J. Comput. Phys.} {\bf 183} (2002) 367.

\bibitem{Vayprep09} J.-L. Vay, {\it et al.},  {\it in preparation.}

\bibitem{LangdonPhysFluid70} A. B. Langdon, C. K. Birdsall, {\it Phys. Fluids} {\bf 13} (1970) 2115.

\bibitem{VayPoP04} J.-L. Vay {\it et al.}, {\it Phys. Plasmas} {\bf 11} (2004) 2928.

\bibitem{VayCPC04} J.-L. Vay, J.-C. Adam, A. H\'eron, {\it Comput. Phys. Comm.} {\bf 164} (2004) 171.

%\bibitem{fonseca08}  R. A. Fonseca {\it et al.}, {\it Plasma Phys. and Control Fusion} {\bf 50} (2008) 124034.

%\bibitem{EsareyAAC04} E. Esarey, {\it et al.}, {\it Proc. 11th Advanced Accelerator Concepts}, Stony Brook, NY (2004) 578.

%\bibitem{lu07} W. Lu {\it et al.}, {\it Phys. Rev. ST Accel. Beams} {\bf 10} (2007) 061301.

%\bibitem{esir01} T. Esirkepov, {\it Comput. Phys. Comm.} {\bf 135} (2001) 144-53.

%\bibitem{MartinsAAC08} S. F. Martins {\it et al.}, {\it Proc. 13th Advanced Accelerator Concepts Workshop}, Santa Cruz, CA (2008) 285.

%\bibitem{vulcan} \url{http://www.clf.rl.ac.uk/Facilities/vulcan/index.htm}

%\bibitem{FawleyAAC08} W. M. Fawley, J.-L. Vay, {\it Proc. 13th Advanced Accelerator Concepts Workshop}, Santa Cruz, CA (2008) 346.

%\bibitem{FawleyPAC09} W. M. Fawley, J.-L. Vay, {\it Proc. Particle Accelerator Conference}, Vancouver, Canada (2009) WE5RFP029.

%\bibitem{Ginger} W. M. Fawley, LBNL Tech. Rpt. LBNL-49625-Rev. 1 (2004); also SLAC Rpt. LCLS-TN-04-3.

%\bibitem{CSR} M. Venturini, {\it et al.}, {\it Phys. Rev. ST Accel. Beams} {\bf 8} (2005) 014202.

\end{thebibliography}



\clearpage
\section*{References\\[-5mm]}

\end{document}